%% file: report.tex
\documentclass[10pt, a4paper]{report}
\usepackage{epsfig,fancybox, wrapfig,calc,amsmath,amsfonts}

\usepackage{atbegshi}
\AtBeginDocument{\AtBeginShipoutNext{\AtBeginShipoutDiscard}}

\input{iodict}
\input{epsf}

\newcommand{\ls}[1]
   {\dimen0=\fontdimen6\the\font
    \lineskip=#1\dimen0
    \advance\lineskip.5\fontdimen5\the\font
    \advance\lineskip-\dimen0
    \lineskiplimit=.9\lineskip
    \baselineskip=\lineskip
    \advance\baselineskip\dimen0
    \normallineskip\lineskip
    \normallineskiplimit\lineskiplimit
    \normalbaselineskip\baselineskip
    \ignorespaces
   }

\def\<#1>{\thinspace{\em#1}\thinspace} 

{\obeyspaces\gdef {\ }}

\newcounter{defenum}
\newcommand{\definition}[1]{\stepcounter{defenum}\item{#1}\\}
\newenvironment{Definitions}%
  {\begin{list}{\textbf{Definition \thedefenum:}}%
    {\settowidth{\labelwidth}{\textbf{Definition 9:}}%
    \settowidth{\labelsep}{\textbf{99}}%
    \setlength{\itemindent}{\labelwidth+\labelsep}
    \setlength{\leftmargin}{0pt}}}%
  {\stepcounter{defenum}\end{list}}

\newlength{\mathboxwidth}

\newcommand{\mathbox}[2]{\settowidth{\mathboxwidth}{#1}%
  \parbox{\mathboxwidth}{#2}}

\title{\huge\textbf{ViPIOS\\VIenna Parallel Input Output System }\\
~\\
\Large{Language, Compiler and Advanced Data Structure Support for Parallel I/O Operations}
~\\
~\\
\normalsize{Project Deliverable}\\
\normalsize{Partially funded by FWF Grant P11006-MAT}\\
\normalsize{Core Project Duration: 1996 - 1998}\\
\normalsize{Deliverable Revised: August 2018}\\
~\\
~\\}

\date{}

\author{Erich~Schikuta \and Helmut~Wanek \and Heinz~Stockinger \and Kurt~Stockinger \and Thomas~F\"urle \and Oliver~Jorns \and Christoph~L\"offelhardt \and Peter~Brezany \and Minh~Dang \and Thomas~M\"uck}

\begin{document}


\sect{}

\maketitle
\tableofcontents
\newpage

\chapter*{Executive Summary}

For an increasing number of data intensive scientific applications, parallel I/O concepts are a major performance issue. Tackling this issue, we develop an input/output system designed for highly efficient, scalable and conveniently usable parallel I/O on distributed memory systems. The main focus of this research is the parallel I/O runtime system support provided for software-generated programs produced by parallelizing compilers in the context of High Performance FORTRAN efforts. Specifically, our design aims for the Vienna Fortran Compilation System.

In our research project we investigate the I/O problem from a runtime system support perspective. We focus on the design of an advanced parallel I/O support, called ViPIOS (VIenna Parallel I/O System), to be targeted by language compilers supporting the same programming model like High Performance Fortran (HPF). The ViPIOS design is partly influenced by the concepts of parallel database technology.

At the beginning of the project we developed a formal model, which forms a theoretical framework on which the ViPIOS design is based. This model describes the mapping of the problem specific data space starting from the application program data structures down to the physical layout on disk across several intermediate representation levels.

Based on this formal model we designed and developed an I/O runtime system, ViPIOS, which provides support for several issues, as
\begin{itemize}
  \item parallel access to files for read/write operations,
  \item optimization of data-layout on disks,
  \item redistribution of data stored on disks,
  \item communication of out-of-core (OOC) data, and
  \item many optimizations including data prefetching from disks based on the access pattern knowledge extracted from the program by the compiler or provided by a user specification.
\end{itemize}

The project was partially funded by the Austrian Science Fund by FWF Grant P11006-MAT in the period 05/1996 - 04/1998.
\vspace{2cm}

Vienna, December 1998\footnote{This report was revised in August 2018 by Erich Schikuta mainly polishing the format and adding some contributions achieved after the core period of the project.}

\chapter{A Short Project History}


\begin{table}
\begin{center}
\begin{tabular}{|l|p{7.5cm}|}
\hline
Name & Responsibilities\\
\hline
Erich Schikuta, Professor & project leader, system design\\
Thomas F\"urle, PhD Student & system design, implementation of basic functionality, UNIX system administrator, caching and prefetching techniques\\
Helmut Wanek, PhD Student & system design, implementation of buffer management and MPI-IO functionality, debugging, formal file model and automatic optimization of I/O operations\\
Heinz Stockinger, Student & overview of research in the field of parallel I/O. Masters Thesis: Glossary on Parallel I/O.\\
Kurt Stockinger, Student & MPI-IO Interface\\
Christoph L\"offelhardt, Student & Special adaptations to overcome MPI client server restrictions\\
Oliver Jorns, Student & HPF interface\\
Peter Brezany, Senior Lecturer & language and compiler support for parallel I/O\\
Minh Dang, PhD Student & Integrating ViPIOS I/O calls into VFC compiler\\
Thomas M\"uck, Professor & basic system design\\
\hline
\end{tabular}
\end{center}
\caption{Contributors}\label{contributors_table}
\end{table}

\begin{table}
\begin{center}
\begin{tabular}{|lp{13.5cm}|}
\hline
 Ref. & Publication\\
\hline
\cite{vipios01} & Erich Schikuta and Thomas F\"{u}rle.
\newblock A transparent communication layer for heterogenous, distributed
  systems.
\newblock In {\em AI'03},
  Innsbruck, Austria, 2003. {IASTED}.\\
\cite{vipios02} & Erich Schikuta and Thomas F\"{u}rle.
\newblock {ViPIOS} islands: Utilizing {I/O} resources on distributed clusters.
\newblock In {\em PDCS'02}, Louisville, {KY,} {USA}, 2002. {ISCA}. \\
\cite{vipios03} & Kurt Stockinger and Erich Schikuta.
\newblock {ViMPIOS,} a {"Truly"} portable {MPI-IO} implementation.
\newblock In {\em PDP'00}, Rhodos, Greece, 2000. {IEEE}.\\
\cite{vipios04} & Kurt Stockinger, Erich Schikuta, Thomas F\"{u}rle, and Helmut Wanek.
\newblock Design and analysis of parallel disk accesses in {ViPIOS}.
\newblock In {\em PCS'99}, Ensenada, Mexico, 1999. {IEEE}.\\
\cite{vipios05} & Thomas F\"{u}rle, Erich Schikuta, Christoph L\"{o}ffelhardt, Kurt Stockinger,
  and Helmut Wanek.
\newblock On the implementation of a portable, client-server based {MPI-IO}
  interface.
\newblock In {\em 5th European {PVM/MPI}}, LNCS 1497/1998, Liverpool, {UK}, 1998.
  Springer.\\
\cite{vipios06} & Erich Schikuta, Thomas F\"{u}rle, and Helmut Wanek.
\newblock {ViPIOS:} the vienna parallel {Input/Output} system.
\newblock In {\em EuroPar'98}, LNCS
  1470/1998, Southampton, {UK}, 1998. Springer Berlin / Heidelberg.\\
\cite{vipios07} & Erich Schikuta, Helmut Wanek, Thomas F\"{u}rle, and Kurt Stockinger.
\newblock On the performance and scalability of client-server based disk {I/O}.
\newblock In {\em {SPAA} Revue at the 10th Annual {ACM} Symposium on Parallel
  Algorithms and Architectures}, Puerto Valarte, Mexico, 1998.\\
\cite{vipios08} & Erich Schikuta, Helmut Wanek, and Thomas F\"{u}rle.
\newblock Design and analysis of the {ViPIOS} message passing system.
\newblock In {\em 6th International Workshop on Distributed Data Processing},
  Akademgorodok, Novosibirsk, Russia, 1998.\\
\cite{vipios09} & Peter Brezany, Alok Choudhary, and Minh Dang.
\newblock Language and compiler support for out-of-core irregular applications
  on distributed-memory multiprocessors.
\newblock In {\em International Workshop on Languages, Compilers, and Run-Time
  Systems for Scalable Computers}. Springer, 1998.\\
\cite{vipios10} & Peter Brezany, Alok Choudhary, and Minh Dang.
\newblock Parallelization of irregular codes including out-of-core data and
  index arrays.
\newblock In {\em Advances in Parallel Computing}, volume~12.
  Elsevier, 1998.\\
\cite{vipios11} & Peter Brezany.
\newblock Automatic parallelization of input/output intensive irregular
  applications.
\newblock In {\em Proceedings of the Second International Conference on
  Parallel Processing and Applied Mathematics}, Zakopane, Poland,
  1997.\\
\cite{vipios12} & Peter Brezany and Minh Dang.
\newblock Advanced optimizations for parallel irregular out-of-core programs.
\newblock In {\em International Workshop on APC}. Springer, 1996.\\
\cite{vipios13} & Peter Brezany, Thomas M\"{u}ck, and Erich Schikuta.
\newblock A software architecture for massively parallel input-output.
\newblock In {\em PARA'96}, LNCS 1184, Lyngby, Denmark,
  1996. Springer Berlin / Heidelberg.\\
\cite{vipios14} & Peter Brezany, Thomas~A. M\"{u}ck, and Erich Schikuta.
\newblock Mass storage support for a parallelizing compilation system.
\newblock In {\em EUROSIM'96}, Delft, The
  Netherlands, 1996. Elsevier.\\
\cite{vipios15} & Peter Brezany, Thomas M\"{u}ck, and Erich Schikuta.
\newblock Language, compiler and parallel database support for {I/O} intensive
  applications.
\newblock In Bob Hertzberger and Giuseppe Serazzi, editors, {\em HPCN-Europe'95}, LNCS 919, Milan, Italy, 1995. Springer Berlin / Heidelberg.\\
\cite{vipios16} & Peter Brezany.
\newblock {\em Input/output intensive massively parallel computing: language
  support, automatic parallelization, advanced optimization, and runtime
  systems}.
\newblock LNCS 1220. Springer Science \&   Business Media, 1997.\\
\hline
\end{tabular}
\end{center}
\caption{Publications}\label{publications_table}
\end{table}

The project titled "Language, Compiler and Advanced Data Structure Support for Parallel I/O Operations"
which was later on also called "Vienna Input Output System (ViPIOS)" - project) started in 1995. 1996 it was
granted funding by the FWF for
two years and its purpose was the design and implementation of a software
system to enhance the parallel I/O capabilities of high performance
computing programs written in HPF. This was achieved by programming an I/O
server which can accomplish HPF I/O requests very efficiently using
multiple disks in parallel. High modularity and portability have also been
a major goal in order to allow for future changes and extensions to the
system. In fact the system design also accounts for topics that have
gained big importance in parallel computing during the duration of the project
(i.e. distributed computing over the Internet and an MPI-IO interface).

Table \ref{contributors_table} lists all the people that took part in the
project with their respective responsibilities and the duration of their work.
The papers published during the course of the project are given in table
\ref{publications_table}, which also contains a reference to a book that
was partly based on this project. The references given point to the bibliography of the report with extended citation.

The following gives only a brieve overview of all the work that has been
conducted as part of the project. More detailed information is to be found
in the later chapters, which are given as references here.

\begin{itemize}
\item Theoretical work\\
The state of the art in parallel I/O has been investigated thoroughly
(\ref{sec_state_of_the_art}). This in the end resulted in the glossary on
parallel I/O given in appendix \ref{Glossary on Parallel I/O}.

The principle architecture and design of the ViPIOS system has been devised
(see \ref{sec_basic_strategies}, \ref{sec_architecture}, \ref{sec_modules},
and \ref{sec_Message_Passing}). The main design considerations besides high
performance and effectiveness have been high modularity, portability and
extensibility. Therefore ViPIOS internally is built on standards only (UNIX,
C and MPI) and offers a variety of interfaces to the user of the system (see
\ref{sec_interfaces}).

Some basic research has also been done in how to automatically optimize
parallel I/O operations. This resulted in a formalization of ViPIOS files and
its operations (see \ref{sec_data} and \ref{sec_file-model}).

\item Practical work\\
First a simple prototype has been built that supports parallel I/O for C
programs on UNIX. It operates according to the basic strategies described in
\ref{sec_Message_Passing}. The implementation was designed as a client server
system using multithreading on both sides (client and server) to achieve
maximum parallelism and thus I/O throughput. I/O operations had to be performed
by calling internal ViPIOS functions directly (ViPIOS proprietary I/O
interface).

An MPI restriction, which does not allow processes to start and stop
dynamically (i.e. all processes that communicate
via MPI have to be started and stopped concurrently) and some limitations to
multitasking and multithreading on different hardware platforms forced the
implementation of three operation modes in ViPIOS (see
\ref{sec_operation_modes}). In {\em library mode} no I/O
server processes are started. ViPIOS only is a runtime library linked to the
application. {\em Dependent mode} needs all the server and client processes
to be started at the same time and {\em independent mode} allows client
processes to dynamically connect and disconnect to the I/O server processes,
which are executed independently. Each of this three modes comes in a threaded
and in a non-threaded version. The non-threaded version only supporting
blocking I/O functionality.

The system was extended by an HPF (see chapter \ref{sec_HPF_Interface}) and
an MPI-IO (see \ref{sec_MPI_Interface}) interface, which allows users to keep
to the standard interfaces they already know. Currently the HPF interface is
only supported by the VFC HPF compiler, which automatically transfers the
application program's FORTRAN Read and Write statements into the appropriate
funtion calls.

The program has been developed on SUN SOLARIS workstations and was ported to
and tested on a cluster of 16 LINUX PCs. Details about the test results can
be found in chapter \ref{Analysis}. The current implementation of ViPIOS
supports most parts of the MPI-IO standard and it is comparable to the
reference MPI-IO implementation ROMIO (both in functionality and in
performance). The advantages of the ViPIOS system are however greater
flexibility (due to the client server approach) and the tight integration
into an HPF compilation system. Flexibility means for instance that it is
possible to read from a persistent file using a data distrubution scheme
different than the one used when the file was written. This is not directly
supported by ROMIO. The client server design also allows for automatic
performance optimizations of I/O requests even in a multiuser environment
(different applications executing concurrently, which pose I/O requests
independently). This generally turns out to be very hard to achieve with a
library approach (because of the communication necessary between different
applications). Though there is no effective automatic performance optimization
yet implemented in ViPIOS the module which will perform this task is already
realized (it is called fragmenter; see \ref{sec_modules}). Currently it only
applies basic data distribution schemes which parallel the data distribution
used in the client applications. A future extension will use a blackboard
algorithm to evaluate different distribution schemes and select the optimal
one.
\end{itemize}

\chapter{Supercomputing and I/O}
In spite of the rapid evolvement of computer hardware the demand for even
better performance seems never ending. This is especially true in scientific
computing, where models tend to get more and more complex and the need for
realistic simulations is ever increasing. Supercomputers and recently clusters
of computers are used to achieve very high performance. The basic idea is to
track the problem down into little parts which can be executed in parallel on
a multitude of processors thus reducing the calculation time.

The use of supercomputers has become very common in various fields of science
like for instance nuclear physics, quantum physics, astronomy, flow dynamics,
meteorology and so on. These supercomputers consist of a moderate to large
number of (eventually specifically designed) processors which are linked
together by very high bandwidth interconnections. Since the design and
production of the supercomputer takes a considerable time the hardware
components (especially the processors) are already dated out when the
supercomputer is delivered. This fact has led to the use of clusters of
workstations (COW) or clusters of PC's. Here off the shelf workstations or
PC's are interconnected by a high speed network (GB-LAN, ATM, etc.). Thus
the most up to date generation of processors can be used easily.
Furthermore these systems can be scaled and updated more easily than
conventional supercomputers in the most cases.
The Beowulf \cite{sterling-beowulf},\cite{www-beowulf} and the Myrinet
\cite{boden-myrinet},\cite{www-myrinet} projects for example show that COW's can indeed
nearly reach the performance of dedicated supercomputing hardware.

One of the most important problems with supercomputers and clusters is the fact
that they're far from easy to program. In order to achieve maximum performance
the user has to know very many details about the target machine to tune her
programs accordingly. This is even worse because the typical user is a
specialist in her research field and only interested in the results of the
calculation. Nevertheless she is forced to learn a lot about computer science
and the specific machine especially. This led to the development of compilers
which can perform the tedious parallelization tasks (semi)automatically. The
user only has to write a sequential program which is transferred to a parallel
one by the compiler.
Examples for such parallel compilation systems are HPF \cite{merlin:HPF2.0},
cite{www:HPF} and C* \cite{rose:C*}.

Finally many of the scientific applications also deal with a very large amount
of data (up to 1 Terabytes and beyond). Unfortunately the development of
secondary and tertiary storage does not parallel the increase in processor
performance. So the gap between the speed of processors and the speed of
peripherals like hard disks is ever increasing and the runtime of applications
tends to become more dependent on the speed of I/O than on the processor
performance. A solution to the problem seems to be the use of a number of
disks and the parallelization of I/O operations. A number of I/O systems and
libraries have been developed to accommodate for parallel I/O. (e.g.
MPI-I/O \cite{corbett:mpi-io2}, PASSION \cite{choudhary:passion}, GALLEY
\cite{nieuwejaar:galley}, VESTA \cite{corbett:jvesta}, PPFS
\cite{elford:ppfs-detail} and Panda \cite{seamons:thesis})
But most of
these also need a good deal of programming effort in order to be used
efficiently. The main idea of the ViPIOS project was to develop a client server
I/O system which can automatically perform near optimal parallel I/O. So the
user simply writes a sequential program with the usual I/O statements. The
compiler transfers this program into a parallel program and ViPIOS
automatically serves the program's I/O needs very efficiently.

The rest of this chapter deals with the automatic parallelization and the
specific problems related to I/O in some more detail. A short summary of the
current state of the art is also given. Chapters \ref{The ViPIOS approach}
and \ref{The ViPIOS design} explain in detail the design considerations and
the overall structure of the ViPIOS System. Chapters \ref{sec_Kernel} to
\ref {sec_HPF_Interface}
describe the current state of the system's implementation and some
benchmarking results are given in chapter \ref{Analysis}.

\section{Automatic Parallelization (HPF)}
The efficient parallelization of programs generally turns out to be very
complex. So a number of tools have been developed to aid programmers in this
task (e.g. HPF-compilers, P3T \cite{fahringer:P3T}).
The predominant programming paradigm today is
the single program - multiple data (SPMD) approach. A normal sequential program
is coded and a number of copies of this program are run in parallel on a number of
processors. Each copy is thereby processing only a subset of the original input
values. Input values and calculation results which have to be shared between
several processes induce communication of these values between the appropriate
processors. (Either in form of message passing or implicitly by using shared
memory architectures.) Obviously the communication overhead is depending
strongly on the underlying problem and on the partitioning of the input data
set. Some problems allow for a very simple partitioning which induces no
communication at all (e.g. cracking of DES codes) other problems hardly allow
any partitioning because of a strong global influence of every data item
(e.g. chaotic systems). Fortunately for a very large class of problems in
scientific computing the SPMD approach can be used with a reasonable
communication overhead.

\section{I/O Bottleneck}
By using HPF a programmer can develop parallel programs rather rapidly. The I/O
performance of the resulting programs however is generally poor. This is due
to the fact that most HPF-compilers split the input program into a host
program and a
node program. After compilation, the host program will be executed on
the host computer as the host process; it handles all the I/O. The
node program will be executed on each node of the underlying hardware
as the node process. The node program performs the actual computation,
whereas input/output statements are transformed into communication
statements between host and node program. Files are read and written
sequentially by the centralized host process. The data is transferred
via the network interconnections to the node processes. In particular,
all I/O statements are removed from the node program. A FORTRAN
READ-statement is compiled to an appropriate READ followed by a SEND-statement
in the host program and
a RECEIVE-statement in the node program. The reason for this behavior is
that
normally on a supercomputer only a small number of the available processors
are actually provided with access to the disks and other tertiary storage
devices. So the host task runs on one of these processors and all the other
tasks have to perform their I/O by communicating with the host task. Since
all the tasks are executed in a loosely synchronous manner there is also a
high probability that most of the tasks will have to perform I/O concurrently.
Thus the host task turns out to be a bottleneck for I/O operations.

In addition to that scientific programs tend to get more demanding with
respect to I/O (some applications are working on Terabytes of input data and
even more) and the performance of I/O devices does not increase as fast as
computing power does. This led to the founding of the Scalable I/O Initiative
\cite{io-initiative}
which tried to address and solve I/O problems for parallel applications. Quite
a view projects directly or indirectly stem from this initiative and a number
of different solutions and strategies have been devised which will be
described shortly in the following section. A quite complete list of
projects in the parallel I/O field as well as a comprehensive bibliography can
be found in the WWW \cite{www-kotz}.

\section{State of the Art}\label{sec_state_of_the_art}
Some standard techniques have been developed to improve I/O performance of
parallel applications. The most important are
\begin{itemize}
\item{two phase I/O}
\item{data sieving}
\item{collective I/O}
\item{disk directed I/O}
\item{server directed I/O}
\end{itemize}
These methods try to execute I/O in a manner that minimizes or strongly
reduces the effects of disk latency by avoiding non contiguous disk accesses
and thereby speeding up the I/O process. More details and even some more
techniques can be found in appendix \ref{Glossary on Parallel I/O}.


In the last years many universities and research teams from different parts
of the world have used and enhanced these basic techniques to produce
software and design proposals to overcome the I/O
bottleneck problem. Basically, three different types of approaches can be
distinguished:

\begin{itemize}

\item{{\em Runtime I/O libraries}}
are highly merged with the language system
by providing a call library for efficient parallel disk accesses. The
aim is that it adapts graciously to the requirements of the problem
characteristics specified in the application program. Typical
representatives are PASSION \cite{thakur:jpassion}, Galley
\cite{nieuwejaar:galley}, or the MPI-IO initiative, which proposes a
parallel file interface for the Message Passing Interface (MPI)
standard \cite{mpi-ioc:mpi-io5,corbett:mpi-overview}.
Recently the MPI-I/O standard has been widely accepted as a programmers
interface to parallel I/O. A portable implementation of this standard is the
ROMIO library \cite{thakur:romio-users}.

Runtime libraries aim for to be tools for the application programmer.
Therefore the
executing application can hardly react dynamically to changing system
situations (e.g. number of available disks or processors) or problem
characteristics (e.g. data reorganization), because the data access
decisions were made during the programming and not during the
execution phase.

Another point which has to be taken into account is the often arising
problem that the CPU of a node has to accomplish both the application
processing and the I/O requests of the application. Due to a missing
dedicated I/O server the application, linked with the runtime
library, has to perform the I/O requests as well. It is often very
difficult for the programmer to exploit the inherent pipelined
parallelism between pure processing and disk accesses by interleaving
them.

All these problems can be limiting factors for the I/O bandwidth. Thus
optimal performance is nearly impossible to reach by the usage of
runtime libraries.

\item{{\em File systems}} are a solution at a quite low level, i.e.
the operating system is enhanced by special features that deal directly with
I/O. All important manufacturers of parallel high-performance computer
systems provide parallel disk access via a (mostly proprietary)
parallel file system interface. They try to balance the parallel
processing capabilities of their processor architectures with the I/O
capabilities of a parallel I/O subsystem. The approach followed in
these subsystems is to decluster the files among a number of disks,
which means that the blocks of each file are distributed across
distinct I/O nodes. This approach can be found in the file systems of
many super-computer vendors, as in Intels CFS (Concurrent File System)
\cite{pierce:pario}, Thinking Machines' Scalable File System (sfs)
\cite{loverso:sfs}, nCUBEs Parallel I/O System
\cite{debenedictis:ncube} or IBM Vesta \cite{corbett:jvesta}.

In comparison to runtime libraries parallel file systems have the
advantage that they execute independently from the application. This
makes them capable to provide dynamic adaptability to the application. Further
the notion of dedicated I/O servers (I/O nodes) is directly supported
and the processing node can concentrate on the application program and
is not burdened by the I/O requests.

However due to their proprietary status parallel file systems do not
support the capabilities (expressive power) of the available high
performance languages directly. They provide only limited disk access
functionality to the application. In most cases the application
programmer is confronted with a black box subsystem. Many systems even
disallow the programmer to coordinate the disk accesses according to
the distribution profile of the problem specification. Thus it is hard or
even impossible to achieve an optimal mapping of the
logical problem distribution to the physical data layout, which
prohibits an optimized disk access profile.

Therefore parallel file systems also can not be considered as a final
solution to the disk I/O bottleneck of parallelized application
programs.

\item{{\em Client server systems}} give a combination of the other two
approaches, which is a dedicated, smart, concurrent executing runtime system,
gathering all available information of the application process both
during the compilation process and the runtime execution. Thus, this
system is able to aim for the static and the dynamic fit properties
\footnote{static fit: Data is distributed across available disks according to
the SPMD data distribution (i.e. the chunk of data which is processed by a
single processor is stored contiguously on a disk; a different processor's data
is stored on different disks depending on the number of disks available).

dynamic fit: Data is redistributed dynamically according to changes of system
characteristics or data access profiles during the runtime of the program.
(i.e. a disk running out of space, too many different applications using the
same disk concurrently and so on.

(See appendix \ref{Glossary on Parallel I/O} for further information.)}.
Initially it can provide the optimal fitting data access profile for the
application (static fit) and may then react to the execution behavior
dynamically (dynamic fit), allowing to reach optimal performance by
aiming for maximum I/O bandwidth.

The PANDA \cite{seamons:panda,seamons:thesis} and the ViPIOS system
are examples for client server systems.
(Note that PANDA is actually called a library by its designers. But since it
offers independently running I/O processes and enables dynamic optimization
of I/O operations during run time we think of it as a client server system
according to our classification)

\end{itemize}

Additionally to the above three categories there are many other proposals
that do not fit exactly into any of the stated schemes. There are many
experimenting test beds and simulation software products that can as well be
used to classify existing file systems and I/O libraries [1]. Those test beds
are especially useful to compare performance and usability of systems.

But while most of these systems supply various possibilities to perform
efficient I/O they still leave the application programmer responsible for the
optimization of I/O operations (i. e. the programmer has to code the calls to
the respective system's functions by hand). Little work has yet been done, to
automatically generate the appropriate function calls by the compiler (though
there are some extensions planned to PASSION). And as far as we know only the
Panda library uses some algorithms to automatically control and optimize the
I/O patterns of a given application.

This is where the ViPIOS project comes in, which implements a database
like approach. The programmer only has to specify what she wants to read or
write (for example by using a simple FORTRAN read or write statement) not how
it should be done. The ViPIOS system is able to decide about data layout
strategies and the I/O execution plan based on information generated at
compile time and/or collected while the run time of the application.

For more information on all the systems mentioned above see appendix
\ref{Glossary on Parallel I/O}, which also lists additional systems not
referenced in this chapter.

\chapter{The ViPIOS Approach}\label{The ViPIOS approach}
ViPIOS is a distributed I/O server
providing fast disk access for high performance applications. It is an I/O
runtime system, which provides efficient access to persistent files by
optimizing the data layout on the disks and allowing parallel read/write
operations. The
client-server paradigm allows clients to issue simple and familiar I/O calls
(e.g. 'read(..)'), which are to be processed in an efficient way by the
server. The application programmer is relieved from I/O optimization tasks,
which are performed automatically by the server. The actual file layout
on disks is solely maintained by the servers which use their knowledge about
system characteristics (number and topology of compute nodes, I/O nodes and
disks available; size and data transfer rates of disks; etc.) to satisfy the
client's I/O requests efficiently.

In order to optimize the file layout on
disk ViPIOS uses information about expected file access patterns which can
be supplied by HPF compilation systems. Since ViPIOS-servers are distributed
on the available processors, disk accesses are effectively parallel. The
client-server concept of ViPIOS also allows for future extensions like
checkpointing, transactions, persistent objects and also support for
distributed computing using the Internet.

ViPIOS is primarily targeted (but not restricted) to networks of workstations
using the SPMD paradigm. Client processes are assumed to be loosely synchronous.

\section{Design Goals}\label{sec_DesignGoals}
The design of ViPIOS followed a data engineering approach,
characterized by the following goals.

\begin{enumerate}
\item {\em Scalability}. Guarantees that the size of the used I/O system,
i.e. the number
of I/O nodes currently used to solve a particular problem, is defined
by or correlated with the problem size. Furthermore it should be
possible to change the number of I/O nodes dynamically corresponding
to the problem solution process. This requires the feature to
redistribute the data among the changed set of participating nodes at
runtime.
The {\em system architecture} (section \ref{sec_architecture})
of VIPIOS is highly distributed and decentralized. This leads to the
advantage that the provided I/O bandwidth of ViPIOS is mainly
dependent on the available I/O nodes of the underlying architecture
only.
\item {\em Efficiency}. The aim of compile time and runtime optimization
is to minimize the number of disk accesses for file I/O. This is
achieved by a suitable data organization (section
\ref{sec_data}) by providing a transparent view of the stored data on
disk to the 'outside world' and by organizing the data layout on disks
respective to the static application problem description and the dynamic
runtime requirements.
\item {\em Parallelism}. This demands coordinated parallel data accesses of
processes to multiple disks. To avoid unnecessary communication and
synchronization overhead the physical data distribution has to reflect
the problem distribution of the SPMD processes. This guarantees that
each processor accesses mainly the data of its local or best suited
disk.
All file data and meta-data (description of files)
are stored in a distributed and parallel form across multiple I/O
devices. In order to find suitable data distributions to achieve maximum
parallelism (and thus very high
I/O bandwidth) ViPIOS may use information supplied by the compilation system
or the application programmer. This information is passed to ViPIOS via hints
(see appendix \ref{sec_Hints}).
If no hints are available ViPIOS uses some general heuristics to find an
initial distribution and then dynamically can adopt to the application's I/O
needs during runtime.
\item {\em Usability}. The application programmer must be able to use the system
without big efforts. So she does not have to deal with details of the
underlying hardware in order to achieve good performance and familiar {\em
Interfaces} (section \ref{sec_interfaces}) are available to program file I/O.
\item {\em Portability}. The system is portable across multiple hardware platforms.
This also increases the usability and therefore the acceptance of the system.
\end{enumerate}

\section{Basic Strategies}\label{sec_basic_strategies}
Naturally ViPIOS supports
the standard techniques (i. e. two phase access, data sieving and collective
operations), which have been adapted to the specific needs of ViPIOS.
In order to meet the design goals described above a number of additional
basic strategies
have been devised and then implemented in ViPIOS.

\subsection{Database like Design}
\begin{figure}
\begin{center}
\includegraphics[width=5cm]{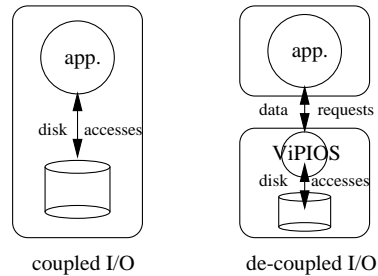}
\caption{Disk access decoupling}\label{decoupling}
\end{center}
\end{figure}

As with database systems the actual disk access operations are {\em decoupled}
from the application and performed by an independent I/O subsystem.
This leads to the situation that an application just sends disk
requests to ViPIOS only, which performs the actual disk accesses in
turn (see figure \ref{decoupling}).

The advantages of this method are twofold:

\begin{enumerate}
\item The application programmer is relieved from the responsibility to
program and optimize all the actual disk access operations. She may therefore
concentrate on the optimization of the application itself relying on the I/O
system to perform the requested operations efficiently.

\item The application programmer can use the same program on all the platforms
supported by the I/O system without having to change any I/O related parts of
the application. The I/O system may use all the features of the underlying
hardware to achieve the highest performance possible but all these features
are well hidden from the programmer. Operations which are not directly
supported by the respective hardware have to be emulated by the I/O system.
\end{enumerate}

So the programmer just has to specify what data she wants to be input or output,
not how that shall actually be performed (i. e. which data item has to be
placed on which disk, the order in which data items are processed and so on).
This is similar to a database approach, where a simple SQL statement for example
produces the requested data set without having to specify any details about
how the data is organized on disk and which access path (or query execution
plan) should be used.

But the given similarities between a database system and a parallel I/O system
also raise an important issue. For database systems an administrator is needed
who has to define all the necessary tables and indexes needed to handle all
the requests that any user may pose. As anyone knows who has already designed
a database this job is far from easy. Special design strategies have been
devised to ensure data integrity and fast access to the data. In the end the
designer has to decide about the database layout based on the requests that
the users of the database are expected to issue.

Now who shall decide which {\em data layout} (see appendix
\ref{Glossary on Parallel I/O}) strategy shall be used. Evidently it must not
be the application programmer but actually the compiler can do an excellent
job here. Remember that the programmer only codes a sequential program which is
transformed by the compiler into a number of processes each of which has only
to process a part of the sequential program's input data. Therefore the
compiler exactly knows which process will need which data items. Additionally
it also knows very much about the I/O profile of the application (i. e. the
order in which data items will be requested to be input or output). All this
information is passed to the I/O server system, which uses it to find a (near)
optimal data layout. Theoretically this can be done automatically because the
I/O profile is completely defined by the given application. In practice a very
large number of possible data layout schemes has to be considered. (One reason
for this is the considerable
number of conditional branches in a typical application. Every branch can
process different data items in a different order thus resulting in a change
of the optimal data layout to be chosen. Though not every branch operation
really affects the I/O behavior of the application the number of possible
layouts to consider still remains very large.)

Different techniques especially designed for searching huge problem spaces
may be used to overcome this problem (e. g. genetic algorithms, simulated
annealing or blackboard methods). These can find a good data layout scheme
in a reasonable time. (Note that it is of no use to find an optimal solution
if the search took longer than the actual execution of the program.)

\subsection{Use of Hints}\label{sec_Hints}
{\em Hints} are the general tool to support ViPIOS
with information for the data administration process.
Hints are data and problem specific information from the "out-side-
world" provided to ViPIOS.

Basically three types of hints can be differntiated,
file administration, data prefetching, and ViPIOS administration hints.

The {\em file administration hints} provide information of the problem
specific data distribution of the application processes (e.g.
SPMD data distribution). High parallelization
can be reached, if the problem specific data distribution of the
application processes matches the physical data layout on disk.

{\em Data prefetching hints} yield better performance by pipelined
parallelism (e.g. advance reads, delayed writes) and file alignment.

The {\em ViPIOS administration hints} allow the configuration of
ViPIOS according to the problem situation respective to the underlying
hardware characteristics and their specific I/O needs (I/O nodes,
disks, disk types, etc.)

Hints can be given by the compile time system, the ViPIOS system administrator
(who is responsible for starting and stopping the server processes, assigning
the available disks to the respective server processes, etc.)
or the application programmer. Normally the programmer should not have to
give any hints but in special cases additional hints may help ViPIOS to find
a suiting data layout strategy. This again parallels database systems where the
user may instruct the system to use specific keys and thus can influence the
query execution plan created by the database system. However the technology of
relational databases is so advanced nowadays that the automatically generated
execution plan can only very rarely be enhanced by a user specifying a special
key to be used. Generally the optimal keys are used automatically. We are quite
confident that a comparable status can be reached for parallel I/O too. But
much research work still has to be done to get there.

Finally hints can be static or dynamic. Static hints are hints that give
information that is constant for the whole application run (e.g. number of
application processes, number of available disks and so on). Dynamic hints
inform ViPIOS of a special condition that has been reached in the application
execution (e.g. a specific branch of a conditional statement has been entered
requiring to prefetch some data, a harddisk has failed). While static hints
may be presented to ViPIOS at any time (i.e. compile time, application startup
and application runtime) dynamic hints only may be given at runtime and are
always sent by the application processes. To generate dynamic hints the
compiler inserts additional statements in the appropriate places of the
application code. These statements send the hint information to the ViPIOS
system when executed.

\subsection{Two-phase data Administration}\label{sec_two_phase}

\begin{figure}
\begin{center}
\includegraphics{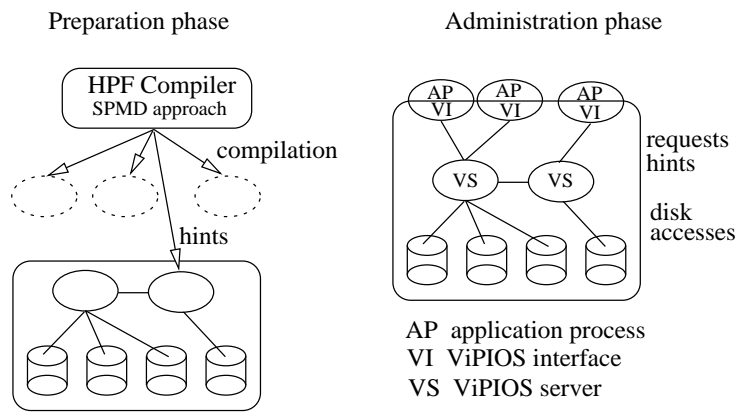}
\caption{Two-phase data administration}\label{twophase}
\end{center}
\end{figure}

The management of data
by the ViPIOS servers is split into two distinct phases, the
preparation and the administration phase (see Figure \ref{twophase}).

The {\em preparation phase}
precedes  the execution of the application processes
(mostly during compilation and startup time).
This phase uses the information collected during the application program
compilation process in form of {\em hints} from the compiler.
Based on this problem specific knowledge the
physical data layout schemes are defined and the disks best suited to actually
store the data are chosen.
Further the data storage areas are prepared,
the necessary main memory buffers allocated, etc.

The following {\it administration phase}
accomplishes the I/O requests of the application
processes during their execution, i.e. the
physical read/write operations and eventually performs necessary
reorganization of the data layout.

The two-phase data administration method aims for putting all the data
layout decisions, and data distribution operations into the
preparation phase, in advance to the actual application execution.
Thus the administration phase performs the data accesses and possible
data prefetching only.

\chapter{The ViPIOS Design}\label{The ViPIOS design}
The system design has mainly been driven by the goals described in
chapter \ref{sec_DesignGoals} and it is therefore built on the following
principles:

\begin{itemize}
\item{Minimum Overhead.} The overhead imposed by the ViPIOS system (e.g. the
time needed to calculate a suitable distribution of data among the available
disks and so on) has to be kept as small as possible. As a rule of thumb an
I/O operation using the ViPIOS system must never take noticeable longer than
it would take without the use of ViPIOS even if the operation can not be speed
up by using multiple disks in parallel.
\item{Maximum Parallelism.} The available disks have to be used in a manner to
achieve maximum overall I/O throughput. Note that it is not sufficient to just
parallelize any single I/O operation because different I/O operations can very
strongly affect each other. This holds true whether the I/O operations have to
be executed concurrently (multiple applications using the ViPIOS system at the
same time) or successively (single application issuing successive I/O requests).
In general the search for a data layout on disks allowing maximum throughput
can be vary time consuming. This is in contradiction with our 'minimum overhead'
principle. So in praxis the ViPIOS system only strives for a very high
througput not for the optimal one. There is no point in calculating the optimal
data layout if that calculation takes longer than the I/O operations would
take without using ViPIOS.
\item{Use of widely accepted standards.} ViPIOS uses standards itself (e.g. MPI
for the communication between clients and servers) and also offers standard
interfaces to the user (for instance application programmers may use MPI-I/O
or UNIX file I/O in their programs), which strongly enhances the systems
portability and ease of use.
\item{High Modularity.} This enables the ViPIOS system to be quickly adopted
to new and changing standards or to new hardware environments by just changing
or adding the corresponding software module.
\end{itemize}

Some extensions to support for future developments in high
performance computing also have been considered like for instance distributed
(Internet) computing and agent technology.

\section{Overall System Architecture}\label{sec_architecture}

\begin{figure}
\begin{center}
\includegraphics[width=11.5cm]{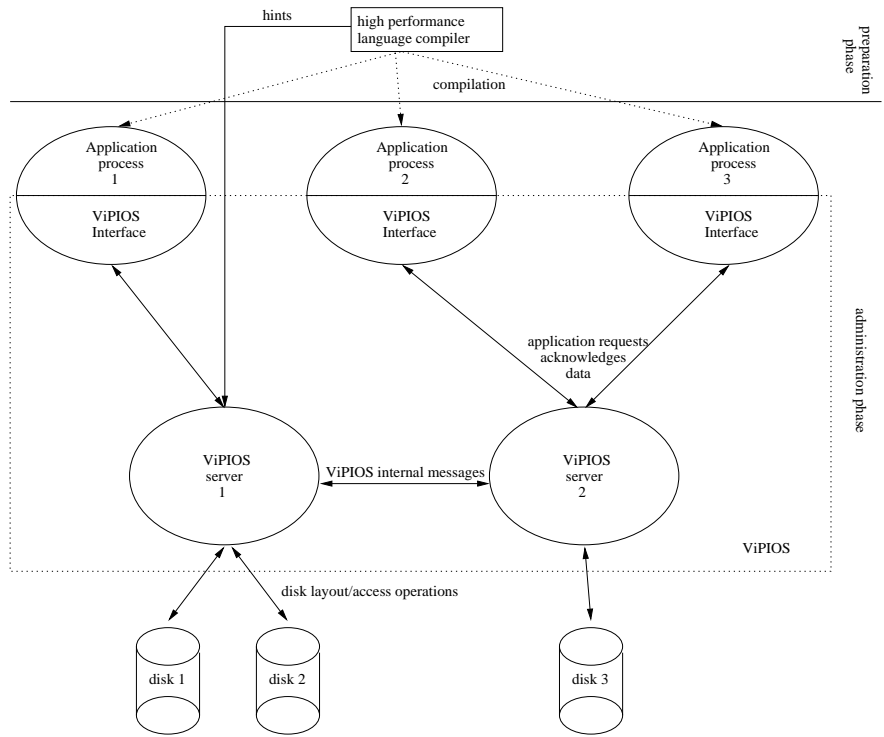}
\caption{ViPIOS system architecture}\label{sys_arch}
\end{center}
\end{figure}

The ViPIOS system architecture is built upon a set of cooperating
server processes, which accomplish the requests of the
application client processes.
Each application process AP is
linked to the ViPIOS servers VS by the {\em ViPIOS interface} VI, which is a
small library that implements the I/O interface to the application and performs
all the communication with the ViPIOS servers (see figure \ref{sys_arch}).

The server processes
run independently on all or a number of dedicated processing
nodes on the underlying cluster or MPP. It is also possible that an application
client and a server share the same processor.
Generally each application process is assigned to exactly one ViPIOS server,
which is called the buddy to this application. All other server processes are
called foes to the respective application. A ViPIOS server can serve any number
of application processes. Hence there is a one-to-many relationship between
servers and the application. (E. g. the ViPIOS server numbered 2 in figure
\ref{sys_arch} is
a buddy to the application processes 2 and 3, but a foe to application process
1.)

While each application process is assigned exactly one ViPIOS server,
a ViPIOS server can serve a number of application processes, i.e. there exists
one-to-many relationship between the servers and the
application.

Figure \ref{sys_arch} also depicts the {\em two phase data administration}
described in chapter \ref{sec_two_phase}:

* The preparation phase precedes the execution of the application processes
(i.e. compile time and application startup time).

* The following administration phase accomplishes the I/O requests posed during
the runtime of the application processes by executing the appropriate physical
read/write operations.

To achieve high data access performance ViPIOS follows the
principle of {\em data locality}.
This means that the data
requested by an application process should be read/written from/to the
best-suited disk.

Logical and physical data locality are to be distinguished.

{\em Logical data locality} denotes to choose the best suited ViPIOS server as
a buddy server for an application process.
This server can be defined by the topological distance and/or the process
characteristics.

{\em Physical data locality} aims to define
the best available (set of) disk(s) for the respective server
(which is called the {\em best disk list, BDL}), i.e.
the disks providing the best (mostly the fastest) data access.
The choice is done on the specific disk characteristics, as
access time, size, toplogical position in the network, and so on.

\section{Modules}\label{sec_modules}

\begin{figure}
\begin{center}
\includegraphics{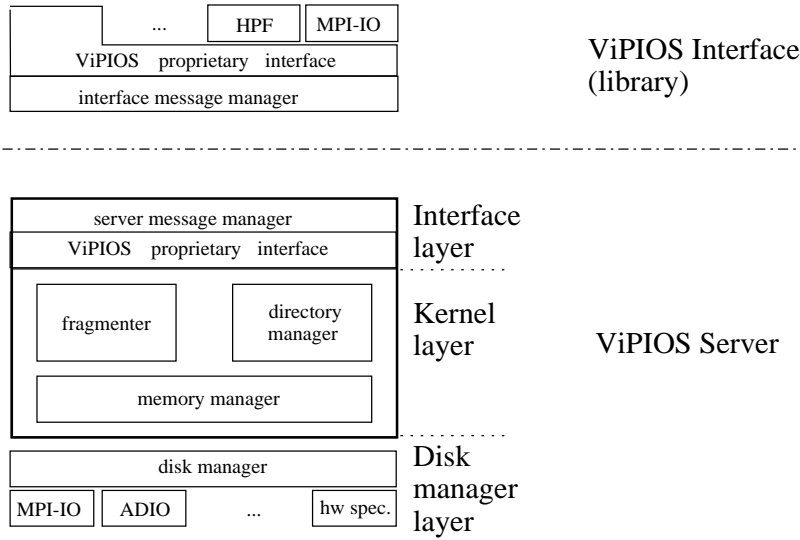}
\caption{Modules of a ViPIOS System}\label{funct_enh}
\end{center}
\end{figure}

As shown in figure \ref{sys_arch} the ViPIOS system consists of the
independently running ViPIOS servers and the ViPIOS interfaces, which are
linked to the application processes. Servers and interfaces themselves are
built of several modules, as can be seen in figure \ref{funct_enh}.

The ViPIOS Interface library is linked to the application and provides the
connection to the "outside world" (i.e. applications, programmers, compilers,
etc.). Different programming interfaces are supported by {\em interface modules}
to allow flexibility and extendability. Currently implemnted are an HPF
interface module (aiming for the VFC, the HPF derivative of Vienna
FORTRAN \cite{chapman94}) a (basic) MPI-IO interface module, and the
specific ViPIOS interface which is also the interface for the
specialized modules. Thus a client application can execute I/O operations by
calling HPF read/write statements, MPI-IO routines or the ViPIOS proprietary
functions.

The interface library translates all these calls into calls to
ViPIOS functions (if necessary) and then uses the interface message manager
layer to send the calls to the buddy server. The message manager also is
responsible for sending/receiving data and additional informations (like for
instance the number of bytes read/written and so on) to/from the server
processes. Note that data and additional information can be sent/received
directly to/from any server process bypassing the buddy server, thereby saving
many additional messages that would be necessary otherwise and enforcing the
minimum overhead principle as stated in chapter \ref{The ViPIOS design}.
(See chapter \ref{sec_Message_Passing} for more details.) The message manager
uses MPI-function calls to communicate to the server processes.

The ViPIOS server process basically contains 3 layers:

\begin{itemize}
\item The {\bf Interface layer} consists of a message manager responsible for
the communication with the applications and the compiler ({\em external
messages}) as well as with other servers ({\em internal messages}). All
messages are translated to calls to the appropriate ViPIOS functions in the
proprietary interface.

\item The {\bf Kernel layer} is responsible for all server specific tasks.
It is built up mainly of three cooperating functional units:

\begin{itemize}

\item The {\bf Fragmenter} can be seen as "ViPIOS's brain". It represents a
smart data administration tool, which models different distribution
strategies and makes decisions on the effective data layout,
administration, and ViPIOS actions.

\item The {\bf Directory Manager} stores the meta information
of the data. Three different modes of operation have been designed, centralized
(one dedicated ViPIOS directory server), replicated (all servers store
the whole directory information), and localized (each server knows the
directory information of the data it is storing only) management.
Until now only localized management is implemented. This is sufficient for
clusters of workstations. To support for distributed computing via the internet
however the other modes are essential (see \ref{sec_Message_Passing}).

\item The {\bf Memory Manager} is  responsible for prefetching, caching and
buffer management.

\end{itemize}

\item The {\bf Disk Manager layer} provides the access to the available and
supported disk sub-systems. Also this layer is modularized to allow
extensibility and to simplify the porting of the system. Available are
modules for ADIO \cite{thakur:abstract}, MPI-IO, and Unix style file systems.

\end{itemize}

\section{Interfaces}\label{sec_interfaces}
To achieve high portability and usability the implementation internally uses
widely spread
standards (MPI, PVM, UNIX file I/O, etc.) and offers multiple modules to
support an application programmer with a variety of existing I/O interfaces.
In addition to that ViPIOS
offers an interface to HPF compilers and also can use different underlying
file systems. Currently the following interfaces are implemented:

\begin{itemize}
\item{User Interfaces}\\
Programmers may express their I/O needs by using

\begin{itemize}
\item{MPI-IO} (see chapter \ref{sec_MPI_Interface}.)
\item{HPF I/O calls} (see chapter \ref{sec_HPF_Interface}.)
\item{ViPIOS proprietary calls} (not recommended though because the programmer
has to learn a completely new I/O interface. See appendix \ref{ViPIOS functions}
for a list of available functions.)
\end{itemize}

\item{Compiler Interfaces}\\
Currently ViPIOS only supports an interface to the VFC HPF compiler (see chapter
\ref{sec_HPF_Interface}).

\item{Interfaces to File Systems}\\
The filesystems that can be used by a ViPIOS server to perform the physical
acceses to disks enclose
\begin{itemize}

\item ADIO (see \cite{thakur:abstract}; this has been chosen because it also
allows to adapt for future file systems and so enhances the portability of
ViPIOS.)

\item MPI-IO (is already implemented on a number of MPP's.)

\item Unix file I/O (available on any Unix system an thus on every cluster of
workstations.)

\item Unix raw I/O (also available on any Unix system, offers faster access but
needs more administrational effort than file I/O. Is not completely implemnted
yet.)

\end{itemize}

\item{Internal Interface}\\
Is used for the communication between different ViPIOS server processes.
Currently only MPI is used to pass messages. Future extensions with respect
to distributed computing will also allow for communication via HTTP.
\end{itemize}


\section{Data Abstraction}\label{sec_data}

\begin{figure}
\begin{center}
\includegraphics{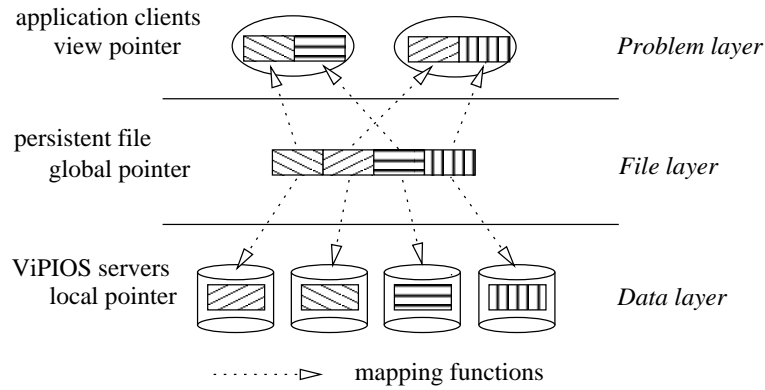}
\caption{ViPIOS data abstraction}\label{abstraction}
\end{center}
\end{figure}

ViPIOS provides a data independent view of the stored data to the
application processes.

Three independent layers in the ViPIOS architecture can be
distinguished, which are represented by file pointer types in ViPIOS.

\begin{itemize}
\item Problem layer. Defines the problem specific data distribution
among the cooperating parallel processes (View file pointer).
\item File layer. Provides a composed view of the persistently stored data
in the system (Global file pointer).
\item Data layer. Defines the physical data distribution among the available disks
(Local file pointer).
\end{itemize}

Thus data independence in ViPIOS separates these layers conceptually
from each other, providing mapping functions between these layers.
This allows {\em logical data independence} between the problem and
the file layer, and {\em physical data independence} between the file
and data layer analogous to the notation in data base systems (\cite
{jardine77, burns86}). This concept is depicted in figure \ref{abstraction}
showing a cyclic data distribution.

In ViPIOS emphasis is laid on the parallel execution of disk accesses.
In the following the supported disk access types are presented.

\begin{figure}[htbp]
\begin{center}
\includegraphics[width=8cm]{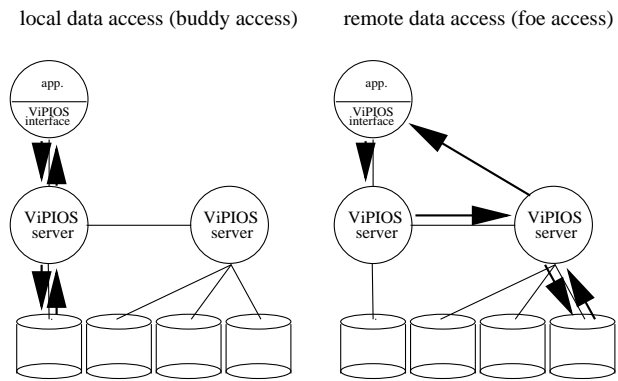}
\end{center}
\caption{Local versus remote data access}\label{dataaccess}
\end{figure}

According to the SPMD programming paradigms parallelism is expressed
by the data distribution scheme of the HPF language in the application
program. Basically ViPIOS has therefore to direct the application
process's data access requests to independent ViPIOS servers only to
provide parallel disk accesses. However a single SPMD process is
performing its accesses sequentially sending its requests to just one
server. Depending on the location of the requested data on the disks
in the ViPIOS system two access types can be differentiated (see figure
\ref{dataaccess}),

\begin{itemize}
\item Local data access,
\item Remote data access
\end{itemize}

\paragraph{Local Data Access.}
The buddy server can resolve the applications requests on its own
disks (the disks of its best disk list). This is also called {\it buddy access}.

\paragraph{Remote Data Access.}
The buddy server can not resolve the request on its disks and has to
broadcast the request to the other ViPIOS servers to find the owner of
the data. The respective server (foe server) accesses the requested
data and sends it directly to the application via the network. This is also
called {\it foe access}.

Based on these access types 3 three disk access modes can be
distinguished, which are called

\begin{itemize}
\item{Sequential,}
\item{Parallel, and}
\item{Coordinated mode.}
\end{itemize}

\paragraph{Sequential Mode.}

The sequential mode of operation allows a single application process
to send a sequential read/write operation, which is processed by a
single VIPIOS server in sequential manner. The read/write operation
consists commonly of processing a number of data blocks, which are
placed on one or a number of disks administrated by the server itself
(disks belonging to the best-disk-list of the server).

\paragraph{Parallel Mode.}

In the parallel mode the application process requests a single
read/write operation. ViPIOS processes the sequential process in
parallel by splitting the operation in independent sub-operations and
distributing them onto available ViPIOS server processes.

This can be either the access of contiguous memory areas (sub-files)
by independent servers in parallel or the distribution of a file onto
a number of disks administrated by the server itself and/or other
servers.

\paragraph{Coordinated Mode.}

The coordinated mode is directly deferred from the SPMD approach by
the support of collective operations. A read/write operation is
requested by a number of application processes collectively. In fact
each application process is requesting a single sub-operation of the
original collective operation. These sub-operations are processed by
ViPIOS servers sequentially, which in turn results in a parallel
execution mode automatically.

\begin{figure}[htbp]
\begin{center}
\includegraphics[width=12cm]{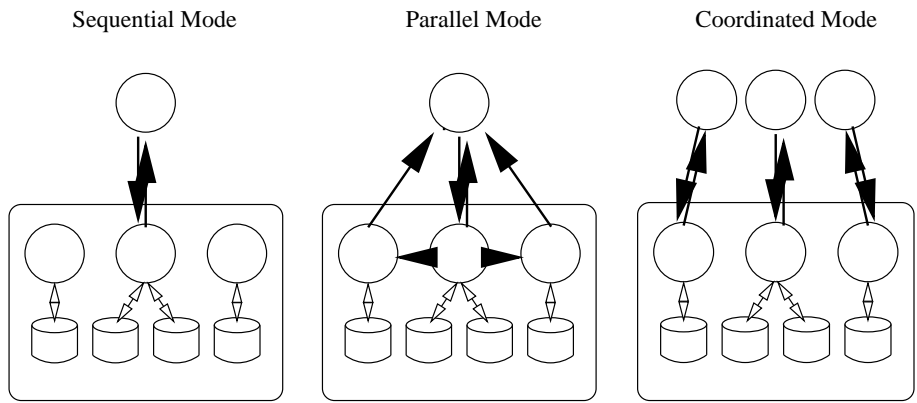}
\end{center}
\caption{Disk Access Modes}\label{accessmode}
\end{figure}

The 3 modes are shown in figure \ref{accessmode}.

\section{Abstract File Model}\label{sec_file-model}
In order to be able to calculate an optimal data layout on disk a formal model
to estimate the expected costs for different layouts is needed. This chapter
presents an abstract file model which can be used as a basis for such a cost
model. The formal model for sequential files and their access operations
presented here is partly based on the works in
\cite{tennent:principles} and \cite{brezany:fileop}.

It is also shown how the mapping functions defined in this model, which
provide logical and physical data abstraction as depicted in figure
\ref{abstraction} are actually implemented in the ViPIOS system.

\input{file_model}

\chapter{ViPIOS Kernel}\label{sec_Kernel}
This chapter describes the actual implementation of the ViPIOS Kernel. It shows
the internal working of the ViPIOS processes and discusses the realization
of the different operation modes, which enable the port of ViPIOS to various
hardware platforms.

\section{The Message Passing System}\label{sec_Message_Passing}
In order to show how a client request actually is resolved by the
ViPIOS server processes some necessary notation is defined first and then
the flow of control and messages for some basic requests (like OPEN,
READ and WRITE) is described.

\subsection{Notation}
In the following some abbreviations are used to denote the various components
of ViPIOS.

\begin{itemize}
\item AP: for an application process (ViPIOS-client) which is in fact an
instance of the application running on one of the compute nodes
\item VI: for the application process interface to ViPIOS (ViPIOS-Interface)
\item VS: for any ViPIOS server process
\item BUDDY: for the buddy server of an AP (i.e. the server process assigned to
the specific AP. See chapter \ref{sec_architecture} for more details.)
\item FOE: for a server, which is foe to an AP (i.e. which is not the BUDDY for
the specific AP. See chapter \ref{sec_architecture} for more details.)
\end{itemize}

For system administration and initialization purposes ViPIOS offers some
special services which are not needed for file I/O operations. These services
include:

\begin{itemize}
\item system services: system start and shutdown, preparation phase routines
(input of hardware topology, best disk lists, knowledge base)
\item connection services: connect and disconnect an AP to ViPIOS.
\end{itemize}

Since these services are relatively rarely used, not every ViPIOS server
process needs to provide them. A ViPIOS server process, which offers system
(connection) services is called a system (connection) controller, abbreviated
SC (CC). Depending on the number of controllers offering a specific service
three different controller operation modes can be distinguished.

\begin{itemize}
\item centralized mode: There exists exactly one controller in the whole
system for this specific service.
\item distributed mode: Some but not all ViPIOS-servers in the system are
controllers for the specific service.
\item localized mode: Every ViPIOS server is a controller for the specific
service.
\end{itemize}

Note that in every ViPIOS configuration at least one system controller and one
connection controller must exist. The rest of this chapter restricts
itself to system and connection controllers in centralized mode, which are
the only ones actually implemented so far. This means
that the terms SC and CC denote one specific ViPIOS server process respectively.
However no assumptions are made whether SC and CC are different processes
or actually denote the same ViPIOS server process. (For distributed computing
via the Internet the other modes for SC and CC could however offer big
advantages and will therefore also be implemented in later versions of ViPIOS.)

An additional service, which is vital for the operation of a ViPIOS system is
the directory service. It is responsible for the administration of file
information (i.e. which part of a file is stored on which disk and where are
specific data items to be read/written). Currently only the localized mode
has been realized, which means that every server process only holds the
information for those parts of the files, which are stored on the disks
administered by that process. Thus each ViPIOS server process currently also
is a directory controller (DC). The directory service differs from the other
services offered by ViPIOS in that it is hidden from the application processes.
So only the server processes can inquire where specific data items can be
found. There is no way for the application process (and thus for the programmer)
to find out which disk holds which data. (For administration purposes however
the system services offer an indirect way to access directory services. An
administrator may inspect and even alter the file layout on disk.)

\subsubsection{Files and Handles}
Applications which use ViPIOS can read and write files by using
ordinary UNIX like functions. The physical files on disks are however
automatically distributed among the available disks by the server processes.
This scattering of files is transparent to the client application and
programmers can therefore apply the well known common file paradigms of the
interface they are using to access ViPIOS (UNIX style, HPF or MPI-IO calls).

The application uses file handles to identify specific files. These handles are
generated by the VI which also administers all the related informations like
position of file pointer, status of I/O operation and so on. This allows for a
very efficient implementation of the Vipios\_IOState function and also reduces
the administration overhead compared to a system where filehandles are managed
by VSs (as will be shown later).

\subsubsection{Basic ViPIOS file access functions}
The AP can use the following operations to access ViPIOS files.

\begin{itemize}
\item Vipios\_Open(Filename, Access mode)\\
Opens the file named 'Filename'. Access mode may be a combination of READ,
WRITE, CREATE, EXCLUSIVE. The function returns a file handle if successful or
an error code otherwise.
\item Vipios\_Read(Filehandle, Number of bytes, buffer)\\
Read a number of bytes from the file denoted by 'Filehandle' into the specified
buffer. Returns number of bytes actually read or an error code. (In case of EOF
the number of bytes read may be less than the requested number. Additional
information can be obtained by a call to the Vipios\_IOState function.)
\item Vipios\_Write(Filehandle, Number of bytes, buffer)\\
Write a number of bytes to the file denoted by 'Filehandle' from the specified
buffer.
\item Vipios\_IRead(Filehandle, Number of bytes, buffer)\\
Immediate read. Same as read but asynchronous (i.e. the function returns
immediately without waiting for the read operation to actually be finished).
\item Vipios\_IWrite(Filehandle, Number of bytes, buffer)\\
Immediate write. Same as write but asynchronous.
\item Vipios\_Close(Filehandle)\\
Closes the file identified by 'Filehandle'.
\item ViPIOS\_Seek(Filehandle, position, mode)\\
Sets the filepointer to position. (The mode parameter specifies if the position
is to be interpreted relative to the beginning or to the end of the file or
to the current position of the filepointer. This parallels the UNIX file seek
function.)
\item Vipios\_IOState(Filehandle)\\
Returns a pointer to status information for the file identified by
'Filehandle'. Status information may be additional error information,
EOF-condition, state of an asynchronous operation etc.
\item Vipios\_Connect(\lbrack System\_ID\rbrack )\\
Connects an AP with ViPIOS. The optional parameter 'System\_ID' is reserved
for future use where an AP may connect to a ViPIOS running on another machine
via remote connections (e.g. internet). The return value is TRUE if the
function succeeded, FALSE otherwise.
\item Vipios\_Disconnect()\\
Disconnects the AP from ViPIOS. The return value is TRUE if the function
succeeded, FALSE otherwise.
\end{itemize}

\subsubsection{Requests and messages}
Requests are issued by an AP via a call to one of the functions declared above.
The VI translates this call into a request message which is sent to the AP's
BUDDY (Except in the case of a Vipios\_Connect call where the message is sent
to the CC which then assigns an appropriate VS as BUDDY to the AP).

According to the above functions the basic {\em message types} are as follows.

CONNECT; OPEN; READ; WRITE; CLOSE; DISCONNECT
\\Note that read and write requests are performed asynchronously by ViPIOS
server processes so that no extra message types for asynchronous operations are
needed. If the application calls the synchronous versions of the read or
write function then the VI tests and waits for the completion of the operation.

ViPIOS-messages consist of a message header and status information. Optionally
they can contain parameters and/or data. The header holds the IDs of the sender
and the recipient of the message, the client ID (=the ID of the AP which
initiated the original external request), the file ID, the request ID and the
message type and class. The meaning of status depends on the type and class of
the message and may for example be TRUE or FALSE for acknowledges or a
combination of access modes for an OPEN message. Number and meaning of
parameters varies with type and class of the message and finally data may be
sent with the request itself or in a seperate message.

\subsection{The execution of I/O Operations}
Figure \ref{fragmenter} shows the modules of a VS which are of interest for
handling requests.

The {\em local directory} holds all the information necessary to map a client's
request to the physical files on the disks managed by the VS. (i. e. wich
portions of a file are stored by this server and how these portions are layout
on the disks.) The {\em fragmenter} uses this information to decompose
({\em fragment}) a request into sub-requests which can be resolved locally and
sub-requests which have to be communicated to other ViPIOS server processes.
The {\em I/O subsystem} actually performs the necessary disk accesses and the
transmission of data to/from the AP. It also sends acknowledge messages to the
AP.

\subsubsection{The request fragmenter}\label{sec:fragmenter}
The fragmenter handles requests differently dependent on their origin. For that
reason we define the following {\em request classes} and the corresponding
{\em message classes}.

\begin{itemize}
\item external requests/messages (ER): from VI to BUDDY
\item directed internal requests/messages (DI): from one VS to another
specific VS
\item broadcast internal requests/messages (BI): from one VS to all other VSs
\item acknowledge messages (ACK): acknowledges the (partial) fulfillment of a
request; can be sent from a VS to another VS or to a VI
\end{itemize}

\begin{figure}
    \centerline{\includegraphics[scale=0.5,angle=270]{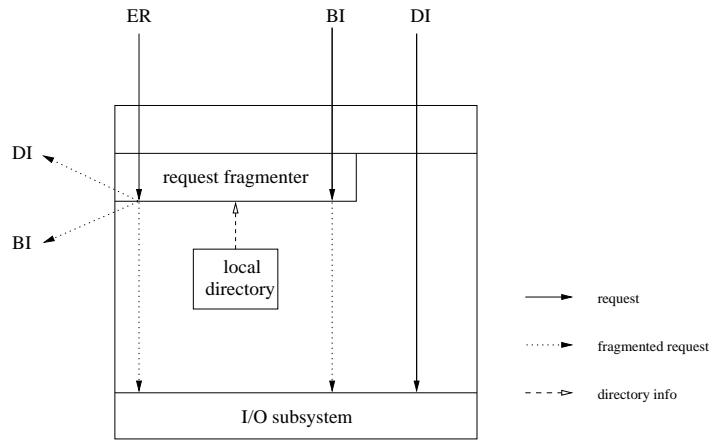}}
    \caption{A ViPIOS-server (VS)}\label{fragmenter}
\end{figure}

Figure \ref{fragmenter} shows how requests are  processed by the fragmenter.
For external requests (ER) the fragmenter uses the VS's local directory
information to determine the sub-request which can be fulfilled locally. It
then passes this part to the VS's I/O subsystem which actually performs the
requested operation.

The remaining sub-requests are committed as internal requests to other VSs. If
the fragmenter already knows which VS can resolve a sub-request (e.g. by hints
about data distribution or if the VS is a directory controller in centralized
or distributed mode) then it sends this sub-request directly to the appropriate
server (DI message). Otherwise the sub-request is broadcast to all the other
VSs (BI message).

Note that only external requests can trigger additional messages to be sent or
broadcast. Internal requests will either be filtered by the fragmenter, if they
have been broadcast (appropriate VS was unknown), or passed directly to the I/O
subsystem, if they have been sent directly (appropriate VS was known in
advance). This design strictly limits the number of request messages that can
be triggered by one single AP's request.

In an optimal configuration files are distributed over VSs such that no
internal requests have to be generated (i. e. every request can be resolved
completely by the BUDDY = {\em Data locality principle}).

\subsubsection{Control and message flow}

\begin{figure}
    \centerline{\includegraphics[scale=0.5,angle=270]{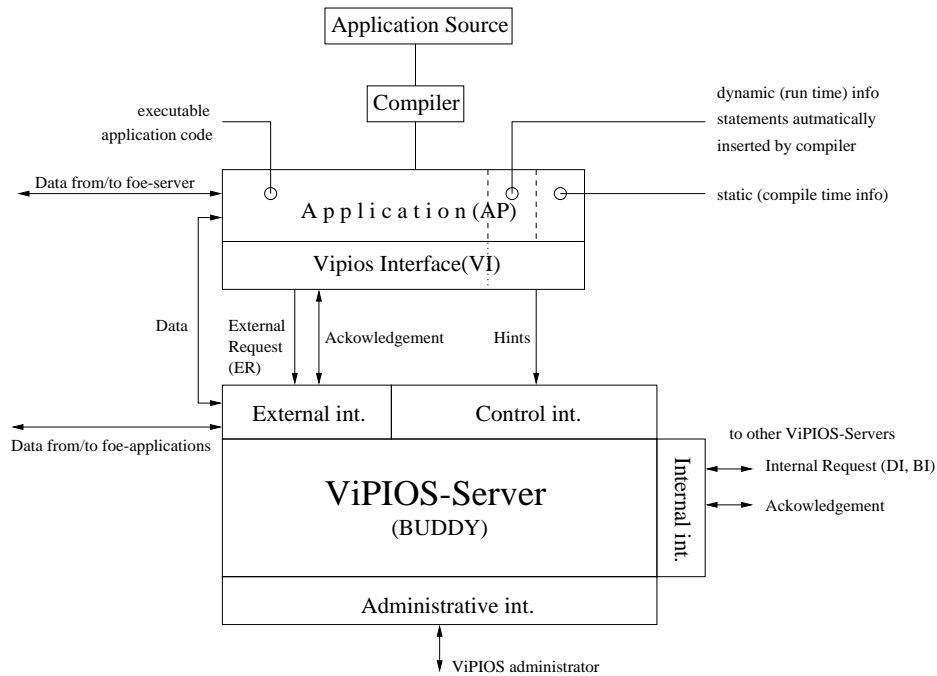}}
    \caption{Overall message flow}\label{flow}
\end{figure}

Figure \ref{flow} depicts the actual message flow in ViPIOS. To keep it simple
only one AP and its associated BUDDY are shown. However the message flow
to/from FOEs is included too. Note that the original application code is
transformed by an HPF compilation system into APs containing static compile
time information (like data distributions etc.) as well as some compiler
inserted statements, which send information to ViPIOS at runtime (hints for
prefetching etc.). These informations are communicated to the BUDDY in the form
of hints and are used to optimize I/O accesses.

The VI sends its requests to the external interface of the BUDDY. To perform
the requested operation the BUDDY's fragmenter may send sub-requests to FOEs
(see \ref{sec:fragmenter}) via the BUDDY's internal interface. Every VS which
resolves a sub-request sends an acknowledge message to the appropriate client's
VI.

The VI collects all the acknowledges and determines if the operation is
completed. If so, it returns an appropriate value to the AP (in case of a
synchronous operation) or sets the state of the operation accordingly (in case
of an asynchronous operation).

Note that in order to save messages all FOEs send their acknowledges directly
to the client's VI bypassing the BUDDY which sent the internal request. This
implies that the VI is responsible for tracking all the information belonging
to a specific file handle (like position of file pointer etc.).

For operations like READ and WRITE the transmission of actual data can be done
in one of the two following ways.

\paragraph{Method 1:}
Data is sent directly with the READ request or with the WRITE acknowledge.\\
In this case the VI has to provide a receive (send) buffer which is large
enough to hold the acknowledge (request) message's header, status and
parameters as well as the data to be read (written). Since the VI actually uses
the same memory as the AP all the buffers allocated by the VI in fact reduce
the memory available to the computing task. Furthermore data has to be copied
between the VI's internal buffer and the AP.

\paragraph{Method 2:}
Data is sent in an additional message following the READ or WRITE acknowledge.\\
The VI uses the AP's data buffer which was provided in the call to Vipios\_read
(Vipios\_write) to receive (send) the data. This can be done because the extra
data message does not have to contain any additional information but the raw
data. All necessary information is already sent with the preceding acknowledge.
This saves the VI from allocating large buffer at the cost of extraneous
messages. (Note that in Figure \ref{flow} data messages are linked directly to
the AP bypassing the VI. This indicates that data transmission is actually
performed using the data buffer of the AP.)

\paragraph{}
The ViPIOS-system decides how data is transmitted for a specific request by
using its knowledge about system characteristics like available memory size and
cost of extra data messages.

\paragraph{}
In addition to the above, every VS supports an administrative interface to
provide for administrative messages (like descriptions of hardware topology,
best disk lists, etc.). In effect the SC gets the administrative messages
provided by the system administrator and then dispatches it to the other VSs.

\section{Operation Modes of ViPIOS}\label{sec_operation_modes}
Unfortunately the client-server architecture that ViPIOS uses can not be
implemented directly on all platforms because of limitations in the underlying
hard- or software (like no dedicated I/O nodes, no multitasking on processing
nodes, no threading, etc.). So in order to support a wide range of different
plattforms ViPIOS uses MPI for portability and offers multiple
{\em operation modes} to cope with various restrictions.

The following 3 different operation modes have been implemented:

\begin{itemize}
\item runtime library,
\item dependent system, or
\item independent system.
\end{itemize}

\paragraph{Runtime Library.} Application programs can be linked with
a ViPIOS runtime module, which performs all disk I/O requests of the
program. In this case \mbox{ViPIOS} is not running on independent servers,
but as part of the application. The interface is therefore not
only calling the requested data action, but also performing it itself.
This mode provides only restricted functionality due to the missing
independent I/O system. Parallelism can only be expressed by the
application (i.e. the programmer).

\paragraph{Dependent System.} In this case ViPIOS is running as an
independent module in parallel to the application, but is started
together with the application. This is inflicted by the MPI-1
specific characteristic that cooperating processes
have to be started at the same time.  This mode allows
smart parallel data administration but objects the
Two-Phase-Administration method by a missing preparation phase.

\paragraph{Independent System.} In this case ViPIOS is running
as a client-server system similar
to a parallel file system or a database server waiting for application
to connect via the ViPIOS interface.
This is the mode of choice to achieve
highest possible I/O bandwidth by exploiting all available data
administration possibilities, because it is the only mode which supports the
two phase data administration method.

\subsection{Restrictions in Client-Server Computing with MPI}
\label{sec_restrictions}

\subsubsection{Independent Mode is not directly supported by MPI-1.}

MPI-1 restricts client-server computing by imposing that all the communicating
processes have to be started at the same time. Thus it is not possible to have
the server processes run independently and to start the clients  at some later
point in time. Also the number of clients can not be changed during execution

\subsubsection{Clients and Servers share MPI\_COMM\_WORLD in MPI-1.}

With \mbox{MPI-1} the global communicator MPI\_COMM\_WORLD is shared by all
participating processes. Thus clients using this communicator for collective
operations will also block the server processes. Furthermore client and server
processes have to share the same range of process ranks. This makes it hard to
guarantee that client processes get consecutive numbers starting with zero,
especially if the number of client or server processes changes dynamically.

Simple solutions to this problem (like using separate communicators for clients
and servers) are offered by some ViPIOS operation modes, but they all require,
that an application program has to be specifically adapted in order to use
ViPIOS.

\subsubsection{Public MPI Implementations (MPICH, LAM) are not Multi Threading
Safe.}

Both public implementations (MPICH \cite{www-MPICH} and LAM \cite{www-LAM}) are
not multi threading save. Thus non-blocking calls (e.g. MPI\_Iread,
MPI\_Iwrite) are not possible without a workaround. Another drawback without
threads is that the servers have to work with busy waits (MPI\_Iprobe) to
operate on multiple communicators.

\subsubsection{Running two or more Client Groups with MPI-2.}
Every new client
group in MPI-2 needs a new intercommunicator to communicate with the ViPIOS
servers. Dynamically joining and leaving a specific already existing group
is not possible. PVM for example offers this possibility with the functions
pvm\_joingroup (...) and pvm\_lvgroup (...).

\subsection{Comparing ViPIOS' Operation Modes}

In the following the advantages and disadvantages of all the operation modes
and their implementation details are briefly discussed.

\subsubsection{Runtime Library Mode}

behaves basically like ROMIO~\cite{thakur:romio-users} or
PMPIO~\cite{fineberg:pmpio}, i.e. ViPIOS is linked as a runtime library to the
application.

\begin{itemize}
    \item{\em Advantage}
        \begin{itemize}
            \item{ready to run solution with any MPI-implementation
            (MPICH, LAM)}
        \end{itemize}
    \item{\em Disadvantage}
        \begin{itemize}
            \item{nonblocking calls are not supported. Optimization like
            redistributing in the background or prefetching is not
            supported}
            \item{preparation phase is not possible, because ViPIOS is
            statically bound to the clients and started together with
            them}
            \item{remote file access is not supported, because there is
            no server waiting to handle remote file access requests, i.e.
            in static mode the server functions are called directly and
            no messages are sent (On systems with multithreading
            capabilities this could be overcome by starting a thread that
            waits for and accomplishes remote file access requests.}
        \end{itemize}
\end{itemize}

\subsubsection{Client Server Modes}

allow optimizations like file redistribution or prefetching and remote file
accesses.

\paragraph{Dependent Mode. }

In Client-Server mode clients and server start at the same time using
application schemes.

\begin{itemize}
\item{\em Advantage}
    \begin{itemize}
        \item{ready to run solution (e.g with MPICH)}
    \end{itemize}
\item{\em Disadvantage}
    \begin{itemize}
        \item{preparation phase is not possible, because the ViPIOS servers
        must be started together with the clients}
        \item{an exclusive MPI\_COMM\_WORLD communicator for clients can
        only be supported in a  patched MPICH version. That patch has been
        implemented but this limits portability)}
    \end{itemize}
\end{itemize}

\paragraph{Independent Mode. }

In order to allow an efficient preparation phase the use of independently
running servers is absolutely necessary.

This can be achieved by using one of the following strategies:

\begin{enumerate}
    \item{MPI-1 based implementations.}\\
        Starting and stopping processes arbitrarily can be simulated with
        MPI-1   by using a number of "dummy" client processes which are
        actually idle and spawn the appropriate client process when needed.
        This simple workaround limits the number of available client
        processes to the number of "dummy" processes started.

        This workaround can't be used   on systems which do not offer
        multitasking because the idle "dummy" process will lock a processor
        completely. Furthermore additional programming effort for waking
        up the dummy proccesses is needed.
        \begin{itemize}
            \item{Advantage}
            \begin{itemize}
                \item{ready to run solution with any MPI-1
                implementation}
            \end{itemize}
            \item{Disadvantage}
            \begin{itemize}
                \item{workaround for spawning the clients necessary,
                because clients cannot be started dynamically}\\
            \end{itemize}
        \end{itemize}
    \item{MPI-2 based implementations.}\\
Supports the connection of independently started MPI-applications with ports.
The servers offer a connection through a port, and client groups, which are
started independently from the servers, try to establish a connection to the
servers using this port. Up to now the servers can only work with one client
group at the same time, thus the client groups requesting a connection to the
servers are processed in a batch oriented way, i.e. every client group is
automatically put into a queue, and as soon as the client group the servers
are working with has terminated, it is disconnected from the servers and the
servers work with the next client group waiting in the queue.
        \begin{itemize}
            \item{Advantages}
            \begin{itemize}
                \item{ready to run solution with any MPI-2
                implementation}
                \item{No workaround needed, because client groups can
                be started dynamically and independently from the
                server group}
                \item{Once the servers have been started, the user can
                start as many client applications as he wants without
                having to take care for the server group}
                \item{No problems with MPI\_COMM\_WORLD. As the server
                processes and the client processes belong to two
                different groups of processes, which are started
                independently, each group has implicitly a separated
                MPI\_COMM\_WORLD}
            \end{itemize}
            \item{Disadvantage}
            \begin{itemize}
                \item{The current LAM version does not support
                multi-threading, which would offer the possibiliy of
                concurrent work on all client groups without busy
                waits}
                \item{LAM Version 6.1 does not work when trying to
                connect processes which run on different nodes}
            \end{itemize}
        \end{itemize}
    \item{Third party protocol for communication between clients and servers
        (e.g. PVM).} \\
        This mode behaves like MPI-IO/PIOFS~\cite{corbett:pfs} or MPI-IO
        for HPSS~\cite{jones:mpi-io}, but ViPIOS uses PVM and/or PVMPI
        (when it is available) for communication between clients and
        servers. Client-client and server-server communication is still
        done with MPI.
        \begin{itemize}
            \item{Advantage}
            \begin{itemize}
                \item{ready to run solution with any MPI-implementation
                and PVM}
                \item{Clients can be started easily out of the shell}
                \item{no problems with MPI\_COMM\_WORLD, because there
                exist two distinct global communicators}
            \end{itemize}
            \item{Disadvantage}
            \begin{itemize}
                \item{PVM and/or PVMPI is additionally needed. Because
                of the wide acceptance of the MPI standard PVM is
                unlikely to be of any future importance. So the system
                should not be used any more.}
            \end{itemize}
        \end{itemize}
\end{enumerate}

\subsection{Sharing MPI\_COMM\_WORLD}

So far, the independent mode using PVM(PI) or MPI-2 is the only ones which
allows to use ViPIOS in a completely transparent way. For the other modes one
of the following methods can be used to simplify or prevent necessary
adaptations of applications.

\begin{enumerate}
    \item{Clients and servers share the global communicator
    MPI\_COMM\_WORLD.}\\
        In this mode ViPIOS offers an intra-communicator MPI\_COMM\_APP for
        communication of client processes and uses another one
        (MPI\_COMM\_SERV) for server processes. This also solves the
        problem with ranking but the application programmer must use
        MPI\_COMM\_APP instead of MPI\_COMM\_WORLD in every MPI function
        call.

    \item{Clients can use MPI\_COMM\_WORLD exclusively.} \\
        This can be achieved  patching the underlying MPI implementation
        and also copes with the ranking problem.
\end{enumerate}

\begin{figure}
\begin{center}
\includegraphics[scale=0.65]{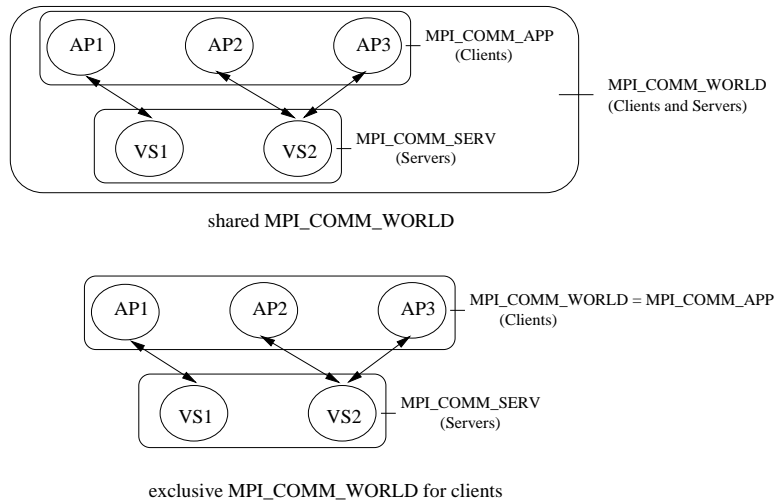}
\caption{shared MPI\_COMM\_WORLD versus exclusive MPI\_COMM\_WORLD}\label{depmode}
\end{center}
\end{figure}

A graphical comparison of this solutions is depicted in Figure \ref{depmode}.

\subsection{Implemented solutions}
Of the approaches described above the following have been implemented so far:
\begin{itemize}
    \item{runtime library mode with MPI-1 (MPICH)}
    \item{dependent mode with MPI-1 with threads (MPICH and patched MPICH)}
    \item{independent mode with the usage of PVM and MPI-1 (MPICH)}
    \item{independent mode with MPI-2 without threads (lam)}
\end{itemize}

\section{Implementation Details of Operation Modes}

\subsection{Dependent Mode with a Shared MPI\/\_\/COMM\/\_\/WORLD}

The first client-server-based implementations were realized in the dependent
mode with a common global communicator MPI\/\_\/COMM\/\_\/WORLD. That means,
the client processes and the server-processes must all be started together as
one single application consisting of client-processes and server-processes, and
all these processes are members of one single MPI\/\_\/COMM\/\_\/WORLD.
Therefore the programmers of ViPIOS-applications must always keep in mind that
the program which they are writing is only one part of the whole system. So
that they may never execute MPI\/\_\/Barrier(MPI\/\_\/COMM\/\_\/WORLD) becaues
MPI would then expect the server-processes to execute the barrier operation
too and the program would be blocked.

\subsection{Dependent Mode with a Separated MPI\/\_\/COMM\/\_\/WORLD}

This modification of ViPIOS has a separate global communicator
MPI\/\_\/COMM\/\_\/WORLD for the client processes. But the client processes and
the server-processes must still be started together concurrently.
However, the programmer of the client processes does no longer have to care
about the ViPIOS server processes. The client progam can be thought of as
running independently and just satisfying its I/O needs by using calls to the
ViPIOS interface library. This approach has been implemented in ViPIOS in two
ways:

\begin{enumerate}
\item{by modification of MPI}
\item{by creating a header file mpi\/\_\/to\/\_\/vip.h, which has to be
included in every source file of a ViPIOS-project just after including mpi.h}
\end{enumerate}

\subsubsection{Modification of MPI}
In the MPICH 1.1 \cite{www-MPICH} implementation of MPI
the internal representation of all the MPI-specific data-types (Communicators,
Groups) is as follows. All these data is stored in an internal list which is
hidden from the user. The only thing which the user can see are pointers to
entries in this list. Each MPI communicator and each MPI group is represented
by one entry in the list and the variables of the types MPI\/\_\/Comm and
MPI\/\_\/Group are nothing else than pointers to these entries. Each entry in
the list has an integer number and the pointers to the entries are just
variables in which the number of these entries is stored. Therefore the types
MPI\/\_\/Comm and MPI\/\_\/Group are just integers. As the global communicator
which contains all processes is automatically stored in the internal list at
position 91 when MPI is initialized, the definition of the constant
MPI\/\_\/COMM\/\_\/WORLD is done in the file mpi.h simply by the line
``\#define MPI\/\_\/COMM\/\_\/WORLD 91''. Therefore the modification of
MPI\/\_\/COMM\/\_\/WORLD was done by substituting the name
MPI\/\_\/COMM\/\_\/WORLD in this line by the name MPI\/\_COMM\/\_UNIVERSAL
and defining MPI\/\_\/COMM\/\_\/WORLD as variable of the type MPI\/\_\/Comm
instead. As soon as ViPIOS is initialized, a communicator containing only the
client-processes is created and stored in the variable
MPI\/\_\/COMM\/\_\/WORLD. Therefore it is important that the programmer does
not access MPI\/\_\/COMM\/\_\/WORLD before initializing ViPIOS.
All the modifications of MPI took place only in the file mpi.h. Therefore it
was {\em not} necessary to recompile MPI. The only thing which has to be done
is substituting the original mpi.h file by the modified one.
(Note that this modification only works for the MPICH version of MPI. Other
implementations may use different representations for MPI\/\_\/COMM\/\_\/WORLD.)

\subsubsection{Creation of a Header file mpi\/\_\/to\/\_\/vip.h}
The modification of MPI\/\_\/COMM\/\_\/WORLD can also be done without any
modification of MPI itself. Instead of modifying mpi.h a header file called
mpi\/\_\/to\/\_\/vip.h can be included immediately after mpi.h in every module
of ViPIOS and in the application modules too. This modifies
the definition of MPI\/\_\/COMM\/\_\/WORLD given in mpi.h after the header
file has been included. So the final effect is the same as if the modified
version of mpi.h had been used.

\subsubsection{Compatibility of mpi\/\_to\/\_\/vip.h to Other MPI
Implementations}
The way of modifying MPI\/\_\/COMM\/\_\/WORLD just explained has only been
applied to MPICH 1.1, but with little modifications of the file
mpi\/\_to\/\_\/vip.h it can also be applied to any other version of MPI.
Whenever MPI\/\_\/COMM\/\_\/WORLD is created with the \#define command this
definition has just do be undone with \#undef and MPI\/\_\/COMM\/\_\/WORLD has
to be redefined as a variable instead. This variable may then be initialized
with the same value which was assigned to the
\#define-name MPI\/\_\/COMM\/\_\/WORLD
in the original mpi.h file in order to avoid having an undefined
\linebreak
MPI\/\_\/COMM\/\_\/WORLD.
All this is done in  mpi\/\_to\/\_\/vip.h and if the value which is assigned
to
\linebreak
MPI\/\_\/COMM\/\_\/WORLD in the original mpi.h file changes in another
MPI implementation, the value with which the variable
MPI\/\_\/COMM\/\_\/WORLD is assigned in the file mpi\/\_to\/\_\/vip.h has to
be changed accordingly.

It is very probable that this will work with the future implementations of MPI
too because the implementors of MPICH are very convinced that defining
MPI\/\_\/COMM\/\_\/WORLD with a \#define construct is the best way of
implementing it. If it is for some reason not possible to initialize the
variable MPI\/\_\/COMM\/\_\/WORLD with the value which was assigned to the
\#define-name MPI\/\_\/COMM\/\_\/WORLD in mpi.h, there is still the posibility
of omitting its initialization. But then it may not be accessed before
initializing ViPIOS (which is not a great problem as it is not recommended to
access it before initializing ViPIOS anyway).

The activities necessary for modifying MPI\/\_\/COMM\/\_\/WORLD done in ViPIOS
itself (i.e. creating of independent communicators for server processes and for
application processes and assignment of the apllication processes' communicator
to the application's MPI\/\_\/COMM\/\_\/WORLD) are completely independent from
the internal implementation of MPI and will never have to be adapted for new
MPI versions.

\subsection{Creation of a Separate MPI\/\_\/COMM\/\_\/WORLD for Fortran}
As Fortran applications have to cooperate with the ViPIOS system,
which is written in C, and it is not possible to declare
variables, which are shared between the C files and the Fortran files, the
manipulation of MPI\/\_\/COMM\/\_\/WORLD for Fortran is more complicated.
MPI\/\_\/COMM\/\_\/WORLD for Fortran is defined in the file mpif.h with the
command "PARAMETER (MPI\/\_\/COMM\/\_\/WORLD=91)". As in mpi.h, the name
MPI\/\_COMM\/\_WORLD has been replaced by the name MPI\/\_COMM\/\_UNIVERSAL.
In the file vipmpi.f, which has to be included into the application with the
USE command,
\linebreak
MPI\/\_\/COMM\/\_\/WORLD is defined as a variable. Moreover, this
file implements the routine MPIO\/\_INIT which has to be called by the
Fortran application in order to initialize ViPIOS. This routine calls via the
Fortran to C interface a C routine which invokes a slightly modified
initialization routine for ViPIOS. This intialization routine returns the value
which has to be assigned to MPI\/\_\/COMM\/\_\/WORLD (a communicator containing
all the client-processes) via a reference parameter back to MPIO\/\_INIT.
MPIO\/\_INIT finally stores it in the variable MPI\/\_\/COMM\/\_\/WORLD. The
whole process is hidden from the application programmer.

\subsection{Independent Mode}
Independent mode means that there are two independently started programs. One
consists of the server processes, and the other one consists of the client
processes. First the server processes must be started which offers a connection
via ports. Then the client application is started which connects to the
server processes. This connection creates intercommunicators, which allow a
communication between the client processes and the
server processes in the same way as it was done in the dependent mode. While
active, the server processes can be connected to by client applications at any
time. Thus different applications can concurrently use the ViPIOS I/O server
processes.
With the MPI 1.1 standard an independent mode of ViPIOS cannot be implemented.
However, it can be done with a MPI 2.0 implementation. Up to now the only
MPI 2.0 implementation, with which the independent mode of ViPIOS has been
tested is LAM 6.1 . Unfortunately it works with LAM 6.1 only if all the
processes are executed on the same node. This is due to instabilities of
LAM 6.1. With an MPI 2.0 implementation which works correctly according to the
MPI 2.0 standard processes could be executed distributed across all the
available processors in independent mode.

For a list of advantages and disadvantages of this MPI-2 based implementation
see chapter \ref{sec_restrictions}.

\subsection{Future Work: Threaded Version of the Independent Mode}
The next step is now to create a threaded version of the independent mode. The
ViPIOS server will then be able to serve more than one ViPIOS client program at
the same time. Every server process will then start one thread for each client
application which connects to the server and each of these threads will then
recieve and process only the requests sent by the one client application, for
which is was started. These threads will then comply each request by starting
another thread whose task is to process just that single request. As soon as
the request is sucessfuly complied the thread terminates. If a client
application closes the connection to the ViPIOS server, the server process
threads whose task was to recieve the requests sent by this client also
terminate.

Unfortunately the attempts to implement this version have failed up to now
because LAM is not thread safe. Some alternatives to LAM have therefore been
considered.

\subsection{Alternatives to LAM}

\textbf{Evaluation of possible alternatives for LAM in order to implement the
independent mode of ViPIOS}

\textbf{Problem:} LAM is instable, not thread-safe and connecting/disconnecting
of processes does not work correctly when it is used to execute a program on
more than one node. An MPICH implementation of the MPI 2.0  standard does not
yet exist, therefore other posibilities have to be found to implement the
independent mode of ViPIOS.

For the ViPIOS client server principle to work the ability to combine two
independently started MPI processes is absolutely necessary. The best solution
would be a connection of the server program with the independently started
client program in a way that an intercommunicator between all the processes of
the server program and all the processes of the client program is created (like
in the LAM implementation described above). Because this does not require any
modifications of the code of the ViPIOS functions (except for
ViPIOS\/\_\/Connect).

\subsubsection{MPI-Glue}
MPI-Glue is an implementation of the MPI 1.1 standard which allows MPI
applications to run on heterogeneous parallel systems. It is especially
designed to combine different homogeneous parallel systems (workstation
clusters where all the workstations are of the same type or supercomputers)
together. In order to be as efficient as possible it imports existing
MPI implementations designed for running MPI on homogeneous parallel systems.
This implementations are used for communication inside one of the homogenous
parallel systems. For communication between different machines MPI-Glue
implements an own, portable MPI based on TCP/IP. MPI-Glue exports all
MPI functions according to the MPI 1.1-Standard to the application and as soon
as any MPI function involving communication is called by the application, it
invokes automatically the required internal function (i.e. when there is a
communication inside a homogeneous parallel system it invokes the
implementation designed for homogeneous parallel systems of that type otherwise
it uses its portable MPI implementation based on TCP/IP to do communication
between two different types of processors, which of course takes much more time
than the communication inside a homogeneous parallel system).

\subsubsection{PACX-MPI}
This system was developed at the computing center of the University of
Stuttgart. It offers the same possibilities as MPI-Glue and it also imports the
system dependent MPI implementations for communication inside homogenous
parallel systems. The difference to MPI-Glue is the way how communication is
performed between two processes running on different platforms. In MPI-Glue the
communication goes directly from one process to another. But in PACX-MPI in
every one of the different homogenous parallel systems which are combined there
exist two additional processes. One has the task to send messages to other
homogeneous parallel systems and the other one recieves messages sent by other
homogeneous parallel systems. If one of the application's processes wants to
send a message to a process running on a different platform, it sends it to the
process, which is to send messages to other parallel systems. That process
sends it to the other system. There the message is recieved by the process
whose task is recieving messages from other homogeneous parallel systems and
this process finally sends the message to the destination process.
Only the two additional processes are able to communicate with other
homogeneous parallel systems using TCP/IP. With PACX-MPI only a small subset of
the MPI functions can be used.

\subsubsection{PVMPI}
PVMPI connects independently started MPI applications, which may run on
inhomogenous platforms using PVM. It creates an intercommunicator which
connects two applications. However, it is {\em not} possible to create an
intercommunicator, which contains all processes of both applications with
MPI\/\_\/Intercomm\/\_\/merge.

\subsubsection{PLUS}
PLUS enables communication between parallel applications using different
models of parallel computation. It does not only allow the communication
between different MPI applications running on different platforms but moreover
the communication between e.g. a MPI application and a PVM application. In
order to make it possible to communicate with processes of another application
in an easy way, the processes of each application can address processes of
another application according to their usual communication scheme. For example
PLUS assigns every process of a remote application, with which a PVM
application has to communicate to a task identifier. As soon as a process of
the PVM application tries to send a message to another process, PLUS tests
whether the task id of the addressed process belongs to a process of a remote
application. If so, PLUS transmits the message to the target process using a
protocol based on UDP. The target process can recieve the message by the scheme
according to its programming model. PLUS uses daemon processes like PACX-MPI to
transmit messages between two applications. With PLUS only a restricted set of
datatypes can be used. As in PVMPI the creation of a global intracommunicator
containing ALL processes of all the applications, which communicate through
PLUS is not possible.

\subsubsection{MPI\/\_\/CONNECT}
MPI\/\_\/CONNECT is a result of optimizing PVMPI. It does not longer need PVM
but uses the metacomputing system SNIPE \cite{www-SNIPE} instead to manage the
message passing between the different parallel systems. SNIPE has the advantage
of a better compatibility to MPI than PVM. MPI\/\_\/CONNECT offers either the
possibility to connect independently started MPI applications via
intercommunicators or starting processes on different parallel systems together
as a single application with a shared MPI\/\_\/COMM\/\_\/WORLD, without the
posibility to start additional tasks later.

\subsubsection{Comparison of the Systems}
In order to group the systems by their most important features in the following
table the systems are classified by two different paradigms. In each system
(except MPI\/\_CONNECT) there is only one of these two paradigms available:

\begin{itemize}
\item{paradigm 1}
All the different homogenous parallel systems start simultanously and have a
common MPI\/\_\/COMM\/\_\/WORLD. No processes can be connected later.
\item{paradigm 2}
All the different homogenous parallel systems are started independently and are
connected dynamically. The communication between them is done via
intercommunicators. No global intracommunicator containing processes of more
than one homogenous parallel system can be created.
\end{itemize}

Table \ref{table_compare} lists the most relevant attributes of the systems
described above.

\begin{table}
\begin{center}
\begin{tabular}{|c|c|p{200pt}|}
\hline
System & paradigm & Portability\\
\hline
MPI-Glue  & 1 & nearly complete MPI 1.1 functionality\\
\hline
PACX-MPI  & 1 & only a small subset of MPI 1.1 functionality\\
\hline
PVMPI  & 2 & As there is no global intracommunicator connecting processes of
different parallel systems, no MPI 1.1  applications can be run on the system
without modification.
However, the local communication (=inside one homogenous parallel system) can
use the whole MPI 1.1 functionality\\
\hline
PLUS  & 2 & As there is no global intracommunicator connecting processes of
different parallel systems, no MPI 1.1  applications can be run on the system
without modification.
However, the local communication (=inside one homogenous parallel system) can
use nearly the whole MPI 1.1 functionality.
Only a restricted subset of the MPI datatypes can be used\\
\hline
MPI\/\_\/CONNECT & both paradigms available & complete MPI 1.1 functionality
\& three extra commands for establishing connections between independently
started MPI programs. Works with MPICH, LAM 6, IBM MPIF and SGI MPI\\
\hline
\end{tabular}
\caption{Comparison of Systems}\label{table_compare}
\end{center}
\end{table}

\subsection{consequences}
After evaluating these systems, it is evident that only PVMPI or
MPI\/\_\/CONNECT can be used to efficiently implement the independent mode of
ViPIOS because only these two systems support the connection of independently
started MPI\/\_\/applications. As MPI\/\_\/CONNECT is an improvement of PVMPI,
it is very likely to be the best choice.

\chapter{The MPI-I/O Interface}\label{sec_MPI_Interface}
\input{vimpios}

\section{MPI}
\input{mpi}

\section{MPI-IO}
\input{mpio_intr}

\input{mpio_interface}

\section{ViMPIOS: Implementation of MPI-IO}
\input{mpio_ini}
\input{file_management}
\input{openclose}
\input{views}

\input{dataaccs}

\input{byte_off}
\input{consist}
\input{advanced_datatypes}

\input{fortran}

\section{Regression Suite}
\input{testmpio}

\chapter{The HPF Interface}\label{sec_HPF_Interface}
\input{interf}

\chapter{Performance Analysis}\label{Analysis}
Some simple performance tests have been performed in order to prove the
efficiency and scalability of the ViPIOS design and the implementation as well.
The results have been compared against other existing I/O systems (i.e. Unix
file I/O and ROMIO) to estimate the overhead of ViPIOS for simple read/write
operations which can not be accelerated by parallelization.
\section{Tests setup}

\begin{figure}
\begin{center}
\includegraphics[scale=0.7]{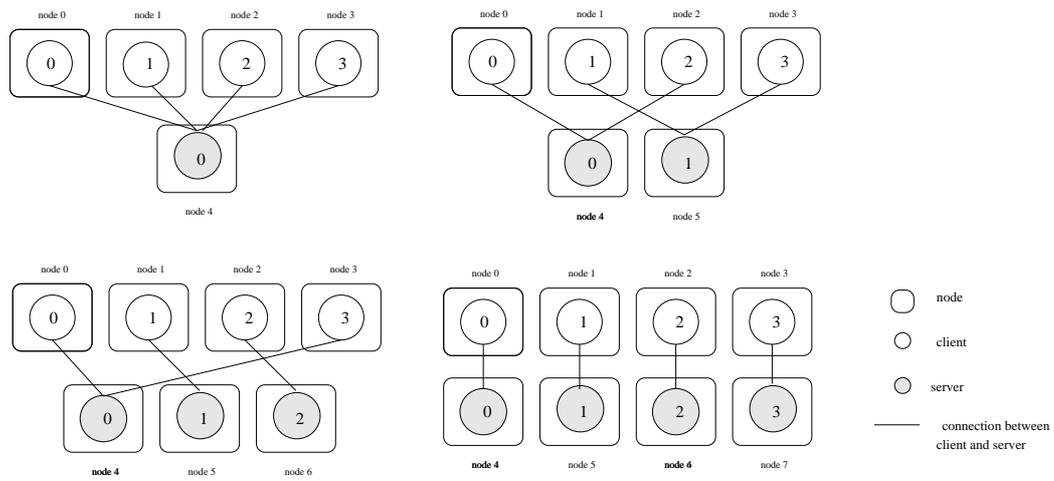}
\caption{Topologies with different number of servers} \label{server_variabel}
\end{center}
\end{figure}
\begin{figure}
\begin{center}
\includegraphics[scale=0.5]{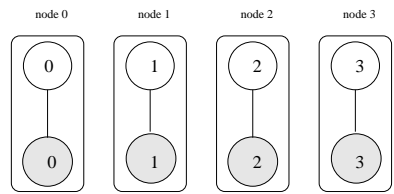}
\caption{Servers and clients reside on the same node} \label{same_node}
\end{center}
\end{figure}
\begin{figure}
\begin{center}
\includegraphics[scale=0.5]{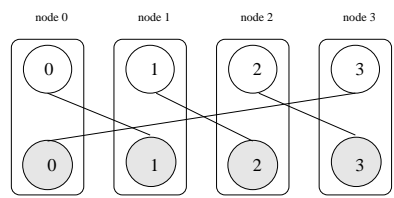}
\caption{Servers and clients reside on different nodes} \label{different_node}
\end{center}
\end{figure}
In order to test ViPIOS we carried out several experiments on a workstation
cluster of 8 LINUX Pentium-133 PCs (nodes) connected by a 100 Mbit Ethernet
network. Each PC
had 32 MB main memory and a hard disk of 2 GByte. Compared to existent super
computers an environment like this offers the following advantages for testing
purposes:
\begin{itemize}
\item low hard- and software costs
\item simple administration
\item availability (the system can be dedicated to the test program so results
are not influenced by changing workloads etc.)
\end{itemize}
Furthermore the Beowulf \cite{www-beowulf} \cite{reschke-design}
\cite{sterling-beowulf} and the Myrinet \cite{www-myrinet} \cite{boden-myrinet}
projects have shown
that whith some enhancements in the network topology such PC-networks can
provide peak performance in excess of 1 GFLOPS and also high disk bandwidths.

First, we ran some scalability tests where we increased the number of server
processes while keeping the number of client processes constant. Moreover we
used dedicated I/O nodes which means that each process (either server or
client) ran on a different node (see figure \ref{server_variabel}).

Second, we experimented with non dedicated I/O nodes which means that on every
node one server and one client process were running concurrently. The results
forced us to differentiate between the following two cases:
\begin{itemize}
\item The client process ran on the same node as its associated server process
as depicted in figure \ref{same_node}.
\item Each client process was connected to a server process on a different node
as depicted in figure \ref{different_node}.
\end{itemize}

\section{Results}
\subsection{Dedicated I/O nodes}
The workload for our first experiment was an 8MB file. This had to be read
by four SPMD client processes (i. e. each client had to read 2 MB without
overlap). In other words, client 0 read the first quarter of the file, client 1
read the second quarter and so on. As depicted in figure \ref{server_variabel}
we increased the number of server processes from one to four.

In order to measure the overall time for reading the whole file, the
concurrently running processes were synchronized before and
after the read request. By imposing these two barriers we guaranteed that the
time to read the whole file corresponded to the reading time of the slowest
process. A short extract from the program code is shown in
figure \ref{code}.

\begin{figure}
\begin{center}
\begin{verbatim}
  /* number of client processes is evaluated */
  MPI_Comm_size(Vipios_comm_app,&nclients);

  MPI_Barrier(Vipios_comm_app);
  /* time is taken after all clients are synchronized */
  start=MPI_Wtime();

  /* each client reads a disjoint part of the file */
  ViPIOS_Seek(fid1, SIZE/nclients*rank, SEEK_SET);
  ViPIOS_Read(fid1,buf,SIZE/nclients);
  MPI_Barrier(Vipios_comm_app);

   /* time is stopped after the last process has finished reading */
   end=MPI_Wtime();
\end{verbatim}
\caption{Program code}\label{code}
\end{center}
\end{figure}

The whole process was iterated 50 times. In order to suppress any caching
effects data was not only read from one file but from 10 different files in a
round robin fashion. Thus, during the first iteration file1 was read, during the
second iteration file2 etc. After 10 iterations file1 was read again. It is
important to state that each server read its data locally from the node it was
running on.
In addition to the mean we measured the maximum, minimum and variance of the
time required for reading the whole file. The results are given in table
\ref{scalability_dedicated}.

\begin{table}
\caption{Scalability results for dedicated I/O nodes}
\label{scalability_dedicated}
\begin{center}
\begin{tabular}{|rrrrrr|}
\hline
clients & servers & max & min & mean & variance \\
\hline
4 & 1 & 4.32 & 2.09 & 3.02 & 0.0694 \\
4 & 2 & 2.15 & 1.20 & 1.78 & 0.0239 \\
4 & 3 & 2.36 & 1.34 & 1.65 & 0.0217 \\
4 & 4 & 1.48 & 0.98 & 1.12 & 0.0115 \\
\hline
\end{tabular}
\end{center}
\end{table}
\begin{figure}
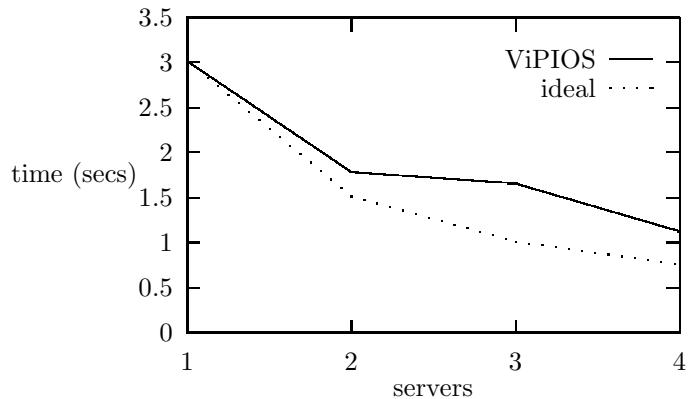

\begin{center}
\include{scalability}
\caption{Scalability results}\label{scalability}
\end{center}
\end{figure}
The graph in figure \ref{scalability} depicts the mean time in seconds to read
the file in relation to the number of server processes.
There seems to be no significant performance gain by increasing the number of
servers from two to three. But remember that clients are synchronized in order
to measure the time it takes to read the 8MB file completely. In the case with
three servers there is one server with two client processes attached. This
server has to read and transfer the same amount of data (4MB) as any server in
the two servers case. So it takes approximately the same time to accomplish
the operation. The other two server-client pairs are idle almost half of the
time. In fact there is a performance gain by introducing a third server
but this will only show up if clients are allowed to work asynchronously.

A comparison with the ideal linear scaling (dotted line in figure
\ref{scalability}) shows that ViPIOS scales well with an increasing number of
servers. Results for superior hardware configurations (MPPs or Beowulf clusters)
are expected to be even better because of increased network bandwidth and less
contention.
%
%
%
%

\subsection{Non dedicated I/O nodes}
Next we studied the results when a server process and the corresponding client
process were running on the same node (see Figure \ref{same_node}). In these
experiments the number of the client processes was always kept equal to the
number of child processes in order to ensure that the work done on every
processing node is the same.
The results were found to be strongly dependent from the topology of the ViPIOS.
If the client and the server connected to it were executing on the same node (
like shown in figure \ref{same_node})
then the performance was very poor. If every client connected to a server which
ran on a different node (figure \ref{different_node}) the performance was
boosted by a factor greater than 30. Further investigation showed that the
reason for this strange behavior was the message passing via MPI (see
\ref{sec_MPI_flaw}). The performance measurements
for the second case are shown in table \ref{constant_workload} for a constant
workload of 8MB. That means that with 2 clients every client read 4MB, with
4 clients every client read 2MB and so on. Table \ref{increasing_workload} gives
the values achieved when every client just read 2MB so that the actual workload
varied from 4MB to 16MB.
\begin{figure}
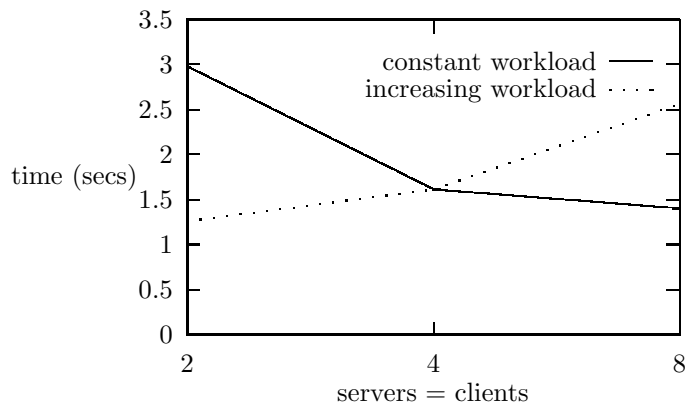

\begin{center}
\include{scale_none_dedicated}
\caption{Non dedicated I/O nodes} \label{non_dedicated_graph}
\end{center}
\end{figure}
As can be clearly seen in figure \ref{non_dedicated_graph} the ViPIOS still
scales
for a constant workload even if the I/O nodes are not dedicated. However the
scaling factor for
increasing the number of processors from 4 to 8 is very small.
The graph for the increasing workload shows that the deviation from the omptimum
(which would be a line parallel to the "server = client" coordinate of course)
increases with the number of servers.

Both of these effects can be easily explained be the increasing network
contention. By using a more elaborate network topology these problems may well
be overridden. Also in a real super-computing environment the interconnection
between the computing nodes will yield a better bandwidth.

\begin{table}
\caption{Constant workload}\label{constant_workload}
\begin{center}
\begin{tabular}{|rrrrrr|}
\hline
clients & servers & max & min & mean & variance \\
\hline
2 & 2 & 3.59 & 2.54 & 2.98 & 0.0734 \\
4 & 4 & 1.96 & 1.22 & 1.61 & 0.0691 \\
8 & 8 & 1.90 & 1.11 & 1.40 & 0.0413 \\
\hline
\end{tabular}
\end{center}
\end{table}

\begin{table}
\caption{Increasing workload}\label{increasing_workload}
\begin{center}
\begin{tabular}{|rrrrrr|}
\hline
clients & servers & max & min & mean & variance \\
\hline
2 & 2 & 2.08 & 0.97 & 1.26 & 0.0624 \\
4 & 4 & 1.96 & 1.22 & 1.61 & 0.0692 \\
8 & 8 & 2.98 & 2.35 & 2.56 & 0.0238 \\
\hline
\end{tabular}
\end{center}
\end{table}

\section{Performance comparison with existing I/O systems}
In order to show that the overhead imposed by message handling and file
administration of the ViPIOS is low we also compared its performance to
UNIX file I/O combined with MPI and to MPI-IO. Performance measurements
have been taken on the PC-network described above. We used MPICH
Version 1.1.0 \cite{www-MPICH} as an MPI implementation and ROMIO Version 1.0.0
\cite{www-ROMIO} as an
implementation of MPI-IO. All the reported values are the means of 50
measurements.

\subsection{ViPIOS compared to UNIX file I/O and MPI}\label{sec_MPI_flaw}
Using normal UNIX file read operations it takes 0.48 seconds to read 2MB from
the local disk. To transfer 2MB of data via MPI from one node to another takes
0.52 seconds. If the sending and the receiving process reside on the same node
then the data transfer via MPI takes at about 50 times as long and the variance
of the duration value gets very big. This may be due to problems in
synchronizing the sender and the receiver processes if they do not run
independently. Here we also find the reason for the bad performance of
ViPIOS when I/O nodes are non dedicated and the client process runs on the same
node as the server process it is connected to.

If we compare the sum of the above times for disk read and data transfer
(1.0 seconds) to the 1.12 seconds ViPIOS takes for the same operation (see
\ref{scalability_dedicated}) we can state that the overhead of the ViPIOS
is reasonably small.

\subsection{ViPIOS compared to ROMIO}
In order to compare ViPIOS to ROMIO we used two test setups. In the first
4 clients had to read  an 8MB file using MPI-I/O operations. Since MPI-I/O
does not directly support access to remote files (files which reside on an
other PC than the client process) the 8MB were distributed in 2MB portions.
So every client just read the 2MB from its local disk. Using ROMIO the read
operation took 0.49 seconds which is in fact very close to the normal
UNIX read (0.48 seconds).

This form of data distribution would be compared
best to a ViPIOS with non dedicated I/O nodes where the server and the
corresponding
client execute on the same node. Unfortunately this is just the configuration
where the performance values of ViPIOS are very bad (39.83 seconds) due to the
behavior of MPI described in the previous section. To overcome this problem we
optimized ViPIOS for that specific topology by using shared memory instead of
MPI to transfer the data from the server to the client. After that ViPIOS just
took 1.41 seconds for the complete read operation. Most of this time (0.84
seconds) was spent for request and acknowledge messages which still were
transmitted via MPI. Further improvements would require to use shared memory
for these messages too. Our time schedule did not allow us to implement that
yet but the results above obviously imply that ViPIOS can handle I/O as quickly
as MPI-I/O even for this specific topology.

In order to have a direct comparison we used a second test setup where the 8MB
file was not distributed
but located in a shared NFS-directory on the disk of the first PC. In this case
the read operation took 3.88 seconds. The comparable ViPIOS configuration (1
dedicated server with 4 clients; see table \ref{scalability_dedicated}
excelled with 3.02 seconds.

%
%
\input{further_tests}

\section{Buffer Management}

Each ViPIOS server process allocates a certain amount of buffer space needed to
accomodate for the data, which has to be read/written to the disks. This buffer
is shared by all the requests that the ViPIOS server process accomplishes
concurrently. A simple buffer manager (first come, first served) has been
implemented, which allots free buffer space to pending requests. If no buffer
space is available at any point in time then open requests, which have no
buffer assigned to them are delayed until another request is completed and
frees its buffer space.

In order to test how crucial the size of the buffer allocated by the ViPIOS
server process is for the performance of the system, we tested the
configuration above (see figure \ref{code}) using different buffer sizes. 5MB
of data were read and the size of the buffer was varied from 5MB to 1KB. We
measured the overall time
that the operation took and also the times needed for the disk access and the
communication between client and servers. (Naturally if the buffer space
available for a request is smaller than the size of the data chunk requested,
then a number of successive disk operations and client server communications
have to be performed.) The program overhead was calculated as the difference
of the overalltime for the operation and the time for messages and disk access
together. This is the time that the program needs for loop control and so on.
The results (see table
\ref{table_buffers}) show that the influences of the buffer size are
neglectable as long as the number of read/write and send/receive operations
for one request does not get too large.

Tables \ref{table_messages} and \ref{table_disk_accesses} give a little more
details. The overall time for communication is split into the times needed
for the different message types. (Each operation included one SEEK and one
READ command sent from the client to the server. The SEEK\_ACK was returned
once by the server. A READ\_ACK and a DATA message containing the actual
data had to be sent whenever the buffer had been filled. So for a 5MB buffer
only one READ\_ACK and one DATA message was needed. For the 1KB buffer 5000
of each of these messages had to be transmitted. Note that the overall
time for READ\_ACK messages roughly doubles if the buffer size halves. But
this is not the case for the DATA messages because while the number of the
messages increases their size decreases at the same time. So the overall time
does not grow that quickly.)
The overall time for disk access also is split into the times needed for the
seek and the read operation. For the read operation the time increases only
slowly (this is again due to the reduced size of the buffer to read while the
number of read operations grows).

\begin{table}
\begin{center}
\begin{tabular}{|r|r|r|r|r|}
\hline
Buffer size & \multicolumn{1}{c|}{Overall time} &
\multicolumn{1}{c|}{Messages} &
\multicolumn{1}{c|}{Disk access} &
\multicolumn{1}{c|}{Program overhead} \\
            & \multicolumn{1}{c|}{mean (sec)} &
\multicolumn{1}{c|}{mean (sec)} &
\multicolumn{1}{c|}{mean (sec)}  &
\multicolumn{1}{c|}{(sec)} \\
\hline
      5 MB  &         1,92 &  0,7233786 &   1,0630203 &        0,1336011 \\
    2.5 MB  &         1,95 &  0,7261519 &   1,0871620 &        0,1333861 \\
      1 MB  &         1,84 &  0,6923981 &   1,0901901 &        0,0574118 \\
    512 KB  &         1,78 &  0,6701392 &   1,0666612 &        0,0431996 \\
    256 KB  &         1,71 &  0,6770387 &   1,0113288 &        0,0178325 \\
    128 KB  &         1,73 &  0,7172205 &   0,9950824 &        0,0164972 \\
     64 KB  &         1,92 &  0,7439646 &   1,1601596 &        0,0178758 \\
     32 KB  &         2,28 &  1,2775443 &   0,9616098 &        0,0378458 \\
     16 KB  &         2,74 &  1,6257222 &   1,0651163 &        0,0447615 \\
      8 KB  &         4,06 &  2,3118020 &   1,5546544 &        0,1953436 \\
      4 KB  &         8,21 &  6,2925069 &   1,6418055 &        0,2756876 \\
      2 KB  &        16,73 & 14,7021435 &   1,6221546 &        0,4057019 \\
      1 KB  &        29,25 & 25,7926610 &   1,6879831 &        1,7651560 \\
\hline
\end{tabular}
\caption{Comparison of Buffer Sizes}\label{table_buffers}
\end{center}
\end{table}

\begin{table}
\begin{center}
\begin{tabular}{|r|r|r|r|r|r|}
\hline
Buffer size &
\multicolumn{1}{c|}{SEEK} &
\multicolumn{1}{c|}{SEEK\_ACK} &
\multicolumn{1}{c|}{READ} &
\multicolumn{1}{c|}{READ\_ACK} &
\multicolumn{1}{c|}{DATA} \\
\hline
      5 MB &  0,0000386 &  0,0001600 &  0,0000300 &  0,0001500 &  0,7230000  \\
    2.5 MB &  0,0000381 &  0,0001598 &  0,0000288 &  0,0002388 &  0,7256864  \\
      1 MB &  0,0000379 &  0,0001607 &  0,0000286 &  0,0004782 &  0,6916927  \\
    512 KB &  0,0000395 &  0,0001586 &  0,0000278 &  0,0008575 &  0,6690558  \\
    256 KB &  0,0000392 &  0,0001586 &  0,0000276 &  0,0017874 &  0,6750260  \\
    128 KB &  0,0000407 &  0,0001611 &  0,0000298 &  0,0038466 &  0,7131423  \\
     64 KB &  0,0000360 &  0,0001616 &  0,0000307 &  0,0079054 &  0,7358308  \\
     32 KB &  0,0000385 &  0,0001592 &  0,0000279 &  0,0132787 &  1,2640401  \\
     16 KB &  0,0000374 &  0,0001518 &  0,0000272 &  0,0236237 &  1,6018821  \\
      8 KB &  0,0000366 &  0,0001594 &  0,0000277 &  0,5675783 &  1,7440000  \\
      4 KB &  0,0000533 &  0,0001966 &  0,0000289 &  1,7435491 &  4,5486790  \\
      2 KB &  0,0000357 &  0,0001519 &  0,0000262 &  4,2165658 & 10,4853639  \\
      1 KB &  0,0000373 &  0,0001572 &  0,0000291 &  9,3677694 & 16,4246678  \\
\hline
\end{tabular}
\caption{Messages detailed}\label{table_messages}
\end{center}
\end{table}

\begin{table}
\begin{center}
\begin{tabular}{|r|r|r|}
\hline
Buffer size &
\multicolumn{1}{c|}{seek} &
\multicolumn{1}{c|}{read} \\
\hline
      5 MB &  0,0000203 &  1,0630000 \\
    2.5 MB &  0,0000425 &  1,0871195 \\
      1 MB &  0,0001155 &  1,0900746 \\
    512 KB &  0,0002340 &  1,0664272 \\
    256 KB &  0,0004866 &  1,0108422 \\
    128 KB &  0,0012508 &  0,9938316 \\
     64 KB &  0,0021597 &  1,1580000 \\
     32 KB &  0,0026852 &  0,9589246 \\
     16 KB &  0,0051107 &  1,0600056 \\
      8 KB &  0,0184110 &  1,5362434 \\
      4 KB &  0,1282281 &  1,5135774 \\
      2 KB &  0,0600973 &  1,5620572 \\
      1 KB &  0,2784313 &  1,4095517 \\
\hline
\end{tabular}
\caption{Disk accesses detailed}\label{table_disk_accesses}
\end{center}
\end{table}

\appendix
\chapter{ViPIOS functions}\label{ViPIOS functions}
\input{vipios_interface}
\input{strided_data_access}


\def\[#1]{\thinspace{\small\bf#1}\thinspace} 
\chapter{Glossary on Parallel I/O}\label{Glossary on Parallel I/O}

\input{intro}
\input{glossar}

\input{gloss-c}
\input{gloss-d-h}

\input{gloss-i-o}
\input{gloss-p}
\input{gloss-r-s}

\input{gloss-t-z}
\input{appendix}

\bibliography{vipios,pario,bib-heinz,bib-brezany,add,wanek,kurt}
\bibliographystyle{plain}

\end{document}

%% file: iodict.tex
\usepackage{calc, fancyheadings}

\newlength{\mytmp}
\newcounter{KWHEAD}
\newcounter{SUBKW}
\newcounter{SUBSUBKW}
\DeclareFontFamily{T1}{phv}{}
\DeclareFontShape{T1}%
{phv}%
{m}%
{n}%
{<-> phvr}%
{}

\newcommand{\keyfont}{\fontfamily{phv}\selectfont}        
\newcommand{\keyreffont}{\bfseries\itshape}   

\newcommand{\beforeskip}{\vspace{1cm}}
\newcommand{\wordref}[1]{{\keyreffont #1}}    
\newcommand{\wrf}[1]{\wordref{#1}}             

\newenvironment{keyword}[1]
  {{\noindent\keyfont #1}%
    \nopagebreak%
    \nopagebreak%
    \nopagebreak%
     \setcounter{KWHEAD}{0}%
     \setcounter{SUBKW}{0}%
     \setcounter{SUBSUBKW}{1}}

  {\begin{list}{}{}}
  {\end{list}}%

\newcommand{\sect}[1]{
  \beforeskip
  \noindent{\keyfont\Huge#1}\\
  \nopagebreak
  \markright{#1}
  \lhead[\fancyplain{}{\bfseries\thepage}]{%
    \fancyplain{}{\bfseries\thepage}}
  \rhead[\fancyplain{}{\bfseries\thepage}]{%
    \fancyplain{}{\keyfont\normalsize\rightmark}}}

%% file: file_model.tex
\begin{Definitions}


\definition{Record}
We define a record as a piece of information in binary representation.
The only property of a record which is relevant to us at the moment is its
size in bytes. This is due to the fact that ViPIOS is only concerned with the
efficient storage and retrieval of data but not with the interpretation of its
meaning.

Let $R$ be the set of all possible records. We then define
\begin{center}
$\mathit{size}: R \to \mathbb{N}$
\end{center}
where $\mathit{size}(\mathit{rec}), \, \mathit{rec} \in R$ denotes the length of the record
in bytes.
In the following the record with size zero is referenced by the symbol $'nil'$.
Further $R_i \subset R$ is the set of all records with size $i$:
\begin{center}
$R_i=\{\mathit{rec} \vert \mathit{rec} \in R \wedge \mathit{size}(\mathit{rec})=i\},
\qquad i \in \mathbb{N}$
\end{center}

\definition{File}
A non empty file $f$ consists of a sequence of records which are all of the same size
and different from $'nil'$.
\begin{center}
$f=<\mathit{rec}_1, \dots, \mathit{rec}_n>, \qquad n \in \mathbb{N}^{+}, \enspace
\mathit{size}(\mathit{rec}_i)=\mathit{size}(\mathit{rec}_j)>0,
\enspace 1 \leq i,j \leq n$
\end{center}
With $F$ denoting the set of all possible files we define the functions
\begin{center}
\mathbox{$\mathit{frec}: F \times \mathbb{N}^{+} \to R$}{$\mathit{flen}: F \to \mathbb{N}$ \\
$\mathit{frec}: F \times \mathbb{N}^{+} \to R$}
\end{center}
which yield the length of (i. e. the number of records in) a file $f$ and specific records
of the file respectively. For a file $f=<\mathit{rec}_1, \dots, \mathit{rec}_n>$ and
$i \in \mathbb{N}^{+}$
\begin{center}
\mathbox{$\mathit{frec}(f,i)= \quad \mathit{rec}_i \quad \text{if} \quad i \le \mathit{flen}(f)$}{
$\mathit{flen}(f)=n$ \qquad and \\
\begin{displaymath}
\mathit{frec}(f,i)=
\begin{cases}
\mathit{rec}_i & \text{if} \quad i \le \mathit{flen}(f), \\
'nil' & \text{otherwise}.
\end{cases}
\end{displaymath}
}
\end{center}
An empty file is denoted by an empty sequence:
\begin{center}
$f=<> \qquad \Leftrightarrow \qquad \mathit{flen}(f)=0$
\end{center}


\definition{Data buffer}
The set $D$ of data buffers is defined by
\begin{center}
$D=\underset{n \in \mathbb{N}}{\bigcup}R^n$
\end{center}
The functions
\begin{center}
\mathbox{$\mathit{drec}: D \times \mathbb{N}^{+} \to R$}{$\mathit{dlen}: D \to \mathbb{N}$ \\
$\mathit{dsize}: D \to \mathbb{N}$ \\
$\mathit{drec}: D \times \mathbb{N}^{+} \to R$
}
\end{center}
give the number of tuple-elements (i. e. records) contained in a data buffer
$d\in D$, \,
its size in bytes and specific records respectively. Thus if $d=(\mathit{rec}_1, \dots,
\mathit{rec}_n), \, n \in \mathbb{N}$ and $i~\in~\mathbb{N}^{+}$ then
\begin{center}
\mathbox{$\mathit{drec}(d,i)= \quad \mathit{rec}_i \quad \text{if} \quad i \le \mathit{dlen}(d)$}{$\mathit{dlen}(d)=n$ \\
$\mathit{dsize}(d)=\sum_{j=1}^{\mathit{dlen}(d)}\mathit{size}(\mathit{rec}_j)$\\
\begin{displaymath}
\mathit{drec}(d,i)=
\begin{cases}
\mathit{rec}_i & \text{if} \quad i \le \mathit{dlen}(d), \\
'nil' & \text{otherwise}.
\end{cases}
\end{displaymath}
}
\end{center}

Of special interest are data buffers which only contain equally sized records. These are
denoted by
\begin{center}
$D_i= \underset{n \in \mathbb{N}}{\bigcup} R_i^n, \qquad i \in \mathbb{N}$
\end{center}
and their size may be computed easily by
$\mathit{dsize}(d)=i*\mathit{dlen}(d)$.

\definition{Access modes}
The set of access modes $M$ is given by:
\begin{center}
$M=\{'read', 'write'\}$
\end{center}

\definition{Mapping functions}
Let $t \in \underset{n \in \mathbb{N}} \bigcup \mathbb{N}^n$ and $t=(t_1, \dots, t_n),
\quad n \in \mathbb{N}$. A mapping function
\begin{center}
$\psi_t: F \to F$
\end{center}
is defined by
\begin{center}

$\psi_t(f) = <\mathit{frec}(f,t_1), \mathit{frec}(f,t_2), \dots,
\mathit{frec}(f,t_n)>$
\end{center}
So for example $\psi_{(2,4,2,6)}(f)$ is the file which contains the records 2,
4, 2 and 6 of the file f in that order. \footnote{Note that $t$ does not have to
be a permutation. So one record of $f$ may be replicated on different positions
in $\psi(f)$.}

The set of mapping functions is denoted by $\Psi$, \, and $\psi^{*}(f)=\psi_{(1,
\dots,\, \mathit{flen}(f))}$ is the mapping function for which f is a fixpoint.

With $()$ denoting the empty tuple the function $\psi_{()}(f)=<>$ for
every file $f \in F$.

\definition{File handle}
The set of file handles $H$ is defined by:

\begin{center}
$H=F \times (\mathcal{P}(M)-\emptyset) \times \mathbb{N} \times \Psi$
\end{center}
where $\mathcal{P}(M)$ is the power set of $M$.

\newpage

To access the information stored in a file handle $\mathit{fh} \in H$ the following
functions are defined:
\begin{center}
\mathbox{$\mathit{mode}: H \to \mathcal{P}(M)-\emptyset$}{
$\mathit{file}: H \to F$ \\
$\mathit{mode}: H \to \mathcal{P}(M)-\emptyset$ \\
$\mathit{pos}: H \to \mathbb{N}$ \\
$\mathit{map}: H \to \Psi$
}
\end{center}

if $\mathit{fh}=(f,m,n,\psi)$, with $f \in F, \, m \in \mathcal{P}(M)-\emptyset$ ,
\, $n \in \mathbb{N}$ \, and $\psi \in \Psi$ then
\begin{center}
\mathbox{$\mathit{mode}(\mathit{fh})=m$}{
$\mathit{file}(\mathit{fh})=f$ \\
$\mathit{mode}(\mathit{fh})=m$ \\
$\mathit{pos}(\mathit{fh})=n$ \\
$\mathit{map}(\mathit{fh})=\psi$
}
\end{center}

\definition{File operations}
Only one file operation (\textbf{OPEN}) directly operates on a file. All other operations
relate to the file using the file handle returned by the \textbf{OPEN} operation. In the
following $f \in F$ is the file operated on, $\mathit{fh} \in H$ is a file handle,
$m \in \mathcal{P}(M)-\empty$ is a set of access modes, $d \in D$ is a data buffer and
$\psi \in \Psi$ is a mapping function. The symbol $'error'$ denotes that the operation
cannot be performed because of the state of parameters. In this case the operation does
not change any of its parameters and just reports the error state. The operations can be
formally described as follows:

\begin{itemize}
\item \textbf{OPEN}$(f,m,\mathit{fh},\psi)$ \qquad is equivalent to: \footnote{Note that
we do not address security aspects in this model. Therefore users are not restricted in
accessing files and the \textbf{OPEN} operation will always succeed.} \\
\begin{center}
$\mathit{fh} \gets (f,m,0,\psi)$
\end{center}


\item \textbf{CLOSE}$(\mathit{fh})$ \qquad is equivalent to:\\
\begin{center}
$\mathit{fh} \gets (<>,\{'read'\},0, \psi_{()})$
\end{center}
Thus every file operation on $\mathit{fh}$ succeeding \textbf{CLOSE} will fail).

\item \textbf{SEEK}$(\mathit{fh},n) \quad n \in \mathbb{N}$ \qquad is equivalent to:\\
($f=\mathit{file}(\mathit{fh}); \, m=\mathit{mode}(\mathit{fh}); \,
\psi=\mathit{map}(\mathit{fh})$)
\begin{displaymath}
\begin{cases}
fh \gets (f,m,n,\psi) & \text{if} \quad \mathit{flen}(\psi(f)) \ge n, \\ \\
'error' & \text{otherwise}.
\end{cases}
\end{displaymath}

\newpage

\item \textbf{READ}$(\mathit{fh},n,d) \quad n \in \mathbb{N}^{+}$ \qquad is equivalent to:
\footnote{Note that the initial content of the data buffer is of no interest. Just its
total size is relevant. This is different to the write operation where the records in the
data buffer have to be compatible with the file written to. The condition assures that we
do not read beyond the end of the file and that the data buffer is big enough to
accommodate for the data read.} \\
($f=\mathit{file}(\mathit{fh}); \, m=\mathit{mode}(\mathit{fh}); \,
p=\mathit{pos}(\mathit{fh}); \psi=\mathit{map}(\mathit{fh})$)
\begin{displaymath}
\begin{cases}
d \gets(\mathit{frec}(\psi(f),p+1), \mathit{frec}(\psi(f),p+2), \dots \\
\qquad \qquad \dots, \mathit{frec}
(\psi(f),p+i)) & \text{if} \quad 'read' \in \mathit{mode}(\mathit{fh})
\wedge\\
fh \gets (f,m,p+i,\psi) &  \quad \wedge i=\text{min}(n, \lfloor
\frac{\mathit{dsize}(d)}{\mathit{size}(\mathit{frec}(f,1))} \rfloor,
\mathit{flen}(\psi(f))-p) > 0, \\ \\
'error' & \text{otherwise}.
\end{cases}
\end{displaymath}

\item \textbf{WRITE}$(\mathit{fh},n,d) \quad n \in \mathbb{N}^{+}$ \qquad is equivalent
to: \footnote{Since files are defined to contain only records which all have the same
size, the data buffer has to hold appropriate records. The \textbf{WRITE} operation as
defined here may be used to append new records to a file as well as to overwrite records
in a file. The length of the file will only increase by the number of records actually
appended.} \\
($f=\mathit{file}(fh); \, m=\mathit{mode}(\mathit{fh}); \, p=\mathit{pos}(\mathit{fh});
\, \psi=\mathit{map}(\mathit{fh})$)
\begin{displaymath}
\begin{cases}
f \gets <\mathit{frec}(f,1), \dots, \mathit{frec}(f,p), \mathit{drec}(d,1),
\dots & \text{if} \quad  'write' \in \mathit{mode}(\mathit{fh}) \wedge n \le
\mathit{dlen}(d) \wedge\\
\qquad \qquad \dots, \mathit{drec}(d,n), \mathit{frec}(f,p+n+1), \dots & \qquad
\wedge (f=<> \wedge (\exists i \in \mathbb{N}^{+} \vert d \in D_i) \vee\\
\qquad \qquad \dots, \mathit{frec}(f, \mathit{flen}(f)) &  \qquad \vee f \ne <>
\wedge d \in D_{\mathit{size}(\mathit{frec}(f,1))}), \\ \\
'error' & \text{otherwise}.
\end{cases}
\end{displaymath}

\item \textbf{INSERT}$(\mathit{fh},n,d) \quad n \in \mathbb{N}^{+}$ \quad is equivalent
to: \footnote{If successful the \textbf{INSERT} operation will always increase the file
size by $n$. \textbf{INSERT}$(fh,n,d)$ is equivalent to \textbf{WRITE}$(fh,n,d)$ iff
$\mathit{pos}(fh)=\mathit{flen}(\mathit{file}(fh))$.} \\
($f=\mathit{file}(\mathit{fh}); \, m=\mathit{mode}(\mathit{fh}); \,
p=\mathit{pos}(\mathit{fh})$)
\begin{displaymath}
\begin{cases}
f \gets <\mathit{frec}(f,1), \dots, \mathit{frec}(f,p), \mathit{drec}(d,1),
\dots & \text{if} \quad 'write' \in \mathit{mode}(\mathit{fh}) \wedge n \le
\mathit{dlen}(d) \wedge \\
\qquad \qquad \dots, \mathit{drec}(d,n), \mathit{frec}(f,p+1), \dots  & \quad \wedge
(f=<> \wedge (\exists i \in \mathbb{N}^{+} \vert d \in D_i) \vee \\
\qquad \qquad \dots, \mathit{frec}(f, \mathit{flen}(f)) &  \quad \vee f \ne <>
\wedge d \in D_{\mathit{size}(\mathit{frec}(f,1))}), \\ \\
'error' & \text{otherwise}.
\end{cases}
\end{displaymath}
\end{itemize}

\end{Definitions}

\subsection{Implementation of a mapping function description}
ViPIOS has to keep all the appropriate mapping functions as part of the
file information of the file. So a data structure is needed to internally
represent such mapping functions. This structure should fulfill the following
two requirements:
\begin{itemize}
\item{} Regular patterns should be represented by a small data structure.
\item{} The data structure should allow for irregular patterns too.
\end{itemize}
Of course these requirements are contradictionary and so a comprimise
actually was implemented in ViPIOS. The structure which will now be described
allows the description of regular access patterns with little overhead yet
also is suitable for irregular access patterns. Note however that the overhead
for completely irregular access patterns may become considerably large. But
this is not a problem since ViPIOS currently mainly targets regular access
patterns and optimizations for irregular ones can be made in the future.

Figure \ref{c_struct} gives a C declaration for the data structure representing
a mapping function.

\begin{figure}
\begin{center}
\begin{verbatim}
struct Access_Desc {
        int no_blocks;
        int skip;
        struct basic_block *basics;
};

struct basic_block {
        int offset;
        int repeat;
        int count;
        int stride;
        struct Access_Desc *subtype;
};
\end{verbatim}
\caption{An according C declaration}\label{c_struct}
\end{center}
\end{figure}


An Access\_Desc basically describes a number (no\_blocks) of independent
basic\_blocks where every basic\_block is the description of a regular access
pattern. The skip entry gives the number of bytes by which the file pointer is
incremented after all the blocks have been read/written.

The pattern described by the basic\_block is as follows: If subtype is NULL
then we have to read/write single bytes otherwise every read/write operation
transfers a complete data structure described by the Access\_Desc block to which
subtype actually points. The offset field increments the file pointer by the
specified number of bytes before the regular pattern starts. Then repeatedly
count subtypes (bytes or structures) are read/written and the file pointer is
incremented by stride bytes after each read/write operation. The number of
repetitions performed is given in the repeat field of the basic\_block
structure.

%% file: vimpios.tex
ViMPIOS (Vienna Message Passing/Parallel Input Output System) is a
por-table, client-server based MPI-IO implementation on the
ViPIOS. At the moment it comprises all ViPIOS routines currently
available. Thus, the whole functionality of ViPIOS plus the
functionality of MPI-IO can be exploited. However, the advantage
of ViMPIOS over the MPI-IO proposed as the MPI-2 standard is the
possibility the assign each server process a certain number of
client processes. Thus, the I/O can actually be done in parallel.
What is more, each server process can access a file scattered over
several disks rather than residing on a single one. The
application programmer need not care for the physical location of
the file and can therefore treat a scattered file as one logical
contiguous file.\\

At the moment four different MPI-IO implementations are available, namely:

\begin{itemize}
\item PMPIO - Portable MPI I/O library developed by NASA Ames Research Center
\item ROMIO - A high-performance, portable MPI-IO implementation developed by Argonne National Laboratory
\item MPI-IO/PIOFS - Developed by IBM Watson Research Center
\item HPSS Implementation - Developed by Lawrence Livermore National Laboratory as part of its Parallel I/O Project
\end{itemize}

Similar to ROMIO all routines defined in the MPI-2 I/O chapter are supported except shared file pointer functions, split collective data access functions, support for file interoperability, error handling, and I/O error classes. Since shared file pointer functions are not supported, the MPI\_MODE\_SEQUENTIAL mode to \textit{MPI\_File\_open} is also not available. \\

In addition to the MPI-IO part the derived datatypes \textit{MPI\_Type\_subarray} and \textit{MPI\_Type\_darray} have been implemented. They are useful for accessing arrays stored in files \cite{ROMIO2}.\\

What is more, changes to the parameters \textit{MPI\_Status} and \textit{MPI\_Request} have been made. ViMPIOS uses the self defined parameter \textit{MPIO\_Status} and \textit{MPI\_File\_Request}. Unlike ROMIO, the parameter textsf{status} can be used for retrieving particular file access information. Thus,  \textit{MPI\_Status} has been modified. The same is true for \textit{MPI\_Request}. Finally, the routines \textit{MPI\_Wait} and \textit{MPI\_Test} are modified to \textit{MPI\_File\_wait} and \textit{MPI\_File\_test}.\\

At the moment, file hints are not supported by ViMPIOS yet. Using file hints would yield following advantages: The application programmer could inform the server about the I/O workload and the possible I/O patterns. Thus, complicated I/O patterns where data is read according to a particular view and written according to a different can be analyzed and simplified by the server. What is more, the server could select the I/O nodes which suit best for the I/O workload. In particular, if one I/O node is idle whereas the other deals with great amount of data transfer, these unbalances could be solved.

%% file: mpi.tex
\subsection{Introduction to MPI}

In this section we will discuss the most important features of the Message Passing Interface (MPI) \cite{MPI}. Rather than describing every function in detail we will focus our attention to the basics of MPI which are vital to understand MPI-IO, i.e. the input/output part of the message passing interface. Thus, the overall purpose of this chapter is to define special MPI terms and explain them by means of the corresponding routines coupled with some examples.\\

The Message Passing Interface is the de facto standard for parallel programs based on the message passing approach. It was developed by the Message Passing Interface Forum (MPIF) with participation from over 40 organizations. MPI is not a parallel programming language on its own but a library that can be linked to a C or FORTRAN program. Applications can either run on distributed-multiprocessors, networks of workstations, or combinations of these. Furthermore, the interface is suitable for MIMD programs as well as for those written in the more restricted SPMD style. A comprehensive overview of parallel I/O terms can be found in \cite{pario}.

\subsection{The Basics of MPI}

The main goal of the standard is to allow the communication of processes whereas the easiest way of interprocess communication is the point-to-point communication where two processes exchange information by the basic operations SEND and RECEIVE. According to \cite{MPI-IO} the six basic functions of MPI are as follows:
\begin{itemize}
\item MPI\_INIT:                initiate an MPI computation
\item MPI\_FINALIZE:    terminate a computation
\item MPI\_COMM\_SIZE:  determine number of processes
\item MPI\_COMM\_RANK:  determine current process' identifier
\item MPI\_SEND:                send a message
\item MPI\_RECV:                receive a message
\end{itemize}

Every program in MPI must be initialized by \textit{MPI\_Init} and terminated by \textit{MPI\_Finalize}. Thus, no other MPI function can be called before \textit{MPI\_Init} or after \textit{MPI\_Finalize}. The syntax of the two functions is:\\
\\
{\sffamily\bfseries int MPI\_Init (int *argc, char *** argv)} \\
{\sffamily\bfseries int MPI\_Finalize (void)} \\

By means of \textit{MPI\_Comm\_rank} the process' identifier can be evaluated. Process numbers start with 0 and have consecutive integer values. In order to find out how many processes are currently running, \textit{MPI\_Comm\_size} is called.

{\sffamily
\begin{tabbing}
{\bfseries int MPI\_Comm\_size (MPI\_Comm comm, int *size)} \\
INOUT   \=parameter     \= description                                  \kill
IN      \> comm         \>  communicator \\
OUT     \> size         \> number of processes in the group of comm
\end{tabbing}
}

{\sffamily
\begin{tabbing}
{\bfseries int MPI\_Comm\_rank (MPI\_comm, int *rank)} \\
INOUT   \=parameter     \= description                                  \kill
IN      \>comm          \>communicator \\
OUT     \>rank          \>rank of the calling process in group of comm
\end{tabbing}
}

In both instructions the argument \textit{comm} specifies a so-called communicator which is used to define a particular group of any number of processes. Suppose 8 processes are currently active and we wish to separate them into two groups, namely \textit{group1} should contain processes with the identifiers from 0 to 3, whereas \textit{group2} consists of the rest of the processes. Thus, we could use a communicator \textit{group1} that refers to the first group and a communicator \textit{group2} that refers to the second group is MPI\_COMM\_WORLD. This MPI predefined communicator includes all processes currently active. \\

On establishing a communication the next step is to explain how information is exchanged by \textit{MPI\_Send} and \textit{MPI\_Recv}. Both instructions execute a blocking message passing rather than a non-blocking one. In a blocking approach a send command waits as long as a matching receive command is called by another process before the actual data transfer takes place.

{\sffamily
\begin{tabbing}
{\bfseries int MPI\_Send (void* buf, int count, MPI\_Datatype, int destination,}\\
{\bfseries int tag, MPI\_Comm comm)} \\
INOUT   \=parameter     \= description                                  \kill
IN      \>buf           \>initial address of send buffer\\
IN      \>count         \>number of elements in the send buffer\\
IN      \>datatype      \>datatype of each send buffer element\\
IN      \>dest          \>rank of destination\\
IN      \>tag           \>message tag\\
IN      \>comm          \>communicator
\end{tabbing}
}

{\sffamily
\begin{tabbing}
{\bfseries int MPI\_Recv (void* buf, int count, MPI\_Datatype datatype, int source,} \\
{\bfseries int tag, MPI\_Comm, MPI\_Status *status)} \\
INOUT   \=parameter     \= description                                  \kill
OUT     \>buf           \>initial address of receive buffer\\
IN      \>count         \>number of elements in the receive buffer\\
IN      \>datatype      \>datatype of each receive buffer element\\
IN      \>source        \>rank of source\\
IN      \>tag           \>message tag\\
IN      \>comm          \>communicator\\
OUT     \>status        \>status object
\end{tabbing}
}

The first three arguments of both instructions are referred to as the message data, the rest is called message envelope. In particular \textit{buf} specifies the initial address of the buffer to be sent. \textit{Count} holds the number of elements in the send buffer, which are defined by the datatype \textit{MPI\_Datatype} (e.g. MPI\_INT, MPI\_FLOAT). The parameter \textit{destination} states the identifier of the process that should receive the message. Similarly, the parameter \textit{source} refers to the process that has sent the message. By means of \textit{tag} a particular number can be related to a message in order to distinguish it from other ones. \textit{Comm} refers to the communicator. Finally, the status information allows checking the source and the tag of an incoming message.\\

The following small program demonstrates how process 0 sends an array of 100 integer values to process 1:

\begin{verbatim}
#include "mpi.h"
int main (int argc,char **argv)
{
  int message[100], rank;
  MPI_Status status;

  /* MPI is initialized */
  MPI_Init(&argc,&argv);

  /* the rank of the current process is determined */
  MPI_Comm_rank(MPI_COMM_WORLD, &rank);

  if (rank==0)
    /*  process 0 sends message with tag 99 to process 1 */
    MPI_Send(message, 100, MPI_INT, 1, 99, MPI_COMM_WORLD);
  else
    /* process 1 receives message with tag 99 from process 0 */
    MPI_Recv(message, 100, MPI_INT, 0, 99, MPI_COMM_WORLD, &status);

  MPI_Finalize();
}
\end{verbatim}

\subsection{Linking and Running Programs}

Compiling and linking of a program is done by \\

\textbf{mpicc -o file\_name file\_name.c} \\

On compiling and linking the program, an executable file is produced which can be executed by the following command: \\

\textbf{mpirun -np  number\_of\_processes  file\_name} \\

-np denotes the number of process and has always be situated before the file name. For example:\\

\textbf{mpirun -np 16 application.c} \\

\subsection{Blocking vs. Non-blocking communication}

The send/receive commands we were discussing so far are so called blocking commands. In other words, the sending process waits until the receiving process has got the message. In contrast, nonblocking communication means that the sending processes do not wait until the operation is complete. Moreover, special functions, namely \textit{MPI\_Wait} and \textit{MPI\_Test}, are used to complete a nonblocking communication. Thus, better performance can be yielded for specific applications, since communication and computation can overlap.\\

Let us again take a look at the most important nonblocking commands before we resume with same examples:

{\sffamily
\begin{tabbing}
{\bfseries int MPI\_Isend (void* buf, int count, MPI\_Datatype datatype, int source, } \\
{\bfseries int tag, MPI\_Comm comm, MPI\_Request *request)} \\
INOUT   \=parameter     \= description                                  \kill
IN      \>buf           \>initial address of send buffer\\
IN      \>count         \>number of elements in the send buffer\\
IN      \>datatype      \>datatype of each send buffer element\\
IN      \>dest          \>rank of destination\\
IN      \>tag           \>message tag\\
IN      \>comm          \>communicator\\
OUT     \>request       \>communication request
\end{tabbing}
}

{\sffamily
\begin{tabbing}
{\bfseries int MPI\_Irecv (void* buf, int count, MPI\_Datatype datatype, int source, } \\
{\bfseries int tag, MPI\_Comm comm, MPI\_Request *request)} \\
INOUT   \=parameter     \= description                                  \kill
OUT     \>buf           \>initial address of receive buffer\\
IN      \>count         \>number of elements in the receive buffer\\
IN      \>datatype      \>datatype of each receive buffer element\\
IN      \>source        \>rank of source\\
IN      \>tag           \>message tag\\
IN      \>comm          \>communicator\\
OUT     \>request       \>communication request
\end{tabbing}
}

The syntax of those instructions is the same as for their blocking counterparts except of the last parameter \textit{request}.

{\sffamily
\begin{tabbing}
{\bfseries int MPI\_Wait (MPI\_Request *request, MPI\_Status *status)}\\
INOUT   \=parameter     \= description                                  \kill
INOUT   \>request       \>request\\
OUT     \>status        \>status object
\end{tabbing}
}

A call to that function returns when the operation identified by \textit{request} is complete. In other words, it waits until a nonblocking send or receive with a matching parameter \textit{request} is executed. \textit{Status} gives information about the completed operation.

{\sffamily
\begin{tabbing}
{\bfseries int MPI\_Test (MPI\_Request *request, int *flag, MPI\_Status *status)} \\
INOUT   \=parameter     \= description                                  \kill
INOUT   \>request       \>communication request\\
OUT     \>flag          \>true if operation completed\\
OUT     \>status        \>status object
\end{tabbing}
}

The function only returns \textit{flag = true} if the operation identified by \textit{request} is complete. Furthermore, the status object contains the information on the completed operation.\\

This example is a modification of the previous one. Rather than using blocking commands, we demonstrate the usage of non-blocking commands.

\begin{verbatim}
#include "mpi.h"
int main (int argc,char **argv)
{
  int message[100], rank;
  MPI_Status status;
  MPI_Request request;

  MPI_Init(&argc,&argv);

  MPI_Comm_rank(MPI_COMM_WORLD, &rank);

  if (rank==0)
  {
    MPI_ISend(message, 100, MPI_INT, 1, 99, MPI_COMM_WORLD, &request);
    /* any computation can be done here */
    MPI_Wait(request,status);
  }
  else
  {
     MPI_IRecv(message, 100, MPI_INT, 0, 99, MPI_COMM_WORLD,
               &request);
     /* any computation can be done here */
     MPI_Wait(request,status);
  }
  MPI_Finalize();
}
\end{verbatim}

\subsection{Derived Datatypes}

Derived datatypes allow the user to define special datatypes that can be any combination of simple datatypes such as MPI\_INT or MPI\_FLOAT. In addition, sections of arrays that need not be contiguous in memory can be referred to as a special datatype. In that section we will give an overview of some important derived datatypes.\\

In order to define a derived datatype, three sets of functions are required, namely a constructor function to construct a derived datatype, e.g.\\ \textit{MPI\_Type\_vector}, a commit function \textit{MPI\_Type\_commit} for applying the new datatype and finally the function \textit{MPI\_Type\_free} that should be applied after the usage of the datatype. \\

The simplest derived datatype is \textit{MPI\_Type\_contiguous}:

{\sffamily
\begin{tabbing}
{\bfseries int MPI\_Type\_contiguous (int count, MPI\_Datatype oldtype, } \\
{\bfseries MPI\_Datatype *newtype)} \\
INOUT   \=parameter     \= description                                  \kill
IN      \>count         \>replication count\\
IN      \>oldtype       \>old datatype\\
OUT     \>newtype       \>new datatype
\end{tabbing}
}

This datatype allows defining a contiguous datatype which consists of \textit{count} elements of a special datatype \textit{oldtype}. The new datatype can be used for further purposes. \\

By means of an example we want to describe the usage of this datatype. Assume that process 0 wants to send an array of 25 integer elements to process 1. In order to use a derived datatype following steps are necessary:

\begin{verbatim}
MPI_Datatype array1;

/* datatype which specifies 25 integer values to contiguous
   locations is created */
MPI_Type_contiguous(25,MPI_INT,&array1);
MPI_Type_commit(&array1);

/* process 0 sends data to processes 1 */
MPI_Send(message, 1, array1, 1, 99, MPI_COMM_WORLD);
MPI_Type_free(&array1);
\end{verbatim}

In our small example we only printed the code for process 0. Taking a look at the third parameter of the send command we notice that a derived datatype is specifed rather than a simple datatype like MPI\_INT. The syntax for the commands which handle the commition and freeing of the derived datatype is given here:

{\sffamily
\begin{tabbing}
{\bfseries int MPI\_Type\_commit (MPI\_Datatype *datatype)} \\
INOUT   \=parameter     \= description                                  \kill
INOUT   \>datatype      \>datatype that is commited
\end{tabbing}
}

{\sffamily
\begin{tabbing}
{\bfseries int MPI\_Type\_free (MPI\_Datatype *datatype)} \\
INOUT   \=parameter     \= description                                  \kill
INOUT   \>datatype      \>datatype that is freed
\end{tabbing}
}

Now assume that process 0 wants to send sections of the array rather than all 25 integer elements. In particular, the first and the last 10 elements shall be sent which means that 2 blocks of data shall be sent whereas 5 elements shall be skipped. A more general derived datatype is \textit{MPI\_Type\_vector}:

{\sffamily
\begin{tabbing}
{\bfseries int MPI\_Type\_vector (int count, int blocklength, int stride, } \\
{\bfseries MPI\_Datatype oldtype, MPI\_Datatype *newtype)} \\
INOUT   \=parameter.....\= description                                  \kill
IN      \>count         \>number of blocks\\
IN      \>blocklength   \>number of elements in each block\\
IN      \>stride        \>number of elements between start of each block\\
IN      \>oldtype       \>old datatype\\
OUT     \>newtype       \>new datatype
\end{tabbing}
}

\textit{Count} holds the number of blocks, \textit{blocklength} specifies the number of elements in each block, and \textit{stride} defines the number of elements between the start of each block whereas stride is a multiple of oldtype. The datatype for our example is as follows:\\
\\
\texttt{MPI\_Type\_vector(2,5,10);}\\

\begin{figure}
\begin{center}
\includegraphics[scale=0.9]{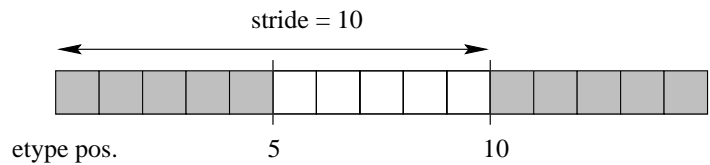}
\caption{Datatype constructor MPI\_TYPE\_VECTOR}
\label{vector1}
\end{center}
\end{figure}

The shape is depicted in Figure \ref{vector1}.\\

A further generalization of the previous datatype is \textit{MPI\_Type\_hvector}. The difference to \textit{MPI\_Type\_vector} is the parameter \textit{stride} which is not given in elements but in bytes. \\

{\sffamily
\begin{tabbing}
{\bfseries int MPI\_Type\_hvector (int count, int blocklength, MPI\_Aint stride, } \\
{\bfseries MPI\_Datatype oldtype, MPI\_Datatype *newtype)} \\
INOUT   \=parameter.....\= description                                  \kill
IN      \>count         \>number of blocks\\
IN      \>blocklength   \>number of elements in each block\\
IN      \>stride        \>number of elements between start of each block\\
IN      \>oldtype       \>old datatype\\
OUT     \>newtype       \>new datatype
\end{tabbing}
}

The next datatype \textit{MPI\_Type\_indexed} allows specifying blocks of different lengths starting at different displacements. Before we present the function we want to give an example of such a particular case.\\

\begin{figure}
\begin{center}
\includegraphics[scale=0.9]{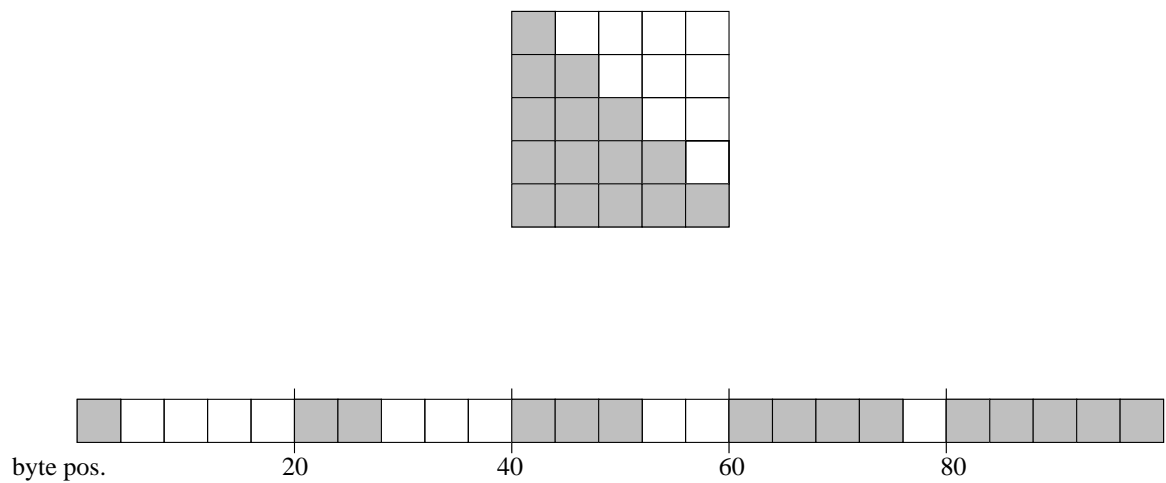}
\caption{Datatype constructor MPI\_TYPE\_INDEXED}
\label{indexed1}
\end{center}
\end{figure}

Assume a 5x5 array of integer values. Further assume that we want to send the lower triangle of that matrix. Thus, the first block to be sent consists of 1 element with the displacement 0. The second block consists of 2 elements with the displacement 6. The third block comprises 3 elements with the displacement 11 etc. The matrix and the corresponding linear file are depicted in Figure \ref{indexed1}.

{\sffamily
\begin{tabbing}
{\bfseries int MPI\_Type\_indexed (int count, int *array\_of\_blocklengths, MPI\_Aint} \\
{\bfseries  *array\_of\_displacements, MPI\_Datatype oldtype, MPI\_Datatype *newtype) } \\
INOUT   \=parameter...................\= description                                  \kill
IN      \>count                         \>number of blocks\\
IN      \>array\_of\_blocklengths       \>number of elements per block\\
IN      \>array\_of\_displacements      \>displacement for each block, in multiples of oldtype\\
IN      \>oldtype                       \>old datatype\\
OUT     \>newtype                       \>new datatype
\end{tabbing}
}

\begin{verbatim}
MPI_Datatype    indexed;
int             a_blocklen[10],
		a_disp[10];

for (i=0; i<5; i++)
{
  a_blocklen[i]=i+1;
  a_disp[i]=i*5;
}

MPI_Type_indexed(4, a_blocklen, a_disp, MPI_INT, &indexed);
MPI_Type_commit(&indexed);
\end{verbatim}

Similar to \textit{MPI\_Type\_hvector} there is also a corresponding \textit{MPI\_Type\_hindexed}. This datatype is identical to the previous except that the displacements are given in bytes and not in multiples of \textit{oldtype}. \\
\\
{\sffamily
{\bfseries int MPI\_Type\_hindexed (int count, int *array\_of\_blocklenghts, } \\
{\bfseries int *array\_of\_displacements, MPI\_Datatype oldtype, MPI\_Datatype *newtype)} \\
}

The most general derived datatype is \textit{MPI\_Type\_struct}. The difference to \textit{MPI\_Type\_indexed} is that each datablock can consist of a different datatype. Thus, it is possible to send an integer, together with a character or double value in one message.

{\sffamily
\begin{tabbing}
{\bfseries int MPI\_Type\_struct (int count, int *array\_of\_blocklengths, MPI\_Aint} \\
{\bfseries  *array\_of\_displacements, MPI\_Datatype *array\_of\_oldtypes,}\\ {\bfseries MPI\_Datatype *newtype) } \\
INOUT   \=parameter...................\= description                                  \kill
IN      \>count                                 \>number of blocks\\
IN      \>array\_of\_blocklengths       \>number of elements per block\\
IN      \>array\_of\_displacements      \>displacement for each block in bytes\\
IN      \>array\_of\_oldtypes           \>old datatype\\
OUT     \>newtype                               \>new datatype
\end{tabbing}
}

Let us again take a look at an example:

\begin{verbatim}
MPI_Datatype    s_types[3]={MPI_INT,MPI_DOUBLE,MPI_CHAR},
		dd_struct;
int             s_blocklen[3]={3,2,16};
MPI_Aint        s_disp[3];

/* set displacements to next free space */
s_disp[0]=0;
s_disp[1]=20;
s_disp[2]=40;

MPI_Type_struct(3, s_blocklen, s_disp, s_types, &dd_struct);
MPI_Type_commit(&dd_struct);
\end{verbatim}

\begin{figure}
\begin{center}
\includegraphics[scale=0.9]{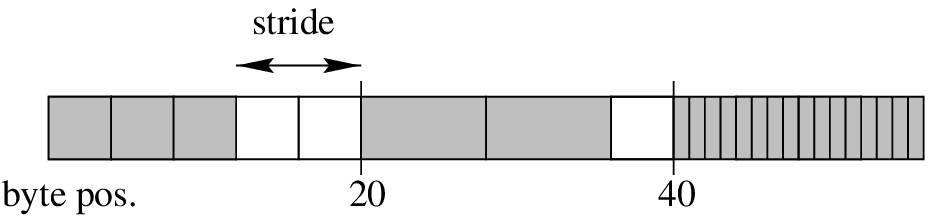}
\caption{Datatype constructor MPI\_TYPE\_STRUCT}
\label{struct1}
\end{center}
\end{figure}

The shape of that derived datatype can be seen in Figure \ref{struct1}. Here the derived datatype consists of 3 non-contiguous blocks of different datatypes, namely MPI\_INT, MPI\_DOUBLE and MPI\_CHAR.

\subsection{Collective Communication}

Unlike the point-to-point communication we stated in the previous chapters, collective communication means that a group of processes can take part in the communication process. In other words, not only one process can send a message to another processes but to several processes with just one command. In contrast, it is also possible for one process to receive messages from all other processes specified in one group. Again, only one receive command is necessary rather than several ones for each message.\\

Let us take a look at one function for collective communication in order to clarify the advantage of a collective communication over a point-to-point communication:

{\sffamily
\begin{tabbing}
{\bfseries int MPI\_Bcast (void* buffer, int count, MPI\_Datatype datatype, } \\
{\bfseries int root, MPI\_Comm comm)} \\
INOUT   \=parameter     \= description                                  \kill
INOUT   \>buffer        \>starting address of buffer\\
IN      \>count         \>number of entries in buffer\\
IN      \>datatype      \>datatype of buffer\\
IN      \>root          \>rank of process to broadcast the message\\
IN      \>comm          \>communicator
\end{tabbing}
}

This routine broadcasts a message from the process identified by the rank \textit{root} to all processes in the group. What is more, a message is also sent to itself.\\

Let us first analyze how this would be done with the conventional point-to-point communication and compare this to the collective version. Assume that our group consists of 8 processes. The message to be sent should be an array of 1000 double values.

\begin{verbatim}
MPI_Comm comm;
double  double_array[1000];
...
if (rank==0)
{
  for (i=0; i<8; i++)
    MPI_Send(double_array, 1000, MPI_DOUBLE, i, 99, comm);
}
else
  MPI_Recv(double_array, 1000, MPI_DOUBLE, 0, 99, comm, &status);
\end{verbatim}

Using collective routines the send-receive process is much shorter since it is not necessary to write separate lines of code for different processes:\\
\\
\texttt{MPI\_Bcast(double\_array, 1000, MPI\_DOUBLE, 0, comm);}\\

\subsection{Communicators}

Communicators are a way of managing the communication among processes in a better way such that several processes can be grouped together and regarded as one homogenous entity. In short, communicators are divided into to kinds:

\begin{itemize}
\item Intra-Communicators
\item Inter-Communicators
\end{itemize}

The first term refers to the communication within a single group of processes whereas the second term refers to the point-to-point communication between two group of processes. For a comprehensive survey of all MPI commands we refer the reader to \cite{MPI}.

%% file: mpio_intr.tex
\subsection{Introduction to MPI-IO}

MPI-IO is a high level interface developed by the MPI-Committee \cite{MPI}. The goal was to create a widely used standard for describing parallel I/O operations within an MPI message passing application. The initial idea was that I/O could be modeled as message passing, namely writing to a file can be regarded as sending a message and reading from a file is equivalent to receiving a message. Thus, MPI-IO is an extension of the MPI 1.1 standard , which did not support parallel file I/O so far for the following reasons \cite{MPI-IO}:

\begin{itemize}
\item not all parallel machines support the same parallel or concurrent file system interface
\item the traditional Unix file system interface is ill suited to parallel computing since multiple processes do not share files at once
\end{itemize}

On giving a short introduction to the I/O problem let us now analyze the most important features of MPI-IO:

\begin{itemize}
\item using derived MPI datatypes yields strided access to the memory and the file
\item non-blocking functions improve the I/O performance by overlapping I/O with computation
\item collective operations may optimize global data access
\item using two types of file pointers, namely individual and shared file pointers, such that exact offsets in the file need not be specified when data is read or written
\item file hints allow specifying the layout of the a file, e.g. number of disks or I/O nodes which hold the information of a striped file
\item filetype constructors can be used to specify array distribution
\item error handling is supported
\end{itemize}

\subsection{First Steps With MPI-IO}

Referring to the introductory chapter we stated that I/O can be modeled as a message passing system. In other words, writing data to a file should be similar to sending a message. In contrast, reading from a file should be modeled as receiving a message. Although MPI-IO supports a large number of different functions for parallel I/O, many programs and applications only use six of them, which are summarized in the following paragraph \cite{MPI-IO}:

\begin{itemize}
\item MPI\_INIT: MPI as well as MPI-IO are initialized
\item MPI\_FILE\_OPEN:	a file is opened
\item MPI\_FILE\_READ:	data is read from a particular location in a file
\item MPI\_FILE\_WRITE:	data is written to a particular location in a file
\item MPI\_FILE\_CLOSE:	a file is closed
\item MPI\_FINALIZE:	MPI as well as MPI-IO are terminated
\end{itemize}

Strictly speaking only four so-called MPI-IO functions are used since MPI\_INIT and MPI\_FINALIZE are already supported by MPI-1.\\

We will now explain the use of these functions by means of two simple examples \cite{MPI-IO}. For introductory purpose we will not go into detail with describing the exact syntax of each function but only mention the most important parameters to focus our attention. We will dedicate a special chapter to the syntax of the functions at a later stage of this thesis.\\

In the first program each process creates its own individual file called \textit{file} followed by an extension which reflects the identifier of the current process. A file can be opened individually by using the parameter MPI\_COMM\_SELF. Furthermore, data is written to the file, which is read back later on.

\begin{verbatim}
#include "mpi.h"
#include "mpio.h"

int main(argc,argv)

int argc;
char *argv[];
{
  int myid;

  MPI_Status status;
  MPI_File fh;

  char filename[12];
 	
  buf= (int *)malloc(50*sizeof(int));	
  for (i=0, i<50, i++)
    buf[i]=i;	

  MPI_Init(&argc,&argv);
  MPI_Comm_rank(MPI_COMM_WORLD,&myid);

  /* open file with filename "file.processid" */	
  sprintf(filename,"%s.%d","ufs:file",myid);

  /* each process opens a separate file */	
  MPI_File_open (MPI_COMM_SELF, filename, MPI_MODE_CREATE |
MPI_MODE_RDWR, MPI_INFO_NULL, &fh );

  /* read data from file */
  MPI_File_read(fh, buf, 50, MPI_INT, &status);

  /* perform computation */	

  /* write data to file */
  MPI_File_write (fh, buf, 50, MPI_INT, &status);

  MPI_File_close(&fh);
	
  free(buf);
  MPI_Finalize();	

}
\end{verbatim}

In the second example each process accesses one common global file rather than its local one. That feature is yielded by the parameter MPI\_COMM\_WORLD. We will not print the whole program code but only the part which differs from the previous example:

\begin{verbatim}
/* each process opens one common file */	
MPI_File_open (MPI_COMM_WORLD, filename, MPI_MODE_CREATE |
MPI_MODE_RDWR, MPI_INFO_NULL, &fh );
\end{verbatim}

\subsection{Some Definitions of MPI-IO}

\textit{File}: An MPI file is an ordered collection of typed data items which can be accessed in a random or sequential way. Furthermore, a communicator (MPI\_COMM\_SELF or MPI\_COMM\_WORLD we discussed in our introductory chapter) on the one hand specifies which group of processes can get access to the I/O operations, on the other hand, it determines whether the access to the file is independent or collective. Since independent I/O requests are executed individually by any of the processes within a communicator group, no coordination among the processes is required. In contrast, the latter case requires each process in a group associated with the communicator to participate in the collective access. \\

\textit{Displacement}: A file displacement defines the beginning of a view (file access pattern) expressed as an absolute byte position relative to the beginning of the file. Furthermore, it can be used to skip head information of a file or to define further access patterns which start at different positions of the file.\\

\textit{Etype}: An etype (elementary datatype) can be regarded as the unit of data access and positioning. In other words, it specifies the data layout in the file. An etype can be any MPI predefined or derived datatype.\\

\textit{Filetype}: A filetype can either be a single etype or a derived MPI datatype of several etypes and describes a template for accessing a file partitioned among processes.\\

\textit{View}: A view is described by a displacement, an etype, and a filetype and defines the current set of data visible and accessible from an open file as an ordered set of etypes. Thus, a view specifies the access pattern to a file.\\

\textit{Offset}: An offset is a position in the file relative to the current view, expressed as a count of etypes.\\

\textit{File pointer}: A file pointer is an implicit offset. On the one hand, MPI provides individual file pointers which are local to each process, on the other hand, file pointers for a group of processes  - so-called shared file pointers - are supported.\\

\textit{File handle}: A file handle can be regarded as a file reference which is created by \textit{MPI\_File\_open} and freed by \textit{MPI\_File\_close}.\\

\begin{figure}
\begin{center}
\includegraphics[scale=0.9]{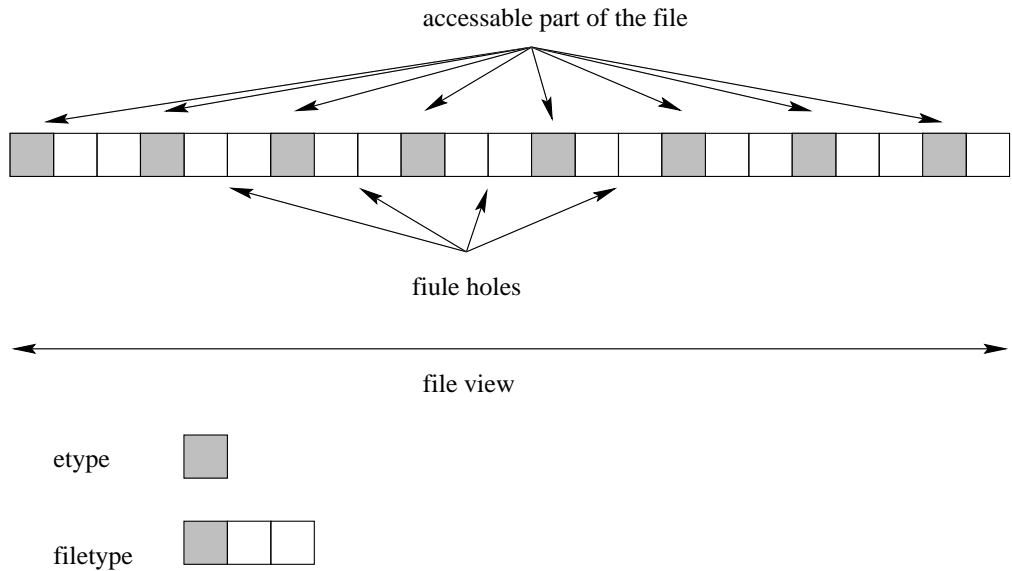}
\caption{File view}
\label{view0}
\end{center}
\end{figure}

By means of an example we will explain the idea of strided access and how this can be achieved by using different file views. Assume a file which holds an array of 24 integer values. Further assume that a process only wants to read or write every third value of the file. We therefore tile the file with a filetype which is a derived MPI datatype. Thus, the values at position (offset) 0,2,5,... can be accessed whereas all the other positions are so-called holes which cannot be accessed by the current process. The view is depicted in Figure \ref{view0}.\\

\begin{figure}
\begin{center}
\includegraphics[scale=0.9]{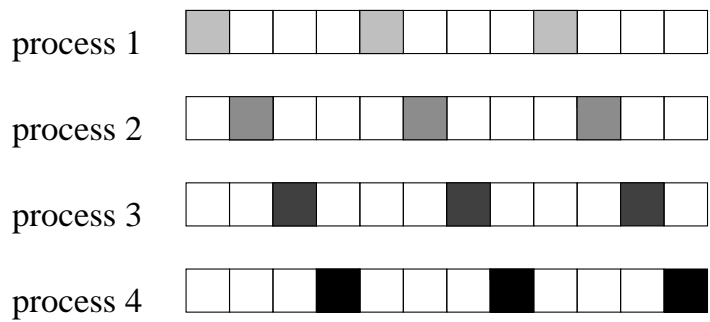}
\caption{File view of 4 processes}
\label{view1}
\end{center}
\end{figure}

In our next example assume that 3 processes access the file in a complementary way. In other words, the file is partitioned among four parallel processes where each process reads or writes at different locations in the file. In particular process 0 accesses the positions 0,3,6,..., process 1 accesses the positions 1,4,7,..., and finally process 2 accesses the positions 2,5,8,... The access patterns of each process are depicted in Figure \ref{view1}.\\

\begin{figure}
\begin{center}
\includegraphics[scale=0.9]{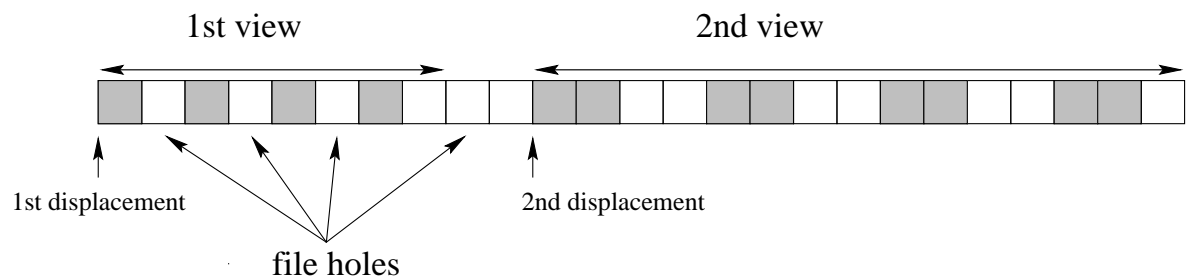}
\caption{Two different file views}
\label{view2}
\end{center}
\end{figure}

Another possibility is to access a file in two different patterns. Thus, we define two tilings (views). In particular, the first part of the file shall be accessed with a stride of 2 elements whereas the second part of the file shall be accessed with a stride of 3 elements. Figure \ref{view2} can make this example clearer:\\

If the file needs to be accessed in the second view, the displacement of the second pattern will skip over the entire first segment.

%% file: mpio_interface.tex
\subsection{MPI-IO routines}

On giving some definitions of the MPI-IO standard we will now describe the routines of that standard and give some examples to get a first idea of how MPI-IO works.

\subsubsection{File Manipulation}

{\sffamily
\begin{tabbing}
{\bfseries int MPI\_File\_open (MPI\_Comm comm, char *filename, int amode,} \\
{\bfseries MPI\_Info info, MPI\_File *fh) } \\
INOUT   \=parameter     \= description                                  \kill 
IN      \>comm          \>communicator\\ 	
IN      \>filename      \>name of file to be opened\\
IN      \>amode         \>file access mode\\
IN      \>info          \>info object \\
OUT     \>fh            \>new file handle (handle)
\end{tabbing}
}

\textbf{Description:}\\
Opens the file \textit{filename} on all processes in the \textit{comm} communicator group.\\

\textbf{Features:}
\begin{itemize}
\item collective routine (all processes must refer to the same file name and use the same access mode)
\item a file can be opened independently by using the MPI\_COMM\_SELF communicator
\item \textit{comm} must be an intra-communicator rather than an inter-communicator
\item \textit{filename} is prefixed to indicate the underlying file system (e.g. "ufs:filename", where ufs stands for Unix file system)
\item \textit{amode} can have the following values which could also be combined together with the bit vector OR: 

\begin{itemize}
\item MPI\_MODE\_RDONLY: file is opened read only
\item MPI\_MODE\_RDWR: reading and writing
\item MPI\_MODE\_WRONLY: write only
\item MPI\_MODE\_CREATE: create the file if it does not exist
\item MPI\_MODE\_EXCL: error is returned if a file is created that already exists
\item MPI\_MODE\_DELETE\_ON\_CLOSE: delete file on close
\item MPI\_MODE\_UNIQUE\_OPEN: file will not be concurrently opened elsewhere. 
Optimization is yielded by eliminating file locking overhead
\item MPI\_MODE\_SEQUENTIAL: file will only be accessed sequentially
\item MPI\_MODE\_APPEND: set initial position of all file pointers to end of file
\end{itemize}
 
\item \textit{info} provides information such as file access patterns and file system specifics. If no info is needed, MPI\_INFO\_NULL can be used.
\end{itemize}

\textbf{Annotation:}\\
Every time the file is accessed the file handle \textit{fh} is used to refer to the file.
Moreover, each open-function must at least contain either MPI\_MODE\_RDONLY, MPI\_MODE\_RDWR or MPI\_MODE\_WRONLY\\

\textbf{Possible errors:}
\begin{itemize}
\item MPI\_MODE\_CREATE or MPI\_MODE\_EXCL are specified together with MPI\_MODE\_RDONLY
\item MPI\_MODE\_SEQUENTIAL is specified together with MPI\_MODE\_RDWR
\item a file which is opened with MPI\_MODE\_UNIQUE is opened concurrently
\item a file which is opened with MPI\_MODE\_SEQUENTIAL is accessed in a non-sequential way 
\end{itemize}

{\sffamily
\begin{tabbing}
{\bfseries int MPI\_File\_close (MPI\_File *fh)} \\
INOUT..\=parameter      \= description                                  \kill
INOUT   \>fh            \>file handle
\end{tabbing}
}

\textbf{Description:}\\
The file defined by \textit{fh} is closed after synchronizing the file, i.e. previous writes to \textit{fh} are transferred to the storage device. Furthermore, the content of the file handle \textit{fh} is destroyed.\\

\textbf{Features:} 
\begin{itemize}
\item collective routine
\item the file is deleted if it was opened with MPI\_FILE\_DELETE\_ON\_CLOSE
\end{itemize}

{\sffamily
\begin{tabbing}
{\bfseries int MPI\_File\_delete (char *filename, MPI\_Info info)} \\
INOUT..\=parameter      \= description                                  \kill
IN      \>filename      \> name of the file \\
IN      \>info          \> file info
\end{tabbing}
}

\textbf{Description: }\\
Deletes the file \textit{filename}.\\

\textbf{Possible errors:}
\begin{itemize}
\item MPI\_ERR\_NO\_SUCH\_FILE: a file is attempted to delete which does not exist
\item MPI\_ERR\_FILE\_IN\_USE, MPI\_ERR\_ACCESS: the file is still opened by any other process
\end{itemize}

{\sffamily
\begin{tabbing}
{\bfseries int MPI\_File\_set\_size (MPI\_File fh, MPI\_Offset size) } \\
INOUT..\=parameter      \= description                                  \kill
INOUT   \>fh            \>file handle\\
IN      \>size          \>size (in bytes) to truncate or expand file
\end{tabbing}
}

\textbf{Description:}\\
Resizes the file defined by \textit{fh}.\\

\textbf{Features:}
\begin{itemize}
\item collective routine (identical value for \textit{size})
\item if \textit{size} is smaller than the current file size, the file is truncated otherwise \textit{size} becomes the new file size 
\item does not effect individual or shared file pointers
\end{itemize}

\textbf{Possible errors:}
\begin{itemize}
\item routine is called if file was opened with MPI\_MODE\_SEQUENTIAL
\item non-blocking requests and split collective operations on \textit{fh} are not completed
\end{itemize}

{\sffamily
\begin{tabbing}
{\bfseries int MPI\_File\_preallocate (MPI\_File fh, MPI\_Offset size) } \\
INOUT..\=parameter      \= description                                  \kill
INOUT   \>fh    \>filehandle\\
IN      \>size  \>size to preallocate file size
\end{tabbing}
}

\textbf{Description:}\\
Storage space is allocated for the first \textit{size} bytes of a file.\\

\textbf{Features:}
\begin{itemize}
\item collective routine (identical value for \textit{size})
\item if \textit{size} is larger than the current file size, the file is increased to \textit{size} otherwise the file size is unchanged
\end{itemize}

\textbf{Possible errors:}
\begin{itemize}
\item routine is called if file was opened with MPI\_MODE\_SEQUENTIAL
\end{itemize}

{\sffamily
\begin{tabbing}
{\bfseries int MPI\_File\_get\_size (MPI\_File fh, MPI\_Offset size) } \\
INOUT..\=parameter      \= description                                  \kill
IN      \>fh    \>file handle\\
OUT     \>size  \>size of the file in bytes
\end{tabbing}
}

\textbf{Description:}\\ 
Returns the current size in bytes of the file defined by \textit{fh}.

{\sffamily
\begin{tabbing}
{\bfseries int MPI\_File\_get\_group (MPI\_File fh, MPI\_Group *group)} \\
INOUT..\=parameter      \= description                                  \kill
IN      \>fh    \>file handle\\
OUT     \>group \>group which opened the file
\end{tabbing}
}

\textbf{Description:}\\ 
Returns a duplicate of the group of the communicator which opened the file defined by \textit{fh}.

{\sffamily
\begin{tabbing}
{\bfseries int MPI\_File\_get\_amode (MPI\_File fh, int *amode)} \\
INOUT..\=parameter      \= description                                  \kill
IN      \>fh    \>file handle\\
OUT     \>amode \>file access mode used to open the file
\end{tabbing}
}

\textbf{Description:}\\ 
Returns the access mode of the file defined by \textit{fh}.

\subsubsection{File Info}

{\sffamily
\begin{tabbing}
{\bfseries int MPI\_File\_set\_info (MPI\_File fh, MPI\_Info info)} \\
INOUT..\=parameter      \= description                                 \kill
INOUT \>fh		\>file handle\\
IN	\>info    	\>info object
\end{tabbing}
}

\textbf{Description:}\\ 
Sets new values for the hints of a file.

{\sffamily
\begin{tabbing}
{\bfseries int MPI\_File\_get\_info (MPI\_File fh, MPI\_Info *info\_used) }\\
INOUT..\=parameter      \= description                                 \kill
IN 	\>fh			\>file handle\\
OUT	\>info\_used	\>new info object (handle)
\end{tabbing}
}

\textbf{Description:}\\ 
Returns the hints of a file.\\

Some examples of file hints:
\begin{itemize}
\item Access\_style: Specifies how a file is accessed whereas a combination of following strings is possible: read\_once, write\_once, read\_mostly, write\_mostly, sequential, reverse\_sequential and random
\item io\_node\_list: Specifies the list of I/O devices to store the file
\end{itemize}

\subsubsection{File Views}

Strided access to a file can be gained by using derived datatypes in combination with views. Recall that a view defines the current set of data visible and accessible from an open file as an ordered set of etypes.\\ 

Let us first present the interfaces before we give a simple example.

{\sffamily
\begin{tabbing}
{\bfseries int MPI\_File\_set\_view (MPI\_File fh, MPI\_Offset disp, MPI\_Datatype etype, } \\ 
{\bfseries MPI\_Datatype filetype, char *datarep, MPI\_Info info)}\\
INOUT..\=parameter   \= description                                  \kill
INOUT	\>fh		\>file handle\\
IN	\>disp		\>displacement\\
IN	\>etype		\>elementary datatype\\
IN	\>filetype	\>filetype\\
IN	\>datarep 	\>data representation\\
IN	\>info		\>info object
\end{tabbing}
}

\textbf{Description:}\\ 
Changes the process' view of the data in the file.\\

\textbf{Features:}
\begin{itemize}
\item collective routine: values for \textit{etype} and \textit{datarep} must be identical on all processes in the group
\item \textit{disp} defines the beginning of the view in bytes, it must have the special value MPI\_DISPLACEMENT\_CURRENT if MPI\_MODE\_SEQUENTIAL was specified
\item \textit{etype} defines the data access and positioning in the file. Thus, every seek which is performed on that file is done in units specified by \textit{etype}. In addition every offset is expressed as a count of \textit{etypes}. More information is given when the command \textit{MPI\_File\_seek} is explained.
\item \textit{filetype} describes the distribution of the data 
\item \textit{datarep} specifies the representation of the data in the file according to three categories:
	\begin{itemize}
\item{native}
\item{internal}
\item{external32}
\end{itemize}
\end{itemize}

\textbf{Annotation:}\\
In the \textit{native} data representation the data is stored in a file exactly as it is in memory \cite{MPI-2}. Since no type conversion is required for that mode, data precision and I/O performance are not lost. With the \textit{internal} data representation type conversion is performed if necessary. Type conversion is always performed with the \textit{external32}.\\

Non-blocking requests and split collective operations (see in a later chapter) on. 
\textit{fh} must be completed

{\sffamily
\begin{tabbing}
{\bfseries int MPI\_File\_get\_view (MPI\_File fh, MPI\_Offset disp, MPI\_Datatype etype, } \\
{\bfseries MPI\_Datatype filetype, char *datarep)} \\
INOUT..\=parameter      \= description                                  \kill
IN      \>fh		\>file handle\\
OUT     \>disp		\>displacement\\
OUT     \>etype	\>elementary datatype\\
OUT     \>filetype	\>filetype\\
OUT     \>datrep  	\>datarepresentation
\end{tabbing}
}

\textbf{Description:}\\ 
Returns the process' view of the data in the file.\\

\textbf{Features:}\\
\begin{itemize}
\item \textbf{Datarep} must be large enough to handle the data representation string. However, the upper limit is defined by MPI\_MAX\_DATAREP\_STRING
\end{itemize}

Example: Assume a file which is an integer array consisting of 100 values starting from 0. Further assume that this file should be accessed in 10 blocks of size 2 with a stride of 10. The corresponding part of the program looks like follows:

\begin{verbatim}
MPI_File_open (MPI_COMM_WORLD, "ufs:file1", MPI_MODE_CREATE | 
    MPI_MODE_RDWR, MPI_INFO_NULL, &fh );

MPI_Type_vector (10,2,10,MPI_INT,&newtype1);
MPI_Type_commit(&newtype1);
MPI_File_set_view (fh, 0, MPI_INT, newtype1, "native", 
                                  MPI_INFO_NULL);
MPI_File_read(fh, buf, 10, MPI_INT, &status);
\end{verbatim}

In that example the displacement is 0. This means that the view begins at position 0 of the file. The displacement can also be used to skip some header information or to specify a second view of a file which begins at a position other than 0 and can be accessed in a different pattern. See Figure \ref{view2}.\\


Example: Again we assume our file which is an integer array consisting of 100 values. We now want to define a view for the second part of the file such that we can access every second value. The new derived datatype and the corresponding view look like follows:

\begin{verbatim}
MPI_Vector (25,1,2, &newtype2);
MPI_Commit (&newtype2);
MPI_File_set_view (fh, 200, MPI_INT, newtype2, "native", 
                                  MPI_INFO_NULL);
\end{verbatim}

Since the displacement is given as an absolute offset in bytes from the beginning of the file, the value 200 sets the file view to the 50th element of the file. This is only true if the size of an integer type is 4 bytes.

\subsubsection{Data Access}

On analyzing different aspects of file manipulation we will now discuss the three fundamental aspects of data access which are as follows:
\begin{itemize}
\item positioning (explicit offset vs. implicit file pointer)
\item synchronism (blocking vs. non-blocking and split collective)
\item coordination (collective vs. non-collective)
\end{itemize}

\textit{Positioning:} All MPI routines that use explicit offsets for accessing data contain \textit{\_at} in their name (e.g. \textit{MPI\_File\_read\_at}). Rather than manipulating individual file pointers or shared file pointers data is accessed at the position defined by an argument. The different positioning methods can be mixed in the same program without effecting each other. This means that changing a file by an explicit offset will not affect the individual file pointer since no file pointer is used or updated. \\

\textit{Synchronism:} Besides blocking I/O routines MPI also supports non-blocking I/O routines which are named \textit{MPI\_File\_ixxx}, where \textit{i} refers to immediate. Similar to MPI, a separate request complete call (\textit{MPIO\_Wait, MPIO\_Test}) is needed to complete the I/O request. However, this does not mean that the data is written to a "permanent" storage. The MPI-IO function that deals with that case is \textit{MPI\_File\_sync} which we will discuss later on.\\

\textit{Coordination:} The collective counterpart to the non-collective data access routine \textit{MPI\_File\_xxx} is \textit{MPI\_File\_xxx\_all}. This means that a collective call is only completed when all other processes that participate in the collective call have completed their tasks. A restricted form of "non-blocking" operations for a collective data access is called \textit{split collective}. Rather than using one routine like \textit{MPI\_File\_xxx\_all}, a pair of \textit{MPI\_File\_xxx\_begin} and \textit{MPI\_File\_xxx\_end}is used. Thus, a single collective operation is separated by two routines where the begin routine can be compared to a non-blocking data access like \textit{MPI\_File\_write} and the end routine acts like a matching routine which completes the operation, e.g. \textit{MPI\_Wait} or \textit{MPI\_Test}. In addition, the counterparts to the \textit{MPI\_File\_xxx\_shared} routines are \textit{MPI\_File\_xxx\_ordered}.

\paragraph{Conventions of Data Access}

Every file which is used by MPI-IO commands is referred to by its file handle \textit{fh}. The parameters \textit{buf}, \textit{count}, and \textit{datatype} specify how the data is stored in memory. Note that similar to a receive, \textit{datatype} must not contain any overlapping regions. Finally, \textit{status} returns the amount of data accessed by a particular process. In detail, the number of datatype entries and predefined elements can be retrieved by calling \textit{MPI\_get\_count} and \textit{MPI\_get\_elements}.

\subsubsection{Data Access With Explicit Offset (\_AT)}

The routines listed in that section can only be used if MPI\_MODE\_SEQUENTIAL was not specified.

\paragraph{Blocking, non-collective:}

{\sffamily
\begin{tabbing}
{\bfseries int MPI\_File\_read\_at (MPI\_File fh, MPI\_Offset offset, void *buf, }\\
{\bfseries int count, MPI\_Datatype datatype, MPI\_status status)}\\
INOUT..\=parameter      \= description                                  \kill
IN      \>fh		\>file handle\\
IN      \>offset		\>file offset\\ 
OUT     \>buf		\>initial address of buffer \\
IN      \>count		\>number of elements in the buffer \\
IN      \>datatype	\>datatype of each buffer element\\
OUT     \>status		\>status object 
\end{tabbing}
}

\textbf{Description:}\\ 
Reads a file beginning at position \textit{offset}. \textit{Status} contains the amount of data accessed by that routine.\\

\textbf{Annotation:}\\
The offset for all explicit offset routines is given in units of \textit{etype} rather than in bytes. Furthermore, \textit{offset} expresses the position relative to the beginning of a file.

{\sffamily
\begin{tabbing}
{\bfseries int MPI\_File\_write\_at (MPI\_File fh, MPI\_Offset offset, void *buf,} \\
{\bfseries int count, MPI\_Datatype datatype, MPI\_status status) } \\
INOUT..\=parameter      \= description                                  \kill
INOUT   \>fh              \>file handle\\
IN      \>offset          \>file offset\\ 
IN      \>buf             \>initial address of buffer \\
IN      \>count           \>number of elements in the buffer \\
IN      \>datatype        \>datatype of each buffer element\\
OUT     \>status          \>status object 
\end{tabbing}
}

\textbf{Description:}\\ 
Writes a file beginning at position \textit{offset}. \textit{Status} contains the amount of data accessed by that routine.

\paragraph{Collective Versions:}

Since the semantic of the collective routines are the same as for their non-collective counterparts, only the synopsis of the routines are printed here:\\

{\sffamily
{\bfseries int MPI\_File\_read\_at\_all (MPI\_File fh, MPI\_Offset offset, void *buf, int count, MPI\_Datatype datatype, MPI\_status status)} \\
}

{\sffamily
{\bfseries int MPI\_File\_write\_at\_all (MPI\_File fh, MPI\_Offset offset, void *buf, int count, MPI\_Datatype datatype, MPI\_status status) } \\
}

\paragraph{Non-blocking Versions:}

Note that non-blocking I/O calls only initiate I/O operations but do not wait for them to complete. Thus, separate request complete calls like MPIO\_WAIT or MPIO\_TEST are needed. The example in the next section will demonstrate the use of non-blocking functions.

{\sffamily
\begin{tabbing}
{\bfseries int MPI\_File\_iread\_at (MPI\_File fh, MPI\_Offset offset, void *buf,} \\
{\bfseries  int count, MPI\_Datatype datatype, MPI\_Request request)}\\
INOUT..\=parameter      \= description                                  \kill
IN      \>fh		\>file handle\\
IN      \>offset	\>file offset\\ 
OUT     \>buf		\>initial address of buffer \\
IN      \>count		\>number of elements in the buffer \\
IN      \>datatype	\>datatype of each buffer element\\
OUT     \>request	\>request object 
\end{tabbing}
}

{\sffamily
\begin{tabbing}
{\bfseries int MPI\_File\_iwrite\_at (MPI\_File fh, MPI\_Offset offset, void *buf,} \\
{\bfseries int count MPI\_Datatype datatype, MPI\_Request request) } \\
INOUT..\=parameter      \= description                                  \kill
INOUT   \>fh              \>file handle\\
IN      \>offset          \>file offset\\ 
IN      \>buf             \>initial address of buffer \\
IN      \>count           \>number of elements in the buffer \\
IN      \>datatype        \>datatype of each buffer element\\
OUT     \>request         \>request object 
\end{tabbing}
}

\subsubsection{Data Access With Individual File Pointers}

One individual file pointer per process per file handle defines the offset of the data to be accessed. Since the semantics are the same as in the previous section we will not go into detail with describing the routines and refer to table.\\

{\sffamily
{\bfseries int MPI\_File\_read (MPI\_File fh, void *buf, int count, MPI\_Datatype datatype, MPI\_Status status)} \\
}

{\sffamily
{\bfseries int MPI\_File\_write (MPI\_File fh, void *buf, int count, MPI\_Datatype datatype, MPI\_Status status)} \\
}

{\sffamily
\begin{tabbing}
{\bfseries int MPI\_File\_seek (MPI\_File fh, MPI\_Offset offset, int whence) } \\
INOUT..\=parameter      \= description                                  \kill
INOUT	\>fh		\>file handle\\
IN	\>offset	\>file offset\\
IN	\>whence	\>update mode (state)
\end{tabbing}
}

\textbf{Description:}\\ 
Updates the individual file pointer according to whence whereas following features are possible:

\begin{itemize}
\item MPI\_SEEK\_SET: file pointer is set to offset
\item MPI\_SEEK\_CUR: file pointer is set to current pointer position plus offset
\item MPI\_SEEK\_END: file pointer is set to the end of the file plus offset
\end{itemize}

\textbf{Annotation:}\\
It is important to mention that the offset is not given in bytes but in units of \textit{etype} defined by the view of the file.

{\sffamily
\begin{tabbing}
{\bfseries int MPI\_File\_get\_position (MPI\_File fh, MPI\_Offset *offset)}\\
INOUT..\=parameter      \= description                                  \kill
IN	\>fh		\>file handle\\
OUT	\>offset	\>offset of file pointer
\end{tabbing}
}

\textbf{Description:}\\ 
Returns the current position of the individual file pointer in \textit{etype} units relative to the current view.

{\sffamily
\begin{tabbing}
{\bfseries int MPI\_File\_get\_byte\_offset (MPI\_File fh, MPI\_Offset offset,}\\
{\bfseries MPI\_Offset *disp) } \\
INOUT..\=parameter      \= description                                  \kill
IN	\>fh		\>file handle\\
IN	\>offset	\>offset of filepointer\\
OUT	\>disp	\>absolute byte position of offset
\end{tabbing}
}

\textbf{Description:}\\ 
Converts a view-relative offset which is given in \textit{etype} units into an absolute byte position relative to the current view.\\

Besides blocking, non-collective routines which we presented here, non-blocking or collective routines are defined in the MPI-IO standard as well.\\

Example: In this example we want to demonstrate that in general each I/O operation leaves the file pointer pointing to the next data item after the last one that is accessed by the operation. In other words, file pointers are updated automatically. Again we use our file which is an array consisting of 100 values. We assume that the first ten values should be stored in the array \textit{buf1[]} and the following 10 values in the array \textit{buf2[]}. Note that after the first non-blocking read the file pointer is adjusted automatically to position 10 in the array. Furthermore, the file pointer is not updated by a routine with an explicit offset.

\begin{verbatim}
MPI_Status status1,status2;
MPIO_Request request1, request2;

MPI_File_open (MPI_COMM_WORLD, "ufs:file1", MPI_MODE_CREATE | 
MPI_MODE_RDWR, MPI_INFO_NULL, &fh );
MPI_File_set_view(fh, 0, MPI_INT, MPI_INT, "native", 
MPI_INFO_NULL);

/* File pointer points to position 0 of the file */
MPI_File_iread(fh, buf1, 10, MPI_INT, &request1);

/* File pointer points to position 10 since 10 values are read by the 
   previous routine */
MPI_File_iread(fh, buf2, 10, MPI_INT, &request2);

/* File pointer points to position 20 of the file since another 10 
   values are read */
MPI_File_read_at(fh, 51, buf3, 10, MPI_INT, &request1);

/* File pointer still points to position 20 since previous read is a 
   routine with explicit offset that does not update the file pointer */
MPI_File_read_(fh, buf4, 10, MPI_INT, &request1);

MPIO_Wait(&request1,&status1);
MPIO_Wait(&request2,&status2);
MPI_File_close(&fh);
\end{verbatim}

On executing the program the arrays contain following values:

\begin{verbatim}
buf1: 0 1 2 3 4 5 6 7 8 9
buf2: 10 11 12 13 14 15 16 17 18 19 
buf3: 50 51 52 53 54 555 56 57 58 59
buf4: 20 21 22 23 24 25 26 27 28 29
\end{verbatim}

\subsubsection{Split Collective Routines}

We have already defined a split collective routine as a restricted form of "non-blocking collective" I/O operations. Before we present the interface routines let us first state the most important semantic rules according to the MPI-IO standard:

\begin{itemize}
\item Each file handle on any MPI process must not have more than one active split collective operation at any time
\item Begin calls are collective and must be followed by a matching collective end call
\item Regular collective operations like MPI\_FILE\_WRITE\_ALL on one process do not match split collective operations on another process.
\item Split collective routines must not be used concurrently with collective routines
\end{itemize}

\begin{verbatim}
MPI_File_write_all_begin(fh,...);
...
/* This collective routine is used concurrently with a split 
   collective routine */
MPI_File_read_all(fh,...);
...
MPI_File_write_all_end(fh,...);
\end{verbatim}

Again the semantics of these operations are the same as for the corresponding collective operations. We therefore only present one example of split collective routines:

{\sffamily
\begin{tabbing}
{\bfseries int MPI\_File\_read\_at\_all\_begin (MPI\_File fh, MPI\_Offset offset, void *buf,}\\
{\bfseries int count, MPI\_Datatype datatype) } \\
INOUT..\=parameter      \= description                                  \kill
IN      \>fh              \>file handle\\
IN      \>offset          \>file offset \\
OUT     \>buf             \>initial address of buffer \\
IN      \>count           \>number of elements in the buffer \\
IN      \>datatype        \>datatype of each buffer element
\end{tabbing}
}

{\sffamily
\begin{tabbing}
{\bfseries int MPI\_File\_read\_at\_all\_end (MPI\_File fh, void *buf, MPI\_status status)} \\
INOUT..\=parameter      \= description                                  \kill
IN      \>fh              \>file handle\\
OUT     \>buf             \>initial address of buffer \\
OUT     \>status          \>status object\\
\end{tabbing}
}

\subsubsection{Data Access With Shared File Pointers}

The offset in the data access routine is described by exactly one shared file pointer per collective MPI\_FILE\_OPEN. Thus, the file pointer is shared among the processes in a particular communicator group. Again, the same semantics are used as in the previous sections with some exceptions:

\begin{itemize}
\item All processes must use the same view.
\item Using collective shared file pointers (\_ORDERED) guarantees a serialized order of multiple calls. In other words, the access to the file is determined by the rank of the processes in the group. In contrast, non-collective shared file pointers (\_SHARED) yield a serialization ordering which is non-deterministic.
\end{itemize}

The functions listed below shall be regarded as a proposal since they are not supported by ViMPIOS yet.

\paragraph{Non-collective Routines:}

{\sffamily
{\bfseries int MPI\_File\_read\_shared (MPI\_File fh, void *buf, int count, MPI\_Datatype datatype, MPI\_Status status) } \\
}

{\sffamily
{\bfseries int MPI\_File\_write\_shared (MPI\_File fh, void *buf, int count, MPI\_Datatype datatype, MPI\_Status status)} \\
}

{\sffamily
{\bfseries int MPI\_File\_seek\_shared (MPI\_File fh, MPI\_Offset offset, int whence)} \\
}

{\sffamily
{\bfseries int MPI\_File\_get\_position\_shared (fh, offset) } \\
}

\paragraph{Collective Routines:}

{\sffamily
{\bfseries int MPI\_File\_read\_ordered (MPI\_File fh, void *buf, int count, MPI\_Datatype datatype, MPI\_Status status) } \\
}

{\sffamily
{\bfseries int MPI\_File\_write\_ordered (MPI\_File fh, void *buf, int count, MPI\_Datatype datatype, MPI\_Status status) } \\
}

\subsubsection{Consistency and Semantics}

We can distinguish three different levels of consistency \cite{MPI-2}:
\begin{itemize}
\item sequential consistency among all processes using a single file handle
\item sequential consistency among all processes using file handle created from a single collective open with atomic mode enabled
\item user imposed sequential consistency
\end{itemize}

Sequential consistency is defined as a set of operations that seem to be performed in some serial order consistent with the program order. In contrast, user-imposed consistency can be yielded due to the program order or calls to MPI\_FILE\_SYNC.

{\sffamily
\begin{tabbing}
{\bfseries int MPI\_File\_set\_atomicity (MPI\_File fh, int flag) } \\
INOUT..\=parameter      \= description                                  \kill 
INOUT \>fh        \>file handle\\
IN    \>flag    	\>true to set atomic mode, false to set non-atomic mode
\end{tabbing}
}

\textbf{Description:}\\ 
Consistency semantics are guaranteed by a collective call of all processes in one group. In detail, any read or write operation to a file can be regarded as an atomic operation.\\

\textbf{Features:} 
\begin{itemize}
\item collective call (values for \textit{fh} and \textit{flag} must be the same for all processes in one group)
\end{itemize}

{\sffamily
\begin{tabbing}
{\bfseries int MPI\_File\_get\_atomicity (MPI\_File fh, int flag) } \\
INOUT..\=parameter      \= description                                  \kill
IN      \>fh      \>file handle\\
OUT     \>flag    \>true if atomic mode, false if non-atomic mode
\end{tabbing}
}

\textbf{Description:}\\ 
Returns either true or false according to the atomicity.

{\sffamily
\begin{tabbing}
{\bfseries int MPI\_File\_Sync (MPI\_File fh)} \\
INOUT..\=parameter      \= description                                  \kill
INOUT   \>fh      \>file handle
\end{tabbing}
}

\textbf{Description:}\\ 
All previous writes to \textit{fh} by the calling process are updated. Furthermore, all updates of other processes are visible as well. Thus, a subsequent read which is executed by the calling process returns the actual data in the file rather than any "dirty read".\\

\textbf{Features:}
\begin{itemize}
\item collective routine
\end{itemize}

Example: This example demonstrates consistency of a file which is written by process 0 and read by process 1. In order to guarantee that process 1 does not read any wrong data the so-called sync-barrier-sync construct is used for the following reason \cite{MPI-2}:

\begin{itemize}
\item MPI\_Barrier ensures that process 0 writes to the file before process 1 reads the file 
\item MPI\_File\_sync guarantees that the data written by all processes is transferred to the storage device. 
\item The second MPI\_File\_sync ensures that all data which has been transferred to the storage device is visible to all processes.
\end{itemize}

\begin{verbatim}

if (myid==0)
/* process 0 write to the file */   
{
  MPI_File_open (MPI_COMM_WORLD, "ufs:file2", MPI_MODE_CREATE | 
    MPI_MODE_RDWR, MPI_INFO_NULL, &fh );
  MPI_File_set_view(fh, 0, MPI_INT, MPI_INT, "native", 
    MPI_INFO_NULL);

  for (i=0; i<1000; i++)                    
    buf1[i]=i;

  MPI_File_write (fh, buf1, 1000, MPI_INT, &status);

  MPI_File_sync(fh);
  MPI_Barrier (MPI_COMM_WORLD);
  MPI_File_sync(fh);

  MPI_File_close(&fh);
}   
else
/* other processes read the updated file */
{
  MPI_File_open (MPI_COMM_WORLD, "ufs:file2", MPI_MODE_CREATE | 
    MPI_MODE_RDWR, MPI_INFO_NULL, &fh );
  MPI_File_set_view(fh, 0, MPI_INT, MPI_INT, "native", 
    MPI_INFO_NULL);

  MPI_File_sync(fh);
  MPI_Barrier (MPI_COMM_WORLD);
  MPI_File_sync(fh);
 
  MPI_File_read (fh, buf1, 1000, MPI_INT, &status);
  MPI_File_close(&fh);
}
\end{verbatim}

Although the second sync seems to be redundant, omitting it would yield an erroneous program.

%% file: mpio_ini.tex
\subsection{Initializing and Finalizing MPI-IO}

On giving the information on MPI-IO and the ViPIOS-interface we
will now have a look at how the ViMPIOS is implemented. In this
section we want to describe the internal handling of establishing
a connection to the ViPIOS server in order to use the
functionalities of MPI-IO. According to \cite{MPI-2} no explicit
routine is required for initializing and finalizing an MPI-IO
session. Thus, a call to \textit{MPI\_Init()} must establish a
connection to the ViPIOS server as well. Moreover, a call to
\textit{MPI\_Finalize()} has to disconnect from the ViPIOS server.
In order to guarantee this, every C application program using
MPI-IO needs to include the header file \textit{vip\_mpio\_init.h}
which looks like follows:

\begin{verbatim}
#define MPI_Init(argc,argv) MPIO_Init_C(argc,argv);
#define MPI_Finalize() MPIO_Finalize();
\end{verbatim}

The routines are defined in the file \textit{vip\_mpio.c}:

\begin{verbatim}
int MPIO_Init_C(int *argc,char ***argv)
{
    return (   ((MPI_Init(argc,argv)==MPI_SUCCESS) &&
                 ViPIOS_Connect(0)) MPI_SUCCESS : -1 );
}

int MPIO_Finalize(void)
{
    return ( (ViPIOS_Disconnect()&&(MPI_Finalize()==MPI_SUCCESS))
              ? MPI_SUCCESS : -1 );
}
\end{verbatim}

Note that this is only true for C applications. A discussion on using Fortran application programs is given in a later section.

%% file: file_management.tex
\subsection{File Management}

In this chapter we will analyze how the file information is handled. In particular, each MPI-IO file contains certain information about the filename, the file handle, the access mode etc. All that information is stored by means of following struct:

\begin{verbatim}
typedef struct {
    MPI_Comm comm;
    char *filename;

    int ViPIOS_fh;
    Access_Desc *view_root,
        *descriptor;
    bool view_is_set;
       
    MPI_Offset disp;
    MPI_Datatype etype;
    MPI_Datatype filetype;

    int contig_filetype; 
    int access_mode;
    int atomicity;
    bool already_accessed;
} File_definition;
\end{verbatim}

The first entry to that struct is the communicator \textit{comm}. When we recall the syntax of the routine \textit{MPI\_File\_open} we find out that each file is opened by a group of processes which are referred to by a communicator. Thus, it is possible to determine whether a file is opened by only one process - this is true if MPI\_COMM\_SELF is used as an argument in \textit{MPI\_File\_open} - or whether the file is opened by a set of processes. This information becomes vital for collective routines which we will discuss later on.\\

The next two parameters are the name of the file \textit{filename} and the file handle \textit{ViPIOS\_fh}. Latter is assigned by the routine \textit{ViPIOS\_Open}. \textit{view\_root} is a pointer to the structure \textit{Access\_Desc} which is defined in the kernel of ViPIOS. The purpose of that structure is to access a file in strides. Furthermore, the variable \textit{view\_is\_set} specifies whether a view to the file currently opened is set or not. We will explain the variables \textit{view\_root} and \textit{descriptor} in more detail when we discuss the implementation of MPI\_FILE\_VIEW. \\

\textit{Disp}, \textit{etype} and \textit{filetype} hold the information about the file view. The variable \textit{contig\_filetype} defines whether the filetype of the view is contiguous. In other words, there are no so-called holes in the file which cannot be accessed by an application. Thus, no additional algorithm is required for computing non-accessible parts of the file.\\

\textit{Access\_mode} contains the access rights to a file, e.g. read only, write only, etc. The last variable \textit{already\_accessed} states whether any file operation was carried out on the file. This becomes important when the displacement of the file view has to be computed.\\

In order to guarantee that an application can open several files at any time,  the information of each file must be administrated carefully, i.e. when an application accesses a file with a certain file handle, the interface must be able to retrieve the information on that particular file. ViMPIOS uses a library which is also applied within the ViPIOS kernel. The idea is as follows: When a new file is opened, a table is created which is a dynamic array of integer values. Each index of that table is a pointer to the structure \textit{File\_Definition} which we analyzed in the previous section. Let us briefly explain how this table is maintained when a file is opened, accessed and closed.\\

In order to access our structure \textit{File\_defintion} a new variable must be defined. This is done in the function \textit{MPI\_File\_open}:

\begin{verbatim}
File_definition help_fh;
\end{verbatim}

In the next step the table which administrates all files must be created:

\begin{verbatim}
static bool first;

if (first) {
    first=FALSE;
    std_tab_init(10,5,sizeof(File_definition),&File_table);
}
\end{verbatim}

\textit{First} is a static variable which determines whether the function \textit{MPI\_File\_open} is called for the first time. According to that result the table is initialized by \textit{std\_tab\_init()}. The first parameter states that storage space for 10 entries to the table is allocated. The second parameter defines how much storage space is allocated at a later stage. In other words, storage space for 10 elements of the size sizeof(File\_definition) is allocated at first. When the table is full, storage space for another 5 elements is allocated etc. The name of the table is denoted by the last parameter of that routine, namely \textit{File\_table}.\\

Let us assume that a file is correctly opened by:

\begin{verbatim}
ViPIOS_Open(filename,amode,&help_fh.ViPIOS_fh);
\end{verbatim}

This means that the ViPIOS server determines a file handle which is stored in \textit{help\_fh.ViPIOS\_fh}.\\

Once storage space of the table is allocated by the routine \textit{std\_tab\_init()} and the structure \textit{File\_definition} is filled with values, i.e. the file name, the access mode etc. are assigned, the file handle and the corresponding file information can be added to the file administration table. This is done by the following routine:

\begin{verbatim}
std_tab_append (&File_table, &help_fh, fh);
n_open_files++;
\end{verbatim}

This routine adds a new element to the table \textit{File\_table}. \textit{help\_fh} is a pointer to the element which has to be inserted. In our case it is the pointer to the struct \textit{File\_defintion} which holds the information on the current file (filename, file handle). The parameter \textit{fh} can be regarded as the index of the table which can be used as a key to retrieve information on the file. Note that \textit{fh} is not the actual file handle which is returned by the \textit{ViPIOS\_Open} call but the index to the table. The actual file handle of the file is retrieved by \textit{help\_fh-\(>\)ViPIOS\_fh} as we will see later on. The variable \textit{no\_open\_files} holds the information about the number of files currently opened.\\

Let us take a look at a small example in order to explain the functionality of the routines discussed so far. Assume that an application program opens three files. As we have already stated at an earlier stage, the file handle for each file is returned by the \textit{ViPIOS\_Open} function call. For our example assume the file handles \textit{help\_fh-\(>\)ViPIOS\_fh} 45,46,47. On initializing the file table, those three values can be added. Thus, index 0 points to the file referred to by file handle \textit{help\_fh-\(>\)ViPIOS\_fh}=45. Index 1 of the file table denotes the file with the file handle 46 and so on. It is important to mention that the file handles which are used by the application are these indices rather than the actual file handles defined by \textit{ViPIOS\_Open}.\\

On explaining the creation of the file table we will now discuss how information about a certain file can be retrieved. Assume that our application program wants to read data from the file with the actual file handle \textit{ViPIOS\_fh}=46. According to our previous explanation the application program uses the file handle \textit{fh} 1, i.e. the second index of our table which points to the file with the file handle 46, rather than the actual file handle \textit{help\_fh.ViPIOS\_fh}. Thus, before the data of the file with the file handle 46 can be read following steps are necessary. The code can be found in \textit{MPI\_File\_read(fh,...)}:

\begin{verbatim}
File_defintion *help_fh;
std_tab_get(File_table, fh, (void **) &help_fh);
\end{verbatim}

The function \textit{std\_tab\_get} returns the element defined by the index \textit{fh}. In our case \textit{fh}=1 which points to the file handle 46. A correct call to \textit{ViPIOS\_Read} looks like follows:

\begin{verbatim}
ViPIOS_Read(help_fh->ViPIOS_fh,...);
\end{verbatim}

Thus, the ViPIOS routine is called with the actual file handle \textit{help\_fh-\(>\)ViPIOS\_fh} rather than the entry to the file table.\\

To sum up, all so-called MPI-IO routines use the index of the file table as their file handle. The actual file handle is derived from the file table and is only used when the ViPIOS routines are called. Moreover, every time an MPI-IO function other than \textit{MPI\_File\_open} is called, the function \textit{std\_get\_table()} must be evoked in order to retrieved the information about a particular file.\\

When the application program closes a file, the entry to the file table must be deleted as well which is done in the routine \textit{MPI\_File\_close}. However, before the file table entry can be deleted, the function \textit{std\_tab\_get()} must be called in order to determine which entry has to be deleted. The index is stored in \textit{fh}. On closing the file with a call to the routine \textit{ViPIOS\_close} the routine \textit{std\_tab\_del()} is called:

\begin{verbatim}
std_tab_del(&File_tab, *fh);
n_open_files--;
\end{verbatim}

This routine simply deletes the element with the index \textit{fh} from the file table and decrements the counter \textit{no\_open\_files} which holds the number of file currently opened.\\

Finally, the whole file table is removed and the allocated storage space is freed when the last open file was closed:

\begin{verbatim}
if (n_open_files==0)
   std_tab_clean(&File_table);
\end{verbatim}

Besides creating, filling and closing the file management table, a further routine is implemented. Before any data access on a file can be performed, a special routine checks whether a corresponding file table entry exists. Thus, following routine is called

\begin{verbatim}
if (std_tab_used(File_table)==0)
{
    printf("\nMPI_File_read: File does not exist!");
    return -1;
}
\end{verbatim}

%% file: openclose.tex
\subsection{File Manipulation}

In this section we want to discuss how the routines for file manipulation are implemented in the ViMPIOS. Let us start off with \textit{MPI\_File\_open}. Besides managing the file table we described in the previous chapter, this routine checks whether the restrictions for the collective mode are obeyed. In particular, files are opened in a collective way when the communicator is not MPI\_COMM\_SELF. Thus, if this is true, a message is broadcast to all members of the communicator group in order to check whether all processes use the same filename and access mode. \\

In addition to closing a file, \textit{MPI\_File\_close} removes a file if the access mode MPI\_MODE\_DELETE\_ON\_CLOSE is set:

\begin{verbatim}
if (amode & MPI_MODE_DELETE_ON_CLOSE) 
    ViPIOS_Remove(filename);
\end{verbatim}

The routines \textit{MPI\_File\_delete}, \textit{MPI\_File\_set\_size} \textit{MPI\_File\_get\_size} and \textit{MPI\_-File\_get\_amode} need no further explanation since they simply call the corresponding ViPIOS interface routines, namely \textit{ViPIOS\_Remove}, \textit{ViPIOS\_File\_-set\_size} and \textit{ViPIOS\_File\_get\_size}. The routine \textit{MPI\_File\_preallocate} has basically the same functionality as the routine \textit{MPI\_File\_set\_size}. The only difference is that the file size is not truncated if the value for preallocating memory space is smaller than the actual file size. \\

The code fragment of  \textit{MPI\_File\_get\_group} for returning the communicator group of the specified file is: 
\begin{verbatim}
int MPI_File_get_group(MPI_File fh, MPI_Group *group)
{
    ...
    return MPI_Comm_group(help_fh->comm, group);
}
\end{verbatim}

Finally, the routine \textit{MPI\_File\_get\_amode} returns the access mode of a particular file which is currently opened:

\begin{verbatim}
int MPI_File_get_amode (MPI_File fh, int *amode)
{
    ...
    *amode=help_fh->access_mode;
    ...
}
\end{verbatim}

%% file: views.tex
\subsubsection{File Views}

In this chapter we want to describe how an MPI-IO view is realized in ViMPIOS. Before any view can be set a derived datatype has to be specified, for example

\begin{verbatim}
MPI_Datatype vector1;

MPI_Type_vector (5,2,10, MPI_INT, &vector1);
MPI_Type_commit(&vector1);
\end{verbatim}

Next the view can be set

\begin{verbatim}
MPI_File_set_view(fh, 0, MPI_INT, vector1, "native",
MPI_INFO_NULL);
\end{verbatim}

In our example we know the our filetype \textit{vector1} is a derived datatype MPI\_TYPE\_VECTOR. Now we want to extract the variables \textit{count}, \textit{blocklen}, \textit{stride} and \textit{oldtype} in order to map the datatype to the ViPIOS access descriptor. In particular we have to search for the structure which stores the information of MPI derived datatypes. One way to do so is to take a look at the MPI implementation of the derived datatype MPI\_TYPE\_VECTOR. However, we will omit printing the whole source code but only analyse its content. The interested reader is refored to the lines of code of the MPI implementation.\\

On checking for bad arguments the derived datatype is checked for being contiguous. This is true if \textit{blocklen} and \textit{stride} have the same value or if \textit{count} is 1. Then, the derived datatype \textit{MPI\_Type\_vector} can be reduced to MPI\_TYPE\_CONTIGUOUS. Moreover, each \textit{MPI\_Type\_vector} is reduced to \textit{MPI\_Type\_vector}. Thus, the stride is evaluated in bytes rather than in multiples of \textit{oldtype}.\\

Since \textit{extent} is a variable of \textit{old\_dtype\_ptr} which in turn is a pointer to the struct \textit{MPIR\_DATATYPE} we have to search for  the definition of that structure. In the source code of MPI it can be found in the header file \textit{datatype.h} which describes all MPI datatypes, i.e. basic datatypes as well as derived datatypes:

\begin{verbatim}
struct MPIR_DATATYPE {
    MPIR_NODETYPE dte_type; /* type of datatype element
                               this is */
    MPIR_COOKIE             /* Cookie to help detect valid
                               item */
    int          committed; /* whether committed or not */
    int          is_contig; /* whether entirely contiguous */
    int              basic; /* Is this a basic type */
    int          permanent; /* Is this a permanent type */
    MPI_Aint        ub, lb; /* upper/lower bound of type */
    MPI_Aint real_ub, real_lb; /* values WITHOUT TYPE_UB/
                               TYPE_LB */
    int             has_ub; /* Indicates that the datatype has
                               a TYPE_UB */
    int             has_lb; /* Indicates that the datatype has
                               a TYPE_LB */
    MPI_Aint        extent; /* extent of this datatype */
    int               size; /* size of type */
    int           elements; /* number of basic elements */
    int          ref_count; /* nodes depending on this node */
    int              align; /* alignment needed for start of
                               datatype */
    int              count; /* replication count */
    MPI_Aint        stride; /* stride, for VECTOR and HVECTOR
                               types */
    MPI_Aint      *indices; /* array of indices, for (H)INDEXED,
                               STRUCT */
    int           blocklen; /* blocklen, for VECTOR and HVECTOR
                               types */
    int         *blocklens; /* array of blocklens for (H)INDEXED,
                               STRUCT */
    struct MPIR_DATATYPE *old_type,
                **old_types,
                *flattened;
    MPI_Datatype self;      /* Index for this structure */
#ifdef FOO
    MPI_Datatype old_type;  /* type this type is built of,
                               if 1 */
    MPI_Datatype *old_types;/* array of types, for STRUCT */
    MPI_Datatype flattened; /* Flattened version, if available */
#endif
};

extern void *MPIR_ToPointer ANSI_ARGS(( int ));

#define MPIR_GET_DTYPE_PTR(idx) \
    (struct MPIR_DATATYPE *)MPIR_ToPointer( idx )

\end{verbatim}

Let us pick out the variables which are important for the datatype \textit{MPI\_Type\_-hvector}. \textit{dte\_type} holds the information about the kind of datatype. In our example the variable contains the data MPIR\_HVECTOR. \textit{committed} states whether the derived datatype was committed in the application program by \textit{MPI\_Type\_commit}. Since the stride in our datatype is greater than the number of elements (\textit{stride}=10 \(>\) \textit{blocklen}=2), our datatype is not contiguous (\textit{is\_contiguous=false}). Furthermore, our datatype is no basic MPI datatype (\textit{basic=false}). \textit{count} holds the number of blocks and \textit{blocklen} the number of elements of each block. We will explain the further variables when we actually need them for our implementation.\\

Since we have found the structure which holds the information about all MPI datatypes we can start our discussion about the implementation of \textit{MPI\_File\_set\_view}. First we define a pointer to that structure for retrieving the information of the access pattern stored in \textit{filetype} of \textit{MPI\_File\_set\_view}:

\begin{verbatim}
struct MPIR_DATATYPE *view;
view=MPIR_GET_PTR(filetype);
\end{verbatim}

Thus, the data access pattern stored in \textit{filetype} which can be any combination of MPI basic or derived datatypes is determined by analyzing the struct MPIR\_DATATYPE.\\

Apart from analyzing the \textit{filetype} we also have to tell the ViPIOS about the file view. This means the information retrieved from MPIR\_DATATYPE has to be mapped to the ViPIOS access descriptor which is handled by the routine \textit{get\_view\_pattern}. Before this function can be called, storage space for the ViPIOS access descriptor has to be allocated and \textit{next\_free} which is defined as \textit{void *next\_free} must be set accordingly:

\begin{verbatim}
descriptor=malloc(1024);
next_free=descriptor;
\end{verbatim}

\subsubsection{The Mapping Function \textit{get\_view\_pattern}}
\textit{int get\_view\_pattern(struct MPIR\_DATATYPE *view, Access\_Desc *descriptor, void **free\_space)}\\

\textit{get\_view\_pattern} is a recursive routine which extracts the information of \textit{filetype}, i.e. the access pattern of the view, and maps it to the ViPIOS access descriptor. The recursion is called as long as the \textit{filetype} is no basic MPI datatype. Furthermore, the return value is 1 if the access pattern is contiguous and 0 otherwise.
First, the function checks whether the \textit{filetype} is one of the following derived datatypes:

\begin{itemize}
\item{MPI\_TYPE\_CONTIGUOUS}
\item{MPI\_TYPE\_HVECTOR}
\item{MPI\_TYPE\_HINDEXED}
\item{MPI\_TYPE\_STRUCT}
\end{itemize}

In the chapter about derived datatypes we stated 6 different derived datatypes rather than 4. The function \textit{get\_view\_pattern} checks only for four different ones since the datatype \textit{MPI\_Type\_vector} is automatically reduced to \textit{MPI\_Type\_hvector}, the same is true for \textit{MPI\_Type\_indexed} and \textit{MPI\_Type\_-hindexed}.\\

\paragraph{MPI\_TYPE\_CONTIGUOUS}

Let us start with the simplest derived datatype and explain how it is mapped to the ViPIOS access descriptor.\\

\textit{MPI\_Type\_contiguous(count,oldtype,newtype)}\\

\textit{count} holds the number of elements of the contiguous datatype. This value is retrieved by \textit{view-\(>\)count}. Thus, it can be mapped as follows:

\begin{verbatim}
descriptor->no_blocks=1;
descriptor->skip=0;
descriptor->basics[0].offset=0;
descriptor->basics[0].repeat=1;
if (!next->basic)
    descriptor->basics[0].count=view->count;
else
    descriptor->basics[0].count=view->count*next->extent;
descriptor->basics[0].stride=0;
\end{verbatim}

Since we use just one homogenous datatype \textit{no\_blocks} and \textit{repeat} are set to 1. What is more, \textit{skip} and \textit{offset} are set to 0. Some more explanation must be given for the parameter \textit{count} which is set according to \textit{next-\(>\)basic} where \textit {next} is an auxiliary variable for retrieving information of \textit {oldtype}:

\begin{verbatim}
struct MPIR_DATATYPE *next;
next=view;
next=next->oldtype;
\end{verbatim}

Assume a file which consists of 100 double values and we wish to define the following view:

\begin{verbatim}
MPI_Type_contiguous(10,MPI_DOULBE,&contig1);
MPI_Type_commit(&contig1);
MPI_File_set_view(....,contig1,...);
\end{verbatim}

On applying the function \textit{get\_view\_pattern} to that example the kind of data-type which is recognized is MPI\_TYPE\_CONTIGUOUS. The kind of datatype of \textit{oldtype} which is MPI\_DOUBLE can only be retrieved by our auxiliary variable \textit{next} which points to the next level in the structure of the filetype. Thus, \textit{next-\(>\)basic} is true since \textit{oldtype} is a basic datatype, namely MPI\_DOUBLE. In other words, \textit{next-\(>\)basic} only returns \textit{true} if the filetype does not consist of nested derived datatypes.\\

Moreover, since every access operation in ViPIOS is done in bytes, \textit{count} must be multiplied by the size of the datatype of \textit{oldtype} which is retrieved by \textit{next-\(>\)extent}. In our example \textit{view-\(>\)count} is 10, thus, the variable \textit{count} of the ViPIOS access descriptor is set to 80 - provided the case that a the datatype double comprises eight bytes. \\

In short, if the view is a nested derived datatype, \textit{count} gets the same values as retrieved from \textit{view-\(>\)count}, i.e. from the MPI struct MPIR\_DATATYPE otherwise \textit{count} is multiplied by the extent of \textit{oldtype}.\\

After this little excursion about the variable \textit{count} of the ViPIOS access descriptor we can now return to the remaining variables.

\begin{verbatim}
if (!next->basic)
{
    descriptor->basics[0].subtype=*free_space;
    get_view_pattern(view->oldtype,descriptor->basics[0].subtype,
                     free_space);
}
else
{
    descriptor->basics[0].subtype=0;
    return 1;
}
\end{verbatim}

If \textit{filetype} is a nested derived datatype, \textit{subtype} points to the next free space allocated for the ViPIOS access descriptor. Furthermore, the recursive function is evoked again to map the next level of \textit{filetype}. \textit{free\_space} is evaluated as follows:

\begin{verbatim}
free_space=(struct basic_block ) ((void) descriptor+sizeof
           (Access_Desc));
descriptor->basics=*free_space;
free_space+=(descriptor->no_blocks*sizeof(struct basic_block));
\end{verbatim}

After the first line of code \textit{free\_space} points to the beginning of the struct \textit{basic\_block} i.e. the first struct of the ViPIOS access descriptor. Since the space for the struct \textit{Access\_Desc} is skipped \textit{descriptor-\(>\)basics} is set to that position as well. In the third line \textit{free\_space} is adjusted, i.e. the space for one basic block is skipped since \textit{no\_blocks} is 1.\\

\paragraph{MPI\_TYPE\_HVECTOR}

On explaining the mapping mechanism of MPI\_TYPE\_CONTIGUOUS we can now resume with the next derived datatype, namely \textit{MPI\_Type\_vector}:\\

\textit{int MPI\_Type\_hvector(count, blocklength, stride, oldtype, *newtype)}\\

The values can be retrieved by \textit{view-\(>\)count}, \textit{view-\(>\)blocklen} and \textit{view-\(>\)stride}.

Similar to the previous datatype only one basic block is required and \textit{count} is set to \textit{view-\(>\)blocklen}. In order to avoid any confusion we refer to the variables of the MPI struct MPIR\_DATATYPE as \textit{view-\(>\)count}, \textit{view-\(>\)blocklen} etc. Thus, when we state, for example, \textit{count} we refer to the ViPIOS access descriptor. Moreover, note that \textit{repeat} corresponds to \textit{view-\(>\)count} and \textit{count} to \textit{view-\(>\)blocklen}. This is the reason why \textit{count} is mapped to \textit{view-\(>\)blocklen} rather than to \textit{view-\(>\)count}:

\begin{verbatim}
descriptor->no_blocks=1;
descriptor->skip=0;
descriptor->basics[0].offset=0;
descriptor->basics[0].repeat=view->count;
...
if (!next->basic)
    descriptor->basics[0].count=view->blocklen;
    get_view_pattern(view->oldtype,descriptor->basics[0].subtype,
                     free_space);
else
    descriptor->basics[0].count=view->blocklen*next->extent;
\end{verbatim}

In addition to assigning the values for \textit{count} the recursive function is called again since the \textit{filetype} is a nested datatype. However, in the program code the recursion is called after the mapping of the remaining values. Here we changed the order of the program code to some extent so that we can more easily explain the mapping technique. \\

\begin{figure}
\begin{center}
\includegraphics[scale=0.9]{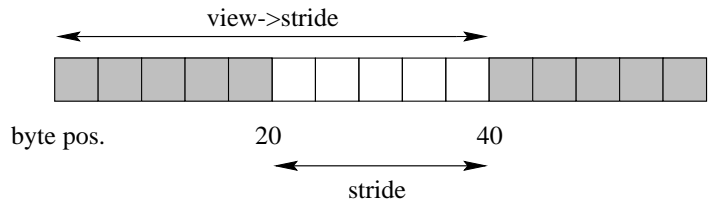}
\caption{MPI\_Type\_hvector}
\label{view7}
\end{center}
\end{figure}

As we have already stated in a previous chapter \textit{stride} and \textit{view-\(>\)stride} are interpreted in a different way. \textit{stride} denotes the gap between two data blocks of the vector whereas \textit{view-\(>\)stride} denotes the number of bytes from the beginning of one block to the beginning of the next one. \textit{stride} is computed as follows:

\begin{verbatim}
stride=view->stride - view->blocklen * next->extent;
\end{verbatim}

where \textit{next-\(>\)extent} holds the extent of \textit{oldtype}. For example,

\begin{verbatim}
MPI_Type_hvector(2,5,40,MPI_INT,&vetor1);
\end{verbatim}

The corresponding view is depicted in Figure \ref{view7}: \\

The stride between the first and the second data block is \textit{view-\(>\)stirde}=40 which is mapped as:\\

\begin{verbatim}
stride=40-5*4=20; \\
\end{verbatim}

since the extent of \textit{oldtype}, i.e. MPI\_INT, is 4.
\textit{Subtype} and the remaining auxiliary variables are adjusted as before.\\

\paragraph{MPI\_TYPE\_HINDEXED}

Unlike the previous datatypes MPI\_TYPE\_HINDEXED is mapped by using several basic blocks since each data block can have a different size:\\

\textit{int MPI\_Type\_hindexed (count, *array\_of blocklengths, *array\_of\_displacements, oldtype, *newtype);}\\

The values can again be retrieved by \textit{view-\(>\)count}, \textit{view-\(>\)blocklens[i]} and \textit{view-\(>\)indices[i]} where \textit{i} ranges from 0 to \textit{view-\(>\)count-1}, i.e. \textit{view-\(>\)blocklen[2]} states the length of the 3rd data block with a stride of \textit{view-\(>\)indices[2]}.\\

\begin{figure}
\begin{center}
\includegraphics[scale=0.9]{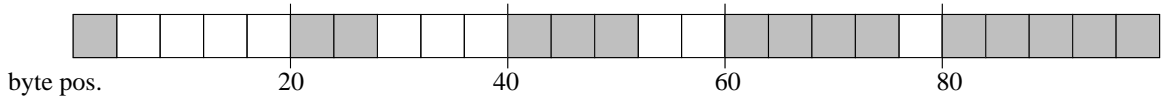}
\caption{MPI\_Type\_indexed}
\label{view8}
\end{center}
\end{figure}

Since we have 5 data blocks of different lengths we use 5 basic blocks (\textit{no\_blocks=
view-\(>\)count} where \textit{count}, i.e. the size, of each block is:
\begin{verbatim}
if (!next->basic)
    descriptor->basics[i].count=view->blocklens[i];
else
    descriptor->basics[0].count=view->blocklens[i]*next->extent;
\end{verbatim}

Next \textit{repeat} is set to 1 and \textit{stride} to 0. Finally, the offset for each basic block must be computed because the array \textit{view-\(>\)indices} denotes the displacement of each data block to the beginning of the datatype rather than the stride between two adjacent blocks. For example, the byte displacement of the second block is \textit{view-\(>\)indices[1]}=20. The gap between the second and the third block is computed by:\\

\textit{gap[2]=view-\(>\)indices[2] - view-\(>\)blocklens[1]*extent\_of\_oldtype - view-\(>\)\\
indices[1]=}\textit{40 - 2*4 - 20 = 12}\\

The offset for the first basic blocks is:

\begin{verbatim}
descriptor->basics[0].offset=view->indices[0];
\end{verbatim}

All remaining offsets are computed similar to the simplified formula given above:

\begin{verbatim}
if (!next->basic)
    descriptor->basics[i].offset=view->indices[i] - descriptor->
    basics[i-1].count*next->extent - view->indices[i-1];
else
    descriptor->basics[i].offset=view->indices[i] - descriptor->
    basics[i-1].count - view->indices[i-1];
\end{verbatim}

This needs some more explanation. Let us start with the fist case were \\ \textit{next-\(>\)basic} is false, i.e. the view is a nested derived datatype. In this case \textit{count} of the previous basic block (\textit{descriptor-\(>\)basics[i-1].count}) is multiplied by the extent of \textit{oldtype} since \textit{count} was not adjusted before. On the other hand, if the view is no nested derived datatype (\textit{next-\(>\)basic=true}) \textit{count} was already adjusted, i.e. the size of that datatype was already multiplied by the extent of \textit{oldtype}(\textit{count=view-\(>\)blocklens[i]*next-\(>\)extent}). Thus, the size of the previous basic blocks needs no longer be multiplied by the extent of \textit{oldtype} in order to compute the correct offset.\\

Recalling our example of the datatype MPI\_TYPE\_HINDEXED we want to compute the offset of the third basic block which is 12, i.e. the stride between the second and the third basic block. Further assume that our view is no nested datatype. Thus, the length of the previous basic blocks is already adjusted (it is 8 rather than 2, i.e. 2*extent of integer=8). The offset is computed according to the second case:\\

\textit{offset=40 - 8 - 20 = 12}\\

The difference to our previous computation is that the length of the previous block is already adjusted (8) and needs no more modification.\\

\paragraph{MPI\_TYPE\_STRUCT}

The most complex derived datatype has the following syntax:\\

\textit{int MPI\_Type\_struct(count, *array\_of\_blocklentghs, *array\_of\_displacements, oldtype, *newtype)}\\

Like we did for MPI\_TYPE\_HINDEXED we use several basic blocks, namely:

\begin{verbatim}
descriptor->no_blocks=view->count;
descriptor->skip=0;
\end{verbatim}

\textit{count} is also defined in the same way. However, the recursive function call differs to some extent. Since each block can consist of different datatypes, the first parameter of the function call is not \textit{view-\(>\)oldtype} but \textit{view-\(>\)oldtypes[i]} where \textit{i} refers to the corresponding block. Thus, we use another auxiliary pointer and adjust it for each basic block:

\begin{verbatim}
next=view;
next=next->oldtypes[i];
...
if (!next->basic)
{
    descriptor->basics[i].count=view->blocklens[i];
    ...
    get_view_pattern(view->oldtypes[i],descriptor->basics[0].
    subtype,free_space);
}
else
    descriptor->basics[0].count=view->blocklens[i]*next->extent;
\end{verbatim}

Again, \textit{repeat} is set to 1 and \textit{stride} to 0. More detailed information must be given about the mapping of \textit{offset}.\\

The offset of the first basic block is:

\begin{verbatim}
descriptor->basics[0].offset=view->indices[0]
\end{verbatim}

whereas the remaining offsets are computed as follows:\\

\textit{descriptor-\(>\)basics[i].offset=view-\(>\)index[i]- view-\(>\)blocklen[i-1]*\\
extent\_of\_oldtypes[i-1]-view-\(>\)indices[i-1]}\\

\begin{figure}
\begin{center}
\includegraphics[scale=0.9]{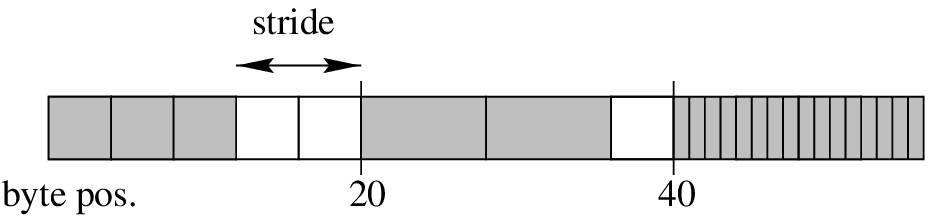}
\caption{MPI\_Type\_struct}
\label{struct1_1}
\end{center}
\end{figure}

Let us explain this by an example. Assume a file which consists of 5 integer, 2 double, and 50 character values. Further assume following derived datatype:
\begin{verbatim}
s_types[3]={MPI_INT,MPI_DOUBLE,MPI_CHAR};
s_blocklens[3]={3,2,16};
s_disps[0]=0;
s_disps[1]=20;
s_disps[2]=40;
\end{verbatim}

The corresponding file view is depicted in Figure \ref{struct1_1}.\\

The values of the displacements are: \textit{view-\(>\)indices[0]}=0, \textit{view-\(>\)indices[1]}=20 and \textit{view-\(>\)indices[2]}=60. The corresponding offsets for each basic block of the ViPIOS access descriptor are computed as follows:\\

\textit{descriptor-\(>\)basics[0].offset=0}\\
\textit{descriptor-\(>\)basics[1].offset=20 - 3*4 - 0 = 8}\\
\textit{descriptor-\(>\)basics[2].offset=60 - 2*8 - 20 = 24}\\

When, for example, the offset of the third basic block is computed, extent of the previous \textit{oldtype} must be retrieved. Thus, we need a further auxiliary variable:
\begin{verbatim}
prev=view;
prev=prev->oldtypes[i-1];
\end{verbatim}

The actual code for computing the offset will be omitted here.\\

Up to now we assume that the \textit{etype} of the view is a basic MPI datatype. However, in our introductory chapter we stated that an \textit{etype} can be a derived datatype as well. Thus, a similar routine for extracting the information of the \textit{etype} is required. The corresponding function is \textit{get\_oldtype}. To start off, the a pointer to the structure which stores the information on MPI datatypes is set in the routine \textit{MPI\_File\_set\_view}. In order to reduce computational overhead, the routine \textit{get\_oldtype} is only called if the \textit{etype} actually is a derived datatype. Moreover, the variable \textit{orig\_oldtype\_etype} gets the basic MPI datatype of the derived datatype. In other words, if the \textit{etype} is the following vector:

\begin{verbatim}
MPI_Type_vector(10,5,20,MPI_INT,&vector}
\end{verbatim}

\textit{orig\_oldtype\_etype} will be MPI\_INT. The remaining lines of code are an extract of the routine \textit{MPI\_File\_set\_view}:

\begin{verbatim}
struct MPIR_DATATYPE *oldtype_etype;
	
oldtype_etype=MPIR_GET_DTYPE_PTR(etype);
orig_oldtype_etype=6;

// only call the function if etype is a derived datatype
if (!oldtype_etype->basic)
    orig_oldtype_etype=get_oldtype(oldtype_etype);
else
    orig_oldtype_etype=oldtype_etype->dte_type;
\end{verbatim}

Let us now take a look at the extraction function. We already know the features of the derived datatype such that we can quickly motivate the following lines of code. First, depending on the derived datatype, the corresponding branches are executed. Thus, the routine is called recursively until the analyzed datatype is a basic MPI datatype which is in turn returned to the calling function, namely \textit{MPI\_File\_set\_view}. Note that the code for the datatypes MPIR\_HINDEXED and MPIR\_STRUCT is more complex because these datatypes can consist of more than one oldtype.

\begin{verbatim}
int get_oldtype (struct MPIR_DATATYPE *old_datatype)
{
    struct MPIR_DATATYPE *next;
    int i;

    next=old_datatype;

    switch (old_datatype->dte_type)
    {
        case MPIR_CONTIG:
        case MPIR_HVECTOR:
        {	
            next=next->old_type;
            if (!next->basic)
                get_oldtype(old_datatype->old_type);
            else
                return next->dte_type;
            break;
        }

        case MPIR_HINDEXED:
        case MPIR_STRUCT:
        {
            for (i=0; i<old_datatype->count; i++)
            {			
                 next=next->old_types[i];
                 if (!next->basic)
                     get_oldtype(old_datatype->old_types[i]);
                 else
                     return next->dte_type;

            }
        }
        default:
            return 6;
    }
}
\end{verbatim}

We still need a mechanism which checks the \textit{etype} and \textit{filetype} for correctness. This means, that the original oldtype of the \textit{etype} must correspond with the original oldtype of the \textit{filetype}. For example, if \textit{etype} is set to MPI\_INT and filetype is set to \textit{MPI\_Type\_vector} its oldtype must be MPI\_INT as well. Thus, one more line must be added to the routine \textit{get\_view\_pattern} which is responsible for verifying these parameters. \\

On analyzing how a view, which is specified by means of derived datatypes, is mapped to the ViPIOS access descriptor we can now resume our discussion on the implementation of the MPI-IO function \textit{MPI\_File\_set\_view}.\\

Since this function is collective, all processes must use the same extent of \textit{etype} and the same data representation. At the moment ViMPIOS only supports the \textit{native} data representation. Thus, if at least two processes are started by the application program, a message is broadcast to check for the same \textit{etypes}. If the \textit{etypes} differ among the processes, the application program is aborted. The last step of the routine \textit{MPI\_File\_set\_view} is to assign following parameters:

\begin{verbatim}
ViPIOS_fh->view_root=root;
ViPIOS_fh->disp=disp;
ViPIOS_fh->etype=etype;
ViPIOS_fh->filetype=filetype;
ViPIOS_fh->info=info;
\end{verbatim}
The last parameter \textit{info} is not implemented yet. Finally, the file pointer is set to position 0 within the file view.

%% file: dataaccs.tex
\subsection{Data Access Routines}

\subsubsection{Read and Write}

Let us start with the routine \textit{MPI\_File\_read}. Unlike we pointed out in the introductory chapter on MPI-IO, the last parameter, namely \textit{status} is of type \textit{MPIO\_Status} rather than \textit{MPI\_Status}. We will discuss its features later in this chapter. \\

On opening the table for the file management and checking the parameters from the application program, the next important step is to interpret the MPI datatype in a way which can be understood by the ViPIOS server. Thus, the routine \textit{convert\_datatype} is called which looks like follows:

\begin{verbatim}
void convert_datatype(MPI_Datatype datatype, int *count)
{
    switch (datatype)
    {
        case MPI_SHORT:
            (*count)*=sizeof(short int);
            break;
        case MPI_INT:
            (*count)*=sizeof(int);
            break;
        case MPI_LONG:
            (*count)*=sizeof(long int);
            break;
        case MPI_UNSIGNED_CHAR:
           (*count)*=sizeof(unsigned char);
            break;
        case MPI_UNSIGNED_SHORT:
            (*count)*=sizeof(unsigned short int);
            break;
        case MPI_UNSIGNED:
            (*count)*=sizeof(unsigned int);
            break;
        case MPI_UNSIGNED_LONG:
            (*count)*=sizeof(unsigned long int);
            break;
        case MPI_FLOAT:
            (*count)*=sizeof(float);
            break;
        case MPI_DOUBLE:
            (*count)*=sizeof(double);
            break;
       default: (*count)*=sizeof(char);
	 }
}
\end{verbatim}

This routine has two main functions. On the one hand, it interprets an MPI datatype to a C datatype, on the other hand, the parameter \textit{count} which states the number of elements to be read is interpreted in byte elements. For example, if the application program prompts to read \textit{count}=10 values of MPI\_DOUBLE,  \textit{count} this is set to 80 provided the case that the size of one double value is 8 bytes. \\

After having adjusted the parameter \textit{count}, the ViPIOS interface can be called in order to read the values. Thus, depending on the filetype of the view, either \textit{ViPIOS\_Read} or \textit{ViPIOS\_Read\_struct} is called. This means that if the file view is contiguous, i.e. there are no holes in the view, the normal \textit{ViPIOS\_Read} is called with following parameters:

\begin{verbatim}
ViPIOS_Read (help_fh->ViPIOS_fh, buf, count,-1);
\end{verbatim}

The first parameter is the entry retrieved from the file management table. The second and the third parameters refer to the buffer and the number of elements to be read. Some more information must be provided for the last parameter. -1 signals the ViPIOS server that the file pointer shall be updated after this call. Thus, if the file pointer was at position 0 before reading the file, it points to position 80 after 80 byte values were read. By means of that parameter we can explain the routine \textit{MPI\_File\_read\_at} which is a so-called routine with an explicit offset. Assume, the file should be read starting from position 100. The corresponding call to the ViPIOS server would be:

\begin{verbatim}
ViPIOS_Read (help_fh->ViPIOS_fh, buf, count,100);
\end{verbatim}

Since routines with explicit offsets shall not interfere with routines without explicit offsets, the file pointer is not updated after calling that routine. \\

After this little excursion to another data access routine we will resume our discussion of \textit{MPI\_File\_read}. We have already mentioned that depending on the file view a special ViPIOS routine is called. Thus, if the file view is not contiguous, following call is made:

\begin{verbatim}
ViPIOS_Read_struct(help_fh->ViPIOS_fh,buf,count,
    help_fh->view_root,help_fh->disp,-1);
\end{verbatim}

The parameter \textit{help\_fh-\(>\)view\_root} is a pointer to the structure which handles the access pattern of the file. \textit{help\_fh-\(>\)disp} holds the information about the displacement of the view. The remaining parameters have the same meaning as we discussed above.\\

Although we have already talked about \textit{MPI\_File\_read\_at} we still need to say something about the offset. Remember that every offset in MPI-IO is interpreted in units of \textit{etype}. Thus, if our \textit{etype} is, for instance, MPI\_DOUBLE, the offset must be multiplied by the extent of MPI\_DOUBLE. Restricting \textit{etypes} to basic MPI datatypes allows us to deploy the function \textit{convert\_datatype} we discussed above. The routines \textit{MPI\_File\_read\_all} and \textit{MPI\_-File\_read\_at\_all} have the same implementation features as their non-collective counterparts. The only exception is a barrier at the end of the routine which synchronizes all processes in the group. This means, that the application can only resume after all processes have executed the collective read operation. \\

Recall that we have stated at the beginning of the section that the parameter \textit{status} is of type \textit{MPIO\_Status} rather than \textit{MPI\_Status}. The ViMPIOS internal structure of \textit{MPIO\_Status} looks like follows:

\begin{verbatim}
typedef struct 
{
    int fid; 
    int count;
}
MPIO_Status;
\end{verbatim}

On calling the ViPIOS interface for any data access routine, for example, \textit{ViPIOS\_Read}, the number of bytes which were actually read are stored by the ViPIOS server in the structure \textit{MPIO\_Status}. Thus,  the file identifier must be stored in the structure which is done in all blocking data access routines:

\begin{verbatim}
status->fid=help_fh->ViPIOS_fh;
\end{verbatim}

The number of bytes which were actually accessed by a particular data access routine can be retrieved by the routine \textit{MPI\_File\_get\_count}. Since we have not discussed its synapses in the introductory chapter on MPI-IO, we will take a look at the whole program code:

\begin{verbatim}
int MPI_File_get_count (MPIO_Status *status,
    MPI_Datatype datatype, int *count)
{
    int res;

    if (status->fid<0)
    {
        printf("\nFile handle is not specified since non-
                 blocking request has not finished correctly!");
        res=0;
    }
   else
        res=ViPIOS_File_get_count(status->fid,count);

     return ( (res==1) ? MPI_SUCCESS :  -1) ;
}
\end{verbatim}

After calling the routine \textit{ViPIOS\_File\_get\_count} with the ViPIOS-file handle stored in \textit {status-\(>\)fid} the parameter \textit{count} holds the number of bytes which were actually accessed.\\

Besides blocking access operations ViMPIOS also supports non-blocking ones. In particular, \textit{MPI\_File\_iread} and \textit{MPI\_File\_iread\_at} as well as \textit{MPI\_File\_iwrite} and \textit{MPI\_File\_iwrite\_at}. However, similar to ROMIO split collective routines are not supported yet. Note that the difference between the parameters of blocking and non-blocking routines is the last parameter, namely \textit{request}, which can be regarded as the index for the particular non-blocking routine. In addition, the routine \textit{MPI\_File\_test} allows testing whether the non-blocking routine has finished. Moreover, \textit{MPI\_File\_wait} is used for waiting until the routine specified by the request number has finished.\\

Similar to the parameter \textit{status} of the blocking data access routines, the type of the parameter \textit{request} of the non-blocking calls is redefined as well. Thus, instead of \textit{MPI\_Request} ViMPIOS uses the type \textit{MPIO\_Request}, which is defined by:

\begin{verbatim}
typedef struct
{
    int  reqid;    /* Request-Id */
    int  fid;      /* File-Id */
}
MPIO_Request;
\end{verbatim}

Moreover, each non-blocking data access stores the ViPIOS-file identifier into that structure which is needed for the routines  \textit{MPI\_File\_wait}, \textit{MPI\_File\_test} and \textit{MPI\_File\_get\_count}:

\begin{verbatim}
request->reqid=help_fh->ViPIOS_fh;
\end{verbatim}

On calling a non-blocking routine, the function \textit{MPI\_File\_test} can be used for checking whether the data access operation is finished which is indicated by the parameter \textit{flag}:

\begin{verbatim}
int MPI_File_test (MPIO_Request *request, int *flag,
    MPIO_Status *status)
{
	 int res;

	 res=ViPIOS_File_Test(request->reqid, flag);

	 if (*flag==TRUE)
	     status->fid=request->reqid;
	 else
        status->fid=-1;

	 return ( (res==1) ? MPI_SUCCESS :  -1) ;
}
\end{verbatim}

If the \textit{flag} is set, i.e. the non-blocking data access operation has finished, the request identifier is stored in the structure of \textit{MPIO\_Status} otherwise the file identifier is set to -1. The reason for this functionality can be explained by looking at the parameters of \textit{MPI\_File\_get\_count}. Recall that the file identifier is required for determining the number of bytes which are accessed. Since this file identifier is retrieved from the parameter \textit{status} which is not used for non-blocking operations the value must be set in the routine \textit{MPI\_File\_test}. Thus, we yield two advantages. On the one hand, we can retrieve the number of bytes actually accessed, the other hand, if the non-blocking function has not finished, i.e. \textit{status-\(>\)fid} is set to -1, the ViPIOS interface \textit{ViPIOS\_File\_get\_count} needs not be called. \\

The routine \textit{MPI\_File\_wait} is implemented in a similar way. However, the difference to the previous one is that it waits until the non-blocking operation specified by the \textit{request} has finished.

\begin{verbatim}
int MPI_File_wait(MPIO_Request *request, MPIO_Status *status)
{
    int res;

    res=ViPIOS_File_Wait(request->reqid);
    status->fid=request->reqid;

    return ( (res==1) ? MPI_SUCCESS :  -1) ;
}

\end{verbatim}

Since the implementation of the routines for writing data to a file - \textit{MPI\_File\_-write}, \textit{MPI\_File\_write\_at}, \textit{MPI\_File\_write\_at\_all} are analogous to the read operations we will omit a comprehensive explanation. \\

\subsubsection{Scatter Functionality}

Recalling the syntax of the data access routines we assumed so far that our data type is a basic MPI data type. However, even more complicated structures can be used in order to simulate a so-called scatter function. In particular this means that a file which is contiguous in memory can be read into the read buffer in a strided way. Let us take a look at an example in order to show the difference to the conventional read operation. Assume that our file resides on the disk in a contiguous way. Further assume that no file view is set. Performing following \textit{read} operation yields a read buffer which exactly corresponds to the data stored on disk. 

\begin{verbatim}
MPI_File_read(fh, buf, 100, MPI_INT, status);
\end{verbatim}

This data access operation simply reads 100 integer values into the read buffer. In order to simulate the scatter mechanism, a derived datatype is used rather than a basic MPI datatpype.:

\begin{verbatim}
MPI_Type_vector(2,5,50,MPI_INT,&vector);
MPI_Type_commit(&vector);

MPI_File_read(fh, buf, 1, vector, status);
\end{verbatim}

Although the parameter \textit{count} is set 1, more than 1 integer value is read. In particluar, 10 integer values are read since the derived datatype comprises 10 elements. However, the read buffer is not filled in contiguously but according to pattern described by the derived datatype. Thus, the first five elements of the read buffer are filled with the values which are read from the file and 45 values are skipped. Finally, the last 5 elements are read into the read buffer. On the whole, the read buffer comprises 55 elements whereas merely 10 integer values stem from the file stored on disk.\\ 

By means of that mechanism an even more complex data access is possible. Assume that you set a file view according to \textit{filetype\_vector} whereas the datatype for the read buffer corresponds to the derived datatype \textit{read\_vector}. Thus, a non-contiguous file can be stored in a different buffer which is non-contiguous as well. Let us demonstrate this case by means of an example:

\begin{verbatim}
MPI_Type_vector(3,10,20,MPI_INT,&filetype_vector);
MPI_Type_commit(&filetype_vector);

MPI_Type_vector(4,3,10,MPI_INT,&read_vector);
MPI_Type_commit(&read_vector);

MPI_File_set_view(fh, 0, MPI_INT, filetype_vector, "native",
    MPI_INFO_NULL);
MPI_File_read(fh, buf, 3, read_vector, status);
\end{verbatim}

The same mechanism can be applied for writing a file.

\subsubsection{Further Data Access Routines}

The routine for updating the file pointer \textit{MPI\_File\_seek} is based on \textit{etype} units as well. Thus, the offset must be converted in the same way as we have stated for the routines with explicit offsets, for example, \textit{MPI\_File\_read\_at}. Next, the actual seek operation is performed where we have to distinguish between two cases. As we have already stated, the file view can either be contiguous or non-contiguous. First case means that the so-called normal seek operation does not suffice. Thus, the ViPIOS server automatically realizes that internally a \textit{ViPIOS\_Seek\_struct} is called. However, the only consequence for the MPI-IO implementation is to call the ViPIOS routine \textit{ViPIOS\_Seek} such that the displacement of the view is added to the offset:

\begin{verbatim} 
if (help_fh->contig_filetype)
{
    
    switch (whence)
    {
        case SEEK_SET:
            res=ViPIOS_Seek (help_fh->ViPIOS_fh, offset+
                             help_fh->disp,whence);
            break;
        ...
     }
}
else
    res=ViPIOS_Seek (help_fh->ViPIOS_fh, offset, whence);
\end{verbatim}

The current position in \textit{etype} units can be retrieved by the routine \textit{MPI\_File\_-get\_position} which in turn calls the routine \textit{ViPIOS\_File\_get\_position}. Since the value which is received from the ViPIOS server is given in bytes rather than in \textit{etype} units, the byte value has to be converted which is done by the routine \textit{byte\_to\_etype}:

\begin{verbatim}
int byte_to_etype(MPI_Datatype datatype, int count)
{
    switch (datatype)
    {
        case MPI_SHORT:
            count/=sizeof(short int);
            break;
        case MPI_INT:
            count/=sizeof(int);
            break;
        case MPI_LONG:
            count/=sizeof(long int);
            break;
        case MPI_UNSIGNED_CHAR:
            count/=sizeof(unsigned char);
            break;
        case MPI_UNSIGNED_SHORT:
            count/=sizeof(unsigned short int);
            break;
        case MPI_UNSIGNED:
            count/=sizeof(unsigned int);
            break;
        case MPI_UNSIGNED_LONG:
            count/=sizeof(unsigned long int);
            break;
        case MPI_FLOAT:
            count/=sizeof(float);
            break;
        case MPI_DOUBLE:
            count/=sizeof(double);
            break;
        default: count/=sizeof(char);
    }
	 return count
}
\end{verbatim}

%% file: byte_off.tex
This routine converts a view relative offset given in \textit{etype} units into an absolute byte position relative to the current view. First, the offset which can be any multiple of \textit{etype} must be converted into bytes. This is handled by the routine \textit{convert\_datatype}. If the \textit{filetype} is contiguous, i.e. the view to the file does not contain any holes, the byte offset can be computed by: \\

\texttt{*disp=offset + ViPIOS\_fh-\(>\)disp;}\\

\begin{figure}
\begin{center}
\includegraphics[scale=0.9]{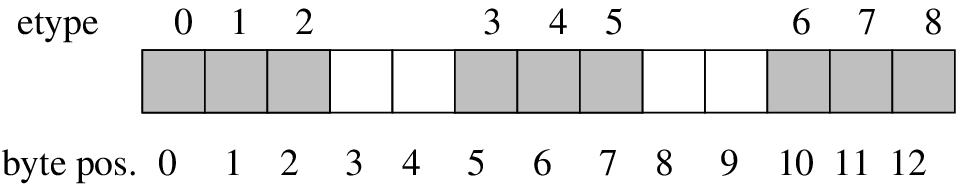}
\caption{File view}
\label{view10}
\end{center}
\end{figure}

Otherwise the ViPIOS access descriptor which holds the information about the file view must be used to compute the byte position. Assume following file view (Figure \ref{view10}).\\

In that example the offset 6 corresponds to the 11th byte position. If the view consists of nested derived datatypes, the byte position cannot be evaluated as straightforward as in that example. The routine \textit{Fill\_access\_descriptor}, on the one hand, computes the extent of the view including all so called holes, i.e. the size for each block is stored in \textit{sub\_count} which is a variable of the ViPIOS access descriptor. On the other hand, it evaluates the number of elements (\textit{sub\_actual} which can actually be accessed, i.e. the size of the view without holes. Thus, the whole structure of the ViPIOS access descriptor is traversed recursively and \textit{sub\_count} as well as \textit{sub\_actual} are evaluated for each basic block.\\

Assume following values for the ViPIOS access descriptor:\\

Level 1: \\
\texttt{no\_blocks=1,\\
offset=0;\\
repeat=2;\\
count=5;\\
stride=5;} \\

Level 2:\\
\texttt{no\_blocks=1;\\
offset=0;\\
repeat=3;\\
count=2;\\
stride=20;}\\

Let us analyze the functionality of the recursive routine by means of the previous example. When the function is called for the first time, \textit{cur\_basic-\(>\)subtype} is \textit{true}, since our view consists of nested datatypes. \textit{sub\_count}=0 and \textit{read\_count}=0. Now the function is called recursively with the values \textit{Fill\_access\_desc(cur\_basic-\(>\)subtype,0,0)}. \\

This time  \textit{cur\_basic-\(>\)subtype} is \textit{false} which means that the function is not called again but \textit{sub\_count, read\_count, cur\_basic-\(>\)sub\_count} and \textit{sub\_actual} are set to 1. Furthermore, following values are computed:\\

\texttt{read\_sum=2 * 5 * 1 = 10;\\
count=0 + 2 * (5 * 1 + 5) - 5  = 15;\\
count=15 + 0 = 15;\\
act\_read=10;}\\

This means that the extent of the inner basic block is 15 whereas only 10 elements can actually be accessed since the stride of that access pattern is 5. However, the variables \textit{sub\_count} and \textit{sub\_actual} are 1 because no more basic blocks exist.\\

When the inner incarnation of the recursive function is completed, the outer incarnation is resumed after the function call. Thus, \textit{cur\_basic-\(>\)sub\_count}=15 and \textit{cur\_basic-\(>\)sub\_actual}=10, i.e. they are assigned the values which are computed by the inner incarnation and hold the information about the extent and the actual number of accessible elements of the sub basic block. Then, the extent of the whole access pattern (view) can be evaluated: \\

\texttt{read\_sum=3 * 2 * 10 = 60;\\
count=0 + 3* (2 * 15 + 20) - 20 = 130;\\
count = 130 + 0 = 130;\\
read\_act=60;}\\

The extent of the outer basic block is 130 whereas 60 byte elements can actually be accessed. \\

On preparing these values we can now derive the absolute byte position from the relative one. This is handled by the routine \textit{Get\_absolute\_byte\_position}. Let us again explain the functionality by means of our previous example. First, the actual size of each basic block is evaluated:\\

\textit{cur\_actual=cur\_basic-\(>\)repeat*cur\_basic-\(>\)count*cur\_basic-\(>\)sub\_actual;}\\

In our example the outer basic block consists of 3 blocks (\textit{repeat}=3) with the size of \textit{count}=2. Since the view is a nested derived datatype, the size of each element is not 1 but \textit{sub\_actual}=10. Thus, \textit{sub\_actual} retrieves the number of elements of the sub basic block. The number of elements of the basic block is:\\

\texttt{cur\_actual= 3 * 2 * 10 = 60;}\\

Next, the start offset is checked whether it lies between the range of the first basic block (\textit{start\_offset<cur\_actual}). If \textit{start\_offset} is greater than \textit{cur\_actual}, the function can only resume if a further basic block exists otherwise the position of  \textit{start\_offset} would lie beyond the range of the view. \\

The next example will demonstrate this case. Assume that our \textit{start\_offset}, i.e. relative byte position,  is 23. The corresponding absolute byte position is 53. Since 23 lies within the range of the first basic block, we resume our computation. In contrast, if our relative offset were greater than 60 the absolute byte offset would be beyond the range of the view. However, the function \textit{Get\_absolute\_byte\_position} assumes that the view is circular, i.e. the new start offset is computed as long as it fits into the actual length of the view:\\

\texttt{start\_offset-=cur\_actual};\\

For example, if the actual length of our view is \textit{cur\_actual}=60, the total extent \textit{sub\_count}=130 and our absolute start offset is 143, then the absolute byte position of 143 is computed as follows:\\

\texttt{start\_offset=143-60=83;\\
start=0 + 3 * (2 * 15 + 20) - 20 + 0 =130 (sub\_count);\\
...
start\_offset=83-60=23;\\
start=130 + 3 * (2 * 15 + 20) - 20 + 0 = 260;\\
...
start=260 + 53 = 313;}\\

Thus, the absolute byte position of \textit{start\_offset}=143 is 303 when we assume that the view of the previous example is circular.\\

Let us now find out how the absolute position of \textit{start\_offset}=23 is evaluated. First, we have to find out the sub block which is referred by the offset.\\

\texttt{blocks=start\_offset / cur\_basic-\(>\)sub\_actual;\\
23 / 10 = 2;}\\

Since our view consists of \textit{repeat*count}=3*2=6 sub blocks, we know that the absolute byte position must be in the range of the third sub block (\textit{blocks}=2 whereas the first sub block is referenced by 0). Next we have to evaluate the byte position of the beginning of the third sub block. This is done by adding the size of the first two sub blocks to the stride between sub block 2 and 3 plus a possible offset:\\

\texttt{*start= blocks*cur\_basic-\(>\)count + (blocks/cur\_basic-\(>\)count)\\
*cur\_basic-\(>\)stride + cur\_basic-\(>\)offset;\\
start= 2*15 + (2/2)*20 + 0 = 50;}\\

Then, the offset within the third sub block is computed:\\

\texttt{start\_offset\%=cur\_basic-\(>\)sub\_actual\\
start\_offset= 23 \% 10 = 3;}\\

Since \textit{cur\_basic-\(>\)subtype} is \textit{true}, i.e. the view consists of a nested derived datatype and consequently further sub blocks exist, the function is called recursively with the values \textit{start\_offset}=3, \textit{start}=50. Then, the new values are evaluated accordingly:\\

\texttt{cur\_actual=2*5*1=10;\\
blocks= 3 / 1 = 3; \\
start= 50 + 31 + (3/5)*5 + 0 = 53;\\
start\_offset= 3 \% 1= 3;}\\

As no further sub blocks exist the variable \textit{start} holds the absolute byte position 53.

%% file: consist.tex
\subsection{File interoperability and Consistency semantics}

The routine \textit{MPI\_File\_get\_type\_extent} returns the extent of datatypes in the file, if the datatype is no NULL type. \\

The routines which handle the file consistency among parallel processes, namely \textit{MPI\_File\_set\_atomicity} and \textit{MPI\_File\_sync} are provided by the interface but are not treated explicitly be the ViPIOS server since every data access operation in ViPIOS is atomic. Thus, non-atomic mode is not supported yet.

%% file: advanced_datatypes.tex
\subsection{Advanced Derived Datatypes}

In our introductory chapter on MPI we have presented several MPI derived data-types. We have already mentioned that they can consist of multiple basic datatypes located either contiguously or non-contiguously in memory. Furthermore, the datatype created by a datatype constructor can be used as an input datatype for other derived datatypes. Thus, it is possible to build nested derived datatypes. The great advantage of such constructions is that noncontiguous data can be accessed with one command rather than reading the first chunk of data, skipping the data not required, reading the next chunk and so forth. In this section we will present two further derived datatypes which are part of the MPI-2 standard but not implemented in MPICH-1.1.

\subsubsection{Subarray Datatype Constructor}

{\sffamily
\begin{tabbing}
  {\bfseries MPI\_Type\_create\_subarray (int ndims, int *array\_of\_sizes,}\\
  {\bfseries int *array\_of\_subsizes, int *array\_of\_starts, int order,}\\
  {\bfseries MPI\_Datatype oldtype, MPI\_Datatype *newtype}\\

INOUT	\=parameter********* \= description                          \kill
IN    \>ndims           	\>number of array dimensions \\
IN    \>array\_of\_sizes  	\>number of elements of oldtype in each \\
	\>				\>dimension of the full array\\	
IN    \>array\_of\_subsizes    	\>number of elements of oldtype in each\\
	\>				\>dimension of the subarray\\
IN 	\>array\_of\_starts		\>starting coordinates of the subarray\\
	\>				\>in each dimension\\
IN    \>order	          	\>array storage order \\
IN 	\>oldtype			\>array element datatype\\
OUT   \>newtype            	\>new datatype
\end{tabbing}
}

\begin{figure}
\begin{center}
\includegraphics[scale=0.9]{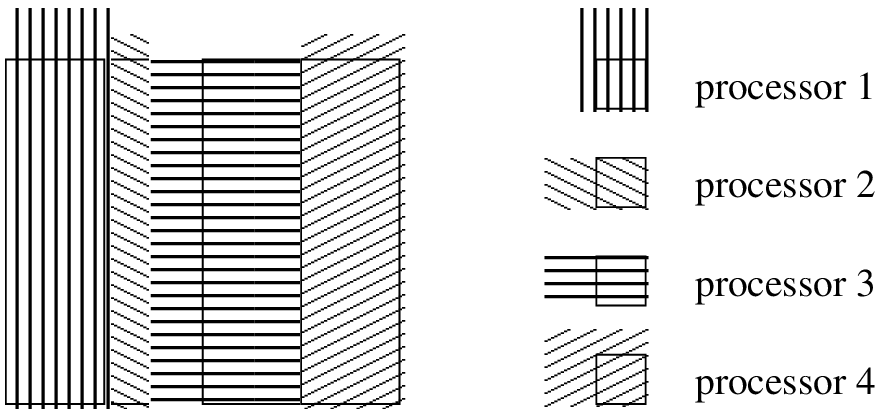}
\caption{Subarray}
\label{subarray1}
\end{center}
\end{figure}

This datatype allows describing an n-dimensional subarray of an n-dimen-sional array whereas the subarray can be located at any position within the array. Thus, a global array can be distributed onto several processors such that each one gets a certain section of the array. Assume a 12x12 array should be distributed onto 4 processes. Further assume that each processor gets 3 consecutive columns of the array as depicted in Figure \ref{subarray1}.\\

By the help of that example we will explain this datatype. \textit{ndims} defines the number of dimensions of the global array. In our case this value is 2. Moreover, this parameter specifies the number of elements of the next three parameters \textit{ array\_of\_sizes[], array\_of\_subsizes[], array\_of\_starts[]}. The size of the global array and the subarray are defined by \textit{array\_of\_sizes[]} and \textit{array\_of\_subsizes}. The location of the subarray within the global array is specified by \textit{array\_of\_starts[]}. Hence, \textit{array\_of\_starts[0]=0, array\_of\_starts[1]=3} describes the subarray of the second processor which starts at the position 0 in the first dimension and at position 2 in the second dimension. Since C and FORTRAN use different orders for addressing arrays, the order can be defined by the parameter \textit{order} and can either be \textit{MPI\_ORDER\_C}, i.e. row-major order, or \textit{MPI\_ORDER\_FORTRAN}, i.e. column-major order. The remaining parameters refer to the datatype of the global array and the name of the created derived datatype.\\

In the following example we present how this datatype can be used with MPI-IO. In our example we assume that the master process, i.e. process with rank 0, firstly writes the data to the file before the new derived datatype can be applied. Thus, on setting the file view, each process can only access the part of the global array according to Figure \ref{subarray1}.\\

\begin{verbatim}

MPI_File     fh;
MPI_Datatype subarray;

int
  rank,
  subarray[4][3],
  array_sizes[2],
  subarray_sizes[2],
  start_pos[2];

MPI_Init(&argc,&argv);
MPI_Comm_rank(MPI_COMM_WORLD,&rank);

/* specify the size of the global array and the subarray */
array_sizes[0]=12;
array_sizes[1]=12;
subarray_sizes[0]=4;
subarray_sizes[1]=3;

/* calculate location of the subarray according to the rank */
start_pos[0]=0;
start_pos[1]=rank*3;	

MPI_Type_create_subarray(2, array_sizes, subarray_sizes,
    start_pos,   MPI_ORDER_C, MPI_INT, &filetype);
MPI_Type_commit(&filetype);

/* each process reads a particular part of the file */   	
MPI_File_open (MPI_COMM_WORLD, "ufs:array_file", MPI_MODE_RDWR,
    MPI_INFO_NULL, &fh );
MPI_File_set_view (fh, 0, MPI_INT, subarray, "native", MPI_INFO_NULL);
MPI_File_read_all(fh, subarray, 4*3, MPI_INT, &status);
MPI_File_close(&fh);

MPI_Type_free(&subarray);
MPI_Finalize();
\end{verbatim}

\subsubsection{Distributed Array Datatype Constructor}

The derived datatype we discussed previously allows accessing different subarrays of a global array where each subarray is regarded as one block. The following derived datatype supports HPF-like distribution patterns like BLOCK-BLOCK distribution or CYCLIC-CYCLIC distribution. Before we give a description of this derived datatype we will give a brief introduction to HPF-distribution patterns. \\

\begin{figure}
\begin{center}
\includegraphics[scale=0.9]{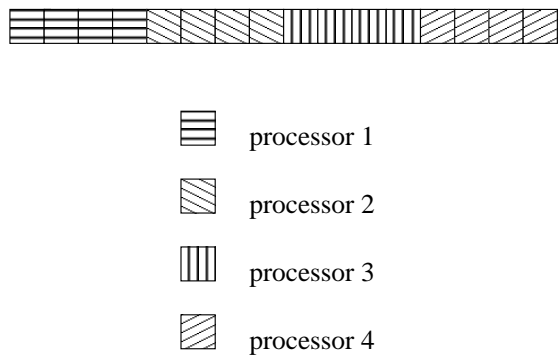}
\caption{BLOCK(4) Distribution}
\label{block4}
\end{center}
\end{figure}

Basically two different distribution patterns are possible, namely BLOCK and CYCLIC. Let us start with a file which can be regarded as a one-dimensional array consisting of 16 elements. Further assume that we have a processor grid of four processors and we wish to distribute the file onto these four processors in the distribution pattern BLOCK(4). This means that each processor gets one contiguous block of the file consisting of 4 elements. The distribution pattern is depicted in Figure \ref{block4}. \\

\begin{figure}
\begin{center}
\includegraphics[scale=0.9]{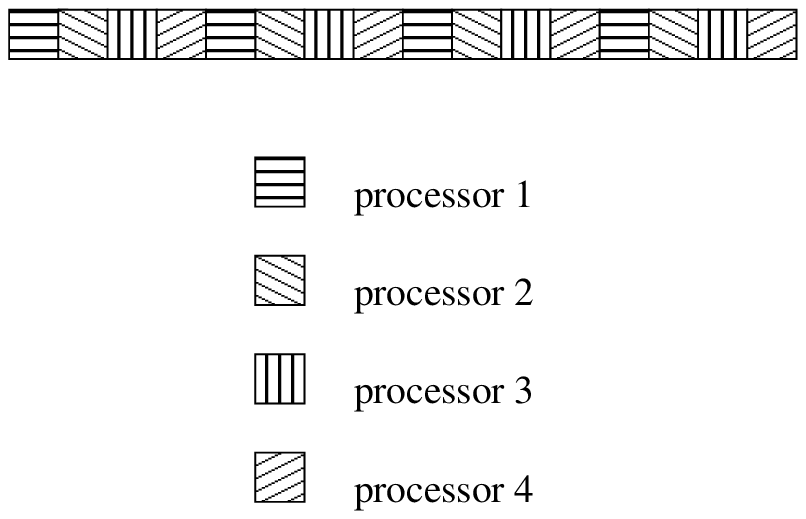}
\caption{CYCLIC(1) Distribution}
\label{cyclic1}
\end{center}
\end{figure}

Another distribution pattern, for instance, is CYCLIC(1). This means that the data is distributed in a round robin fashion onto the four processors. In this case, processor 1 gets the first element of the file, processor 2 gets the second element of the file and so on. When each processor has got one element, this process is repeated again. Thus, in the second run processor 1 gets the fifth element, processor 2 the sixth and so on. Figure \ref{cyclic1} demonstrates the result of this distribution pattern.\\

Now assume that our file corresponds to a 2-dimensional array rather than a 1-dimensional one. In particular, our file should be an 8x8 array. Thus, more complex distribution patterns are possible by combining the ones we discussed so far. For example, BLOCK-CYCLIC means that the distribution pattern in the first dimension is BLOCK whereas the distribution pattern in the second dimension is CYCLIC. In order to get familiar with these patterns, we will give some examples. \\

\begin{figure}
\begin{center}
\includegraphics[scale=0.9]{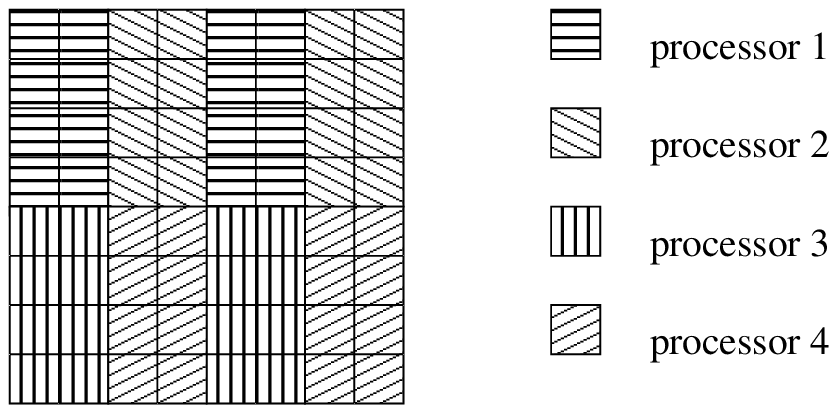}
\caption{BLOCK(4), CYCLIC(2) Distribution}
\label{b4c2}
\end{center}
\end{figure}

In our first example we assume an 8x8 array which shall be distributed in the first dimension according to BLOCK(4) and in the second one according to CYCLIC(2). Further assume that our processor grid consists of 4 processors such that each dimension comprises two processors. The result can be seen in Figure \ref{b4c2}.\\

\begin{figure}
\begin{center}
\includegraphics[scale=0.9]{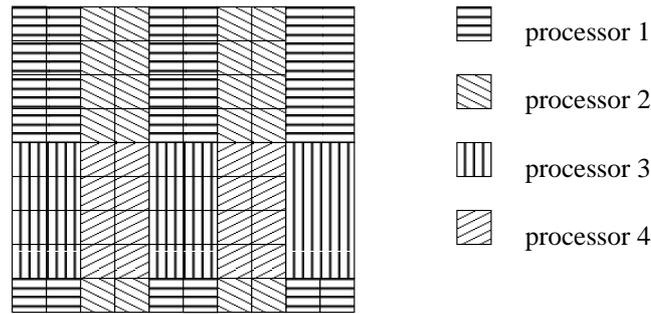}
\caption{BLOCK(4), BLOCK(2) Distribution With Irregular Patterns}
\label{b4c2_1}
\end{center}
\end{figure}

In the next example we want to distribute a 9x10 array according to the pattern CYCLIC(2), CYCLIC(2). In contrast to our previous example not each processor gets the same number of elements. The result can be seen in Figure \ref{b4c2_1}.\\

On giving a brief introduction to HPF-like distribution patterns, we can now resume our discussion on the derived datatype for distributed arrays.

{\sffamily
\begin{tabbing}
  {\bfseries MPI\_Type\_create\_darray (int size, int rank, int ndims}\\
  {\bfseries int *array\_of\_gsizes, int *array\_of\_distribs,}\\
  {\bfseries int *array\_of\_dargs, int *array\_of\_psizes, int order}\\
  {\bfseries MPI\_Datatype oldtype, MPI\_Datatype *newtype)}\\

INOUT	\=parameter********* \= description                          \kill
IN    \>size	           	\>size of process group \\
IN    \>rank	           	\>rank in process group \\
IN    \>ndims           	\>number of array dimensions \\
	\>				\>as well as processor grid dimensions\\
IN    \>array\_of\_gsizes  	\>number of elements of oldtype in each \\
	\>				\>dimension of the global array\\	
IN    \>array\_of\_distribs    	\>distribution of array in each dimension\\
IN 	\>array\_of\_dargs		\>distribution argument in each dimension\\
IN 	\>array\_of\_psizes		\>size of process grid in each dimension \\
IN    \>order	          	\>array storage order \\
IN 	\>oldtype			\>array element datatype\\
OUT   \>newtype            	\>new datatype
\end{tabbing}
}

By means of \textit{MPI\_Type\_create\_darray} a datatype can be created that corresponds to the distribution of an \textit{ndims}-dimensional array onto an \textit{ndims}-dimensional array of logical processes. \textit{size} states the number of  processes in the group. \textit{ndims} specifies the number of dimensions of the array to be distributed as well as the number of dimensions of the processor grid. The size of each processor grid is given in \textit{array\_of\_psizes[]}. It is important to note the following equation must be satisfied:

\begin{displaymath}
\sum_{i=0}^{ndims-1}
array \ of \ psizes[i]=size
\end{displaymath}\\

\begin{figure}
\begin{center}
\includegraphics[scale=0.9]{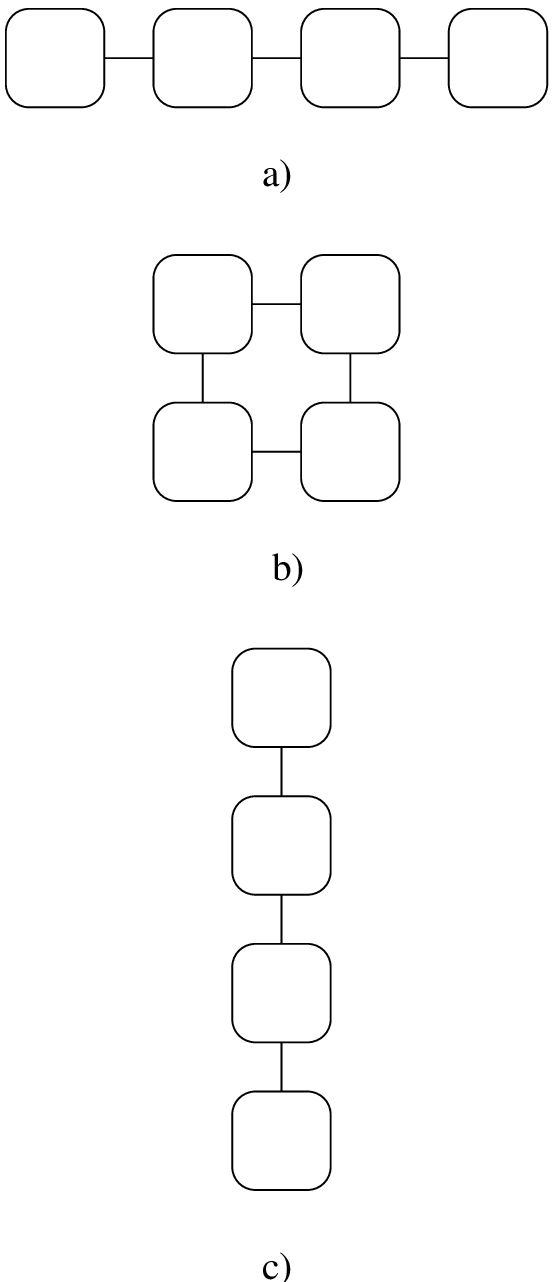}
\caption{Processor Grid}
\label{processor_grid}
\end{center}
\end{figure}

Let us explain this with an example. Suppose that our global array is a 2-dimensional 16x16 array and we wish to distribute it onto \textit{size}=4 processes. Since our global array consists of \textit{ndims}=2 dimensions, our processor grid consists of two dimensions as well. Thus, we have three different possibilities for constructing our processor grid. The first dimension could consist of \textit{array\_of\_psizes[0]}=1 processor and the second of \textit{array\_of\_psizes[1]}=4 (1*4=4) processors. In this case the processor grid corresponds to shape a in Figure \ref{processor_grid}. In shape b both dimensions comprise 2 processors. Shape c is the opposite of shape a, i.e. dimension 1 consists of 4 processors whereas dimension 2 consists of 1 processor.\\

This processor grid serves as the basis for the distribution pattern specified by the array \textit{array\_of\_distribs[]}. In particular, each dimension of the global array can be distributed in three different ways:

\begin{itemize}
  \item MPI\_DISTRIBUTE\_BLOCK- corresponds to BLOCK distribution
  \item MPI\_DISTRIBUTE\_CYCLIC- corresponds to CYCLIC distribution
  \item MPI\_DISTRIBUTE\_NONE - this dimension is not distributed
\end{itemize}

The distribution argument of each dimension is stored in \textit{array\_of\_dargs[]} and can either be a number or the constant MPI\_DISTIRBUTE\_DFLT\_DARG whereas following assumption must be satisfied:\\

\textit{array\_of\_dargs[i] * array\_of\_psizes[i]\(>\)=array\_of\_gsizes[i]}.\\

Let us again take a look at one example in order to see the functionality of that datatype. In particular we wish to take a look at the code fragment of an application program which distributes a 9x10 array onto four processes in the pattern CYCLCIC(2), CYCLIC(2) as we have seen in Figure \ref{b4c2_1}. Assume that the array storage order is C.

\begin{verbatim}
MPI_Comm_rank(MPI_COMM_WORLD, &rank);
MPI_Comm_size(MPI_COMM_WORLD, &nprocs);

ndims=2
order=MPI_ORDER_C

/* size of the array to be distributed */
array_of_gsizes[0]=9;
array_of_gsizes[1]=10;

/* distribution and distribution argument */
array_of_distribs[0]=MPI_DISTRIBUTE_CYCLIC;
array_of_distribs[1]=MPI_DISTRIBUTE_CYCLIC;
array_of_dargs[0]=2;
array_of_dargs[1]=2;

/* processor grid */
array_of_psizes[0]=2;
array_of_psizes[1]=2;

MPI_Type_create_darray(nprocs, rank, ndims, array_gsizes,
    array_of_distribs, array_of_dargs, array_of_psizes,
    order, MPI_INT; &newtype);
MPI_Type_commit(&newtype);
\end{verbatim}

\subsubsection{Implementation of the Subarray Datatype Constructor}

On getting used to the functionality of the derived datatypes we can now turn to describing the implementation of \textit{MPI\_Type\_create\_subarray}. The basic information for the implementation of this and the following derived datatype can be found in ROMIO.\\

First, the input parameters to that derived datatype must be checked for being correct. In particular none of the entries for the parameters \textit{ndims, array\_of\_sizes[], array\_of\_subsizes[]} must be less than 1. Furthermore, \textit{array\_of\_subsizes[i]} must not be less than \textit{array\_of\_sizes[i]}. \textit{array\_of\_starts[i]} must not be less than 0 or \textit{array\_of\_starts[i] \(>\) (array\_of\_sizes[i]-array\_of\_sub-sizes[i]}.
Next the parameters \textit{oldtype} and \textit{order} are checked. \\

According to \textit{order} the datatype is built by using the existing derived datatypes of MPI-1 we discussed in a previous chapter. Lets take a look at the implementation of the \textit{order = MPI\_ORDER\_C}. Recall that in this case the last dimension of the array changes fastest since the ordering used by C arrays is row-major. For example, assume a 4x4 array. Thus, we address it by the indices [0][0], [0][1],[0][2],[0][3],[1][0] and so forth. The code fragment will help explaining the algorithm:

\begin{verbatim}
if (order == MPI_ORDER_C)
{
    // dimension ndims-1 changes fastest
    if (ndims == 1)
        MPI_Type_contiguous(array_of_subsizes[0], oldtype, &tmp1);
    else
    {
        MPI_Type_vector(array_of_subsizes[ndims-2],
        array_of_subsizes[ndims-1],
        array_of_sizes[ndims-1], oldtype, &tmp1);
	
        size = array_of_sizes[ndims-1]*extent;

        for (i=ndims-3; i>=0; i--) {
            size *= array_of_sizes[i+1];
        MPI_Type_hvector(array_of_subsizes[i], 1, size, tmp1, &tmp2);
        tmp1 = tmp2;
    }
}
\end{verbatim}

Next, the dimension \textit{ndims} of the derived data is checked. If \textit{ndims} is 1, we can reduce the whole datatype to \textit{MPI\_Type\_contiguous} since the file can only be accessed in a contiguous block with the size of \textit{array\_of\_subsizes[]}. \\

\begin{figure}
\begin{center}
\includegraphics[scale=0.9]{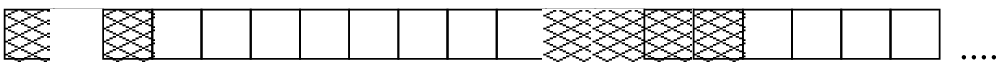}
\caption{Vector}
\label{vector12_3_12}
\end{center}
\end{figure}

If \textit{ndims} is greater than 1, the derived datatype can be built with the datatype \textit{MPI\_Type\_vector}. Recall the example from the previous section where we distributed a 12x12 array onto 4 processes such that each process gets 3 consecutive columns of the array. Since the parameters to \textit{MPI\_Type\_vector} are \textit{count, blocklength} and \textit{stride}, the section for the first process can be described by \textit{MPI\_Type\_vector(12,3,12)}. The pattern is depicted in Figure \ref{vector12_3_12}.\\

In general, \textit{count} corresponds to \textit{array\_of\_subsizes[ndims-2]}, \textit{blocklength} to \textit{array\_of\_subsizes[ndims-1]} and \textit{stride} to \textit{array\_of\_sizes[ndims-1]}. If \textit{ndims} is of an order higher than 2, a further derived datatype, namely \textit{MPI\_Type\_hvector}, is used for describing the remaining dimensions. \\

Next, we have to add the displacement for each process. For example, process 2 should read the forth, fifth and sixth columns which is defined by \textit{disp[1]=array\_of\_starts[1]=3}. Since each process is supposed to access a different block of the array, the start displacement \textit{disps[1]} is different for each process. Finally, the derived datatype \textit{MPI\_Type\_struct} is used for describing the section which should be accessed by each process. In particular, this datatype consists of three blocks whereas the second block contains the information of the access pattern we have described by \textit{MPI\_Type\_contiguous}, \textit{MPI\_Type\_vector} or \textit{MPI\_Type\_hvector}. The first block is of the type \textit{MPI\_LB} and the third of \textit{MPI\_UB}. \textit{MPI\_LB} and \textit{MPI\_UB} are so-called "pseudo-datatypes" which are used to mark the lower bound and the upper bound of a datatype. Since the extent, i.e. the span from the first to the last byte, of both datatypes is 0, they neither effect the size nor count of a datatype. The reason for using these datatypes is to define explicitly the lower and upper bound of a derived datatype. For more information we refer the reader to \cite{MPI}. \\

The implementation is shown in the following code fragment. However, a detailed description is given in the next section when we discuss the implementation of the derived datatype \textit{MPI\_Type\_create\_subarray}.

\begin{verbatim}
// add displacement and UB 	
disps[1] = array_of_starts[ndims-1];
size = 1;
for (i=ndims-2; i>=0; i--) {
    size *= array_of_sizes[i+1];
    disps[1] += size*array_of_starts[i];
}

disps[1] *= extent;
disps[2] = extent;
for (i=0; i<ndims; i++)
    disps[2] *= array_of_sizes[i];

block_lens[0] = block_lens[1] = block_lens[2] = 1;

types[0] = MPI_LB;
/* datatype we described previously */
types[1] = tmp1;
types[2] = MPI_UB;

MPI_Type_struct(3, block_lens, disps, types, newtype);
\end{verbatim}

\subsubsection{Implementation of the Distributed Array Datatype Constructor}

In this section we will describe the implementation of our last derived datatype, namely \textit{MPI\_Type\_create\_subarray}. On checking the input parameters for correctness we can start the implementation according to the order of the array. Similar to the previous section we will concentrate on the C order whereas we will not go into detail with describing the FORTRAN order since the assumptions are only slightly different.\\

Before any distribution pattern can be taken into account each process must be mapped to the processor grid. This is done according to the formula given in \cite{MPI-2}:

\begin{verbatim}
procs=size; /* number of processes in that process group */
tmp_rank=rank;
for (i=0; í<ndims; i++)
{
    procs = procs / array_of_psizes[i];
    coords[i] = tmp_rank / procs;
    tmp_rank = tmp_rank % procs;
}
\end{verbatim}

Let us demonstrate this functionality of this formula by means of the process with rank=2. We assume that our processor grid consists of 2 processors per dimension as depicted in Figure \ref{processor_grid}. On applying the correct values to the formula, namely \textit{procs=4}, \textit{tmp\_rank}=2 and \textit{ndims}=2, we yield the result \textit{coord[0]=1} and \textit{coord[1]=0}. Thus, the process with rank 2 corresponds to the processor with the coordinates [0,1], i.e. the processor in the lower left corner of the processor grid.\\

Next, each dimension of the array is analyzed separately starting from the highest dimension. Since we will concentrate on C-order, the last dimension changes fasted. We have already stated that \textit{MPI\_Type\_create\_darray} allows three different ways of distributing the array, namely MPI\_DISTRIBUTE\_-BLOCK, MPI\_DISTRIBUTE\_CYCLIC and MPI\_DISTRIBUTE\_NONE. Since latter case can be reduced to a BLOCK distribution on only one process, i.e. no distribution is actually performed, we merely have to discuss the routines which handle the first and the second case.\\

\paragraph{BLOCK-Distribution}

Let us start off with the routine for the BLOCK distribution. First, the distribution argument \textit{darg} is checked.  If it is set to MPI\_DISTRIBUTE\_DFLT\_-DARG, the block size of the corresponding process is set to:

\begin{verbatim}
if (darg== MPI_DISTRIBUTE_DFLT_DARG)
    blksize= (global_size + nprocs-1) / nprocs;
else
    blksize=darg;
\end{verbatim}

according to \cite{MPI-2} whereas \textit{global\_size} corresponds to \textit{array\_of\_gsizes[i]}, i.e. the size of the array in that particular dimension, otherwise \textit{blksize} is set to the size of the distribution argument.\\

\begin{figure}
\begin{center}
\includegraphics[scale=0.9]{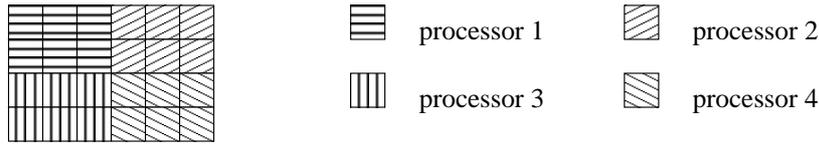}
\caption{4x6 Array With BLOCK(2),BLOCK(3) Distribution}
\label{array46}
\end{center}
\end{figure}

Next, we must check whether each process gets the same number of elements. For example, if wish to distribute a 1-dimesional array consisting of 6 elements onto 2 processors according to the pattern BLOCK(4). Thus, the process with rank 0 gets the first four blocks, whereas process 1 merely gets 2 elements. The following code fragment handles this case:

\begin{verbatim}
j=global_size-blksize*rank;
mysize=min(blksize,j);
if ( mysize < 0 )
    mysize=0;
\end{verbatim}

Now we are ready to implement the block-distribution by means of derived datatypes we have already discussed in the chapter on MPI. In particular, we have to bear in mind the dimension of the array we are currently analyzing. Assume a 4x6 array with a distribution pattern of BLOCK(2), BLOCK(3) as depicted in Figure \ref{array46}. This 2-dimensional array can be linearized to a 1-dimensional array as presented in the same figure. The distribution pattern of the last dimension, i.e. dimension 1, BLOCK(3) can be described by the derived datatype \textit{MPI\_Type\_contiguous}. All other dimensions must be described by \textit{MPI\_Type\_hvector} with the correct stride. What is more, the \textit{oldtype} of this datatype must be the datatype of the previous dimension, i.e. a nested derived datatype is built.\\

\begin{figure}
\begin{center}
\includegraphics[scale=0.9]{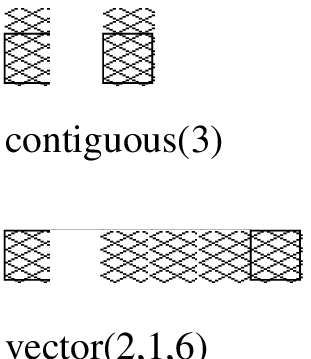}
\caption{Vector Distribution Pattern}
\label{block_example1}
\end{center}
\end{figure}

In our example the distribution pattern of dimension 1 is BLOCK(3). Thus, following datatype is used:

\begin{verbatim}
MPI_Type_contiguous(3, MPI_INT, dim1);
\end{verbatim}

Since dimension 0 is distributed according to BLOCK(2), this can be described by using the derived datatype \textit{MPI\_Type\_vector} whereas \textit{oldtype} is the datatype we created before. The stride is computed as the length of the array of the next dimension, i.e. \textit{array\_of\_gsize[1]}=6. Thus, the vector is:

\begin{verbatim}
MPI_Type_hvector(2,1,6,dim1,dim0);
\end{verbatim}

The two steps which are necessary, for simulating a BLOCK(2),BLOCK(3) distribution are shown in Figure \ref{block_example1}\\

In order to apply this algorithm for any dimension, the stride must be set accordingly, namely to:

\begin{verbatim}
for (i=nidms-1; i>dim; i--)
    stride *= array_of_gsizes[i];
\end{verbatim}

Thus, the formula is set accordingly for a n-dimensional array.\\

Finally, the start offset for each process must be set. Recall that the size of the first dimension of the array to be distributed is 4 and the distribution pattern is BLOCK(2). Thus, the first process in the grid of dimension 0 should read the first and the second elements whereas the second process should start accessing the third element. In general, this is handled by the following code:

\begin{verbatim}
*st_offset = block_size * rank;
\end{verbatim}

Let us now return to the point after calling the procedure for handling the block distribution. The loop for analyzing the distribution pattern, i.e. BLOCK, CYCLIC or NONE is repeated until the last dimension. However, each call returns a new value for the \textit{filetype} and the corresponding offset for each process as we described above. It is important to mention that each \textit{newtype} of the derived datatype, i.e. the distribution pattern of the particular dimension is used as the \textit{oldtype} of the next dimension. In other words, the derived datatype of one dimension is used for building the derived datatype of the next dimension and, thus, creating a nested derived datatype which describes the distribution pattern of all dimensions. A short abstract of the code shall demonstrate this:

\begin{verbatim}
if (order==MPI_ORDER_C)
{
    for (i=ndims-1, i>=0, i--)
    {
        switch(array_of_distribs[i]
        {
            case MPI_DISTRIBUTE_BLOCK:
                Block_distribution(...,type_old, &type_new,
                                   st_offsets+i);
		    break;
            case MPI_DISTRIBUTE_CYCLIC:
                Cyclic_distribution(...,type_old, &type_new,
                                   st_offsets+i);
                break;
		...
         }
        /* use the new datatype as the basis for the next
           datatype * /
        type_old = type_new;
    }	
	
    /* add displacement */
    disps[1]=st_offsets[ndims-1];
}
\end{verbatim}

After this loop we know the access pattern (\textit{type\_new}) of the particular process  and its start position \textit{disps[1]}. The easiest way to combine all this information is by using the derived datatype \textit{MPI\_Type\_struct} with the following values:

\begin{verbatim}
disps[0]=0;
disps[2]= multiple of array_of_gsizes;
block_lens[0]=block_lens[1]=block_lens[2]=1;
types[0]=MPI_LB;
types[1]=type_new;
types[2]=MPI_UB;

MPI_Type_struct(3, block_lens, disps, types, newtype);
\end{verbatim}

In short, the distribution pattern is made up by the second block of the derived datatype. The first and the third block are made up by the datatypes MPI\_LB and MPI\_UB.

\paragraph{CYCLIC-Distribution}

This section will be dedicated to the description of the second distribution pattern, namely CYCLIC. Similar to the previous section, we will merely take a look at the implementation of the C order since the implementation of the FORTRAN order only slightly differs. \\

\begin{figure}
\begin{center}
\includegraphics[scale=0.9]{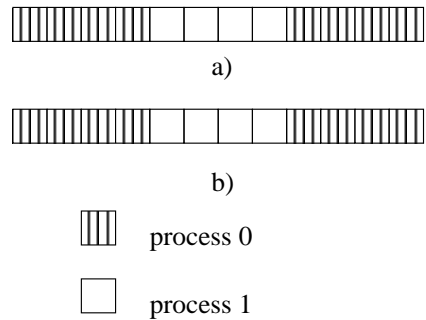}
\caption{Regular And Irregular Distribution Patterns}
\label{regular_irregular}
\end{center}
\end{figure}

First, the number of blocks per process is calculated according to a slightly modified version of the formula given in \cite{MPI-2}:

\begin{verbatim}
nblocks = ( array_of_gsizes[dim] + blksize-1 ) / blksize;
count = nblocks / nprocs;
left_over= nblocks - count * nprocs;
if (rank < left_over)
  count = count + 1;
\end{verbatim}

For example, if we wish to distribute a 1-dimesional array of 12 elements onto 2 processes with the distribution pattern CYCLIC(4) the first process gets two blocks whereas the second process only gets one block consisting of 4 elements (see Figure \ref{regular_irregular} a).\\

Next, the size of the last block is calculated. In our previous example, the size of each block was four elements. However, this is not true for an array comprising, for instance, 14 elements. Thus, each process gets two blocks but the last block of process 2 consists of only two elements rather than four (see Figure \ref{regular_irregular} b):\\

\begin{verbatim}
/* check whether any irregular block exists */
if ( (remaining_elements=array_of_gsizes[dim] %
     (nprocs*blksize)) != 0)
{
    /* compute the size of the last block */
    last_blksize=remaining_elements-blksize*rank;
        if ( (last_blksize < blksize) && (last_blksize > 0) )
        {
            count--;
            evoke_struct=1;
        }
}
\end{verbatim}

Again, \textit{blksize} refers to the distribution argument. In our case it is 2 - CYCLIC(2). Note that if the size of the last block is less than the remaining ones, \textit{count} is decremented by 1 and \textit{evoke\_struct} is set to 1. Thus, the last block is treated by means of the special derived datatype as described later on. \\

The regular blocks are treated in the following way. Let us explain each step by means of the introductory example where we have taken a look at the source code of an application program. Recall that the array size of the 2-dimensional array is 9x10, the distribution pattern is CYCLIC(2), CYCLIC(2) and the processor grid comprises 2 processes per dimension.\\

Since in the C order the last dimension changes fastest, we handle this case first. The second dimension consists of 10 elements. Thus, process 0 gets \textit{count}=3 blocks of the size \textit{blksize}=2 and process 1 gets 2 blocks of \textit{blksize}=2.

\begin{verbatim}
if (dim == ndims-1)
{
    stride = nprocs*blksize*orig_extent;
    MPI_Type_hvector(count, blksize, stride, type_old, type_new);
}
\end{verbatim}

Since the stride is given in bytes, it must be multiplied by the datatype of elements of the array. All other dimensions cannot be treated as straightforwardly. Therefore, we have to split up the access pattern into two derived datatypes, one which builds a sub block using the block size as the first parameter, and a second one which uses the number of blocks, i.e. \textit{count}, as the first parameter. Moreover, it is a nested datatype based on the previously created sub block. In addition, the strides for these two derived datatypes must be computed. Let us list the code fragment before we explain its meaning:

\begin{verbatim}
/* compute sub_stride and stride */
sub_stride = orig_extent;
stride = nprocs*blksize*orig_extent;

for (i=ndims-1; i>dim; i--)
{
    sub_stride *= array_of_gsizes[i];
    stride *= array_of_gsizes[i];
}
			
/* datatypes for sub_block and block */
MPI_Type_hvector (blksize, 1, sub_stride, type_old, &sub_block);
MPI_Type_hvector(count, 1, stride, sub_block, type_new);
\end{verbatim}

\begin{figure}
\begin{center}
\includegraphics[scale=0.9]{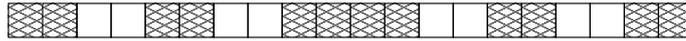}
\caption{Distribation Pattern of Dimension 1 }
\label{dd_dim1}
\end{center}
\end{figure}

We will demonstrate our approach for the first process. Assume that we have already built the derived datatype for the second dimension. Since the first dimension consists of 9 elements, process 1 gets 3 blocks whereas the last blocks is smaller than the previous ones. Thus, \textit{count} is reduced to 2. What is more, the block size of the regular blocks is 2 as well. Provided with that information we can now build the first derived datatype, which represents the sub block. Since our array comprises only two dimensions \textit{sub\_stride} is set to \textit{array\_of\_gsizes[2]}=10. For simplicity let us restrict to the number of elements and thus not regard the actual number of byte elements, which would be necessary for the derived datatype. Thus, the pattern of the derived datatype which describes the sub block looks like follows (Figure \ref{dd_dim1}):\\

\begin{figure}
\begin{center}
\includegraphics[scale=0.9]{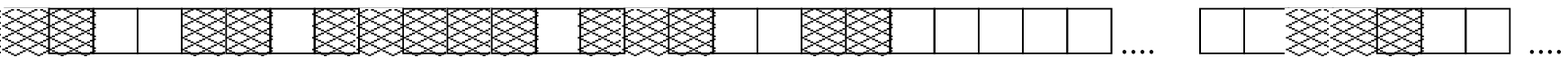}
\caption{Distrubution Pattern of Both Dimensions}
\label{dd_complete}
\end{center}
\end{figure}

Next the derived datatype which describes the whole access pattern is built. This is done by creating a derived datatype \textit{MPI\_Type\_hvector} such that the number of blocks corresponds to \textit{count}. What is more, the block size is 1 and the oldtype is the newtype from our previously created sub block. Finally, the stride must be set accordingly. First, the stride is multiplied by the number of processes of the particular processor grid dimension - in our case it is two - times the block size. Next, it is multiplied by the sizes of the array to be distributed. On building both derived datatypes the access pattern for process 1 depicted in Figure \ref{dd_complete} is yielded.\\

Since the first dimension consists of 9 elements rather than 8, we still have to add the last block. Recall that its block size is smaller than the remaining one and thus has to be treated separately:

\begin{verbatim}
if ( dim == ndims-1)
{
    types[0] = *type_new;
    types[1] = type_old;
    disps[0] = 0;
	
    if (count == 0 )
        disps[1]= blksize*(nprocs-1)*sub_stride;
    else
        disps[1] = count*stride;

    blklens[0] = 1;
    blklens[1] = last_blksize;
}

else
{
    // sub_type is added!!!
    MPI_Type_hvector (last_blksize, 1, sub_stride, type_old,
                      &sub_block_struct);

    types[0] = *type_new;
    types[1] = sub_block_struct;
    disps[0] = 0;

    if (count == 0 )
        disps[1]= blksize*(nprocs-1)*sub_stride;
    else
        disps[1] = count*stride;

    blklens[0] = 1;
    blklens[1] = 1;
}

MPI_Type_struct(2, blklens, disps, types, &type_tmp);
*type_new = type_tmp;

\end{verbatim}

\begin{figure}
\begin{center}
\includegraphics[scale=0.9]{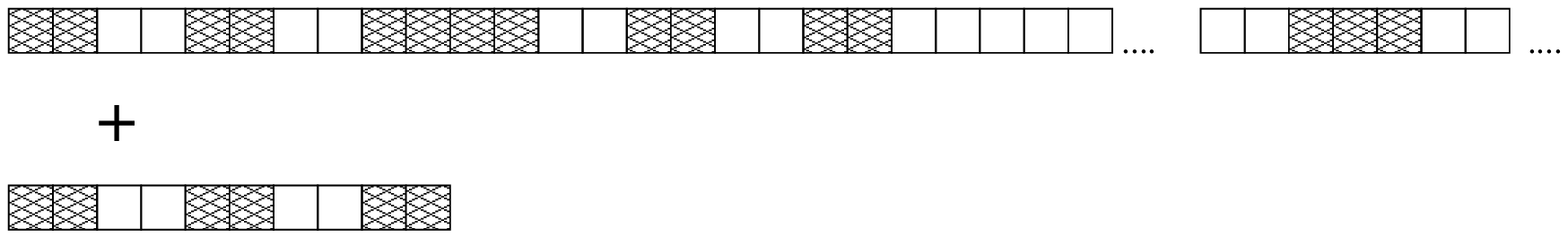}
\caption{Pattern of an Irregular Distribution}
\label{dd_complete_irreg}
\end{center}
\end{figure}

Since we are analyzing the first dimension, we have to take a look at the else branch of the outer if-statement. Again, a derived datatype for a sub block is build. Rather than taking \textit{blksize} as the first argument, \textit{last\_blksize} is used now. Next, the previously created datatype for the regular pattern is combined with the irregular pattern by means of \textit{MPI\_Type\_struct}. Note that the displacement of the irregular pattern depends on the number of blocks \textit{count} of the regular pattern. In our case the displacement can simply be computed by multiplying the number of blocks by the stride. The new pattern is depicted in Figure \ref{dd_complete_irreg}.\\

\begin{figure}
\begin{center}
\includegraphics[scale=0.9]{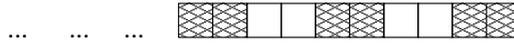}
\caption{Irregular Distribution Pattern of One Block}
\label{dd_irregular}
\end{center}
\end{figure}

Imagine that the first dimension of the processor grid consists of \textit{nprocs}=4 processes rather than 2. Further assume that we have already built the derived datatype for the second dimension as shown so far. Thus, assuming a 9x10 array with a distribution pattern CYCLIC(2) the first three processes get one block consisting of two elements whereas the last process gets one block of only one element. In this case \textit{count} is 0 and \textit{last\_blksize} is 1. Figure \ref{dd_irregular} demonstrates the effect of the new stride. Since \textit{sub\_stride} corresponds to the second dimension of the array, i.e. 10, \textit{stride} is set to 2*3*10=60.\\

After having built the derived datatypes for the regular as well as the irregular patterns, the last step is to add the lower and upper boundary as we described for the BLOCK distribution. Although we have only discussed the implementation for a 2-dimensional array, this algorithm works for n-dimensions as well. \\

%% file: fortran.tex
\subsection{Fortran-Interface}

In this section we want to state the most important points about converting a Fortran application program using MPI-IO to the C-interface.\\

The first thing to be mentioned is that every procedure call in Fortran is \textit{call by reference}. The name of a procedure can only consist of lower case letters and must be terminated with an underscore "\_". Next, strings must be treated with special care since there is no "null" which indicates the end of a character array. Thus, this indicator must be added to the character array.\\

In the following code fragment we will show the implementation of the MPI-IO routine \textit{MPI\_File\_open}. Note that the routine is a procedure rather than a function as it is true for the C implementation. Thus, the return value from the C function is stored in the parameter \textit{*ierror}. In order to distinguish between input and output parameters, input parameters are indicated by \textit{const}. Since the name of the file to be opened is a character array, it must be converted accordingly. First, the routine \textit{std\_str\_f\_to\_c} is called which in turn calls the routine \textit{strn\_copy0}. On returning from both routines, the Fortran filename is converted to a character array which can correctly be interpreted by the C routine and does the actual file opening. 

\begin{verbatim}
#define 		STRING80_LEN 80
typedef char	STRING80 [STRING80_LEN+1];

void mpi_file_open_ (const int *comm, const char *filename,
    const int *amode, const int *info, int *fh, int *ierror)
{
    STRING80 c_filename;
        
    /* convert a c-array into a fortran */
    std_str_f_to_c (filename, c_filename, STRING80_LEN);

    /* call the correspoding function from the c interface */
    *ierror=MPI_File_open (*comm, c_filename, *amode, *info, fh);
}

static void std_str_f_to_c (const char *fstr, char cstr [], size_t max)
{
    register int i;

     /* do nothing */
     for (i=max-1; i >= 0 && fstr[i] == ' '; i--) ;

    (void) strncpy0 (cstr, fstr, i+1);
}

static char *strncpy0 (char *s1, const char *s2, size_t n)
{
    (void) strncpy (s1, s2, n);

    *(s1+n) = '\0';

    return (s1);
}
\end{verbatim}

A further point to bear in mind when mapping from Fortran to C is the different interpretation of datatypes. For example, following MPI datatypes are supported by Fortran, which have no direct C MPI datatypes as their counterparts, namely:

\begin{itemize}
	\item MPI\_INTEGER
	\item MPI\_REAL
	\item MPI\_DOUBLE\_PRECISION
	\item MPI\_COMPLEX
	\item MPI\_LOGICAL
	\item MPI\_CHARACTER
	\item MPI\_BYTE
\end{itemize}

Thus, the converting function, which is used in the Fortran MPI-IO interface is:

\begin{verbatim}
int MPI_Type_f2c (MPI_Datatype datatype)
{
    switch (datatype)
    {
        case MPI_CHARACTER:
            return MPI_CHAR;
            break;
		
        case MPI_INTEGER:
            return MPI_INT;
            break;
			
        case MPI_REAL:
            return MPI_FLOAT;
            break;

        case MPI_DOUBLE_PRECISION:
            return MPI_DOUBLE;
            break;

        case MPI_LOGICAL:
            return MPI_UNSIGNED;
            break;						

        default: return MPI_BYTE;
    }
}
\end{verbatim}

This datatype conversion is used, for example, in each routine for reading or writing a file. Let us therefore take a look at the mapping function for reading:

\begin{verbatim}
void mpi_file_read_ (const int *fh,void *buf,const int *count, 
        const MPI_Datatype  *datatype, int *status, int *ierror )
{
    int                     res;
    MPIO_Status   status_c;

    MPI_Datatype datatype_c;

    datatype_c = MPI_Type_f2c(*datatype);

    *ierror=MPI_File_read(*fh, buf, *count, datatype_c, &status_c);

    *status=status_c.fid;
}
\end{verbatim}

The forth parameter, namely  \textit{datatype}, is converted from the Fortran MPI representation to the C MPI representation via the routine \textit{MPI\_Type\_f2c}.\\

We still need to focus out attention to the parameter \textit{status} which is of type \textit{MPIO\_Status} in the C interface. Since all objects in FORTRAN are of type integer, the C interface is called with the parameter \textit{status\_c} rather than the value received from the FORTRAN application program. On returning from the C routine, the file identifier from the structure \textit{status\_c} is assigned to the parameter \textit{status}. Thus, the FORTRAN application program merely deals with the file identifier of the particular data access routine rather than with the whole structure. The same conversion mechanism is also true for the remaining blocking data access routines. \\

The implementation for the non-blocking routines is analogous. However, since these routines do not contain the parameter \textit{status}, the conversion mechanism is done for the parameter \textit{request}:

\begin{verbatim}
void mpi_file_iread_ (const int *fh,void *buf,const int *count,
     const MPI_Datatype  *datatype, int *request, int *ierror )
{
    int     res;
    MPIO_Request    request_c;
    MPI_Datatype     datatype_c;

    datatype_c = MPI_Type_f2c(*datatype);

    *ierror=MPI_File_iread(*fh, buf, *count, datatype_c, &request_c);

    *request=request_c.reqid;
}
\end{verbatim}

Thus, the application program only deals with the request identifier rather than the whole structure which must be considered in the routine for checking whether the outstanding non-blocking operation has finished:

\begin{verbatim} 
 void mpi_file_test_ ( int *request, int *flag, int *status, int *ierror)
{
    MPIO_Status 	status_c;
    MPIO_Request	request_c;

    request_c.reqid=*request;	

    *ierror = MPI_File_test (&request_c, flag, &status_c);
    *status=status_c.fid;
}
\end{verbatim}

This routine firstly assigns the request identifier received from the FORTRAN application to the structure \textit{request\_c}. Later, the value from the structure \textit{status\_c} is assigned to the parameter \textit{status} which in turn can be used by the FORTRAN application. Similar conversions are made for the routines \textit{mpi\_file\_wait\_} and \textit{mpi\_file\_get\_count\_}. \\

On analyzing the peculiarities of the Fortran to C interface we have to discuss the handling for connecting and disconnecting from the ViPIOS server. Before the functionalities of  MPI-IO can be used, MPI-IO has to be initialized. This is done by the routine \textit{MPIO\_Init}, which is located in a Fortran module. Thus, every Fortran application program must include the module \textit{vipmpi}. This routine, on the one hand, establishes the connection to the ViPIOS server, on the other hand, it manipulates the MPI\_COMM\_WORLD communicator in a way such that all client processes can use one MPI\_COMM\_WORLD without the interference of the server processes. In contrast to C application programs the header file similar to \textit{vip\_mpio\_init.h} is not required. However, the header files \textit{vip\_mpio\_f2c.h} and \textit{vip\_mpio\_def.h} are still needed.\\

We want to conclude our discussing with a small Fortran application which uses the derived datatype MPI\_TYPE\_DARRAY for distributing a 4x5x6- array onto several processes according to the distribution pattern BLOCK(2), CYCLIC(3) and CYCLIC(2) whereas the processor grid consists of 2 processors per dimension. Besides the code for the application program a graphical interpretation of the distribution array is given as well. Note that all data objects in Fortran are integer values. This becomes clear when we take a look at the declaration section of the example program. For instance, the \textit{newtype} of a derived datatype in Fortran  is of the type \textit{integer} rather than MPI\_Datatype as we know it from the C application programs. 

\begin{verbatim}
      program main
      USE vipmpi	
      implicit none

      include 'mpif_vip.h'
      include 'vip_mpio_f2c.h'       
      include 'vip_mpio_def_f2c.h'

      character	filename*81

      integer newtype, i, ndims, array_of_gsizes(3)
      integer order, intsize, nprocs,fh,ierr
      integer array_of_distribs(3), array_of_dargs(3)
      integer array_of_psizes(3)
      integer readbuf(1024), writebuf(1024)
      integer mynod, array_size, bufcount

      call MPI_INIT(ierr)
      call MPIO_INIT(ierr)	

      call MPI_COMM_SIZE(MPI_COMM_WORLD, nprocs, ierr)
      call MPI_COMM_RANK(MPI_COMM_WORLD, mynod, ierr)

      ndims = 3
      order = MPI_ORDER_FORTRAN
      filename = 'ufs:file_create'
      
c    specify the size of the array to be distributed
      array_of_gsizes(1)=4
      array_of_gsizes(2)=5
      array_of_gsizes(3)=6

c    distribution pattern of each dimension
      array_of_distribs(1) = MPI_DISTRIBUTE_BLOCK
      array_of_distribs(2) = MPI_DISTRIBUTE_CYCLIC
      array_of_distribs(3) = MPI_DISTRIBUTE_CYCLIC

c    distribution argument of each dimension
      array_of_dargs(1) = 2
      array_of_dargs(2) = 3
      array_of_dargs(3) = 2

      do i=1, ndims
           array_of_psizes(i) = 0
      end do

c    create processor array
      call MPI_DIMS_CREATE(nprocs, ndims, array_of_psizes, ierr)

      call MPI_TYPE_CREATE_DARRAY(nprocs, mynod, ndims,
     $     array_of_gsizes, array_of_distribs, array_of_dargs,
     $     array_of_psizes, order, MPI_INTEGER, newtype, ierr)

      call MPI_TYPE_COMMIT(newtype, ierr)

      array_size = array_of_gsizes(1) * array_of_gsizes(2) *
     $     array_of_gsizes(3)

c     write the array to the file

      call MPI_FILE_OPEN(MPI_COMM_WORLD,
     $     filename,
     $     MPI_MODE_CREATE+MPI_MODE_RDWR+MPI_MODE_UNIQUE_OPEN,
     $     MPI_INFO_NULL, fh, ierr)
 
      call MPI_FILE_SET_VIEW(fh, 0, MPI_INTEGER, newtype, "native",
     $     MPI_INFO_NULL, ierr)
      call MPI_FILE_WRITE_ALL(fh, writebuf, bufcount, MPI_INTEGER,
     $     status, ierr)
      call MPI_FILE_CLOSE(fh, ierr)

      call MPI_TYPE_FREE(newtype, ierr)
      call MPI_FINALIZE(ierr)
      call MPIO_FINALIZE(ierr)

      stop
      end

\end{verbatim}

%% file: testmpio.tex
\subsection{Program Testmpio}

Up to now we were discussing the theoretical background of MPI-IO and the portable implementation on ViPIOS. Since an extensive testing phase is part of each software engineering process, we will present one test program from the University of California and Lawrence Livermore National Laboratory written in April 1998. This so-called "regression suite" verifies a lot of different MPI-IO routines by simulating different cases of application programs. For example, functions for testing collective open and close of files, independent reads and writes with file and buffer types, file control etc. Besides checking the functionalities of the MPI-IO implementation, the time for the verification procedure is taken. Since ViMPIOS does not support shared file pointers, special error handling routines and different representation modes, the last routines could not be tested.\\

The structure of the test program is as follows. The main function takes the input parameters, for example the user path for storing the files, and calls the routine \textit{dotest()}, which in turn calls the different test routines which we stated before. Moreover, the time for the whole process is taken. \\

On giving a brief introduction do the regression suite we will now anaylise each routine separately.

\subsubsection{test\_manycomms}
Some files are opened with a couple of different communicators. Starting with a group of processes which is split into two sub groups, reading and writing is analyzed. \\

In particular, following interface routines are tested:\\

\textit{MPI\_Barrier, MPI\_Comm\_Free, MPI\_Comm\_split
}\\

\textit{Purpose}:\\
This routine checks whether several files can be opened using nested communicators. Thus, for example, files are opened with the communicators 1,2,3,4 and 5. Then, the files are closed in a different way. For instance, the file which is opened by communicator 3 is closed first etc. In order to accomplish this task, a communicator group is split into several sub groups which operate on different nested files. What is more, consistency semantics are obeyed by using atomic access operations (\textit{MPI\_File\_set\_atomicity} and synchronization points (\textit{MPI\_File\_sync}). However, for the time being, ViPIOS merely operates with atomic modes. What is more, the function for synchronizing concurrent file I/O is still a task to be fulfilled in the near future.

\subsubsection{test\_openmodes}
Different open modes are checked. First, the file is opened with the access mode MPI\_MODE\_CREATE and MPI\_MODE\_WRONLY. Thus, reading this file is supposed to fail. Later, the modes MPI\_MODE\_READ\_ONLY and MPI\_MODE\_DELETE\_ON\_CLOSE are checked.\\

\textit{Purpose:}\\
This routines checks whether the open modes are consistent. In other words, a file which is opened, for example, in the \textit{write only} mode, should not be readable and vice versa.

\subsubsection{test\_manyopens}
First, a couple of files are opened and filled with data which are read back in a different order later on. Next,  further files are opened with different access modes such that each file has a distinct entry to the file table. \\

\textit{Purpose:}\\
Similar to the function \textit{test\_manycomms} the behavior of  the file table is tested whether it can cope with several files which are opened at the same time and accessed in a different way.

\subsubsection{test\_openclose}
MPI files are opened and closed in a collective way. Thus, it checks whether the MPI-IO recognizes errors with using wrong access modes for different files. What is more, a file is tried to close which was not opened before.\\

\textit{Purpose:}\\ 
This routine merely check the basic functionality of opening and closing a file.

\subsubsection{test\_readwrite}
Independent I/O calls are tested here. MPI\_BYTE is chosen for the data access buffer as well as for the filetypes. Each node writes some bytes to a separate part of the same file which is read back later on in order to compare the results. Besides checking whether the correct number of bytes are read or written, the inbuffer and and outbuffer are compared, i.e. the input string must be identical to the output string. No file view is set in that routine.\\

\textit{Purpose:}\\
This routine checks the basic functionalities of blocking data access operations.

\subsubsection{test\_rdwr}
Collective I/O routines are tested with derived filetypes and buffer types. Thus, on the one hand a view is set, on the other hand a contiguous derived data type for the read and write buffer is used.\\

\textit{Purpose:}\\
Besides checking the features of file views, the functionality of so-called data scattering is tested. 

\subsubsection{test\_filecontrol}
This routine tests different MPI-IO routines like \textit{MPI\_File\_get\_position}, \textit{MPI\_-File\_set\_size} or \textit{MPI\_File\_get\_byte\_offset}. Moreover, different file views are set. This test is concluded with splitting the communicator group into two sub groups which separately operate on the files.\\

\textit{Purpose:}\\ 
The most important MPI-IO routines are checked. 

\subsubsection{test\_localpointer} 
First, each node accesses the file in a contiguous way but with different displacements. Later, some data is skipped and a couple of byte values are written after this hole. Different read and write operations are performed in order to check the holes in the file. In the second run a file with a more complicated file view is opened in order to perform similar tests. \\

\textit{Purpose:}\\
The behavior of setting file views is analyzed in detail. Besides setting correct file views, erroneous views are set to check whether the implementation recognizes these inconsistencies. A possible incorrect assignment would be to use an etype of the type MPI\_INT   and a filetype of the type MPI\_DOUBLE.

\subsubsection{test\_collective} 
Collective I/O with explicit interleaving is tested. What is more, each process writes data of different size.\\

\textit{Purpose:}\\
This routine checks the behavior of writing to a file according to different access patterns and different number of byte elements.

\subsubsection{test\_nb\_rdwr}
This routine does the same as \textit{test\_rdwr}. Except of using blocking routines it tests non-blocking operations.

\subsubsection{test\_nb\_localpointer}
This routine does the same as \textit{test\_nb\_localpointer}. Except of using blocking routines it tests non-blocking operations.

%% file: interf.tex
This chapter describes the interface between HPF (High Performance Fortran)
and ViPIOS (Vienna Parallel Input Output System). First a quick introduction
to the relevant HPF features is given. Then the implementation of the interface
is discussed in detail.

\section{HPF (High Performance Fortran)}

HPF has been developed to support programmers in the development of parallel
applications. It uses the SPMD paradigm for transferring a sequential program
to a parallel one, which can be executed on SIMD and MIMD architectures.
Basically the same (sequential) program is executed on every processor
available. But each processor only works on a subset of the originial input
data. The result of the whole computation has to be composed from all the
results of the single processors.

HPF itself is an extension to FORTRAN 90 and supplies the programmer with the
functionality needed to generate SPMD programs. The programmer has to supply
the sequential version of the program (in FORTRAN 90) and also to define how the
data is to be distributed among the various processors. The HPF compiler then
automatically generates the according parallel program by inserting the communication statements necessary to distribute the data and to coordinate the different processes.

Any HPF specific statement (i.e. the ones which are not FORTRAN 90 statements)
starts with the string \texttt{!HPF\$}. So all these statements are treated as
a comment by a FORTRAN 90 compiler and the sequential version of the program
can be easily compiled and tested. After the \texttt{!HPF} token the HPF
compiler expects an HPF directive. The most important directives are those
for the definition of data distribution, which are discussed in the following.

\subsection{HPF-directives}\label{sec_directives}

\subsubsection{\texttt{PROCESSORS}}

This directive allows to define an abstract processor array. The number of
processors defined in an abstract processor array can vary from the number of
physically existing processors. (If the number of physical processors is less
than the number of physical processors, then two or more tasks will be executed
on specific processors. The number
of tasks executing allways corresponds to the number of logical processors. Each
task is supposed to run on one of these logical processors.)

The example code \texttt{!HPF\$ PROCESSORS PROCS(3,4)} declares an
abstract (i.e logical) processor array with three processors in the first
dimension and four processors in the second dimension. The reason for the
availiability of multidimensional abstract processor arrays is that the
datastructures used in high performance computing mostly are arrays of higher
dimension. The mapping of data elements to specific processors can therefore
very often be done elegantly by using an appropriate logical view on the
processors available. Note that the abstract processor array does not have
to correlate to the physical topology of processors and their interconnections
in the targeted hardware architecture. It is the responsibility of the HPF
compilation system to map the logical processor array to the physical one
supplied by the hardware.

\subsubsection{Distribution Formats}
While distribution formats are not directives on their own, they are needed
to specify the data distribution in the \texttt{DISTRIBUTE} directive, which
is explained in the next chapter. The following distribution formats are
supported by HPF:

\paragraph{\texttt{BLOCK / BLOCK(blocksize)}}
Generally data becomes divided into blocks of equal size. Each data element
\emph{(1...N)} is assigned to one  corresponding processor \emph{(1...P)}.
Using \texttt{BLOCK} without any further particular blocksize in brackets
causes calculation of blocksize for each processor depending on the size of
data and the number of processors in the according dimension of the processor
array. If the number of data elements can be
divided into commensurate blocks of size \emph{(N/P)} (i.e. N is divisible by
P), then each of the \emph{P} blocks is assigned to the corresponding
processor (i.e. the first block to the first processor and so on). If the
number of data is not divisible by the number of processors, then commensurate
blocks of size \emph{$b = \lceil N/P\rceil$} are assigned to the first
\emph{1...(P-1)} processors.
The remaining \emph{N-P} elements are assigned to the last processor \emph{P}.

\begin{figure}
  \centerline{\includegraphics[scale=0.6]{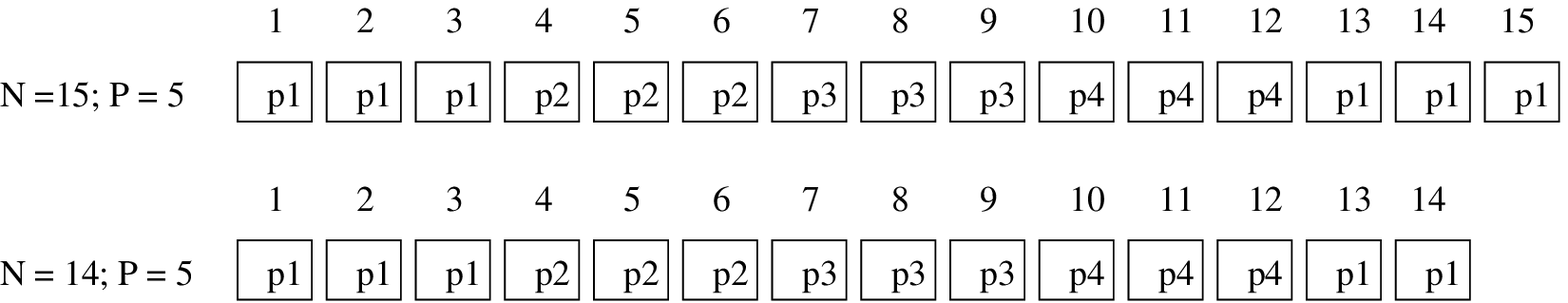}}
  \caption{\texttt{BLOCK} distribution} \label{block}
\end{figure}

Figure \ref{block} shows two examples how data elements are assigned to
processors. In the first case \emph{N} is divisible by \emph{P}. In the case
where \texttt{N = 14} the last processor ($p_1$) gets assigned the remaining
seventh element.

If the optional parameter blocksize is given then all the blocks are calculated
to be this size (except the last one if \emph{N} is not divisible by blocksize).
The
blocks are then distributed onto the processors in a cyclic fashion (i.e. The
first block to the first processor , ... the Pth block to the Pth processor,
the \emph{(P+1)th} block to the first processor and so on until all the blocks
are
assigned a processor.)

\paragraph{\texttt{CYCLIC / CYCLIC(blocksize)}}

CYCLIC without particular blocksize causes each element to be assigned to
a processor in ascending order (i.e. the first element to the first processor,
the second element to the second processor and so on). If the number of elements
exceeds the number of processors the elements are allocated to processors
cyclically (i.e. the \emph{(P+1)th} element is assigned to processor one again,
the \emph{(P+2)th} to procesor two and so on.
If a blocksize is given the resulting distribution is often also called
\texttt{BLOCK\_CYCLIC}. In this case not single data elements but blocks of
data elements of the specified size are assigned to the processors in the
same cyclic fashion.

\begin{figure}
  \centerline{\includegraphics[scale=0.6]{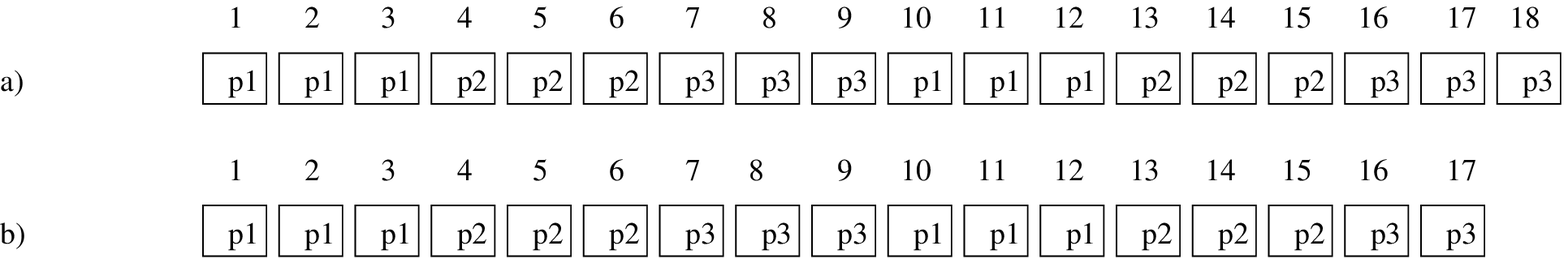}}
  \caption{\texttt{CYCLIC / CYCLIC(blocksize)} distribution} \label{cyclic}
\end{figure}

Figure \ref{cyclic} a) shows how the elements wrap around the processor
array. Example b) in figure \ref{cyclic} shows how blocks of blocksize \emph{$b = \lceil N/P
\rceil$} are assigned to each processor. Except the last block assigned to
processor $p_3$, which holds the remaining \emph{N-P = 2}
elements.

\paragraph{\texttt{GEN\texttt{\_}BLOCK}}
The generic block distribution allows for blocks of arbitrary size, which may
vary from processor to processor. Thus processor one may for instance be
assigned to 10 elements, processor two to 7 elements and processor three to 154
elements. This enables completely irregular distributions to be realized.
However this distribution strategy is not implemented yet in all the HPF
compilation systems and is also not supported by the ViPIOS HPF interface now.

\paragraph{INDIRECT}
Is similar to generic block but the block sizes can be given by pointers that
actually point to the actual size. So while in generic block the block sizes
are known at compile time and are constant during runtime, the indirect
distribution allows for variable sized blocks, the size of which can be
changed at runtime. This distribution too is not implemented yet.

\paragraph{*}

This distribution format causes data not to become distributed at all. This
means that data elements are replicated (i.e. every processor gets a copy
of the data elements).

\paragraph{\texttt{DISTRIBUTE ONTO}}

Data mapping is achieved by the distribution directive
\texttt{DISTRIBUTE}. The kind of distribution is specified by distribution
formats like \texttt{BLOCK}, \texttt{CYCLIC} and  \texttt{*}, which are
explained above.

Each dimension of an array can be distributed independently, as shown in
the following example.

\begin{verbatim}
!HPF$ PROCESSORS PROCS(3,4)  	
	INTEGER, DIMENSION (14,17) :: B
!HPF$ DISTRIBUTE (CYCLIC(3),BLOCK) ONTO PROCS :: B
\end{verbatim}

The chosen test array \texttt{B} is a two-dimensional array
\texttt{INTEGER, DIMENSION (14,17) :: B}. \\ \texttt{!HPF DISTRIBUTE
(CYCLIC(3),BLOCK) ONTO PROCS :: B} distributes each dimension of
array \texttt{B} depending on the  distribution format onto the processor
array. The chosen formats are \texttt{CYCLIC(3)} (\texttt{BLOCK\_CYCLIC}) for the first dimension and \texttt{BLOCK} for the second dimension.

\begin{figure}
  \centerline{\includegraphics[scale=0.6]{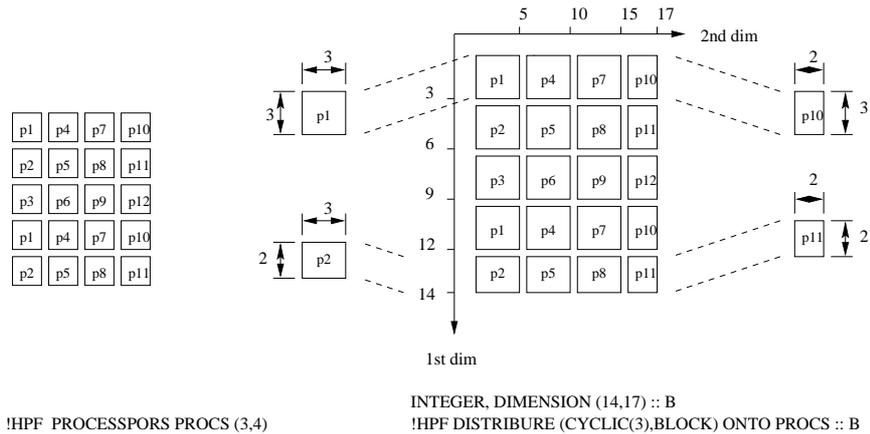}}
  \caption{processor-array and data mapping onto processors}
\label{procs}
\end{figure}

\subsection{VFC Vienna Fortran Compiler}

The \emph{VFC} compiler system performs a source-to-source translation from \emph{HPF} to \emph{Fortran 90} SPMD source code with special \emph{ViPIOS} calls. System libraries are used to perform I/O operations and create a runtime descripor. This descriptor contains all neccessary information to perform data distribution through the \emph{ViPIOS} system and is explained in detail in the following section.

\subsubsection{The runtime descriptor}

The runtime descriptor is a special datastructure containing all information about the processor array the data become distributed onto, the number of dimensions of the data, its extensions in each dimension and its type. Further information are explained in detail:

\begin{figure}
  \centerline{\includegraphics[scale=0.6]{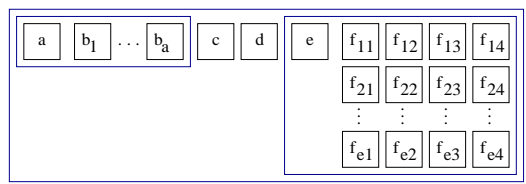}}
  \caption{Interface runtime descriptor} \label{rtdsc}
\end{figure}

\begin{center}
    \begin{tabular}{p{2cm}p{3cm}p{6cm}} \hline
	field		& arguments & description  \\ \hline
	a		& \emph{1...N}	&	number of dimensions of the
processor array \\
 	$b_1...b_a$ 	& \emph{$(1...N)_1$} ... \emph{$(1...N)_a$}	&
number of processors of the processor array for each dimension \\
 	c		& & type of passed data \\
	d		& & size (in Byte) of each element \\
	e		& & umber of dimensions of the data-array  \\ \hline
	each dim.& & \\

	$f_{e1}$	&	& global length of dimension \\
	$f_{e2}$	&	& local length of dimension \\
	$f_{e3}$	&	& distribution (BLOCK, CYCLIC, BLOCK\_CYCLIC,
...) \\ 	$f_{e4}$	&	& distribution argument
for all kinds of distribution except GEN\_BLOCK blocksize is equal
to $f_{e4}$. otherwise $f_{e4}$ contains the number of
processors.\\ \hline
\end{tabular} \end{center}

The elements of the runtime descriptor are shown in
figure \ref{rtdsc}.

\section{ViPIOS-HPF-interface}

The interface (which is contained in the file  \texttt{vip\_test\_rt.c}
transfers all the information contained in the runtime descriptors into the
format used by the ViPIOS system. ViPIOS uses two datastructures to describe
data distribution. These are the structures \texttt{Access\_Desc} and \texttt{basic\_block}.

\subsection{The datastructures \texttt{Access\_Desc} and \texttt{basic\_block}}
\label{sec_datastructures}

The \emph{HPF}-\emph{ViPIOS}-interface uses the recursive datastructurs
\texttt{Access\_Desc} and \texttt{basic\_block}.
\texttt{Access\_Desc} is defined for each dimension once. Like
\texttt{Access\_Desc} the structure \texttt{basic\_block} contains several
variables needed to perform read and write operations. Depending on the
distribution directive the structure \texttt{basic\_block} appears once or
twice in each dimension.

The datastructure is given by the following code fragment defined in
\texttt{vip\_int.h}:

\begin{verbatim}
typedef struct
{
        int     no_blocks;
        struct basic_block      *basics;
        int     skip;
}
Access_Desc;

struct basic_block
{
        int offset;
        int repeat;
        int count;
        int stride;
        Access_Desc     *subtype;
        int sub_count;
        int sub_actual;
};
\end{verbatim}

\begin{center}
    \begin{tabular}{p{3.5cm}p{8cm}}
	\emph{struct Access\_Desc} & \\
	\texttt{no\_blocks}	& number of subsets \texttt{basic\_block} \\
	\texttt{struct basic\_block}	& pointer to subsetstructure
	\texttt{struct basic\_block} \\
  	\texttt{skip} 		& number of elements to skip after R/W operation \\
\\
 	\emph{struct basic\_block} & \\
	\texttt{offset}	& number of elements to skip to set filehandle to
startposition  \\
	\texttt{repeat} & how many times element blocks appear \\ 	
	\texttt{count} & number of elements of one block \\
	\texttt{stride}	& number of elements between two blocks \\
	\texttt{* subtype} & pointer to next \texttt{Access\_Desc} if further
dimension \\
	\texttt{sub\_count} & not necessary in this interface \\
	\texttt{sub\_actual} & not necessary in this interface \\
\end{tabular} \end{center}

Each component of these structures describes how data has to be mapped onto
the processor array.
The following sections explain the values of all the components of
\texttt{Access\_Desc} and \texttt{basic\_block}
for the simple example, which we already have used previously.

\begin{verbatim}
!HPF$ PROCESSORS PROCS(3,4)  	
	INTEGER, DIMENSION (14,17) :: B
!HPF$ DISTRIBUTE (CYCLIC(3),BLOCK) ONTO PROCS :: B
\end{verbatim}

\begin{figure}
  \centerline{\includegraphics[scale=0.6]{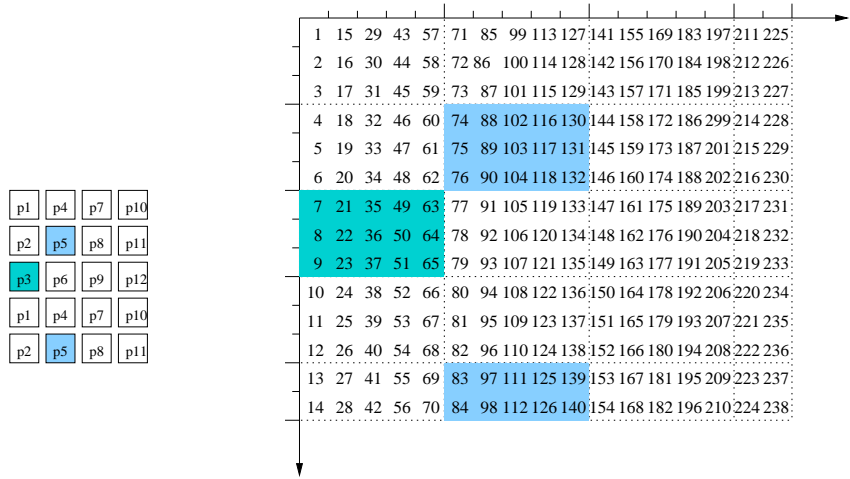}}
  \caption{data mapping} \label{mapp}
\end{figure}

Figures \ref{procs} and \ref{mapp} illustrate how data is distributed among
the processors in this case.
In this example the processors
\texttt{3} and \texttt{5} reveal two common constellations of the
datastructures' component values. The component values for these two processors are therefore calculated and explained in detail in the following.

\subsubsection{\texttt{Access\_Desc} and \texttt{basic\_block} arguments - its
values for specific processors}
\subsubsection*{processor \texttt{3}}
\begin{figure}
  \centerline{\includegraphics[scale=0.6]{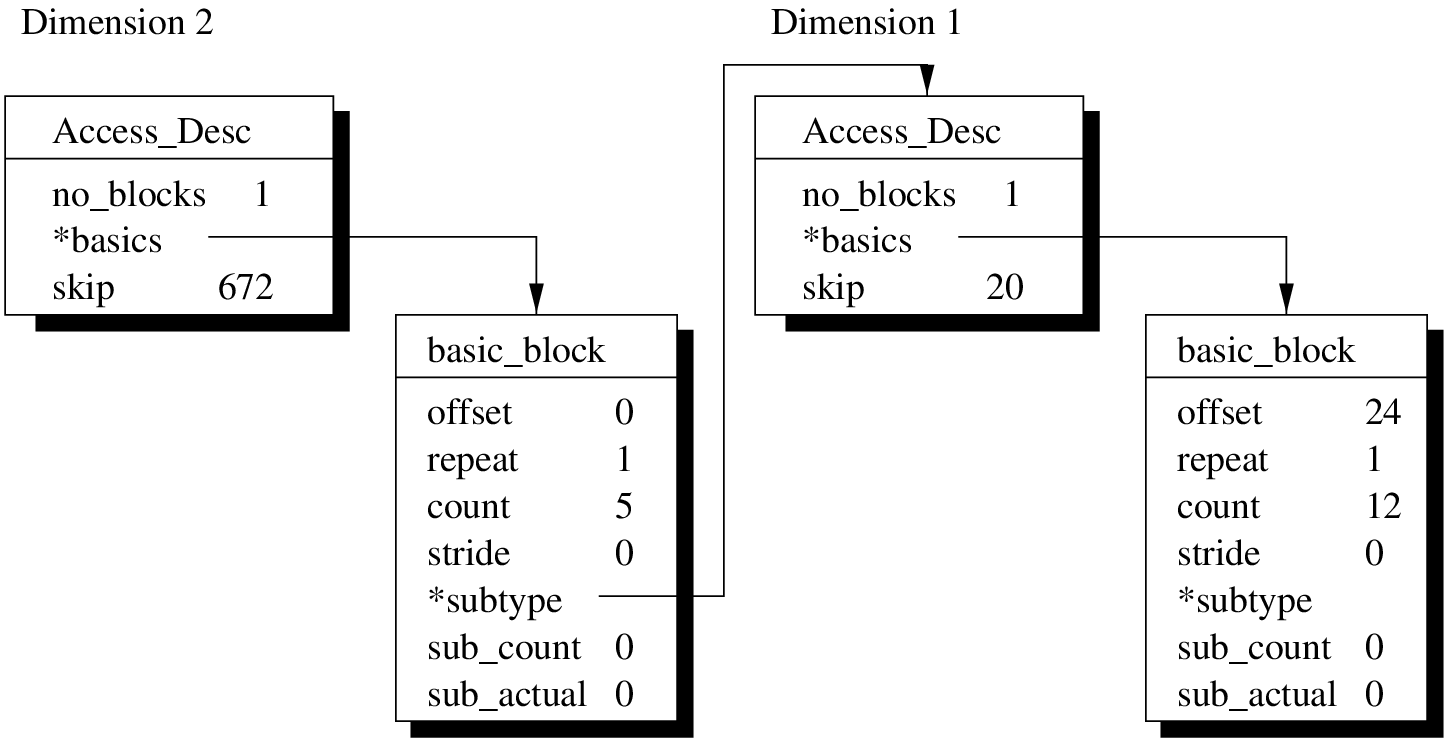}}
  \caption{processor \texttt{3}: data structures \texttt{Access\_Desc} and
\texttt{basic\_block}}
\label{lnkone}
\end{figure}
\begin{figure}
  \centerline{\includegraphics[scale=0.6]{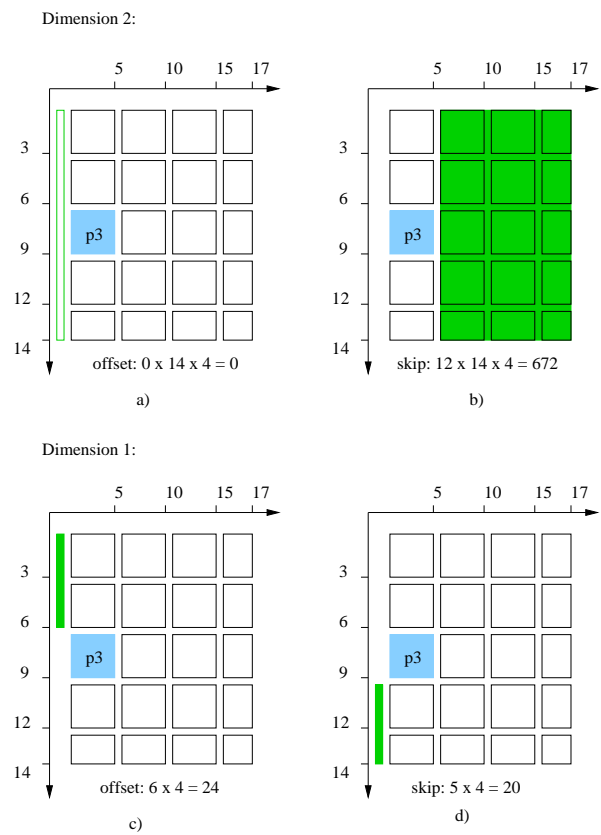}}
  \caption{processor \texttt{3}: \texttt{skip} and \texttt{offset} for each
dimension}
\label{dimone}
\end{figure}
In this example data mapping for processor \texttt{3} creates a structure
consisting of \texttt{Access\_Desc} and \texttt{basic\_block} which are linked
together as shown in figure \ref{lnkone}. For each dimension structure
\texttt{Access\_Desc} and \texttt{basic\_block} are defined separately. The
structure describes the second dimension first because the internal interface
of \emph{ViPIOS} operates in recursive fashion.
Argument \texttt{no\_blocks} declares the number of structures
\texttt{basic\_block} used to describe data mapping for each dimension - in
this case one \texttt{basic\_block}.
The next argument \texttt{*basics} is a pointer to this structure.
One of the arguments of \texttt{basic\_block} is \texttt{offset}. It stores the
the number of bytes the file pointer has to skip from the beginning of the
file to the first position where read or write operations start. For the second
dimension \texttt{offset} is zero as shown in figure \ref{dimone}.
\texttt{skip} describes how many bytes the file pointer has to skip after read
or write operations. As shown in figure \ref{dimone} b) 672 bytes have to be
skipped along the axis of the second dimension. Along the first dimension
\texttt{skip} is set to 20 bytes (see figure \ref{dimone} c). Processor
\texttt{3} occures only for one time along the second dimension -
\texttt{repeat}s value is \texttt{1}.  \texttt{count} (for dimensions higher
than one) indicates the number of elements assigned to a processor but not the
number of bytes. In the first dimension \texttt{count} describes the number of
bytes to be read or written. As the example shows five elements are assigned to
processor \texttt{3} in the second dimension. For the first dimension the value
of \texttt{count} is three bytes.
If more than one processor would appear in the second dimension
(\texttt{repeat > 1}) the value for \texttt{count} would not
change. On the other hand in dimension one the value of \texttt{repeat} is
always one even if a processor appears more than one time. In this case however
the value of \texttt{count} is the summ of all bytes assigned to a processor in
this dimension  The argument \texttt{stride} stores the number of bytes to skip
between processor elements. If \texttt{repeat}s value is set to one the value
of \texttt{stride} must be zero. Only if \texttt{repeat} is greater than one
\texttt{stride} differes from zero. In the case a processor appears
more than one time in dimemsion two the first dimension would be taken
into consideration for calculation of \texttt{stride} as it is done for
\texttt{skip} (see also figure \ref{dimone} b).

\subsubsection*{processor \texttt{5}}
\begin{figure}
  \centerline{\includegraphics[scale=0.6]{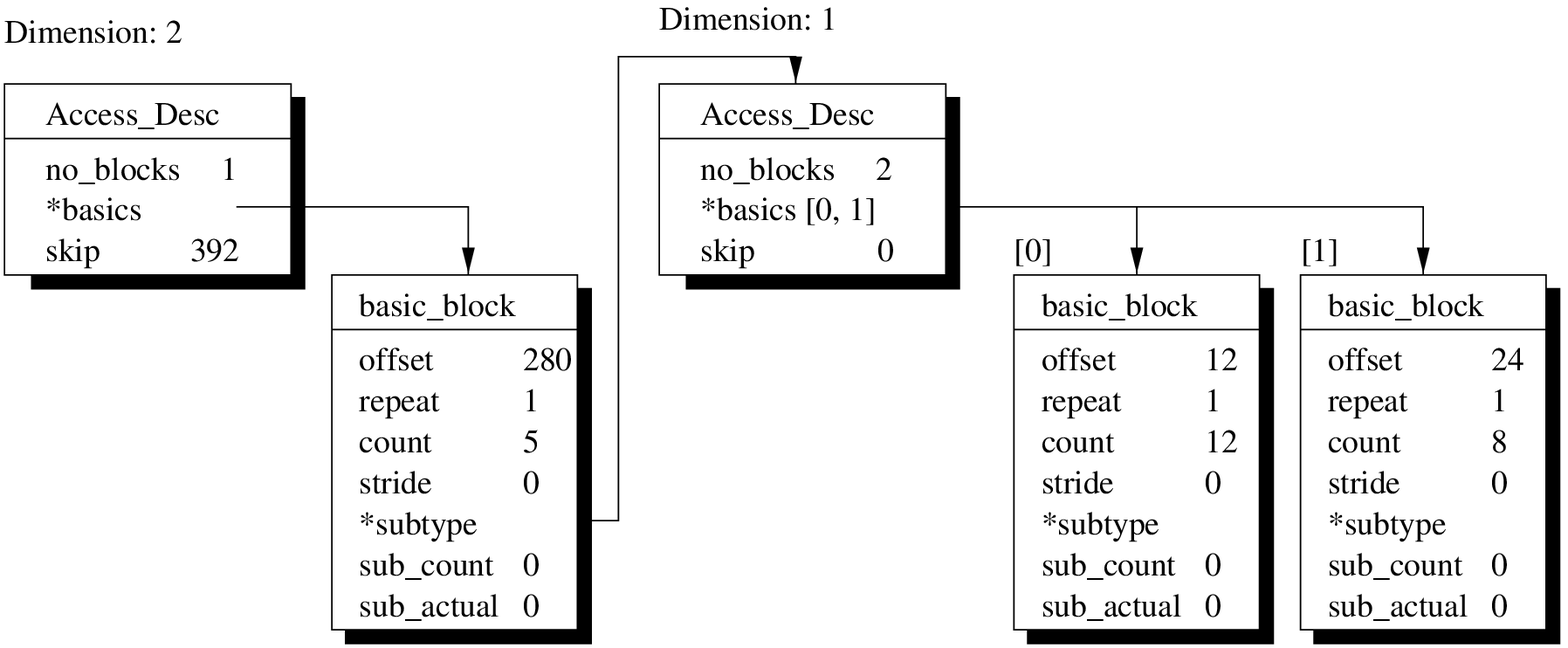}}
  \caption{processor \texttt{5}: data structures \texttt{Access\_Desc} and
\texttt{basic\_block}}
\label{lnktwo}
\end{figure}

\begin{figure}
  \centerline{\includegraphics[scale=0.6]{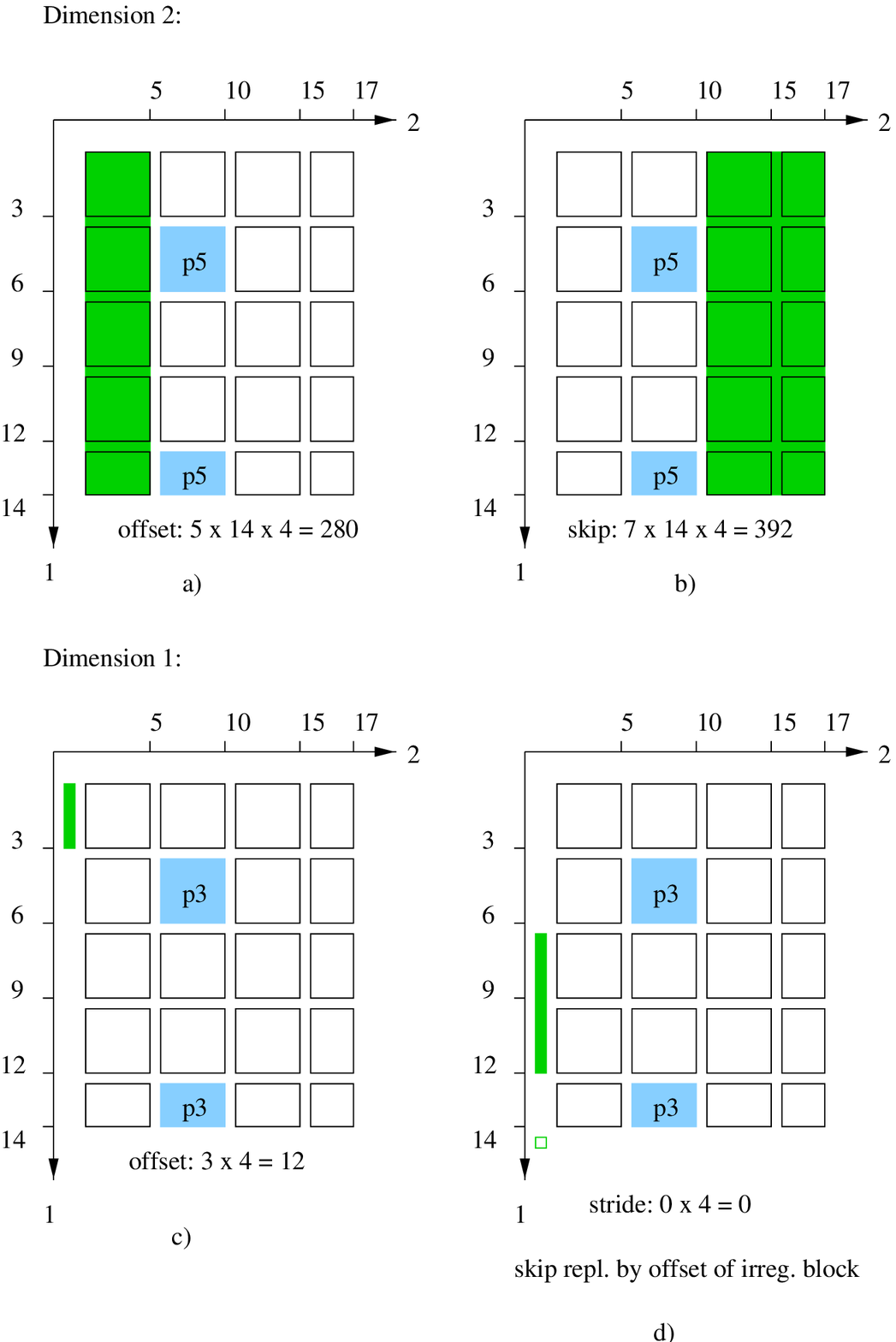}}
  \caption{processor \texttt{5}: \texttt{skip} \texttt{offset} and
\texttt{stride} for each dimension}
\label{dimtwo}
\end{figure}

Figure \ref{lnktwo} shows the datastructure for processor \texttt{5}.
\texttt{no\_blocks} defines an array of two structures of
\texttt{basic\_block} in dimension one. Unlike data mapping for processor
\texttt{3} data mapping for this processor is divided into two blocks. The
first block describes the regular block the second one the irregular block. It
is important to distinguish these two because of their different blocklength
expressed by \texttt{count}. In figure \ref{dimtwo} data mapping for processor
\texttt{5} and the different length of data blocks along the axis of the first
dimension can be seen.
An important detail relating to argument \texttt{skip}: This argument
is defined for the regular block as well as for the irregular block for each
dimension in \texttt{Access\_Desc}  If an irregular block is defined
\texttt{skip} is always set to zero. Instead of this \texttt{offset} of the
irregular block repleaces \texttt{skip} of the regular block.

%
%
\subsection{Interface functions}

This section gives a detailed description of the interface functions of
\texttt{vip\_test\_rt.c}. It is seperated into two parts. One gives a quick
overview of the functions input arguments, its functionality and return values
in tabular form. The second part of the description gives a more detailed
description of the functionality based on the example \emph{HPF} code, which
already has been used  in sections \ref{sec_directives} and
\ref{sec_datastructures}.

\subsubsection{Functions for all operations}

These functions are used to establish or complete connections between clients and server. They are indpendent from the applications demand.

\begin{verbatim}
int VIPIOS_connect (const char * system_name, int connection) {
        return ( ViPIOS_Connect (connection) ? 0 : -1);
}
\end{verbatim}
\begin{tabular}{p{2cm}p{2cm}p{6cm}}
input var & connection & to establish a connection between an application and the ViPIOS system. \\
return val &  0 or (-1) & true (connection established) or false (connection not established)
\end{tabular}
%
%
\begin{verbatim}
int VIPIOS_disconnect(int connection) {
        return  ( ViPIOS_Disconnect () ? 0 : -1);
}
\end{verbatim}
\begin{tabular}{p{2cm}p{2cm}p{6cm}}
input var & connection & disconnects the application from the ViPIOS system \\
return val &  0 or (-1) & true (disconnected) or false (not dsconnected)
\end{tabular}

%
\subsubsection{Functions for operations on binary data}

To perform operations dealing with binary data the following functions are used:

\begin{verbatim}
int VIPIOS_open_binary
(int connection, const char *filename, int status, int io_stat) {
    int fd, flags;
    switch (status) {
        case 0: flags = 0;                break;
        case 1: flags = MPI_MODE_CREATE;  break;
        case 2: flags = 0;                break;
        case 3: flags = MPI_MODE_CREATE;  break;
    }
    switch (io_stat) {
        case 0:
            flags |= MPI_MODE_RDWR;       break;
        case 1: flags |= MPI_MODE_RDONLY; break;
        case 2: flags |= MPI_MODE_WRONLY;
    }
    return (ViPIOS_Open ( filename, flags, &fd) ? fd : -1);
}
\end{verbatim}
\begin{tabular}{p{2cm}p{2cm}p{6cm}}
input var	& *filename 	& the name of the file e.g. /tmp/test\_file1
\\ 		& status	& \texttt{0} old, \texttt{1} new, \texttt{2}
unknown, \texttt{3} replace, \texttt{4} scratch \\
		& io\_stat	& file access mode: \texttt{0}
read \& write \texttt{1} read (only), \texttt{2} write (only).\\
return val&  fd or (-1) & file descriptor or false (connection not established)
\end{tabular}

Depending on the value of the passed argument \texttt{status} which gives information about the file to be opened, \texttt{flags} becomes assigned a specific \texttt{MPI\_MODE} value. The bits of the number of \texttt{flags} which is neccessary as argument in \texttt{ViPIOS\_Open (filename, flags, \&fd)} are furthermore bound up with one of the \texttt{MPI} modi depending on \texttt{io\_stat}. This flag describes the status of the file while using I/O-operations.

\subsubsection*{read and write binary arrays}

\begin{verbatim}
int VIPIOS_read_binary_array (int connection, int fd, const void * data,
const int * array_dist) {
        Access_Desc * descriptor = prep_for_set_structure (array_dist);
        return = ViPIOS_Read_struct (fd, data, 0, descriptor, 0, -1);
}

int VIPIOS_write_binary_array (int connection, int fd, const void * data,
const int * array_dist) {
        Access_Desc * descriptor = prep_for_set_structure (array_dist);
        return = ViPIOS_Write_struct (fd, data, 0, descriptor, 0, -1);
}
\end{verbatim}

\begin{tabular}{p{2cm}p{2cm}p{6cm}}
input var	& connection & to establish a connection between an application and the ViPIOS system. \\
		& fd		& file descriptor \\
		& data	 	& data to distribute \\
 		& array\_dist	& runtime descriptor \\
return val	& 0 or (-1)	& true read/write operation performed or not
\end{tabular} \\

Both functions \texttt{VIPIOS\_read\_binary\_array} and \texttt{VIPIOS\_write\_binary\_array} use the same list of passed arguments. The only difference depending on if binary array shall be read or written is the call of \texttt{ViPIOS\_Read\_struct(fd, data, 0, descriptor, 0, -1)} either or \texttt{ViPIOS\_Write\_struct(fd, data, 0, descriptor, 0, -1)}. The argument \texttt{descriptor} in this case is a pointer to the data structure \texttt{Access\_Desc} which becomes initialized in \texttt{prep\_for\_set\_structure(array\_dist)}. The passed argument is a reference to the runtime descriptor delivered by the runtime system of \emph{VCPC}.

\subsubsection*{\texttt{Access\_Desc * prep\_for\_set\_structure (const int
*array\_dist)}}

\begin{tabular}{p{2cm}p{2cm}p{6cm}}
input var	& array\_dist	& runtime descriptor \\
return val	& descriptor	& pointer to data structure \texttt{Access\_Desc}
\end{tabular} \\

\texttt{void *next\_free} is used as an auxiliary pointer to write the data structure \texttt{Accesss\_Desc}. Its inital address is the memory address of \texttt{descriptor}.
Contiguous memory allocation \texttt{descriptor = malloc (1024)} for the
datastructure \texttt{Access\_Desc} gurarantees  faster processing while
operations on \texttt{Access\_Desc} are performed:

\begin{verbatim}
    void *next_free;        /* will point to next free space (in rec) */
    descriptor = malloc (1024); /* initial free space for Access_Desc */
    next_free  = descriptor;    /* init points to Access_Desc */
\end{verbatim}

Some definitions made to make operations using specific information from the runtime descriptor easier to use.
\texttt{distance} points to the first element of the runtime descriptor where the information about the first dimension of the data to distribute are stored.
\texttt{array\_dimension} stores the number or dimensions of data.

\begin{verbatim}
    distance = array_dist[0]+4;
    array_dimension = array_dist[array_dist[0]+3];
    proc_nbr_array_count = array_dist[0];
    max_proc_array_dim = array_dist[0];
\end{verbatim}

\begin{figure}
  \centerline{\includegraphics[scale=0.6]{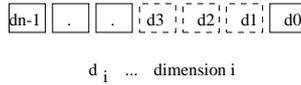}}
  \caption{process grid coordinates} \label{dimcontr}
\end{figure}

The processor grid coordinates calculated by the following code fragment are stored in \texttt{dim\_contr[i + 1]}.
The maximum number of dimensions of the data does not may exceed the number of digits of this array. Otherwise the size of the array defined by \texttt{int dim\_contr[5]} in this case must become changed to a higher value.
Each coordinate $c_i$ of the data is stored at index position $n - 1$ of \texttt{dim\_contr}.
The organisation of process grid coordinates in \texttt{dim\_contr} is shown
in figure \ref{dimcontr}.\\

Depending on the current process number the corresponding processor grid coordinates can be calculated. To find out the actual process id \texttt{MPI\_Comm\_rank( MPI\_COMM\_WORLD, \&proc\_nbr);} is used. The process id is stroed at the reference \texttt{\&proc\_nbr}.

\begin{verbatim}
    result = MPI_Comm_rank(MPI_COMM_WORLD, &proc_nbr);
    aux = proc_nbr;

    for (i = 0; i < array_dimension; i++) {
        total_array_size = total_array_size / array_dist[array_dimension - i];
        dim_contr[i + 1] = aux / total_array_size;
        aux = proc_nbr % total_array_size;
    }
\end{verbatim}

If the processor grid coordinates for the specific process are calculated and stored in \texttt{dim\_contr} the next step is to set up the datastructure \texttt{Access\_Desc}. With the informations of the runtime descriptor, the pointer to the allocated storage for \texttt{Access\_Desc} and the processor grid coordinates as arguments the function \texttt{set\_structure} is called. It sets up the datastructure \texttt{Access\_Desc} by the informations calculated in \texttt{prep\_for\_set\_structure} before.
\begin{verbatim}
    set_structure (descriptor, &next_free, array_dist);
    return descriptor;
\end{verbatim}
At last the result value of \texttt{prep\_for\_set\_structure} is a reference to the complete datastructure \texttt{Access\_Desc} as a result of \texttt{set\_structure}.
%
%
\subsubsection*{\texttt{void set\_structure (Access\_Desc * descriptor, void **free\_space, int * array\_dist)}}

\begin{tabular}{p{2cm}p{2cm}p{6cm}}
input var	& descriptor	& pointer to data structure \texttt{Access\_Desc}  \\
		& *free\_space	& auxiliary pointer to pointer of address of \texttt{Access\_Desc} \\
		& array\_dist	& runtime descriptor \\
return val	& descriptor	& reference of input var descriptor
\end{tabular} \\

Depending on how each dimension has to become distributed the function \emph{set\_structure} provides several blocks for: \\
\\ \
\texttt{* ... no distribution}, \\
\texttt{BLOCK(n) ... block distribution} and \\
\texttt{CYCLIC(n) ... cyclic distribution}.\\

In each block all specific data for \texttt{Access\_Desc} are calculated seperate.
Before these data and its calculation becomes explained in detail the following description points out variables used in \texttt{Access\_Desc} which are distribution independent:

\begin{verbatim}
    descriptor->no_blocks = 1;
    down_count = array_dimension - (step_dim_count + 1);
\end{verbatim}

\texttt{no\_blocks} is set to \texttt{1} initially. The variable \texttt{down\_count} becomes assigned the current number of dimension for each iteration of \texttt{set\_structure}. If more than one dimension has to be calculated the value of \texttt{down\_count} decrements for each dimension.
The following description shows the different algorithms needed for the calculation of each value in \texttt{Access\_Desc} depending on the distribution.

\subsubsection{no distribution} \label{nodistr}

\begin{verbatim}
    descriptor->skip = 0;
\end{verbatim}
Because all elements shall become read in one dimension there are no data to skip. Thats why \texttt{skip} is set to \texttt{0}.
\begin{verbatim}
    *free_space = (struct basic_block *) ((void *)descriptor) +
        sizeof(Access_Desc);
    descriptor->basics = *free_space;
    *free_space += sizeof (struct basic_block);
\end{verbatim}
\texttt{*free\_space} (used as auxilliary pointer) gets assigned the address of the next structure after \texttt{Access\_Desc} - it points to  \texttt{basic\_block}. Eventually \texttt{descriptor->basics} of \texttt{Access\_Desc} gets assigned the address of \texttt{basic\_block} by the address stored in \texttt{*free\_space}. At last \texttt{*free\_space += sizeof (struct basic\_block)} assignes the next free address after the last structure \texttt{basic\_block} for the next dimension - structure \texttt{Access\_Desc} and the following structure(s) \texttt{basic\_block}.
\begin{verbatim}
    descriptor->basics->offset = 0;
\end{verbatim}
Like \texttt{descriptor->skip} also \texttt{descriptor->basics->offset} is set to \texttt{0}.
\begin{verbatim}
    descriptor->basics->repeat = array_dist[distance +
        (4 * step_dim_count) + 1];
    descriptor->basics->stride = 0;
    descriptor->basics->count = 1 * type_size;
\end{verbatim}
The number of elements of the global length in each dimension mark \texttt{
descriptor->basics->repeat}. This describes how many times elements of size \texttt{descriptor->basics->count = 1 * type\_size} has to be read.
In case of no distribution there is no stride of data necessary.
%
%
\subsubsection{BLOCK(n) distribution} \label{blocksec}

\begin{verbatim}
    global_length =  array_dist[distance +
        ( 4 * (array_dimension - 1 - step_dim_count))];
    local_length = array_dist[distance +
        ( 4 * (array_dimension - 1 - step_dim_count)) + 1];
    argument = array_dist[distance +
        ( 4 * (array_dimension - 1 - step_dim_count)) + 3];
\end{verbatim}

Definitions like \texttt{global\_length} \texttt{local\_length} and \texttt{argument} are made to reduce expense if informations from the runtime-descriptor are used more than once. The values of these three definitions depend on the current dimension obtained by \texttt{step\_dim\_count}.

Relating to the fifth processor (see figure \ref{dimtwo}) for the first dimension \texttt{skip} becomes calculated as follows:

\begin{verbatim}
    descriptor->skip  = global_length -
        (argument * (dim_contr[step_dim_count + 1] + 1)) + (argument - local_length);
\end{verbatim}

The \texttt{global\_length} for this dimension is still calculated by the lines before. It is set to \texttt{17} in this example. \texttt{argument} for cyclic distribution of this dimension from the runtime-descriptor gives information about how many elements each processor becomes assigned. In this case \texttt{argument} is set to \texttt{3} as the distribution directive of the corresponding \emph{HPF} code demands. The value of the array \texttt{dim\_contr} in position \texttt{step\_dim\_count + 1} stores the processor coordinate of the corresponding dimension. In this case it is set to \texttt{1}.
\texttt{(dim\_contr[step\_dim\_count + 1] + 1)} which results in \texttt{2} marks the second processor in this dimension. Multiplied with the number of blocks each processor gets assigned (\texttt{argument}) and furthermore subtracted from the global length \texttt{skip} results in \texttt{8}. This
means that \texttt{8} elements have to be skipped for dimension 1 while data for
processor 5 is read or written.

There is one special case if the argument exceeds the local length (see figure \ref{dimtwo} the rightmost elements in the matrix)). This is the case where the number of elements assigned to a processor is smaller than the number of elements a processor could get assigned. (Elements 15 to 17 in the second dimension) For the last processor in figure \ref{dimtwo} the calculation of \texttt{descriptor->skip} would result in \texttt{-3} by the calculation described above. For this case \texttt{argument - local\_length} is a neccessary correction. It results to \texttt{3} and marks \texttt{descriptor\_skip} to \texttt{0}. In all other cases the correction \texttt{argument - local\_length} results to zero because of equal values.

The next operations multiply the number of elements of all remaining dimensions to the calculated value in \texttt{skip}. Therefore \texttt{down\_count} contains the number of the following dimensions from \texttt{step\_dim\_count}. In this case \texttt{down\_count} is \texttt{1}. \\
In figure \ref{dimtwo} c) seven elements are skipped along dimension 2. These seven elements become multiplied with the global length of the remaining dimensions, in this case only one dimension (dimension 1) which has \texttt{14} elements.
\begin{figure}
  \centerline{\includegraphics[scale=0.6]{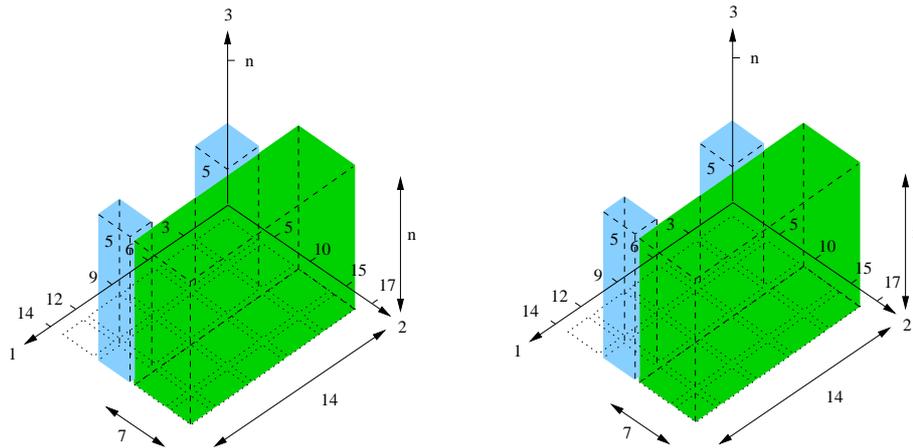}}
  \caption{three dimensional skip}
\label{3dskip}
\end{figure}
\\
Assuming a third dimension for the example data array the elements of \texttt{skip} would form a three dimensional block as shown in figure \ref{3dskip}. Dimension \texttt{1} and \texttt{2} and its distributions are the same as used in the example before. The only difference is the extended third dimension wich is not distributed (distribution \texttt{*} to keep graphical presenation simple). Furthermore it shows all data elements assigned to processor \texttt{5} for a three dimensions.

\begin{verbatim}
    down_count = array_dimension - (step_dim_count + 1);
    for (u = 0; u < down_count; u++)
        descriptor->skip *= array_dist[distance + (4 * u)];
    descriptor->skip *= type_size;
\end{verbatim}

At last \texttt{descriptor->skip} which now contains all elements to be skipped becomes multiplied with \texttt{type\_size} - to optain the proper size for \texttt{integer} for example.

All variables such as \texttt{no\_blocks} and \texttt{skip} in \texttt{Access\_Desc} are calculated now. To be able to assign values to \texttt{offset}, \texttt{repeat}, \texttt{count}, \texttt{stride} and \texttt{sub\_type} of the next datastructure \texttt{basic\_block} for this dimension the auxilliary pointer \texttt{*free\_space} becomes set to the next addresses like in section \ref{nodistr}.

\begin{verbatim}
    *free_space = (struct basic_block *) ((void *)descriptor) + sizeof(Access_Desc);
    descriptor->basics = *free_space;
    *free_space += sizeof (struct basic_block);
\end{verbatim}

\texttt{*free\_space} points at the next address after the last structure \texttt{basic\_block}. That is why all following descriptions relate to variables defined in \texttt{basic\_block}.

\begin{verbatim}
    descriptor->basics->offset = 1;
\end{verbatim}

\texttt{offset} is set to \texttt{1} initially. This is neccessary for the following calculations.
Depending on if the coordinate for the current dimension in the processor array is set or not the variable \texttt{offset} becomes calculated or zero. It is zero in case of if the processor coordinate for this dimension (\texttt{dim\_contr[step\_dim\_count + 1]}) is zero. Otherwise the number of blocks as offset (expressed by the processor coordinate) times the number of elements (\texttt{argument}) for each block (\texttt{argument  * dim\_contr[step\_dim\_count+1}) gives the number of element to offset till the first element becomes read or written in this dimension.
If an offset is set the remaining dimensions are also multiplied too.

\begin{verbatim}
    if (dim_contr[step_dim_count + 1]) {

        down_count = array_dimension - (step_dim_count + 1);
        descriptor->basics->offset = argument  *
            dim_contr[step_dim_count+1];

        for (i = 0; i < down_count; i++)
            descriptor->basics->offset *=
                array_dist[distance + (4 * i)];
    }
    else
        descriptor->basics->offset = 0;

    descriptor->basics->offset *= type_size;
\end{verbatim}

Each block appears only once in each dimension if distribution is \texttt{BLOCK}.
\begin{verbatim}
    descriptor->basics->repeat = 1;
\end{verbatim}

Because there is no gap between each block \texttt{stride} is zero.

\begin{verbatim}
    descriptor->basics->stride = 0;
\end{verbatim}

The value of the last variable \texttt{count} depends on the present dimension. If the last dimension is not reached \texttt{count} becomes assigned the value of the local length of the corresponding dimension. \texttt{count} assignes only the number of elements read or written.
If the last dimension is reached the local length of the last dimension is assigned too. In this case the value of \texttt{count} also becomes multiplied with \texttt{type\_size}. This means that only if the last dimension is reached \texttt{count} contains the size of the block (number of elements) times the size of an element.
\begin{verbatim}
    if (step_dim_count + 1 < array_dimension) {
        descriptor->basics->count = array_dist[distance +
            (4 * down_count) +1];
    }
    else {
        descriptor->basics->count = array_dist[distance +
            1] * type_size;
    }
\end{verbatim}
%
%
\subsubsection{CYCLIC distribution}

\begin{verbatim}
    global_length =  array_dist[distance +
        ( 4 * (array_dimension - 1 - step_dim_count))];
    local_length = array_dist[distance +
        ( 4 * (array_dimension - 1 - step_dim_count)) + 1];
    argument = array_dist[distance +
        ( 4 * (array_dimension - 1 - step_dim_count)) + 3];
\end{verbatim}

Like in the beginning of section \ref{blocksec} where the case of \texttt{BLOCK} distribution was explained the same definitions for \texttt{global\_length} \texttt{local\_length} and \texttt{argument} are made here too. \\

The following definitions are made to distinguish two kinds of data - regular and irregular data. Regular data is given if the length of each block is equal to the blocklength given by the runtime descriptor which is expressed by \texttt{argument}. Irregular data consists of only one block which in turn is formed by the last elements of the respective dimension. The number of these elements is less than \texttt{argument}. As shown in figure \ref{dimtwo} cyclic data distribution for processor 5 creates two blocks with different sizes. The first block (regular) consists of three elements (4, 5, 6) along the vertical axis (row order major). The second block (irregular) of processor 5 consists of only two elements (13, 14).
For each kind of these two blocks (regular and irregular) a datastructure \texttt{basic\_block} is definend. See figure \ref{lnktwo} where the datastructure for both dimensions is shown graphically.

The calculations are as follows.\\

\texttt{nbr\_occ\_elem\_mod} is zero if no irregular block exists. This is the case if the local length is divisible exactly.
Depending on if an irregular block exists or not the global length is \texttt{nbr\_occ\_elem\_mod} or zero. The global length for regular and irregular blocks can be distinguished by index \texttt{0} or \texttt{1}.

\begin{verbatim}
    nbr_occ_elem_mod = local_length % argument;
    mem_glob_length[0] = global_length - nbr_occ_elem_mod;
    mem_glob_length[1] = nbr_occ_elem_mod;
\end{verbatim}

The following calculations for \texttt{skip} depend on if \texttt{local\_length} is greater, equal or smaller than \texttt{argument}. \\
The first case \texttt{local\_length > argument} does not include that an irregular block exists. It only means that an irregular block is possible. \\
First an auxiliary variable \texttt{temp} stores the global length without the elements of the irregular block. In this example \texttt{temp = 14 - (14 mod 3)}. After this the whole number of regular blocks becomes calculated by \texttt{temp / argument}. In the next step \texttt{temp} becomes the an indicator for computing the proper size of \texttt{skip} of the regular block by \texttt{4 mod 3 = 1}. \texttt{temp} represents the number of all the blocks of
the regular part (i.e. the blocks which are all of equal size) it is therefor
also referred to as an auxiliary skip value.

\begin{verbatim}
    /* (local length > argument) irregular block possible  */
    if (local_length > argument) {
        temp = global_length - (global_length % argument);
        temp = temp / argument;
        temp = temp % array_dist[array_dist[0]-step_dim_count];
\end{verbatim}

\texttt{temp} compared to the present processor coordinate allowes to calculate the proper number of blocks to skip as follows:

\begin{verbatim}
        if (dim_contr[step_dim_count+1] < temp) {
        /* subcount actual proc coordinate from available number of procs */
            descriptor->skip = temp - (dim_contr[step_dim_count+1] + 1);
        }
\end{verbatim}

If the current processor coordinate is smaller than the calculated auxiliary skip the remaining elements

example:\\
xyzxy|z\\
01201 2\\
temp = 17\\
argument = 3\\
(i)  temp = 15\\
(ii) temp = 15/3 = 5\\
(iii) temp = 5\%3 = 2  = auxskip\\
respective to processor two. It has coordinate 1 and thus \\
\texttt{descriptor-> skip = 0}, since 2 - (1+1)\\

The second case is given if the the value of auxiliary skip \texttt{temp} is equal to the present processor coordinate. Therefore \texttt{skip} is the number of processors in the processor array minus the present processor. \\
At last if auxiliary skip \texttt{temp} is less than the processor coordinate

\begin{verbatim}
        else {
            if (dim_contr[step_dim_count+1] == temp) {
	        descriptor->skip = array_dist[array_dist[0] - step_dim_count]
		-1;
            }
	    else {
 	        descriptor->skip = array_dist[array_dist[0]-step_dim_count]-
		(dim_contr[step_dim_count + 1] + 1);
	        printf ("lod tmp %d\n", descriptor->skip);
	    }
        }
\end{verbatim}

If the processor coordinate is greater than \texttt{temp} \texttt{skip} becomes the value of the number of processors minus the processor coordinate for the present dimension.

example:\\
xyzxy|z\\
01231 2\\
temp = 17\\
argument = 3\\
(i)  temp = 15\\
(ii) temp = 15/3 = 5\\
(iii) temp = 5\%4 = 1 = auxskip\\
respective to processor three, which has coordinate 2 and thus \\
\texttt{descriptor-> skip = 0}, since 4 - (2)\\

The number, which is stored in \texttt{skip} as a result of the operations till this point is the number of processors followed by the present processor.

To obtain the number of elements to skip, the present value of \texttt{skip} is
multiplied with \texttt{agrument}.

\begin{verbatim}
        descriptor->skip *= argument;
\end{verbatim}

At last if an irregular block exists \texttt{skip} becomes multiplied with \texttt{type\_size}. That is if \texttt{nbr\_occ\_elem\_mod} is not zero. Furthermore the value of \texttt{no\_blocks} is \texttt{2} because of two structures \texttt{basic\_block}.\\
If no irregular block exists \texttt{no\_blocks} is \texttt{1}. If a block with less elements than \texttt{agument} exists (This block must be assigned to another processor. Otherwise an irregular block for the present processor would exist.) the number of these elements become added to \texttt{skip}.

\begin{verbatim}
        if (nbr_occ_elem_mod) { /* irregular block is given */
            descriptor->skip *= type_size;
            descriptor->no_blocks = 2;
        }
        else { /* no irregular block */
            descriptor->no_blocks = 1;
            descriptor->skip += (global_length % argument);
            descriptor->skip *= type_size;
        }
    } /* fi local_length > argument */
\end{verbatim}

The next possible state is given if the \texttt{local\_length} is equal to \texttt{argument}. In this case no irregular block for the present processor is possible. Of course a possible irregular block of another processor is taken into consideration. \\
From the \texttt{global\_length} where the irregular block (if it exists) is included the present processor coordinate which becomes multiplied with the number of element of each block \texttt{argument} becomes subtracted. So \texttt{skip} contains the number  of all elements from the rightmost present processor to the right border. \\
\texttt{no\_blocks} is \texttt{1} again.

\begin{verbatim}
    else { /* local length == arguement */

        if (local_length == argument) {
            descriptor->skip =
                global_length - ((dim_contr[step_dim_count + 1] +1) * argument);
            descriptor->skip *= type_size;
            descriptor->no_blocks = 1;
        }
\end{verbatim}

The last case is given if the \texttt{local\_length} is smaller than \texttt{argument}.
In this case the block assigned to the present processor must be the last elements in the proper dimension. \texttt{skip} must be zero. \texttt{mem\_glob\_length[0]} must be redefined.

\begin{verbatim}
            else {  /* local length < arguement -> must be last element */
                if (local_length < argument) {
                    mem_glob_length[0] = global_length;
                    descriptor->skip = 0;
                    descriptor->no_blocks = 1;
                }
            }
        }
\end{verbatim}

The auxiliary pointer \texttt{*free\_space} points to the next free address after \texttt{Access\_Desc}. That's where \texttt{descriptor->basic} points to through \texttt{descriptor->basics = *free\_space;}.
The next free address is the address after the last structure \texttt{basic\_block}. Depending on \texttt{descriptor->no\_blocks} (one or two blocks) \texttt{*free\_space} points one or two blocks after \texttt{Access\_Desc}.

\begin{verbatim}
        /* point to basic_block */
        *free_space = (struct basic_block *) ((void *)descriptor) +
            sizeof(Access_Desc);
        descriptor->basics = *free_space;
        *free_space += descriptor->no_blocks * sizeof (struct basic_block);
\end{verbatim}

The following calculations relate to variables stored in \texttt{basic\_blocks}.
For each block (regular and irregular) the values of \texttt{offset}, \texttt{repeat}, \texttt{stride} and \texttt{count} are calculated. \\
The index \texttt{i} of the slope indicates which block becomes calculated at the moment. \\
The first variable \texttt{offset} for the regular block is the number of elements assigned to processors before the first element assigned to the present processor. If the first processor gets assigned the first elements too \texttt{offset} is zero. \\
For the irregular block \texttt{offset} becomes assigned the value of \texttt{skip}. Supplementary \texttt{skip} becomes set to zero. The calculation of \texttt{skip} in \texttt{Access\_Desc} is used as \texttt{offset} in this case.

\begin{verbatim}
    for ( i = 0; i < descriptor->no_blocks; i++) {
        descriptor->basics[i].offset = 1; /* default */
        down_count = array_dimension - (step_dim_count + 1);

        if ( i == 0) { /*  calculation regular block */
            if (dim_contr[step_dim_count + 1]) {
                descriptor->basics[i].offset =
                    argument * dim_contr[step_dim_count + 1] * type_size;
            }
            else {
                descriptor->basics[i].offset = 0;
            }
        }
        else { /* offset equ to skip of regular block */
            descriptor->basics[i].offset = descriptor->skip;	
            descriptor->skip = 0;
        }
\end{verbatim}

Calculation of \texttt{repeat} is divided into several steps. First variable \texttt{temp} stores the number of blocks with length \texttt{argument}. Outgoing from this number \texttt{aux\_repeat} can be calculated. \texttt{aux\_repeat} becomes assigned the number of how many times the present processor becomes assigned data elements with length \texttt{argument} ignoring the difference between regular and irregular blocks. \\
This is corrected by the next statements. If an irregular block exists and the \texttt{local\_length} is greater or equal \texttt{argument} (data assignment is \texttt{repeat}ed more than one time) the variable \texttt{temp} becomes assigned the global length of the regular part first. This value divided through the number of processors of the processor array for the corresponding dimension results in  a value which can be compared to processor coordinate. If they are equal the number of \texttt{aux\_repeat} must become decremented.

The last step is to assign the proper value of \texttt{repeat} to \texttt{descriptor->basics[i].repeat}. This depends on if \texttt{i} is zero (regular block) or one (irregular block).
If \texttt{i} is zero the value of \texttt{aux\_repeat} can be assigned to \texttt{repeat}. If \texttt{i} is one \texttt{repeat} becomes assigned the value \texttt{1} because an irregular block becomes repeated only one time.

\begin{verbatim}	
        temp = global_length / argument;

        if (global_length % argument) {
            temp += 1;
        }
        aux_repeat = ((temp - dim_contr[step_dim_count + 1] -1) /
            array_dist[array_dist[0] - step_dim_count]) + 1;

        if ((global_length % argument) && (local_length >= argument)) {
            temp = (global_length - (global_length % argument)) / argument;
            temp = temp % array_dist[array_dist[0] - step_dim_count];
            if (dim_contr[step_dim_count + 1] == temp )
                aux_repeat -= 1;	
            }

        if ( (!i) )  /* calc only neccessary in regular block  */
            descriptor->basics[i].repeat = aux_repeat;
        else /* irregular block */
            descriptor->basics[i].repeat = 1;
\end{verbatim}

Outgoing from calculation of \texttt{aux\_repeat} before \texttt{stride} can be calculated. If \texttt{aux\_repeat} is greater than \texttt{1} (\texttt{stride exists})) \texttt{stride} is the number of blocks between two blocks assigned to the proper processor. In any case this is the number of processors of the processor array minus one.
If \texttt{aux\_repeat} is equal or lower \texttt{1} only one block of elements becomes assigned to one processor. That is why \texttt{stride} is zero.

\begin{verbatim}
        if (aux_repeat > 1)
            descriptor->basics[i].stride =
                array_dist[array_dist[0]-step_dim_count] - 1;
        else
            descriptor->basics[i].stride = 0;
	
        descriptor->basics[i].stride *= (argument * type_size);
\end{verbatim}

Considering the first case - the calculation of \texttt{count} of the regular block the variable \texttt{count} ist \texttt{1} if the passed \texttt{argument} is \texttt{1}. In this case each element becomes assigned to a processor. \\
If \texttt{argument} is not equal \texttt{1} two cases can be distinguished. In the first case the \texttt{local\_length} is smaller than \texttt{argument}. That is when \texttt{count} becomes assigned the value of \texttt{local\_length} (the length of one block with blocklength \texttt{local\_length} smaller than \texttt{argument}). If \texttt{local\_length} is greater than \texttt{argument} the variable \texttt{count} becomes assigned the value of \texttt{argument}. \texttt{argument} is the maximum blocklength in the regular block. \\
If the first dimension is reached the calculated value of \texttt{count} becomes multiplied with \texttt{type\_size}. This is only done if the values of the first dimension are calculated. \\
Because of the irregular block consists of only one block \texttt{count} is the the number of elements forming this irregular block. In this case there is no distinction between the first dimension and all further dimensions. Each calculated value of \texttt{count} for the irregular block becomes multiplied with \texttt{type\_size}.

\begin{verbatim}

        if (i == 0) { /* regualar block */

            if (argument == 1) /* cyclic(1) */
                descriptor->basics[0].count = 1;
            else	
	
            if (local_length < argument)
                descriptor->basics[0].count = local_length;
            else
                descriptor->basics[0].count = argument;

            if ((step_dim_count + 1) >= array_dimension)
                descriptor->basics[0].count *= type_size;
        }
        else { /* irregular block */

            descriptor->basics[i].count = local_length % argument;
            descriptor->basics[i].count *= type_size;
        }
\end{verbatim}

At last the the calculated values \texttt{skip}, \texttt{offset} and \texttt{stride} become multiplied with all \texttt{global\_length}s of the remaining dimensions. \\
\texttt{sub\_count}s and \texttt{sub\_actual}s values are zero. These two variables are not neccessary for this interface. Instead of its values are set to zero (for the regular and irregular block).

\begin{verbatim}
        down_count = array_dimension - (step_dim_count + 1);

        for (u = 0; u < down_count; u++) {
            descriptor->skip *= array_dist[distance + ( 4 * u)];
            descriptor->basics[i].offset *=
                array_dist[distance + (4 * u)];
            descriptor->basics[i].stride *=
                array_dist[distance + (4 * u)];
        }
        descriptor->basics->sub_count = 0;
        descriptor->basics->sub_actual = 0;
        descriptor->basics[i].sub_count = 0;
        descriptor->basics[i].sub_actual = 0;

    } /* i: [0 ... descriptor->no_blocks] */
} /* switch */
\end{verbatim}

%% file: scalability.tex
\setlength{\unitlength}{0.240900pt}
\ifx\plotpoint\undefined\newsavebox{\plotpoint}\fi
\sbox{\plotpoint}{\rule[-0.200pt]{0.400pt}{0.400pt}}%
\begin{picture}(1049,629)(0,0)
\font\gnuplot=cmr10 at 10pt
\gnuplot
\sbox{\plotpoint}{\rule[-0.200pt]{0.400pt}{0.400pt}}%
\put(220.0,113.0){\rule[-0.200pt]{184.288pt}{0.400pt}}
\put(220.0,113.0){\rule[-0.200pt]{4.818pt}{0.400pt}}
\put(198,113){\makebox(0,0)[r]{0}}
\put(965.0,113.0){\rule[-0.200pt]{4.818pt}{0.400pt}}
\put(220.0,183.0){\rule[-0.200pt]{4.818pt}{0.400pt}}
\put(198,183){\makebox(0,0)[r]{0.5}}
\put(965.0,183.0){\rule[-0.200pt]{4.818pt}{0.400pt}}
\put(220.0,254.0){\rule[-0.200pt]{4.818pt}{0.400pt}}
\put(198,254){\makebox(0,0)[r]{1}}
\put(965.0,254.0){\rule[-0.200pt]{4.818pt}{0.400pt}}
\put(220.0,324.0){\rule[-0.200pt]{4.818pt}{0.400pt}}
\put(198,324){\makebox(0,0)[r]{1.5}}
\put(965.0,324.0){\rule[-0.200pt]{4.818pt}{0.400pt}}
\put(220.0,395.0){\rule[-0.200pt]{4.818pt}{0.400pt}}
\put(198,395){\makebox(0,0)[r]{2}}
\put(965.0,395.0){\rule[-0.200pt]{4.818pt}{0.400pt}}
\put(220.0,465.0){\rule[-0.200pt]{4.818pt}{0.400pt}}
\put(198,465){\makebox(0,0)[r]{2.5}}
\put(965.0,465.0){\rule[-0.200pt]{4.818pt}{0.400pt}}
\put(220.0,536.0){\rule[-0.200pt]{4.818pt}{0.400pt}}
\put(198,536){\makebox(0,0)[r]{3}}
\put(965.0,536.0){\rule[-0.200pt]{4.818pt}{0.400pt}}
\put(220.0,606.0){\rule[-0.200pt]{4.818pt}{0.400pt}}
\put(198,606){\makebox(0,0)[r]{3.5}}
\put(965.0,606.0){\rule[-0.200pt]{4.818pt}{0.400pt}}
\put(220.0,113.0){\rule[-0.200pt]{0.400pt}{4.818pt}}
\put(220,68){\makebox(0,0){1}}
\put(220.0,586.0){\rule[-0.200pt]{0.400pt}{4.818pt}}
\put(475.0,113.0){\rule[-0.200pt]{0.400pt}{4.818pt}}
\put(475,68){\makebox(0,0){2}}
\put(475.0,586.0){\rule[-0.200pt]{0.400pt}{4.818pt}}
\put(730.0,113.0){\rule[-0.200pt]{0.400pt}{4.818pt}}
\put(730,68){\makebox(0,0){3}}
\put(730.0,586.0){\rule[-0.200pt]{0.400pt}{4.818pt}}
\put(985.0,113.0){\rule[-0.200pt]{0.400pt}{4.818pt}}
\put(985,68){\makebox(0,0){4}}
\put(985.0,586.0){\rule[-0.200pt]{0.400pt}{4.818pt}}
\put(220.0,113.0){\rule[-0.200pt]{184.288pt}{0.400pt}}
\put(985.0,113.0){\rule[-0.200pt]{0.400pt}{118.764pt}}
\put(220.0,606.0){\rule[-0.200pt]{184.288pt}{0.400pt}}
\put(45,359){\makebox(0,0){time (secs)}}
\put(602,23){\makebox(0,0){servers}}
\put(220.0,113.0){\rule[-0.200pt]{0.400pt}{118.764pt}}
\put(855,541){\makebox(0,0)[r]{ViPIOS}}
\put(877.0,541.0){\rule[-0.200pt]{15.899pt}{0.400pt}}
\put(220,538){\usebox{\plotpoint}}
\multiput(220.00,536.92)(0.733,-0.500){345}{\rule{0.686pt}{0.120pt}}
\multiput(220.00,537.17)(253.576,-174.000){2}{\rule{0.343pt}{0.400pt}}
\multiput(475.00,362.92)(7.221,-0.495){33}{\rule{5.767pt}{0.119pt}}
\multiput(475.00,363.17)(243.031,-18.000){2}{\rule{2.883pt}{0.400pt}}
\multiput(730.00,344.92)(1.705,-0.499){147}{\rule{1.460pt}{0.120pt}}
\multiput(730.00,345.17)(251.970,-75.000){2}{\rule{0.730pt}{0.400pt}}
\put(855,496){\makebox(0,0)[r]{ideal}}
\multiput(877,496)(20.756,0.000){4}{\usebox{\plotpoint}}
\put(943,496){\usebox{\plotpoint}}
\put(220,538){\usebox{\plotpoint}}
\multiput(220,538)(15.960,-13.269){16}{\usebox{\plotpoint}}
\multiput(475,326)(19.995,-5.567){13}{\usebox{\plotpoint}}
\multiput(730,255)(20.552,-2.901){13}{\usebox{\plotpoint}}
\put(985,219){\usebox{\plotpoint}}
\end{picture}

%% file: scale_none_dedicated.tex
\setlength{\unitlength}{0.240900pt}
\ifx\plotpoint\undefined\newsavebox{\plotpoint}\fi
\sbox{\plotpoint}{\rule[-0.200pt]{0.400pt}{0.400pt}}%
\begin{picture}(1049,629)(0,0)
\font\gnuplot=cmr10 at 10pt
\gnuplot
\sbox{\plotpoint}{\rule[-0.200pt]{0.400pt}{0.400pt}}%
\put(220.0,113.0){\rule[-0.200pt]{184.288pt}{0.400pt}}
\put(220.0,113.0){\rule[-0.200pt]{0.400pt}{118.764pt}}
\put(220.0,113.0){\rule[-0.200pt]{4.818pt}{0.400pt}}
\put(198,113){\makebox(0,0)[r]{0}}
\put(965.0,113.0){\rule[-0.200pt]{4.818pt}{0.400pt}}
\put(220.0,183.0){\rule[-0.200pt]{4.818pt}{0.400pt}}
\put(198,183){\makebox(0,0)[r]{0.5}}
\put(965.0,183.0){\rule[-0.200pt]{4.818pt}{0.400pt}}
\put(220.0,254.0){\rule[-0.200pt]{4.818pt}{0.400pt}}
\put(198,254){\makebox(0,0)[r]{1}}
\put(965.0,254.0){\rule[-0.200pt]{4.818pt}{0.400pt}}
\put(220.0,324.0){\rule[-0.200pt]{4.818pt}{0.400pt}}
\put(198,324){\makebox(0,0)[r]{1.5}}
\put(965.0,324.0){\rule[-0.200pt]{4.818pt}{0.400pt}}
\put(220.0,395.0){\rule[-0.200pt]{4.818pt}{0.400pt}}
\put(198,395){\makebox(0,0)[r]{2}}
\put(965.0,395.0){\rule[-0.200pt]{4.818pt}{0.400pt}}
\put(220.0,465.0){\rule[-0.200pt]{4.818pt}{0.400pt}}
\put(198,465){\makebox(0,0)[r]{2.5}}
\put(965.0,465.0){\rule[-0.200pt]{4.818pt}{0.400pt}}
\put(220.0,536.0){\rule[-0.200pt]{4.818pt}{0.400pt}}
\put(198,536){\makebox(0,0)[r]{3}}
\put(965.0,536.0){\rule[-0.200pt]{4.818pt}{0.400pt}}
\put(220.0,606.0){\rule[-0.200pt]{4.818pt}{0.400pt}}
\put(198,606){\makebox(0,0)[r]{3.5}}
\put(965.0,606.0){\rule[-0.200pt]{4.818pt}{0.400pt}}
\put(220.0,113.0){\rule[-0.200pt]{0.400pt}{4.818pt}}
\put(220,68){\makebox(0,0){2}}
\put(220.0,586.0){\rule[-0.200pt]{0.400pt}{4.818pt}}
\put(603.0,113.0){\rule[-0.200pt]{0.400pt}{4.818pt}}
\put(603,68){\makebox(0,0){4}}
\put(603.0,586.0){\rule[-0.200pt]{0.400pt}{4.818pt}}
\put(985.0,113.0){\rule[-0.200pt]{0.400pt}{4.818pt}}
\put(985,68){\makebox(0,0){8}}
\put(985.0,586.0){\rule[-0.200pt]{0.400pt}{4.818pt}}
\put(220.0,113.0){\rule[-0.200pt]{184.288pt}{0.400pt}}
\put(985.0,113.0){\rule[-0.200pt]{0.400pt}{118.764pt}}
\put(220.0,606.0){\rule[-0.200pt]{184.288pt}{0.400pt}}
\put(45,359){\makebox(0,0){time (secs)}}
\put(602,23){\makebox(0,0){servers = clients}}
\put(220.0,113.0){\rule[-0.200pt]{0.400pt}{118.764pt}}
\put(855,541){\makebox(0,0)[r]{constant workload}}
\put(877.0,541.0){\rule[-0.200pt]{15.899pt}{0.400pt}}
\put(220,533){\usebox{\plotpoint}}
\multiput(220.00,531.92)(0.993,-0.500){383}{\rule{0.894pt}{0.120pt}}
\multiput(220.00,532.17)(381.145,-193.000){2}{\rule{0.447pt}{0.400pt}}
\multiput(603.00,338.92)(6.436,-0.497){57}{\rule{5.193pt}{0.120pt}}
\multiput(603.00,339.17)(371.221,-30.000){2}{\rule{2.597pt}{0.400pt}}
\put(855,496){\makebox(0,0)[r]{increasing workload}}
\multiput(877,496)(20.756,0.000){4}{\usebox{\plotpoint}}
\put(943,496){\usebox{\plotpoint}}
\put(220,290){\usebox{\plotpoint}}
\multiput(220,290)(20.581,2.687){19}{\usebox{\plotpoint}}
\multiput(603,340)(19.585,6.870){20}{\usebox{\plotpoint}}
\put(985,474){\usebox{\plotpoint}}
\end{picture}

%% file: further_tests.tex
\section{Further Tests on ViPIOS and ViMPIOS}

In the previous chapter we were discussing some tests concerning the scalability of ViPIOS. In 
particular we tried to show that the performance of the overall system increases when the number of 
sever processes is increased as. However, the upper limit for the number of server processes is 
determined by the number of client processes. The reason for this statement is quite obvious when 
we bear in mind that each client process can only be handled by one sever process rather than 
multiple ones. \\

In this chapter we will show some test results which we have already analyzed in the previous 
chapter. However, rather than reading file sizes of 2 MB, we will now take a look at larger file sizes. The reason for presenting these results in addition to the one we have discussed in the previous chapter is that during the first and this second test phase some changes to the ViPIOS kernel where made which could have some impacts of the overall performance. In addition to that, this workstation cluster was increased by 8 Linux PCs such that at present our system comprises 16 PC clustered together over an Ethernet network.\\

Later in this chapter we will compare the performance of ViMPIOS with ROMIO where we will notice 
that the performance of ROMIO is about 20 times better than ViMPIOS for small file sizes. The 
advantage can be explained by the internal caching and prefetching mechanism of ROMIO (see 2-
Phase-Method). For larger file sizes this method does not improve the performance any more. Thus, 
the performance of ROMIO is merely three times higher for file sizes beyond 30 MB.\\

By means of further test series will analyze the behavior of derived datatypes. We first try to write 
files in a certain way, let's say BLOCK, BLOCK, and read them back at a later stage CYCLIC, 
CYCLIC.\\

We will conclude this test phase by describing the performance of what we may call \textit{sparse derived 
datatypes}. A sparse derived datatype is a derived datatype which has a large \textit{holes} between 
adjacent data blocks. Since ROMIO uses the 2-Phase-Method, we might expect ViMPIOS to 
outperform the counterpart in this special case.\\

\subsection{Scalability Results With Larger Files}

In this sub section we want to present to scalability results for a 60 MB file. Following assumptions 
are made for the analysis: The number of client processes is kept constant throughout each test phase 
whereas the number of server processes ranges from 1 to 4. According to our experience, which we 
derived from the tests in the previous section, we would expect the system to have a significant 
performance gain when the number of server processes is increased. The results are depicted in the Table \ref{scalability_dedicated_II}.\\

\begin{table}
\caption{Scalability results II for dedicated I/O nodes} 
\label{scalability_dedicated_II}
\begin{center}
\begin{tabular}{|rrrrrr|}
\hline
clients & servers & max & min & mean & variance \\
\hline
4 & 1 & 88.79 & 81.61 & 86.08 & 10.1245 \\
4 & 2 & 52.00 & 46.12 & 49.04 & 5.7647 \\
4 & 3 & 51.91 & 45.89 & 48.52 & 6.3144 \\
4 & 4 & 28.80 & 25.30 & 26.51 & 2.6367 \\
\hline
\end{tabular}
\end{center}
\end{table}

Since the curve resembles where much the one depicted in Figure \ref{scalability}, we will not preset it here one more but refer the reader to the previous chapter.

\subsection{ROMIO vs. ViMPIOS}

\subsubsection{Read-Write-Behavior}

In the introduction of this chapter we stated that ROMIO's performance for writing a file and reading 
it back later on is much higher for small files. The larger the files get, the better ViMPIOS converges 
to the results of ROMIO although ViPIOS does not support any internal caching or prefetching 
algorithm to optimize the system yet. In the table we give the results for both systems. However, if the 
file size lies beyond 200 MB, ViPIOS still works erroneously. Thus, some more research work must 
be invested in order to verify this drawback. The results can be seen in Table \ref{romio_vs_vimpios}. 

In order to compare the two systems directly the second column gives the ratio of the mean times for ROMIO and ViMPIOS. The further colums state the test results of ROMIO whereas the last columns contain the equivalent information on ViMPIOS. To sum up the results very briefly we can state state ROMIO's performance  outranges ViMPIOS at on order of magnitude of 21 for file sizes of about 16 MB. However, this gap shrinks to a factor of merely 2.5 for file sizes lager than 16 MB. In our table we only depict the values for three different file sizes.\\

\begin{table}
\caption{Performance of ROMIO vs ViMPIOS} 
\label{romio_vs_vimpios}
\begin{center}
\begin{tabular}{|rrrrrrrr|}
\hline
File size & R. vs. V. & max & min & mean &  max & min & mean \\
\hline
16 & 1:21 & 5.19 & 1.36 & 1.95 & 43.15 & 41.17 & 41.09\\
40 & 1:2.50 & 167.23 & 42.79 & 105.94 & 339.36 & 239.09 & 289.31\\
100 & 1:2.50 & 406.80 & 154.62 & 279.60 & 702.12 & 696.36 & 698.81\\
\hline
\end{tabular}
\end{center}
\end{table}

\subsubsection{Derived Datatypes}

This kind of tests is supposed to be the most promising ones for ViMPIOS due to the 
completely different approaches of solving this problem. We have already stated that ROMIO uses 
the 2-Phase-Method for reading and writing data. Thus, the whole range of data including so-called 
holes is read rather than accessing only the data, which is actually needed. Since ViPIOS´ approach 
handles this problem by recursively accessing the data according to the file view (compare the 
chapter on the implementation of derived datatypes), i.e. no \textit{unnecessary} data is accessed and therefore I/O overhead is decreased. Unfortunately, no test results are available yet to some internal 
inconsistencies of the system.

%% file: vipios_interface.tex
\section{ViPIOS Interface}
In this section we describe the interface of the ViPIOS which
serves as the basis for the MPI-IO implementation we will describe
in the next chapters. The first part deals with establishing a
connection to the ViPIOS and how to disconnect at a later stage.
In the second part we summarize the commands to perform file
manipulation with the ViPIOS.

\subsection{Connection and Disconnecting}
Before an application program can use the functionality of the ViPIOS a connection must be established first.\\

{\sffamily
\begin{tabbing}
{\bfseries bool ViPIOS\_Connect(int ViPIOS\_System\_ID)}\\
INOUT   \=parameterxxxxxxxxxxx  \= description              \kill
IN  \>ViPIOS\_System\_ID    \>system identifier

\end{tabbing}
}

\textbf{Description:}\\
Initializes the ViPIOS and establishes a connection between an application program and the ViPIOS.\\

\textbf{Example:}\\
\texttt{ViPIOS\_Connect(0);}\\

{\sffamily
    {\bfseries bool ViPIOS\_Disconnect(void)}\\
}

\textbf{Description:}\\
Disconnects the application program from the ViPIOS.\\

{\sffamily
    { \bfseries{bool ViPIOS\_Shutdown(void)} }\\
}

\textbf{Description:}\\
Shuts the ViPIOS down. All processes are closed and the ViPIOS FAT is written back. This function can only be used by an administrative interface.

\subsection{File Manipulation}
{\sffamily
\begin{tabbing}
{\bfseries bool ViPIOS\_Open(const char filename[], int flags, int *fid)}\\
INOUT \=parameter     \= description                                  \kill
IN    \>filename        \>name of the file\\
IN    \>flags       \>file access mode (R, W, RW)\\
INOUT \>fid             \>file identifier\\
\end{tabbing}
}

{\textbf{Description:}}\\
Opens an existing or creates a new file.\\

\textbf{Example:}\\
\texttt{ViPIOS\_Open ("matrix", 'r', \&fid1);}\\

{\sffamily
\begin{tabbing}
{\bfseries bool ViPIOS\_Close(int fid)}\\
INOUT \=parameter     \= description                                  \kill
IN \>fid            \>file identifier\\
\end{tabbing}
}

\textbf{Description:}\\
Closes an open file.\\

{\sffamily
\begin{tabbing}
{\bfseries bool ViPIOS\_Remove(const char filename[])}\\
INOUT \=parameter     \= description                                  \kill
IN \>filename           \>name of the file\\
\end{tabbing}
}

\textbf{Description:}\\
Removes an existing ViPIOS file.\\

{\sffamily
\begin{tabbing}
{\bfseries int ViPIOS\_File\_set\_size (int fid, int size) } \\
INOUT..\=parameter      \= description                                  \kill
INOUT   \>fid           \>file identifier\\
IN      \>size          \>size (in bytes) to truncate or expand file
\end{tabbing}
}

\textbf{Description:}\\
Resizes the file defined by \textit{fid}.\\

{\sffamily
\begin{tabbing}
{\bfseries int ViPIOS\_File\_get\_size (int fid, int *size) } \\
INOUT..\=parameter      \= description                                  \kill
IN      \>fid   \>file identifier\\
OUT     \>size  \>size of the file in bytes
\end{tabbing}
}

\textbf{Description:}\\
Returns the current size in bytes of the file defined by \textit{fid}.

\subsection{Data Access}

\paragraph{Blocking Routines}

{\sffamily
\begin{tabbing}
{\bfseries bool ViPIOS\_Read (int fid, void *buffer, int count, int offset)} \\
INOUT \=parameter     \= description                                  \kill
IN  \>fid               \>file identifier assigned in ViPIOS\_Open \\
OUT \>buffer        \>initial address of buffer\\
IN  \>count     \>number of bytes to read from file\\
IN  \>offset        \>byte offset
\end{tabbing}
}

{\textbf{Description}}\\
Reads data from an open file denoted by the file identifier into \textit{buffer}. The last parameter \textit{offset} states whether the operations is a so-called routine with \textit{explicit offset} or not. Further information is given in the next routine.\\

\textbf{Example}\\
\texttt{ViPIOS\_Read (fid1, buf, 15,-1);}\\

{\sffamily
\begin{tabbing}
{\bfseries int  ViPIOS\_Read\_struct (int fid, void *buffer, int len,}\\
{\bfseries Access\_Desc *desc, int offset, int at)}\\
INOUT \=parameter     \= description                                  \kill
IN  \>fid               \>file identifier assigned in ViPIOS\_Open \\
OUT \>buffer        \>initial address of buffer\\
IN  \>len           \>number of bytes to read from file\\
IN  \>desc      \>initial address of access descriptor\\
IN  \>offset        \>displacement of the file\\
IN  \>at            \>offset relative to the displacement
\end{tabbing}
}

{\textbf{Description}}\\
Reads data from an open file in a strided way according to the file access pattern, i.e. the file view, specified by \textit{desc}. \textit{offset} is similar to the parameter \textit{disp} in the routine \textit{MPI\_File\_set\_view} and specifies the start position of the strided access. The last parameter \textit{at} allows distinguishing between data access with explicit offset and data access with an individual file pointer. The value -1 means that the file is read from the current position. Any value greater than 0 sets the file pointer to the specified position. However, since data access with explicit offsets should not interfere with data access with individual file pointers (see corresponding section in the chapter about MPI-IO), the file pointer is not updated after the read operation. Thus, the file pointer is only updated if the value of the parameter \textit{at} is set to -1. A detailed description of how the strided access is accomplished and how the parameter \textit{desc} is used is given in the next section.\\

\textbf{Example:}\\
\texttt{ViPIOS\_Read\_struct (fid1,buf,40,view\_root,20,-1);}\\

40 byte values starting from the position 20 are read according to the file access pattern defined by \textit{view\_root}. Since the last parameter (\textit{at}) is set to -1, the data access with an individual file pointer is simulated. Thus, the file pointer is updated.\\

\texttt{ViPIOS\_Read\_struct (fid1,buf,40,view\_root,20,80);}\\

Here, data access with explicit offset is simulated. The file is read from position 20 relative to the beginning of the file. Furthermore, the file is read from position 80 relative to the file access pattern. In contrast to the previous example the file pointer is not updated. The exact meaning of the two different offsets becomes clear when we take a look at the implementation of the MPI-IO routines.\\

{\sffamily
\begin{tabbing}
{\bfseries bool ViPIOS\_Write (int fid, const void *buffer, int count, int offset)} \\
INOUT \=parameter     \= description                                  \kill
IN  \>fid               \>file identifier assigned in ViPIOS\_Open \\
IN  \>buffer        \>initial address of buffer\\
IN  \>count     \>number of bytes read from file\\
IN  \>offset        \>byte offset
\end{tabbing}
}

{\textbf{Description:}}\\
Writes data contained in \textit{buffer} to an open file denoted by the file identifier. The last parameter \textit{offset} states whether the operations is a so-called routine with \textit{explicit offset} or not.\\

\textbf{Example:}\\
\texttt{ViPIOS\_Write (fid1, buf, 15,-1);}\\

{\sffamily
\begin{tabbing}
{\bfseries int  ViPIOS\_Write\_struct (int fid, const void *buffer, int len,}\\
{\bfseries Access\_Desc *desc, int offset, int at)}\\
INOUT \=parameter     \= description                                  \kill
IN  \>fid               \>file identifier assigned in ViPIOS\_Open \\
IN  \>buffer        \>initial address of buffer\\
IN  \>len           \>number of bytes to read from file\\
IN  \>desc      \>initial address of ViPIOS access descriptor\\
IN  \>offset        \>displacement of the file\\
IN  \>at            \>offset relative to the displacement
\end{tabbing}
}

{\textbf{Description:}}\\
Writes data in a strided way according to \textit{desc} to an open file starting from position \textit{disp}. The parameters \textit{offset} and \textit{at} have the same meaning as in \textit{ViPIOS\_Read\_struct} we have analyzed above.\\

\textbf{Example:}\\
\texttt{ ViPIOS\_Write\_struct(fid1,buf,40,view\_root,120,-1); }\\

\paragraph{Non-Blocking Routines}

{\sffamily
\begin{tabbing}
{\bfseries int  ViPIOS\_Iread  (int fid, void *buffer, int count,}\\
{\bfseries int offset, int *req\_id)}\\
INOUT \=parameter     \= description                                  \kill
IN  \>fid               \>file identifier assigned in ViPIOS\_Open \\
OUT \>buffer        \>initial address of buffer\\
IN  \>count     \>number of bytes to read from file\\
IN  \>offset        \>byte offset\\
IN  \>req\_id       \>identifier of the request
\end{tabbing}
}

{\textbf{Description}}\\
Reads data from an open file denoted by the file identifier into \textit{buffer} in a non-blocking way.\\

\textbf{Example}\\
\texttt{ViPIOS\_Iread (fid1, buf, 15,-1,\&req\_id);}\\

{\sffamily
\begin{tabbing}
{\bfseries int  ViPIOS\_Iread\_struct (int fid, void *buffer, int len,}\\
{\bfseries Access\_Desc *desc, int offset, int at, int *req\_id)}\\
INOUT \=parameter     \= description                                  \kill
IN  \>fid               \>file identifier assigned in ViPIOS\_Open \\
OUT \>buffer        \>initial address of buffer\\
IN  \>len           \>number of bytes to read from file\\
IN  \>desc      \>initial address of access descriptor\\
IN  \>offset        \>displacement of the file\\
IN  \>at            \>offset relative to the displacement\\
IN  \>req\_id           \>identifier of the request

\end{tabbing}
}

{\textbf{Description}}\\
Reads data from an open file denoted by the file identifier into \textit{buffer} in a non-blocking way.\\

{\sffamily
\begin{tabbing}
{\bfseries int  ViPIOS\_Iwrite  (int fid, void *buffer, int count,}\\
{\bfseries int offset, int *req\_id)}\\
INOUT \=parameter     \= description                                  \kill
IN  \>fid               \>file identifier assigned in ViPIOS\_Open \\
IN  \>buffer        \>initial address of buffer\\
IN  \>count     \>number of bytes to read from file\\
IN  \>offset        \>byte offset\\
IN  \>req\_id       \>identifier of the request
\end{tabbing}
}

{\textbf{Description:}}\\
Writes data contained in \textit{buffer} to an open file denoted by the file identifier in a non-blocking way. \\

\textbf{Example:}\\
\texttt{ViPIOS\_Write (fid1, buf, 15,-1,\&req\_id);}\\

{\sffamily
\begin{tabbing}
{\bfseries int  ViPIOS\_Iwrite\_struct (int fid, const void *buffer, int len,}\\
{\bfseries Access\_Desc *desc, int offset, int at, int *req\_id)}\\
INOUT \=parameter     \= description                                  \kill
IN  \>fid               \>file identifier assigned in ViPIOS\_Open \\
IN  \>buffer        \>initial address of buffer\\
IN  \>len           \>number of bytes to read from file\\
IN  \>desc      \>initial address of ViPIOS access descriptor\\
IN  \>offset        \>displacement of the file\\
IN  \>at            \>offset relative to the displacement\\
IN  \>req\_id           \>request identifier
\end{tabbing}
}

{\textbf{Description:}}\\
Writes data in a strided way according to \textit{desc} to an open file starting from position \textit{disp}.\\

{\sffamily
\begin{tabbing}
{\bfseries int ViPIOS\_File\_Test (int req\_id, int *flag)} \\
INOUT \=parameter     \= description                                  \kill
IN  \>req\_id               \>identifier of the request\\
OUT \>flag          \>flag
\end{tabbing}
}

{\textbf{Description:}}\\
This routine checks whether an outstanding non-blocking routine has finished. The result is given in \textit{flag}.\\

\textbf{Example:}\\
\texttt{ ViPIOS\_File\_Wait (req\_id, \&flag);}\\

int  ViPIOS\_File\_Wait     (int req\_id);

{\sffamily
\begin{tabbing}
{\bfseries int ViPIOS\_File\_Test (int req\_id)} \\
INOUT \=parameter     \= description                                  \kill
IN  \>req\_id               \>identifier of the request
\end{tabbing}
}

{\textbf{Description:}}\\
This routine checks waits until an outstanding non-blocking routine has finished.\\

\textbf{Example:}\\
\texttt{ ViPIOS\_File\_Wait (req\_id);}\\

\paragraph{Further Access Routines}

{\sffamily
\begin{tabbing}
{\bfseries bool ViPIOS\_Seek (int fid, int offset, int offset\_base)}\\
INOUT \=parameterxxxxxx \= description                                  \kill
IN  \>fid               \>file identifier assigned in ViPIOS\_Open \\
IN  \>offset        \>absolute file offset\\
IN  \>offset\_base  \>update mode
\end{tabbing}
}

\textbf{Description:}\\
Updates the file pointer of a file according to \textit{offset\_base}, whereas following features are possible:

\begin{itemize}
\item SEEK\_SET: pointer is set to \textit{offset}
\item SEEK\_CUR: pointer is set to the current pointer position plus \textit{offset}
\item SEEK\_END: pointer is set to the end of file
\end{itemize}

\textbf{Example:}\\
\texttt{ViPIOS\_Seek (fid1, 50, SEEK\_SET);}\\

The file pointer is set to position 50 of the file denoted by the file identifier.\\

{\sffamily
\begin{tabbing}
{\bfseries bool ViPIOS\_Seek\_struct (int fid, int offset, int offset\_base, }\\
{\bfseries Access\_Desc *desc)}\\
INOUT \=parameterxxxxxxx\= description                                  \kill
IN  \>fid               \>file identifier assigned in ViPIOS\_Open \\
IN  \>offset        \>absolute file offset\\
IN  \>offset\_base  \>update mode\\
IN  \>desc      \>initial address of ViPIOS access descriptor
\end{tabbing}
}

\textbf{Description:}\\
Updates the file pointer of a file according to \textit{offset\_base} within a predefined file access pattern rather than merely in a contiguous way.\\

\textbf{Example:}\\
\texttt{ViPIOS\_Seek (fid1, 50, SEEK\_SET, view\_root);}\\

The file pointer is set to position 50 of the file according the file access pattern, i.e. file view.\\

{\sffamily
\begin{tabbing}
{\bfseries int ViPIOS\_File\_get\_position (int fid, int *pos)}\\
INOUT..\=parameter      \= description                                  \kill
IN  \>fid       \>file identifier\\
OUT \>pos       \>position of file pointer
\end{tabbing}
}

\textbf{Description:}\\
Returns the current position of the individual file pointer in bytes relative to the beginning of the file.

\newpage
\section{How to use the ViPIOS}
\subsection{Quick Start}
On describing the interface of the ViPIOS we will now explain all steps which are necessary to use the ViPIOS runtime library from an application program written in MPI. We assume that the ViPIOS server has already been compiled and the library \textit{libvipios.a} resides in the same directory. This library contains the interface we described in the previous section. \\
\\
First, the application program must be compiled and linked with the ViPIOS library. The syntax is the same as for a usual C or FORTRAN compiler. For example,\\
\\
\texttt{gcc -o vip\_client application1.c libvipios.a}\\
\\
Thus, the application program \textit{application1.c} is treated as a client process called \textit{vip\_serv}.\\

Next, the application schema must be written. This is a text file which describes how many server and client processes you want to use and on which host they should run. A possible application schema \textit{app-schema} for one server and one client process is:
\begin{verbatim}
vipios2 0 /home/usr1/vip_serv
vipios1 1 /home/usr1/vip_client
\end{verbatim}
In that example the server process \textit{vip\_serv} is started on the host called \textit{vipios2} whereas the client process \textit{vip\_client} is started on the host \textit{vipios1}.

\subsection{An Example Program}
Let us first analyze a simple program which opens a file called \textit{infile}, reads the first 1024 bytes of the file and stores them in a file called \textit{outfile}. Further assume that the program is run by one server and one client process.\\
\\
The client program \textit{application1.c} looks like follows:
\begin{verbatim}
#include <stdio.h>
#include "mpi.h"
#include "vip_func.h"

void main ( int argc, char **argv )
{
  int   i,fid1, fid2;
  char  outfile [15], buf[1024];

  MPI_Init (&argc, &argv);

  ViPIOS_Connect (0);
  ViPIOS_Open ("infile", 'r', &fid1);
  ViPIOS_Read (fid1, (void *) buf, 1024);
  ViPIOS_Close(fid1);

  ViPIOS_Open (outfile, 'w', &fid2);
  ViPIOS_Write (fid2, (void *) buf, 1024);
  ViPIOS_Close(fid2);

  ViPIOS_Disconnect();
  ViPIOS_Shutdown();
}
\end{verbatim}

The next step is to specify the number of servers and clients which should be involved in the computation. As we stated before we want to run 1 server and 1 client process. Thus, we define a text file called \textit{app11-schema} which contains following information:\\
\begin{verbatim}
vipios1 0 /home/usr1/vip_serv
vipios2 1 /home/usr2/kurt/vip_client
\end{verbatim}
We assume that the server and the client program reside in the specified directories. Furthermore, we see that the server process \textit{vip\_serv} is started on \textit{vipclus1} and the client process on \textit{vipios2}. We are now ready to start the application program as we described previously.

%% file: strided_data_access.tex
\subsubsection{Strided Data Access}

In this section we will describe how a file can be accessed in strided way rather than in contiguous chunks as it is true, for example, for \textit{ViPIOS\_Read}. The approach analyzed here refers to the routines \textit{ViPIOS\_Read\_struct, ViPIOS\_Write\_struct ViPIOS\_Seek\_struct}. Thus, it is possible to define a certain view to a file similar to the function \textit{MPI\_File\_set\_view}. As a consequence, the application program can access the data as if it were contiguous although it is physically scattered across the whole file. Let us now analyze the underlying data structure and how it can be applied for accessing a file in a strided way:

\begin{verbatim}
typedef struct
{
  int                   no_blocks;
  struct basic_block    *basics;
  int                   skip;
}
Access_Desc;

struct basic_block
{
  int                    offset;
  int                    repeat;
  int                    count;
  int                    stride;
  Access_Desc            *subtype;
  int                    sub_count;
  int                    sub_actual;
};
\end{verbatim}

Basically, the data structure which we refer to as the "ViPIOS access descriptor" consists of two structs whereas the first struct \textit{Access\_Desc} defines the number of blocks \textit{no\_blocks} and a displacement \textit{skip} which can be used similar to the parameter \textit{disp} in \textit{MPI\_File\_set\_view}. The second struct \textit{basic\_block} specifies each block.\\

\begin{figure}
\begin{center}
\includegraphics[scale=0.9]{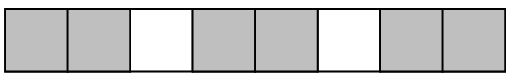}
\caption{File view}
\label{view3}
\end{center}
\end{figure}

In order to understand the functionality of this data structure let us assume that we wish to access a file according to the view in the Figure \ref{view3}.\\

Our file consists of 8 elements of datatype \textit{byte} and we wish to access the file in three blocks of  two elements with the stride of one element. Recalling the chapter of derived datatypes we could describe this access pattern by means of a vector, namely

\begin{verbatim}
MPI_Type_vector(3,2,3,MPI_BYTE,&vector1);
\end{verbatim}

How can this datatype be mapped to our data structure ViPIOS access descriptor? Assume that the file view can be described by one basic block. Thus, we set \textit{no\_blocks} to 1. Since the basic block starts at position 0, i.e. no file information is skipped, \textit{skip} is set to 0. \textit{basics} is a pointer to the structure \textit{basic\_block}. Thus, the first basic block can be referenced by \textit{basics[0]}. Further basic blocks could be referenced in the same way, e.g. basic block 8 is referenced by \textit{basics[7]}.\\

Now we can describe the basic block. Since we wish to access the file at position 0 we have to set \textit{offset} to 0 as well. \textit{Repeat=3} states how many data blocks our view consists of. This variable corresponds to the first parameter in \textit{MPI\_Type\_vector}. Furthermore, each block comprises 2 elements. Thus, \textit{count} is set to 2. Note that \textit{count} is always given in bytes because every access operation in ViPIOS is made in units of bytes.\\

Finally, the variable \textit{stride} has to be filled with a value. Unlike the stride of the \textit{MPI\_Type\_vector} the stride of the ViPIOS access descriptor specifies the gap between each data block rather than the number of elements between the start of each data block. This means that \textit{stride} is set to 1 rather than to 3. The remaining variables \textit{sub\_count} and \textit{sub\_actual} will not be explained any further since they are not important for that part of the impelementation.\\

\begin{figure}
\begin{center}
\includegraphics[scale=0.7]{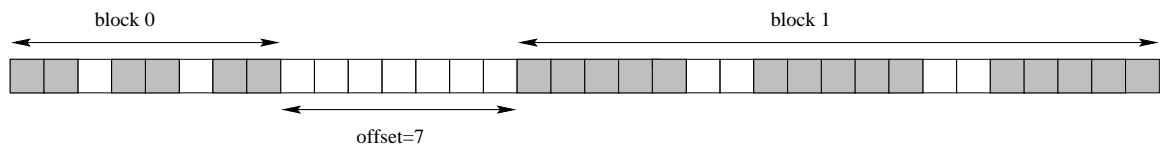}
\caption{Two Different File Views}
\label{view4}
\end{center}
\end{figure}

Now we could raise the question about the purpose of the variable \textit{no\_blocks} in the ViPIOS access descriptor when our view can fully be described by one basic block. The answer to that question is that we also wish to access files in a more heterogeneous way. Assume that the first part of the file should be accessed in the way described in the previous section whereas the second part should be accessed differently. The whole view is depicted in picture Figure \ref{view4}.\\

The access pattern of the second part of the file corresponds to the vector

\begin{verbatim}
MPI_Type_vector(3,5,7,MPI_BYTE,&vector2);
\end{verbatim}

Since the file is accessed in two different patterns we set \textit{no\_blocks} to 2. The second basic block is defined as: \textit{repeat=3}, \textit{count=5}, \textit{stride=2}. Since the gap between the first access pattern (basic block 0) and the second access pattern (basic block 1) is 7 elements, we set \textit{offset} to 7.\\

\begin{figure}
\begin{center}
\includegraphics[scale=0.9]{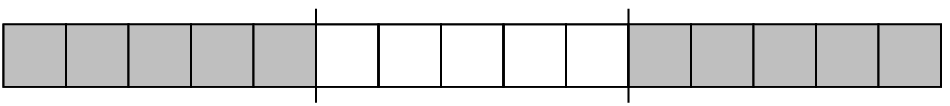}
\caption{Level 1}
\label{view5}
\end{center}
\end{figure}

Taking a look at the definition of the ViPIOS access descriptor we notice the pointer \textit{subtype} which is a pointer to the struct \textit{Access\_Desc}. Thus, it can be used for more complex views. Up to now we used basic MPI datatypes for accessing our files. In particular, the parameter \textit{oldtype} of \textit{MPI\_Type\_vector} was MPI\_BYTE. Now assume that we define a nested datatype such that \textit{oldtype} of the second derived datatype is in turn a derived datatype rather than a basic one. For example,

\begin{verbatim}
MPI_Type_vector (2,5,10, MPI_BYTE, &level1);
MPI_Type_vector (3,2,3,level1, &level2);
\end{verbatim}

The derived datatype \textit{level1} describes the pattern depicted in Figure \ref{view5}.\\

\begin{figure}
\begin{center}
\includegraphics[scale=0.7]{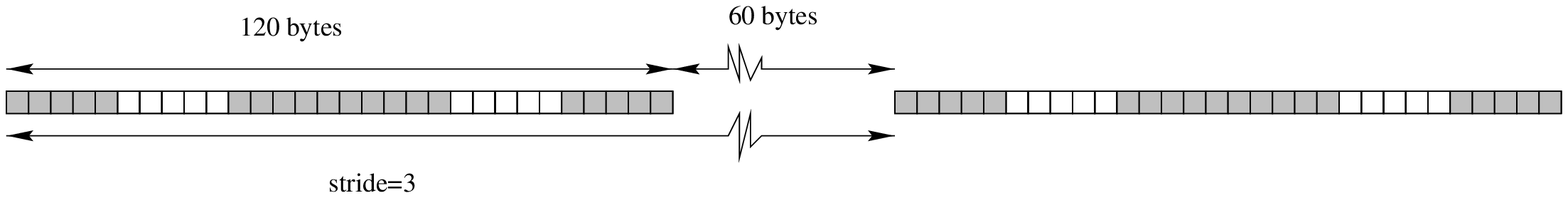}
\caption{Level 2}
\label{view6}
\end{center}
\end{figure}

Since \textit{oldtype} of the derived datatype \textit{level2} is the derived datatype \textit{level1}, the access pattern described by \textit{level2} is depicted in Figure \ref{view6}.\\

\textit{level2} consists of 3 blocks where the size of each block is 2. In order to describe this nested access pattern with the ViPIOS access descriptor \textit{subtype} is needed which points to another level of access pattern. More information is given when we describe the implementation of \textit{MPI\_File\_set\_view}.

%% file: intro.tex
\section{Introduction}


In the last decades VLSI technology has improved and, hence, the
power of processors. What is more, also storage technology has
produced better main memories with faster access times and bigger
storage spaces. However, this is only true for main memory, but
not for secondary or tertiary storage devices. As a matter of
fact, the secondary storage devices are becoming slower in
comparison with the other parts of a computer.\\

Another driving forces for the development of big main memories
are scientific applications, especially Grand Challenge
Applications. Here a big amount of data is required for
computation, but it does not fit entirely into main memory. Thus,
parts of the information have to be stored on disk and transferred
to the main memory when needed. Since this transfer - reading from
and writing to disk, also known as input/output (I/O) operations -
is very time consuming, I/O devices are the bottleneck for compute
immense scientific applications. Recently much effort was devoted
to I/O, in particular universities and research laboratories in
the USA, Austria, Australia, Canada and Spain. There are many
different approaches to remedy the I/O problem. Some research
groups have established I/O libraries or file systems, others deal
with new data layout, new storage techniques and tools for
performance checking.\\

This dictionary  gives an overview of all the available research
groups in David Kotz's homepage (see below) and depicts the
special features and characteristics. The \emph{abstracts} give
short overviews, \emph{aims} are stated as well as \emph{related
work} and \emph{key words}. The \emph{implementation platform}
shall state which platforms the systems have been tested on or
what they are devoted to. \emph{Data access strategies} illustrate
special features of the treatment of data (e.g.: synchronous,
locking, ...). \emph{Portability} includes whether a system is
established on top of another one (e.g.: PIOUS is designed for
PVM) or if it supports interfaces to other systems. Moreover, many
systems are applied to real world applications (e.g.: fluid
dynamics, aeronautics, ...) or software has been developed (e.g.:
matrix multiplication). All this is stated at \emph{application}.
Small examples or code fragments can illustrate the interface of
special systems and are inserted at \emph{example}.\\

What is more, some special features are depicted and discussed
explicitly. For instance, a Two-Phase Method from PASSION is
explained in detail under a specific key item. Furthermore, also
platforms like the Intel Paragon are discussed. To sum up, the
dictionary shall provide information concerning research groups,
implementations, applications, aims, functionalities, platforms
and background information. The appendix gives a detailed survey
of all the groups and systems mentioned in the Dictionary part and
compares the features.\\

This dictionary is intended for people having background
information regarding operating systems in general, file systems,
parallel programming and supercomputers. Moreover, much work is
based on C, C++ or Fortran, and many examples are written in these
programming languages. Hence, a basic knowledge of these languages
and the general programming paradigms is assumed.\\

Most of the facts presented in this dictionary are based on a web
page of David Kotz, Dartmouth College, USA, on parallel I/O. The
internet made it possible to obtain all necessary information
through ftp-downloading instead of using books or physical
libraries (the web could be referred to as a logical library).
This is also the place to thank David Kotz for his work, because
without it, this dictionary would never have been written. The URL
for the homepage is:

\begin{center}
\texttt{http://www.cs.darmouth.edu/pario}
\end{center}

Note that most of the names listed at \emph{people} are only from
people participating in writing the papers, i.e. normally the
names listed in the papers of the bibliography are listed in this
work.\\


%% file: glossar.tex

This glossary is based on "Dictionary on Parallel Input/Output" \cite{Dictionary} by the same author. The glossary presented here gives only a brief survey. For more detailed facts refer to \cite{Dictionary}.\\

\begin{keyword}{ADIO (Abstract-Device Interface for Portable Parallel-I/O)}\\
Since there is no standard \wrf{API} for parallel I/O, ADIO is supposed to provide a strategy for implementing \wrf{API}s (not a standard) in a simple, portable and efficient way \cite{thakur:abstract} in order to take the burden of the programmer of choosing from several different \wrf{API}s. Furthermore, it makes existing applications portable across a wide range of different platforms. An \wrf{API} can be implemented in a portable fashion on top of ADIO, and becomes available on all file systems on which ADIO has been implemented.\\
\end{keyword}

\begin{keyword}{ADOPT (A Dynamic scheme for Optimal PrefeTching in parallel file systems)}\\
ADOPT is a dynamic \wrf{prefetching} scheme that is applicable to any distributed system, but major performance benefits are obtained in {distributed memory} I/O systems in a parallel processing environment~\cite{singh:adopt}. Efficient accesses and \wrf{prefetching} are supposed to be obtained by exploiting access patterns specified and generated from users or compilers.\\

{I/O node}s are assumed to maintain a portion of memory space for \wrf{caching} blocks. This memory space is partitioned into a current and a prefetch cache by ADOPT. In particular, the current cache serves as a buffer memory between {I/O node}s and the disk driver whereas a prefetch cache has to save prefetched blocks. Prefetch information is operated and managed by ADOPT at the {I/O node} level.\\

The two major roles of the I/O subsystem in ADOPT are receiving all prefetch and disk access requests and generating a schedule for disk I/O. Finally, ADOPT uses an Access Pattern Graph (APGraph) for information about future I/O requests.\\
\end{keyword}

\begin{keyword}{Agent Tcl}\\	
A transportable {agent} is a named program that can migrate from machine to machine in a heterogeneous network~\cite{Agent1}. It can suspend its execution at an arbitrary point, transport to another machine and resume execution on the new machine. What is more, such {agent}s are supposed to be an improvement of the conventional {client-server} model. Agent Tcl is used in a range of information-management applications. It is also used in serial workflow applications (e.g. an {agent} carries electronic forms from machine to machine). It is used for remote salespersons, too.\\

Transportable {agent}s are autonomous programs that communicate and migrate at will, support the {peer-to-peer} model and are either clients or servers. Furthermore, they do not require permanent connection between machines and are more fault-tolerant.\\
\end{keyword}

\begin{keyword}{Alloc Stream Facility (ASF)}\\
Alloc Stream Facility (ASF) is an application-level I/O facility in the \wrf{Hurricane File System} (\wrf{HFS}). It can be used for all types of I/O, including disk files, terminals, pipes, networking interfaces and other low-level devices~\cite{krieger:asf-tr}. An outstanding difference to \wrf{UNIX I/O} is that the application is allowed direct access to the internal buffers of the I/O library  instead of having the application to specify a buffer into or from which the I/O data is copied. The consequence is another reduction in copying.\\
\end{keyword}

\begin{keyword}{ANL (Argonne National Laboratory) - Parallel I/O Project}\\
ANL builds and develops a testbed and parallel I/O software, and applications that will test and use this software. The I/O system applies two layers of high-performance networking. The primary layer is used for interconnection between {compute node}s and {I/O server}s whereas the second layer connects the {I/O server}s with \wrf{RAID} arrays.\\
\end{keyword}

\begin{keyword}{Application Program Interface (API)}\\ 
There are two complementary views for accessing an {OOC} data structure with a {global view} or a {distributed view}~\cite{bennett:jovian}. A {global view} indicates a copy of a globally specified subset of an {OOC} data structure distributed across disks. The library needs to know the \wrf{distribution} of the {OOC} and in-core data structures as well as a description of the requested data transfer. The library has access to the {OOC} and in-core data \wrf{distribution}s. With a {distributed view}, each process effectively requests only the part of the data structure that it requires, and an exchange phase between the {coalescing processes} is needed ({all-to-all communication} phase).\\
\end{keyword}

\begin{keyword}{array distribution}\\
An array  distribution can either be regular (block, cyclic or block-cyclic) or irregular (see \wrf{irregular problems}) where no function specifying the mapping of arrays to processors can be applied. Data that is stored on disk in an array is distributed in some fashion at compile time, but it does not need to remain fixed throughout the whole existence of the program. There are some reasons for redistributing data:
\begin{itemize}
\item Arrays and array sections can be passed as arguments to subroutines.
\item If the \wrf{distribution} in the subroutine is different from that in the calling program, the array needs to be redistributed.\\
\end{itemize}
\end{keyword}

\begin{keyword}{automatic data layout}\\
A \wrf{Fortran D} program can be automatically transferred into an SPMD node program for a given {distributed memory} target machine by using data layout directives. A good data layout is responsible for high performance. An algorithm partitions the program into code segments, called phases.\\
\end{keyword}

\begin{keyword}{Bridge parallel file system}\\
This file system provides three interfaces from a high-level UNIX-like interface to a low-level interface that provides direct access to the individual disks. A prototype was built on the {BNN Butterfly}.\\
\end{keyword}



%% file: gloss-c.tex
\begin{keyword}{C*}\\
C* supports \wrf{data parallel} programming where a sequential program operates on parallel arrays of data, with each virtual processor operating on one parallel data element. A computer multiplexes physical processors among virtual processors to support the parallel model. A variable in C* has a shape describing the rectangular structure and defining the logical configuration of parallel data in virtual processors.\\ 
\end{keyword}

\begin{keyword}{Cache Coherent File System (CCFS)}\\
The Cache Coherent File System (CCFS) is the successor of \wrf{ParFiSys} and consists of the following main components~\cite{carretero:case}:
\begin{itemize}
\item {Client File Server (CLFS)}: deals with user requests providing the file system functionality to the users, and interfacing with the CCFS components placed in the {I/O node}s; it is sited on the clients
\item {Local File Server (LFS)}: interfaces with I/O devices to execute low level I/O and is sited on disk nodes
\item {Concurrent Disk System (CDS)}: deals with {LFS} low level I/O requests and executes real I/O on the devices and is sited on the disk nodes\\
\end{itemize}
\end{keyword}

\begin{keyword}{caching}\\
In order to avoid or reduce the latency of physical I/O operations, data can be cached for later use. Read operations are avoided by \wrf{prefetching} and write operations by postponing or avoiding writebacks. Additionally, smaller requests can be combined and large requests can be done instead. One important question is the location of the cache. \wrf{PPFS} employs three different levels: server cache (associated with each I/O sever), client cache (holds data accessed by user processes), and global cache (in order to enforce consistency).\\
\end{keyword}

\begin{keyword}{CAP Research Program}\\
CAP is an integral part of parallel computing research at the Australian National University (ANU). A great deal of work is devoted to the \wrf{Fujitsu AP1000} since CAP is an agreement between ANU and the High Performance group of Fujitsu Ltd.\\
\end{keyword}

\begin{keyword}{CHANNEL}\\ 
\wrf{PASSION} introduces CHANNEL as modes of communication and synchronization between \wrf{data parallel} tasks~\cite{avalani:channels}. A CHANNEL provides a uniform one-directional communication mode between two \wrf{data parallel} tasks, and concurrent tasks are plugged together. This results in a {many-to-many communication} between processes of the communicating tasks. The general semantics of a CHANNEL between two tasks are as follows:
\begin{itemize}
\item Distribution Independence: If two tasks are connected via a CHANNEL, they need not have the same data \wrf{distribution}, i.e. whereas task 1 employs a cyclic fashion, task 2 can use a block fashion and the communication can still be established. Hence, both data \wrf{distribution}s are independent.
\item Information Hiding: A task can request data from the CHANNEL in its own \wrf{distribution} format. This is also true if both tasks use different data \wrf{distribution} formats.
\item Synchronization: The task wanting to receive data from the CHANNEL has to wait for the CHANNEL to get full before it can proceed.\\
\end{itemize}
\end{keyword}

\begin{keyword}{CHAOS}\\
CHAOS deals with efficiently coupling multiple \wrf{data parallel} programs at runtime. In detail, a mapping between data structures in different \wrf{data parallel} programs is established at runtime. Languages such as \wrf{HPF}, C an {pC++} are supported. The approach is supposed to be general enough for a variety of data structures~\cite{CHAOS1}.\\

Firstly, the implementation used asynchronous, one sided {message passing} for inter-\ application data transfer with the goal to overlap data transfer with computation. Secondly, optimized messaging schedules were used. The number of messages  transmitted has to be minimized. Finally, buffering was used to reduce the time spent waiting for data. The data transfer itself can be initiated by a consumer or a producer \wrf{data parallel} program. Furthermore, the inter-application data transfer is established via a library called {Meta-Chaos}. \wrf{PVM} is the underlying messaging layer, and each \wrf{data parallel} program is assigned to a distinct \wrf{PVM} group.\\

{Meta-Chaos} is established to provide the ability to use multiple specialized parallel libraries and/or languages within a single application, i.e. one can use different libraries in one program in order to run operations on distributed data structures in parallel. \\
\end{keyword}

\begin{keyword}{CHARISMA (CHARacterize I/O in Scientific Multiprocessor Applications)}\\
CHARISMA is a project to characterize I/O in scientific multiprocessor applications from a variety of production parallel computing platforms and sites~\cite{kotz:workload}. It recorded individual read and write requests in live, multiprogramming workloads. It turned out that most files were accessed in  complex, highly regular patterns.\\
\end{keyword}

\begin{keyword}{checkpointing}\\
Checkpointing allows processes to save their state from time to time so that they can be restarted in case of failures, or in case of swapping due to resource allocation. What is more, a checkpointing mechanism must be both space and time efficient. Existing checkpointing systems for {MPP}s checkpoint the entire memory state of a program. Similarly, existing checkpointing systems work by halting the entire application during the construction of the checkpoint. Checkpoints have low latency because they are generated concurrently during the program's execution. \\
\end{keyword}

\begin{keyword}{ChemIO (Scalable I/O Initiative)}\\
ChemIO is an abbreviation for High-Performance I/O for Computational Chemistry Applications. The {Scalable I/O Initiative} will determine application program requirements and will use them to guide the development of new programming language features, compiler techniques, system support services, file storage facilities, and high performance networking software~\cite{bagrodia:sio-character}.\\

Key results are:
\begin{itemize}
\item implementation of scalable I/O algorithms in production software for computational chemistry applications
\item dissemination of an improved understanding of scalable parallel I/O systems and algorithms to the computational chemistry community
\end{itemize}

The objectives of the Application Working Group of a the {Scalable I/O Initiative} include:
\begin{itemize}
\item collecting program suites that exhibit typical I/O requirements for \wrf{Grand Challenge Applications} on massively parallel processors
\item monitoring and analyzing these applications to characterize parallel I/O requirements for large-scale applications and establish a baseline for evaluating the system software and tools  developed during the {Scalable I/O Initiative}
\item modifying, where initiated by the analysis, the I/O structure of the application programs to improve performance
\item using the system software and tools from other working groups and representing the measurement and analysis of the applications to evaluate the proposed file system, network and system support software, and language features
\item developing instrumented parallel I/O benchmarks
\end{itemize}
\end{keyword}

\begin{keyword}{clustering}\\
A file is divided into segments which reside on a particular server. This can be regarded as a collection of records. What is more, each file must have at least one segment.\\
\end{keyword}

\begin{keyword}{CM-2 (Connection Machine)}\\
CM-2 is a \wrf{SIMD} machine where messages between processors require only a single cycle.\\
\end{keyword}

\begin{keyword}{CMMD I/O System}\\
This system provides a {parallel I/O interface} to parallel applications on the Thinking Machines {CM-5}, but the {CM-5} does not support a parallel file system. Hence, data is stored on large high-performance \wrf{RAID} systems.\\
\end{keyword}

\begin{keyword}{coding techniques}\\ 
Magnetic disk drives  suffer from three primary types of failures: transient or noise-related failures (corrected by repeating the offending operation or by applying per sector error-correction facilities), media defect (usually detected and corrected in factories) and catastrophic failures like head crashes. Redundant arrays can be used to add more security to a system. The scheme is restricted to leave the original data unmodified on some disks ({information disk}s) and define redundant encoding for that data on other disks ({check disk}s).\\
\end{keyword}

\begin{keyword}{concurrency algorithms}\\
concurrency algorithms can be divided into two classes:
\begin{itemize}
\item {Client-distributed state (CDS)} algorithms are optimistic and allow the {I/O daemon}s to schedule in parallel all data accesses which are generated by a given file pointer. This method can lead to an invalid state that forces rollback: a file operation may have to be abandoned and re-tried. {CDS} algorithms distribute global state information in the form of an operation "{commit}" or "{abort}" message, sent to the relevant {I/O daemon}s by the client. In \wrf{PIOUS} this model is realized with a transaction called volatile. {CDS} algorithms provide the opportunity to efficiently multicast global state information.
\item {Server-distributed state (SDS)} algorithms are conservative, allowing an {I/O daemon} to schedule data access only when it is known to be consistently ordered with other data accesses. {SDS} never leads to an invalid state, because global state information is distributed in the form of a token that is circulated among all {I/O daemon}s servicing a file operation.
\end{itemize}
\end{keyword}

\begin{keyword}{concurrency control}\\
Sequential consistency ({serializability}) dictates that the results of all read and write operations generated by a group of processes accessing storage must be the same as if the operations had occurred within the context of a single process. It should gain the effect of executing all data accesses from one file operation before executing any data accesses from the other one. This requires global information: each {I/O daemon} executing on each {I/O node} must know that it is scheduling data access in a manner that is consistent with all other {I/O daemon}s. Concurrency control algorithms can be divided into two classes: client-distributed and server-distributed. See also \wrf{concurrency algorithms}.\\
\end{keyword}

\begin{keyword}{Concurrent File System (CFS)}\\ 
CFS is the file system of the \wrf{Intel Touchstone Delta} and provides a UNIX view of a file to the application program~\cite{bordawekar:delta-fs}. Four I/O modes are supported:
\begin{itemize}
\item Mode 0: Here each node process has its own file pointer. It is useful for large files to be shared among the nodes.
\item Mode 1: The {compute node}s share a common file pointer, and I/O requests are serviced on a first-come-first-serve basis.
\item Mode 2: Reads and writes are treated as global operations and a global synchronization is performed.
\item Mode 3: A synchronous ordered mode is provided, but all write operations have to be of the same size.
\end{itemize}
\end{keyword}

\begin{keyword}{Cray C90}\\
The Cray C90 is a {shared memory} platform.\\
\end{keyword}

\begin{keyword}{CVL (C Vector Library)}\\
CVL (also referred to as {DartCVL}) is an interface to a group of simple functions for mapping and operating on vectors. The target machine is a \wrf{SIMD} computer. In other words, CVL is a library of low-level vector routines callable from C. The aim of CVL is to maximize the advantages of hierarchical virtualization~\cite{Dart18}.\\
\end{keyword}

%% file: gloss-d-h.tex
\begin{keyword}{data parallel}\\ 
A data parallel program applies a sequence of similar operations to all or most elements of a large data structure. \wrf{HPF} is such a language. A program written in a data parallel style allows advanced compilation systems to generate efficient code for most {distributed memory machines}.
\end{keyword}

\begin{keyword}{data prefetching}\\
The time taken by the program can be reduced if it is possible to overlap computation with I/O in some fashion. A simple way of achieving this is to issue an asynchronous I/O request for the next {slab} immediately after the current {slab} has been read. As for prefetching, data is prefetched from a file, and on performing the computation on this data the results are written back to disk. This is repeated again afterwards. Prefetching can pre-load the cache to reduce the cache miss ratio, or reduce the cost of a cache miss by starting the I/O early.\\
\end{keyword}

\begin{keyword}{data reuse}\\
The data already fetched into main memory is reused instead of read again from disk. As a result, the amount of I/O is reduced, here are.\\
\end{keyword}

\begin{keyword}{data sieving}\\
Normally data is distributed in a {slab} and not concentrated on a special address. Direct reading of data requires a lot of I/O requests and high costs. Therefore, a whole {slab} is read into a temporary buffer and the required data is extracted from this buffer and placed in the {ICLA}.\\

All routines support the reading/writing of regular sections of arrays which are defined as any portion of an array that can be specified in terms of its lower bound, upper bound and stride in each dimension. For reading a strided section, instead of reading only the requested elements, large contiguous chunks of data are read at a time into a temporary buffer in main memory. This includes unwanted data. The useful part of it is extracted from the buffer and passed to the calling program. A disadvantage is the high memory requirement for the buffer.\\
\end{keyword}

\begin{keyword}{DDLY (Data Distribution Layer)}\\
DDLY is a run-time library providing a fast high-level interface for writing parallel programs. DDLY is not yet ready, but is supposed to support automatic and efficient data partitioning and \wrf{distribution} for \wrf{irregular problems} on {message passing} environments~\cite{trabado:io}. Addidionally, parallel I/O for both regular and \wrf{irregular problems} should be provided.\\ 
\end{keyword}

\begin{keyword}{disk-directed I/O}\\
Disk-directed I/O can dramatically improve the performance of reading and writing large, regular data structures between {distributed memory} and distributed files~\cite{kotz:expand-tr}, and is primarily intended for the use in multiprocessors~\cite{kotz:int-ddio}.\\
 
In a traditional UNIX-like interface, individual processors make requests to the file system, even if the required amount of data is small. In contrast, a {collective-I/O interface} supports single joint requests from many processes. Disk-directed I/O can be applied for such an environment. In brief, a collective request is passed to the {I/O processor}s for examining the request, making a list of disk blocks to be transferred and  sorting the list. Finally, they use double-buffering and special remote memory messages to pipeline the data transfer. This strategy is supposed to optimize disk access, use less memory and has less CPU and {message passing} overhead.\\

It is distinguished between a sequential request to a file, which is at a higher offset than the previous one, and a consecutive request, which is a sequential request that begins where the previous one ended. In a {simple-strided access} a series of requests to a node-file is done where each request has the same size and the file pointer is incremented by the same amount between each request. Indeed, this would correspond to reading a column of data from a matrix stored in row-major order. A group of requests that is part of this {simple-strided} pattern is defined as a {strided segment}. {Nested patterns} are similar to {simple strided access}, but it is composed of {strided segment}s separated by regular strides in the file.\\
\end{keyword}

\begin{keyword}{Disk Resistent Arrays (DRA)}\\
DRA extend the programming model of \wrf{Global Arrays} to disk. The library contains details concerning data layout, addressing and I/O transfer in disk array objects. The main difference between DRA and \wrf{Global Arrays} is that DRA reside on disk rather than in main memory.\\
\end{keyword}

\begin{keyword}{distribution}\\
The term distribution determines in which segment the record of a file resides and where in that segment. It is equivalent to a one-to-one mapping from file record number to a pair containing segment number and segment record number.\\
\end{keyword}

\begin{keyword}{distributed computing}\\
Distributed computing is a process whereby a set of computers connected by a network is used collectively to solve a single large program. {Message passing} is used as a form of interprocess communication.\\
\end{keyword}

\begin{keyword}{EXODUS}\\
EXODUS an \wrf{object-oriented database} effort and serves as the basis for \wrf{SHORE}. EXODUS provides a {client-server} architecture and supports multiple servers and transactions~\cite{carey:shore}.The programming language E, a variant of C++, is included in order to support a convenient creation and manipulation of persistent data structures. Although EXODUS has good features such as transactions, performance and robustness, there are some important drawbacks: storage objects are untyped arrays of bytes, no type information is stored, it is a {client-server} architecture, it lacks of support for access control, and existing applications built around UNIX files cannot easily use EXODUS.\\
\end{keyword}

\begin{keyword}{Express}\\
Express is a toolkit that allows to individually address various aspects of concurrent computation. Furthermore, it includes a set of libraries for communication, I/O and parallel graphics.\\
\end{keyword}

\begin{keyword}{ExtensibLe File Systems (ELFS)}\\
ELFS is based on an object-oriented appraoch, i.e. files should be treated as typed objects~\cite{karpovich:elfs}. Ease-of-use can be implemented in a way that a user is allowed to manipulate data items in a manner that is more natural than current file access methods available. For instance, a 2D matrix interface can be accessed in terms of rows, columns or blocks. In particular, the user can express requests in a manner that matches the semantic model of data, and does not have to take care of the physical storage of data, i.e. in the object-oriented approach the implementation details are hidden. Ease of development is supported by encapsulation and inheritance as well as code reuse, extensibility and modularity.\\
\end{keyword}

\begin{keyword}{file level parallelism}\\
A conventional file system is implemented on each of the processing nodes that have disks, and a central controller is added which controls a transparent striping scheme over all the individual file systems~\cite{nieuwejaar:galley}. The name file level parallelism stems from the fact that each file is explicitly divided across the individual file systems. Moreover, it is difficult to avoid arbitrating I/O requests via the controller ({bottleneck}). \wrf{HiDIOS} has introduced a {disk level parallelism} (parallel files vs. parallel disks).\\
\end{keyword}

\begin{keyword}{file migration}\\
The amount of data gets larger and larger, hence, storing this data on a magnetic disk is not always feasible. Instead, tertiary storage devices such as tapes and optical disks are used. Although the costs per megabyte of storage are lowered, they have longer access times than magnetic disks. A solution to this situation is to use file migration systems that are used by large computer installations to store more data than that which would fit on magnetic disks.\\
\end{keyword}

\begin{keyword}{FLEET}\\
FLEET is a FiLEsystem Experimentation Testbed for experimentation with new concepts in parallel file systems.\\
\end{keyword}

\begin{keyword}{Fortran D} \\
Fortran D  is a version of Fortran that provides data decomposition specifications for two levels of parallelism (how should arrays be aligned with respect to each other, and how should arrays be distributed onto the parallel machine). Furthermore, a Fortran D compilation system translates a Fortran D program into a {Fortran 77} SPMD node program. A consequence can be a reduction or hiding of communication overhead, exploited parallelism or the reduction of memory requirements.\\  
\end{keyword}

\begin{keyword}{Fujitsu AP1000}\\
The AP1000 is an experimental large-scale \wrf{MIMD} parallel computer with configurations range from 64 to 1024 processors connected by three separate high-bandwidth communication networks. There is no {shared memory}, and the processors are typically controlled by a host like the {SPARC} Server. A processor is a {SPARC} 25MHz, 16MB RAM processor. Programs are written in C or Fortran. \wrf{HiDIOS} is a parallel file system implemented on the AP1000.\\
\end{keyword}

\begin{keyword}{Galley}\\
Galley is a parallel file system intended to meet the needs of parallel scientific applications. It is based on a three-dimensional structuring of files. Furthermore, it is supposed to be capable of providing high performance I/O.\\
	
It was believed that parallel scientific applications would access large files in large consecutive chunks, but results have shown that many applications make many small regular, but non-consecutive requests to the file system. Galley is designed to satisfy such applications. The goals are:
\begin{itemize}
\item efficiently handle a variety of access sizes and patterns
\item allow applications to explicitly control parallelism in file access
\item be flexible enough to support a variety of interfaces and policies, implemented in libraries
\item allow easy and efficient implementations of libraries
\item scale to many compute and {I/O processor}s
\item minimize memory and performance overhead\\
\end{itemize}
\end{keyword}

\begin{keyword}{Global Arrays (GA)}\\
Global Arrays are supposed to combine features of {message passing} and {shared memory}, leading to both simple coding and efficient execution for a class of applications that appears to be fairly common~\cite{Global}. Global arrays are also regarded as "A Portable 'Shared Memory' Programming Model for Distributed Memory Computers". GA also support the NUMA (Non-Uniform Memory Access) {shared memory} paradigm. What is more, two versions of GA were implemented: a fully distributed one and a mirrored one. See also \wrf{Disk Resistent Arrays(DRA)}.

In comparison to common models, GA are different since they allow {task-parallel} access to distributed matrices. Furthermore, GA support three distinctive environments:
\begin{itemize}
\item {distributed memory}, {message passing} parallel computers with interrupt-driven communication (Intel Gamma, \wrf{Intel Touchstone Delta}, \wrf{Intel Paragon}, {IBM SP1})
\item networked workstation clusters with simple {message passing}
\item {shared memory} parallel computers (\wrf{KSR-2}, {SGI})\\
\end{itemize}
\end{keyword}

\begin{keyword}{Global Placement Model (GPM)}\\
In \wrf{PASSION} there are two models for storing and accessing data: the {Local Placement Model (LPM)} and the Global Placement Model (GPM). For many applications in supercomputing main memory is too small, therefore, main parts of the available data are stored in an array on disk. The entire array is stored in a single file, and each processor can directly access any portion of the file. In a GPM a global data array is stored in a single file called {Global Array File (GAF)}. The file is only logically divided into local arrays, which saves the initial local file creation phase in the {LPM}. However, each processors' data may not be stored contiguously, resulting in multiple requests and high I/O latency time.\\
\end{keyword}

\begin{keyword}{GPMIMD (General Purpose MIMD)}\\ 
A general purpose multiprocessor I/O system has to pay attention to a wide range of applications that consist of three main types: normal UNIX users, databases and scientific applications. Database applications are characterized by a multiuser environment with much random and small file access whereas scientific applications support just a single user having a large amount of sequential accesses to a few files.\\

The main components are {processing nodes (PN)}, network, input/output  nodes ({ION}) and disk devices. In order to describe a system, four parameters can be used: number of {I/O node}s, number of controllers, number of disks per controller, and degree of synchronization across disks of a controller. Additionally, another two concepts must be considered: file \wrf{clustering} and {file striping}. A declustered file is distributed across a number of disks such that different blocks of the same file can be accessed in parallel from different disks. In a stripped file a block can be read from several disks simultaneously.\\ 
\end{keyword}

\begin{keyword}{Grand Challenge Applications}\\
{Massively parallel processors (MPPs)} are more and more used in order to solve Grand Challenge Applications which require much computational effort. They cover fields like physics, chemistry, biology, medicine, engineering and other sciences. Furthermore, they are extremely complex, require many Teraflops of communication power and deal with large quantities of data. Although supercomputers  (see \wrf{supercomputing applications}) have large main memories, the memories are not sufficiently large to hold the amount of data required for Grand Challenge Applications. High performance I/O is necessary if a degrade of the entire performance of the whole program has to be avoided. Large scale applications often use the Single Program Multiple Data (SPMD) programming paradigm for \wrf{MIMD} machines. Parallelism is exploited by decomposing of the data domain.\\
\end{keyword}

\begin{keyword}{HiDIOS (High performance Distributed Input Output System)}\\
HiDIOS (part of the \wrf{CAP Research Program}) is a parallel file system for the \wrf{Fujitsu AP1000} multicomputer~\cite{tridgell:hidios}. What is more, HiDIOS is a product of the ACSys \wrf{PIOUS} project. HiDIOS uses a {disk level parallelism} (instead of the \wrf{file level parallelism}) where a parallel disk driver is used which combines the physically separate disks into a single large parallel disk by stripping data cyclically across the disks. Even the file system code is written with respect to the assumption of a single large, fast disk.\\

Requests are placed in request queues, which are thereafter processed by a number of independent threads. After request processing the manager can return and, hence, can receive further requests while previous ones may be blocked waiting for disk I/O. The {meta-data} system makes it possible to immediately service {meta-data} manipulation (such as file creation, renaming) without disk I/O.\\
\end{keyword}

\begin{keyword}{HPF (High Performance Fortran)}\\
High Performance Fortran is an extension to {Fortran 90} with special features to specify data \wrf{distribution}, alignment or \wrf{data parallel} execution, and it is syntactically similar to \wrf{Fortran D}. HPF was designed to provide language support for machines like \wrf{SIMD}, \wrf{MIMD} or vector machines. Moreover, it provides directives like \texttt{ALIGN} and \texttt{DISTRIBUTE} for distributing arrays among processors of {distributed memory machines}. Here an array can either be distributed in a block or cyclic fashion.\\

HPF is also supposed to make programs independent of single machine architectures. Although HPF can reduce communication cost and, hence, increase the performance, this is only true for regular but not for \wrf{irregular problems}.\\
\end{keyword}

\begin{keyword}{Hurricane File System (HFS)}\\		
The Hurricane File System is developed for large-scale {shared memory} multicomputers~\cite{krieger:hfs}. HFS is a part of the \wrf{Hurricane operating system}. The file system consists of three user level system servers: the {Name Server, Open File Server (OFS)} and {Block File Server (BFS)}.
\begin{itemize}
\item The {Name Sever} manages the name space and is responsible for authenticating requests to open files.
\item The {OFS} maintains the file system state kept for each open file.
\item The {BFS} controls the system disks, is responsible for determining to which disk an operation is destined and directs the operation to the corresponding device driver.
\item {Dirty Harry (DH)} collects dirty pages from the memory manager and makes requests to the {BFS} to write the pages to disk.
\item The \wrf{Alloc Stream Facility (ASF)} is a user level library. It maps files into the application's address space and translates read and write operations into accesses to mapped regions.
\end{itemize}

Each of those file system servers maintains a different state. Whereas the {Name Server} maintains a logical directory state (e.g. access permission and directory size) and directory contents, the {OFS} maintains logical file information (length, access permission, ...) and the per-open instance state. Finally, the {BFS} maintains the block map for each file. Obviously, these states are different from each other and independent, consequently, there is no need for different servers within a cluster to communicate in order to keep the state consistent.\\
\end{keyword}

\begin{keyword}{Hurricane operating system}\\ 
Hurricane is a micro-kernel and single storage operating system that supports mapped file I/O. A mapped file system allows that the application can map regions of a file into its address space and access the file by referencing memory in mapped regions. Moreover, main memory can be used as a cache of the file system. Another feature of Hurricane is a facility called {Local Server Invocations (LSI)} that allows a fast, local, cross-address space invocation of server code and data, and results in new workers being created in the server address space. {LSI} also simplifies {deadlock} avoidance.\\
\end{keyword}

%% file: gloss-i-o.tex
\begin{keyword}{I/O problem}\\
The I/O problem (also referred to as the {I/O bottleneck problem}) stems from that fact that the processor technology is increasing rapidly, but the performance and the access time of secondary storage devices such as disks and floppy disk drives have not improved to the same extend. Disk seek times are still low, and I/O becomes an important {bottleneck}. The gap between processors and I/O systems is  increased immensely, which is especially obvious and tedious in multiprocessor systems. However, the I/O subsystem performance can be increased by the usage of several disks in parallel. As for the \wrf{Intel Paragon} XP/S, \wrf{RAID}s are supported.\\
\end{keyword}

\begin{keyword}{in-core communication}\\
In-core communication can be divided into two types: demand-driven and producer-driven:
\begin{itemize}
\item {demand-driven}: The communication is performed when a processor requires off-processor data during the computation of the {ICLA}. A node sends a request to another node to get data.
\item {producer-driven}: When a node computes on an {ICLA} and can determine that a part of this {ICLA} will be required by another node later on, this node sends that data while it is in its present memory. The producer decides when to send the data. This method saves extra disk access, but it requires knowledge of the data dependencies so that the processor can know beforehand what to send.\\
\end{itemize}
\end{keyword}

\begin{keyword}{Intel iPSC/860 hypercube}\\
The Intel iPSC/860 is a {distributed memory}, {message passing} \wrf{MIMD} machine, where the {compute node}s are based on Intel i860 processors that are connected by a hypercube network. {I/O node}s are connected to a single {compute node} and handle I/O. What is more, {I/O node}s are based on the Intel i386 processor.\\
\end{keyword}

\begin{keyword}{Intel Paragon}\\
The Intel Paragon (also referred to as Intel Paragon XP/S) multicomputer has its own operating system \wrf{OSF/1} and a special file system called \wrf{PFS (Parallel File System)}. The Intel Paragon is supposed to address \wrf{Grand Challenge Applications}. In particular, it is a {distributed memory} multicomputer based on Intel's teraFLOPS architecture. More than a thousand heterogeneous nodes (based on the Intel i860 XP processors) can be connected in a two-dimensional rectangular mesh. Furthermore, these nodes communicate via {message passing} over a high-speed internal interconnect network. A \wrf{MIMD} architecture supports different programming styles including SPMD and \wrf{SIMD}. However, it does not have {shared memory}. \wrf{SPIFFI} is a scalable parallel file system for the Intel Paragon.\\
\end{keyword}

\begin{keyword}{Intel Touchstone Delta}\\
The Intel Touchstone Delta System is a {message passing} multicomputer consisting of processing nodes that communicate across the two dimensional mesh interconnecting network. It uses Intel i860 processors as the core of communication nodes. In addition, the Delta has 32 Intel 80386 processors as the core of the {I/O node}s where each {I/O node} has 8 Mbytes memory that serves as I/O cache. Furthermore, other processor nodes such as service nodes or ethernet nodes are used.\\
\end{keyword}

\begin{keyword}{irregular (unstructured) problems}\\
Basically, in irregular problems data access patterns cannot be predicted until runtime. Consequently, optimizations carried out at compile-time are limited. However, at run-time data access patterns of nested loops are usually known before entering the loop-nest, which makes it possible to utilize various preprocessing strategies.\\
\end{keyword}

\begin{keyword}{Jovian}\\
Jovian is an I/O library that performs optimizations for one form of {collective-I/O}~\cite{bennett:jovian}. It makes use of a Single Program Multiple Data (SPMD) model of computation. Jovian distinguishes between {global} and {distributed view}s of accessing data structures.  In the {global view} the I/O library has access to the in-core and {out-of-core} data \wrf{distribution}s. What is more, {application process}es requesting I/O have to provide the library with a globally specified subset of the data structure. In contrast, in the {distributed view} the {application process } has to convert local in-core data indices into global {out-of-core} ones before making any I/O request. The library consists of two types of processes: {application processes (A/P)} and {coalescing processes (C/P)} (similar to server processes in a DBMS). At link time there is no distinction between {A/P}s and {C/P}s. The name {C/P} stems from the fact that coalescing I/O requests into a larger one can increase I/O performance. A user can determine which process will run the application and which will perform coalescing of I/O requests.\\
\end{keyword}

\begin{keyword}{Kenal Square (KSR-2)}\\
KSR-2 is a non-uniform access {shared memory machine}.\\
\end{keyword}

\begin{keyword}{Linda}\\
Linda is a concurrent programming model with the primary concept of a {tuple space}, an abstraction via which cooperating processes communicate.\\
\end{keyword}

\begin{keyword}{loosely synchronous}\\
In a loosely synchronous model all the participating processes alternate between phases of computation and I/O. In particular, even if a process does not need data, it still has to participate in the I/O operation. What is more, the processes will synchronize their requests ({collective communication}).\\
\end{keyword}

\begin{keyword}{mapped-file I/O}\\
A contiguous memory region of an application's address space can be mapped to a contiguous file region on secondary storage. Accesses to the memory region behave as if they were accesses to the corresponding file region.\\ 
\end{keyword}

\begin{keyword}{metacomputing}\\
Metacomputing defines an aggregation of networked computing resources, in particular networks of workstations, to form a single logical parallel machine. It is supposed to offer a cost-effective alternative to parallel machines for many classes of parallel applications. Common metacomputing environments such as \wrf{PVM}, \wrf{p4} or \wrf{MPI} provide interfaces with similar functions as those provided for parallel machines. These functions include mechanisms for interprocess communication, synchronization and \wrf{concurrency control}, fault tolerance, and dynamic process management. Except of \wrf{MPI-IO}, they do not support file I/O or serialize all I/O requests.\\
\end{keyword}

\begin{keyword}{MIMD (Multiple Instruction Stream Multiple Data Stream)}\\
MIMD is a more general design than \wrf{SIMD}, and it is used for a broader range of application. Here each processor has its own program acting on its own data. It is possible to brake a program into subprograms which can be distributed to the processors for execution. Several problems can occur. For example, the scheduling of the processors and their synchronization. What is more, there will also be a need for more flexible communication than in a \wrf{SIMD} model. MIMD appears in two forms. First, with a private memory for each process - also referred to as {distributed memory} - and, second, with a {shared memory}. A {distributed memory} approach uses {message passing} for interprocess communication.\\
\end{keyword}

\begin{keyword}{MPI (Message Passing Interface)}\\
In the last years, many vendors have implemented their own variants of the {message passing} paradigm, and it turned out that such systems can be efficiently and portably implemented~\cite{MPI}. Message Passing Interface (MPI) is the de facto standard for {message passing}. MPI does not include one existing {message passing} system, but makes use of the most attractive features of them. The main advantage of the {message passing} standard is said to be 'portability and ease-of-use'. MPI is intended for writing {message passing} programs in C and {Fortran77}. MPI has gained some new features which are expressed in \wrf{MPI-2}.\\
\end{keyword}

\begin{keyword}{MPI-2}\\
MPI-2 is the product of corrections and extensions to the original \wrf{MPI} Standard document~\cite{mpi2-io}. Although some corrections were already made in Version 1.1 of \wrf{MPI}, MPI-2 includes many other additional features and substantial new types of functionality. In particular, the computational model is extended by dynamic process creation and one-sided communication, and a new capability in form of parallel I/O is added (\wrf{MPI-IO}). (Note that every time when \wrf{MPI} is mentioned this dictionary refers to Version 1.0. Thus, if a passage refers to MPI-2, it explicitly uses the term MPI-2.)\\
\end{keyword}

\begin{keyword}{MPI-IO}\\
Despite the development of \wrf{MPI} as a form of interprocess communication, the \wrf{I/O problem} has not been solved there. (Note: \wrf{MPI-2} already includes I/O features.) The main idea is that I/O can also be modeled as {message passing}: writing to a file is like sending a message while reading from a file corresponds to receiving a message~\cite{mpi-ioc:mpi-io5}. Furthermore, MPI-IO supports a high-level interface in order to support the partitioning of files among multiple processes, transfers of global data structures between process memories and files, and optimizations of physical file layout on storage device.\\
\end{keyword}

\begin{keyword}{MPL (Mentat Programming Language)}\\
Mentat is an object oriented parallel processing system. MPL is a programming language based on C and used to program the machines MP-1 and MP-2.\\
\end{keyword}

\begin{keyword}{Multipol}\\
Multipol is a publicly available library of distributed data structures designed for irregular applications (see \wrf{irregular problems}). Furthermore, it contains a thread system which allows overlapping communication latency with computation~\cite{Berkeley3}.\\ 
\end{keyword} 

\begin{keyword}{nCUBE}\\
The proposed file system for the nCUBE is based on a two-step mapping of a file into the {compute node} memorie, where the first step provides a mapping from subfiles stored on multiple disks to an abstract data set, and the second step is mapping the abstract data set into the {compute node} memories. One drawback is that it does not provide an easy way for two {compute node}s to access overlapping regions of a file.\\
\end{keyword}

\begin{keyword}{Network-Attached Peripherals (NAP)}\\
NAP make storage resources directly available to computer systems on a network without requiring a high-powered processing capability. This makes it possible for a single network-attached control system such as {HPSS (High-Performance Storage System)} to manage access to the storage devices without being required to handle the transferred data. In particular, {HPSS} is capable of coordinating concurrent I/O operations over a non-blocking network fabric to achieve very high aggregate I/O throughput.\\
\end{keyword}

\begin{keyword}{OSF/1}\\
OSF/1 is the operating system for the \wrf{Intel Paragon} multicomputer.\\
\end{keyword}

%% file: gloss-p.tex
\begin{keyword}{p4}\\
p4 is a library of macros and subroutines developed at \wrf{ANL} for programming parallel machines. It supports {shared memory} and {distributed memory}, where the former is based on monitors and the later is based on {message passing}. Like in \wrf{PVM}, p4 offers a {master-slave} programming model.\\
\end{keyword}

\begin{keyword}{Pablo}\\
Pablo is a massively parallel, {distributed memory} performance analysis environment to provide performance data capture, analysis, and presentation across a wide variety of scalable parallel systems~\cite{Pablo1}. Pablo can help to identify and remove performance {bottleneck}s at the application or system software level. The Pablo environment includes software performance instrumentation, graphical performance data reduction and analysis, and support for mapping performance data to both graphics and sound. In other words, Pablo is a toolkit for constructing performance analysis environments.\\
\end{keyword}

\begin{keyword}{Panda (Persistence AND Arrays)}\\
Panda is a library for input and output of multidimensional arrays on parallel and sequential platforms. Panda provides easy-to-use and portable array-oriented interfaces to scientific applications, and adopts a server-directed I/O strategy to achieve high performance for collective I/O operations~\cite{seamons:thesis}. 

Panda combines three techniques in order to obtain performance:
\begin{itemize}
\item storage of arrays by subarray chunks in main memory and on disk
\item high-level interfaces to I/O subsystems
\item use of \wrf{disk-directed I/O} to make efficient use of disk bandwidth
\end{itemize}
Array chunking can improve the locality of computation on a processor,
and improve I/O performance. High-level interfaces are considered to be flexible, easier to be used by programmers and give applications better portability.\\
\end{keyword}

\begin{keyword}{ParFiSys (Parallel File System)}\\ 
ParFiSys was developed to provide I/O services for a \wrf{General Purpose MIMD machine (GPMIMD)}~\cite{carretero:parfisys}. It was named \wrf{CCFS} in earlier projects. ParFiSys tries to realize the concept of  "minimizing porting effort" in the following way: 
\begin{itemize}
\item standard {POSIX} interface
\item parallel services are provided transparently, and the physical data \wrf{distribution} across the system is hidden
\item a single name space allows all the user applications to share the same directory tree\\
\end{itemize}
\end{keyword}

\begin{keyword}{PARTI (Parallel Automated Runtime Toolkit at ICASE)}\\
PARTI is a subset of the \wrf{CHAOS} library and specially considers \wrf{irregular problems} that can be divided into a sequence of concurrent computational phases. The primitives enable the \wrf{distribution} and retrieval of globally indexed, but irregularly distributed data sets over the numerous local processor memories. What is more, it should efficiently execute unstructured and block structured problems on {distributed memory} parallel machines~\cite{CHAOS5}. The PARTI primitives can be used by parallizing compilers to generate parallel code from programs written in \wrf{data parallel} languages.\\
\end{keyword}

\begin{keyword}{Partial Redundancy Elimination (PRE)}\\
PRE is a technique for optimizing code by suppressing partially redundant computations, and is used in optimizing compilers for performing common subexpression eliminiation and strength reduction~\cite{PRE}. An {Interprocedural Partial Redundancy Elimination algorithm (IPRE)} is used for optimizing placement of communication statements and communication preprocessing statements in {distributed memory} compilations. In this environment the communication overhead can be decreased by message aggregation. In other words, each processor requests a small number of large amounts of data. The optimization is obtained by placing a preprocessing statement to determine the communicated data. The information is stored in a {communication-schedule}. The developed {IPRE} algorithms is applicable on arbitrary recursive programs.\\
\end{keyword}

\begin{keyword}{Partitioned In-core Model (PIM)}\\
This is one of the three basic models of \wrf{PASSION} for accessing {out-of-core} arrays. It is a variation of the \wrf{Global Placement Model}. An array is stored in a single global file and is logically divided into a number of partitions, each of which can fit in the main memory of all processors combined. Hence, the computation problem is rather an in-core problem than an {out-of-core} one.\\
\end{keyword}

\begin{keyword}{PASSION (Parallel And Scalable Software for Input-Output)}\\		PASSION is a runtime library that supports a \wrf{loosely synchronous} {SPMD} programming model of parallel computing~\cite{choudhary:passion-paragon}. It assumes a set of disks and {I/O node}s which can either be dedicated processors or some of the {compute node}s can also serve as {I/O node}s. Each of these processors may either share the set of disks or have its local disk. What is more, PASSION considers the \wrf{I/O problem} from a language and compiler point of view. \wrf{Data parallel} languages like \wrf{HPF} and {pC++} allow writing parallel programs independently of the underlying architecture. Such languages can only be used for \wrf{Grand Challenge Applications} if the compiler can automatically translate {out-of-core (OOC)} \wrf{data parallel} programs. In PASSION, an {OOC} \wrf{HPF} program can be translated to a {message passing} node program with explicit parallel I/O.\\

PASSION distinguishes between an in-core and an {out-of-core} program. Whereas in an in-core program the entire amount of data (e.g. elements of a {distributed array} in a {distributed memory machine}) fits in the local main memory of a processor, large programs and large data do not fit entirely in the main memory and have to be stored on disk. Such data arrays are referred to as {Out-of-core Local Array }. Unfortunately, many massively parallel machines such as {CM-5}, \wrf{Intel iPSC/860}, \wrf{Intel Touchstone Delta} or \wrf{nCUBE}-2 do not support {virtual memory} otherwise the {OCLA} can be swapped in and out of disk automatically,  and the \wrf{HPF} compiler could also be used for {OOC} programs.\\
\end{keyword}

\begin{keyword}{PFS (Parallel File System)}\\
PFS is the file system for \wrf{Intel Paragon}'s operating system \wrf{OSF/1}. In general, \wrf{OSF/1} provides two forms of parallel I/O:
\begin{itemize}
\item PFS gives high-speed access to a large amount of disk storage, and is optimized for simultaneous access by multiple nodes. Files can be accessed with parallel and non-parallel calls.
\item Special I/O system calls, called parallel I/O calls, give applications better performance and more control over parallel file I/O. These calls are compatible with the \wrf{Concurrent File System} (CFS) for \wrf{Intel iPSC/860 hypercube}.\\
\end{itemize}
\end{keyword}

\begin{keyword}{PIOFS (IBM AIX Parallel File System)}\\
PIOFS is a {parallel file system} for the IBM SP2. It uses UNIX like read/write and logical partitioning of files. Furthermore, logical views can be specified (subfiles). PIOFS is capable of scaling I/O performance as the underlying machine scales in compute performance. What is more, applications can be parallized in two different ways: logically or physically. Physically means that a file's data is spread across multiple server nodes whereas logically refers to the partitioning of a file into subfiles. Other features: faster job performance, {scalability}, portability and application support, and  file \wrf{checkpointing}.\\
\end{keyword}

\begin{keyword}{PIOUS (Parallel Input-OUtput System)}\\
Since in \wrf{metacomputing} environments I/O facilities are not sufficient for a good performance, the virtual, {parallel file system} PIOUS was designed to incorporate true parallel I/O into existing \wrf{metacomputing} environments without requiring modification to the target environment, i.e. PIOUS executes on top of a \wrf{metacomputing} environment\cite{moyer:application}. What is more, parallel applications become clients of the PIOUS task-parallel application via library routines. In other words, PIOUS supports parallel applications by providing coordinated access to file objects with guaranteed consistency semantics.\\
\end{keyword}

\begin{keyword}{Portable Parallel File System (PPFS)}\\
PPFS is a file system designed for experimenting with I/O performance of parallel scientific applications that use a traditional UNIX file system or a vendor-specific {parallel file system}. PPFS is implemented as a user level I/O-library in order to obtain more experimental flexibility. In particular, it is a library between the application and a vendor's basic system software. Furthermore, the correct usage of PPFS requires some assumptions: The underlying file system has to be a standard UNIX file system, which allows the file system to be portable across a wide range of UNIX systems without changing the kernel or the device drivers~\cite{PPFS1}. Additionally, PPFS has to sit on top of a {distributed memory} parallel machine. It is assumed that applications are based on a {distributed memory} {message passing} model.\\ 
\end{keyword}

\begin{keyword}{PVM (Parallel Virtual Machine)}\\
PVM is a software tool allowing a heterogeneous collection of workstations and supercomputers to function as a single high-performance parallel machine, i.e. a workstation cluster can be viewed as a single parallel machine (see also \wrf{metacomputing})~\cite{PVM1}. PVM can be used in both parallel and \wrf{distributed computing} environments. A {message passing} model is used to exploit \wrf{distributed computing} across the array of processes or processors. Moreover, data conversion and task scheduling are also handled across the network.\\

PVM should link computing resources. What is more, the parallel platform can also consist of different computers on different locations (heterogeneity). PVM makes a collection of computers appear as a large virtual machine. The principles upon which PVM is based are: user-configured host pool, translucent access to hardware, process-based computation, explicit {message passing} model, heterogeneity support and multiprocessor support.\\
\end{keyword}

%% file: gloss-r-s.tex
\begin{keyword}{RAID (Redundant Array of Inexpensive Disks)}\\
RAID (Redundant Array of Inexpensive Disks - due to the destructiveness of the term "inexpensive", RAID is also known as Redundant Array of Independent Disks) organizes multiple independent disks into a large, high-performance logical disk, stripes data across multiple disks and accesses them in parallel to achieve high data transfer and higher I/O rates. What is more, disk arrays increase secondary storage throughput. However, these large disk arrays have also a major drawback: they are highly vulnerable to disk failures. An array with x disks is x-times more likely to fail. A solution to this problem is to employ a redundant disk array and error-correcting codes to tolerate disk failures. Even this model has a disadvantage: all write operations have to update the redundant information, which reduces the performance of writes in the disk array.\\

Another drawback of a RAID system is that the throughput is decreased for small writes. What is more, such small data requests are especially important for on-line transaction processing. Thus, a powerful technique called {parity logging} is proposed for overcoming this problem.\\
\end{keyword}

\begin{keyword}{RAID-I}\\
RAID-I ("RAID the First") is a prototype \wrf{RAID} level 5 system. It was designed to test workstation based-{file server}s concerning high bandwidth and high I/O rates. It is based on {Sun 4/280} workstations with 128 MB RAM and  28 5 1/4 inch SCSI disks and four dual-string SCSI controllers. The most serious reason why RAID-I was ill-suited for high-bandwidth I/O was the memory contention.\\
\end{keyword}

\begin{keyword}{RAID-II}\\
RAID-II ("RAID the second") is a scalable high-bandwidth network {file server} and is designed for heterogeneous computing environments of diskless computers, visualization workstations, multimedia platforms and UNIX workstations. It should support the research and development of storage architectures and file systems. It is supposed to run under {LFS}, the {Log-Structured File System}. What is more, {LFS} is specially optimized to serve as a high-bandwidth I/O and crash recovery file system.\\
\end{keyword}

\begin{keyword}{raidSim}\\
A \wrf{RAID} simulator, raidSim, is an event-driven simulator for both modeling non-redundant and redundant disk arrays. It does neither model the CPU, host disk controllers nor I/O busses, but only disks.\\
\end{keyword}

\begin{keyword}{RAMA}\\
RAMA is a parallel file system that is intended primarily as a cache or storage area for data stored on tertiary storage. Furthermore, RAMA uses hashing algorithms to store and retrieve blocks of a file.\\
\end{keyword}

\begin{keyword}{RAPID (Read Ahead for Parallel-Independent Disks)}\\
RAPID is a fully parallel file system testbed that allows implementations of various buffering and \wrf{prefetching} techniques to be evaluated. The architectural model is a medium to large scale \wrf{MIMD} {shared memory} multiprocessor with memory distributed among processor nodes. Some results show that \wrf{prefetching} often reduces the total execution time. As a matter of fact, the hit ratio is only a rough indicator of overall performance of a \wrf{caching} system since it tends to be optimistic and ignores \wrf{prefetching} overhead~\cite{kotz:prefetch}.\\
\end{keyword}

\begin{keyword}{read ahead}\\ 
Communication between clients and servers (or in distributed systems in general) is one of the main overheads in a file system. Hence, I/O requests are packaged, and level locks and resources are managed in groups. Read ahead reads new blocks in advance when a minimum threshold is reached. {Flush ahead} is the opposite of read ahead and frees clean blocks in order to satisfy write requests as soon as possible. \\
\end{keyword}

\begin{keyword}{Remote Memory Servers}\\
The memory server model extends the memory hierarchy of multicomputers by introducing a remote memory layer whose latency lies somewhere between local memory and disk. A memory server is a multicomputer node whose memory is used for fast backing storage and logically lies between the local physical memory and fast stable storage such as disks.\\
\end{keyword}

\begin{keyword}{ROMIO}\\
ROMIO is a high-performance, portable implementation of \wrf{MPI-IO}. A key feature component is an internal abstract I/O device layer called \wrf{ADIO}~\cite{thakur:romio-users}. \\
\end{keyword}

\begin{keyword}{Scalable I/O Facility (SIOF)}\\
SIOF is a project to enable I/O performance to scale with the computing performance of parallel computing systems and achieve terascale computing.\\ 
\end{keyword}
		
\begin{keyword}{Scotch Parallel Storage System (SPFS)}\\
	Parallel storage systems are constructed as testbeds for the development of advanced parallel storage subsystems and file systems for parallel storage. Scotch has been developing a portable, extensible framework, {RAIDframe}, applicable to simulation and implementation of novel \wrf{RAID} desigin order to advance parallel storage subsystems~\cite{gibson:scotch1}. The key features are the separation of mapping, operation semantics, \wrf{concurrency control} and error handling. The file system research is based on \wrf{prefetching} and \wrf{caching} techniques. {Transparent Informed Prefetching (TIP)} and the Scotch Parallel File System (SPFS) are the results of the work. The benefit of {TIP} is its ability to increase the I/O concurrency of a single-threaded application.\\
\end{keyword}

\begin{keyword}{shared file pointer}\\
A shared file pointer is much more powerful than a traditional private or local UNIX file pointer since it can simplify the coding and increase the performance of parallel applications. As for a shared file pointer, it is ensured that a file is read sequentially even if many processes share the same file. Additionally, it reduces the number of disk seeks and increases the effectiveness of \wrf{prefetching}.\\
\end{keyword}

\begin{keyword}{Shared Virtual Memory}\\
Shared Virtual Memory implements coherent shared memory on a multicomputer without physically shared memory The {shared memory} system presents all processors with a large coherent {shared memory} address space. Any processor can access any memory location at any time. See also \wrf{GPM}.\\
\end{keyword}

\begin{keyword}{SHORE (Scalable Heterogeneous Object REpository)}\\
SHORE is a persistent object system (under development) that represents a merger of \wrf{object-oriented database (OODB)} and file system technologies~\cite{carey:shore}. The work is based on \wrf{EXODUS}, an earlier object-oriented data base effort.\\
\end{keyword}

\begin{keyword}{SIMD (Single Program Multiple Data)}\\
\wrf{SPMD} is a model for large-scale scientific and engineering applications. The same program is executed an each processor, but the input data to each of the programs may be different.\\

The most widely used classification is the one where the von Neumann model is viewed as a Single Stream  of Instructions controlling a Single Stream of Data ({SISD}). One instruction produces one result and, hence, there is a Single Instruction Stream and a Single Data Stream. One step towards parallelism leads to the SIMD model, another step ends up with Multiple Instruction Streams (\wrf{MIMD}). In the classical example of a parallel SIMD model, a number of identical {processing element}s receive the same instruction broadcast by a higher instance. Each {processing element} performs the instruction on its own data item. In other words, a SIMD instruction means that a Single Instruction causes the execution of identical operations on Multiple pairs of Data. Furthermore, this is the simplest conceptual model for a vector computer. The synchronization can be obtained by using a broadcast command that keeps the processes in a lockstep, and the processes need to talk to each other for synchronization purpose. Additionally, they need not store their own programs, which results in a smaller design and a bigger amount of processes.\\
\end{keyword}

\begin{keyword}{SPIFFI (Scalable Parallel File System)}\\
SPIFFI is a high-performance parallel file system that stripes files across multiple disks~\cite{freedman:spiffi}. SPIFFI provides applications with a high-level flexible interface including one individual  and three \wrf{shared file pointer}s.\\
\end{keyword}

\begin{keyword}{STARFISH}\\
STARFISH is a parallel file system simulator which ran on top of the Proteus parallel architecture simulator, which in turn ran on a {DEC-5000} workstation.\\
\end{keyword}

\begin{keyword}{supercomputing applications}\\
Supercomputing applications are generating more and more data, but I/O systems cannot keep abreast, i.e. they become less able to cope with the amount of information in a sensible amount of time. The solution requires correct matching of bandwidth capability to application bandwidth requirements, and using of buffering to reduce the peak bandwidth that I/O systems have to handle.\\

Conventional file systems use \wrf{caching} for reducing I/O bandwidth requirements. Thus, the number of requests can be decreased, and the system performance is increased. Another method of reducing I/O is the usage of {delayed write}s. A {write-behind cache policy} is required, which allows a program to continue executing after writing data to the cache without waiting for the data to be written to the disk.\\

The environment of a supercomputer (e.g. {Cray Y-MP} 8/832) is different from a conventional one. It is characterized by a few large processes that consume huge amounts of memory and CPU time. Jobs are not interactive, but submitted in batch and run whenever the scheduler can find enough resources.\\ 

Supercomputers are ideal for applications that require the manipulation of large arrays of data. They are especially applied in fields like fluid dynamics, structural dynamics or seismology.\\
\end{keyword}

%% file: gloss-t-z.tex
\begin{keyword}{task parallel program}\\ 
A task parallel program consists of a set of (potentially dissimilar) parallel tasks that perform explicit communication and synchronization. {Fortran M (FM)} and {CC++} are examples of such a language.\\
\end{keyword}

\begin{keyword}{TOPs (The Tower of Pizzas)}\\
TOPs is a portable software system providing fast parallel I/O and buffering services~\cite{tan:pizzas}.\\
\end{keyword}

\begin{keyword}{TPIE (Transparent Parallel I/O Environment)}\\
TPIE is designed to allow programmers to write high performance I/O-efficient programs for a variety of platforms~\cite{vengroff:tpie-man}. The work on TPIE is still in progress.\\
\end{keyword}

\begin{keyword}{Two-Phase Method (TPM)}\\
\wrf{PASSION} introduces a Two-Phase Method which consists of the following two phases~\cite{thakur:ext2phase}:
\begin{itemize}
\item READ DATA (processes cooperate to read data in large chunks)
\item DISTRIBUTE DATA (interprocess communication is used so that each processor gets the data it requested)
\end{itemize}
\end{keyword}

\begin{keyword}{UNIX I/O}\\
The UNIX I/O facility can be applied in a uniform way to a large variety of I/O services, including disk files, terminals, pipes, networking interfaces and other low-level devices. Many application programs use higher-level facilities, because they have more specified features. An example is the standard I/O library \texttt{stdio} for C, which is an application-level facility. The functions correspond to the UNIX functions.\\
\end{keyword}

\begin{keyword}{Vesta}\\
Vesta is a {parallel file system} providing  parallel file access to application programs running on multicomputers with parallel I/O subsystems. A file can be divided into partitions (multiple disjoint sequences)~\cite{corbett:vesta-di}. Furthermore, Vesta allows a direct access from a {compute node} to the {I/O node} without referencing any centralized metadata. Consequently, Vesta is based on a {client-server} model, which allows libraries to be implemented on top of Vesta.\\ 
\end{keyword}

\begin{keyword}{Vienna Fortran (VF)}\\
Vienna Fortran is a \wrf{data parallel} language which supports the {SPMD} model of computation. Furthermore, it provides explicit expressions for data mapping. The corresponding compiler is called {Vienna Fortran Compilation Systems (VFCS)}.\\
\end{keyword}

\begin{keyword}{VIP-FS (VIrtual Parallel File System)}\\
VIP-FS is a straight-forward interface to parallel I/O~\cite{delrosario:vipfs-tr}. It is virtual because it is implemented using multiple individual standard file systems integrated by a {message passing} system. VIP-FS makes use of {message passing} libraries to provide a parallel and distributed file system which can execute over multiprocessor machines or heterogeneous network environments. \\
\end{keyword}

\begin{keyword}{ViPIOS (Vienna Parallel Input-Output System)}\\
The ViPIOS represents a mass-storage sub-system for highly I/O intensive scientific applications on massively parallel supercomputers. 
The ViPIOS is based on a client-server approach combining the advantages
of parallel I/O runtime libraries and parallel file systems. I/O bandwidth is maximized by exploiting the concept of data locality, by
optimizing the data layout on disk and, thus, allowing efficient
parallel read/write accesses \cite{brezany:architecture}. What is more,
the ViPIOS is influenced by the concepts of parallel database technology.\\

The ViPIOS chooses the file layout as close as possible to the problem
specification
in focus to reach data local accesses. It is not guaranteed
that the physical distribution is equal to the problem distribution. Thus,
the physical data layout is transparent to the application processes and
provided by different view layers (represented by file pointers accordingly)
to the application programmer. A prototype implementation is ready; the performance analysis shows promising results.\\
\end{keyword}

\begin{keyword}{Vulcan multicomputer}\\
This computer located at the IBM T. J. Watson Research Center is a {distributed memory}, {message passing} machine, with nodes connected by a multistage packet-switched network. The nodes include {compute node}s as well as {storage node}s.\\
\end{keyword}

%% file: appendix.tex

\section{Parallel I/O products}

\begin{figure}
\begin{center}
\epsfig{figure=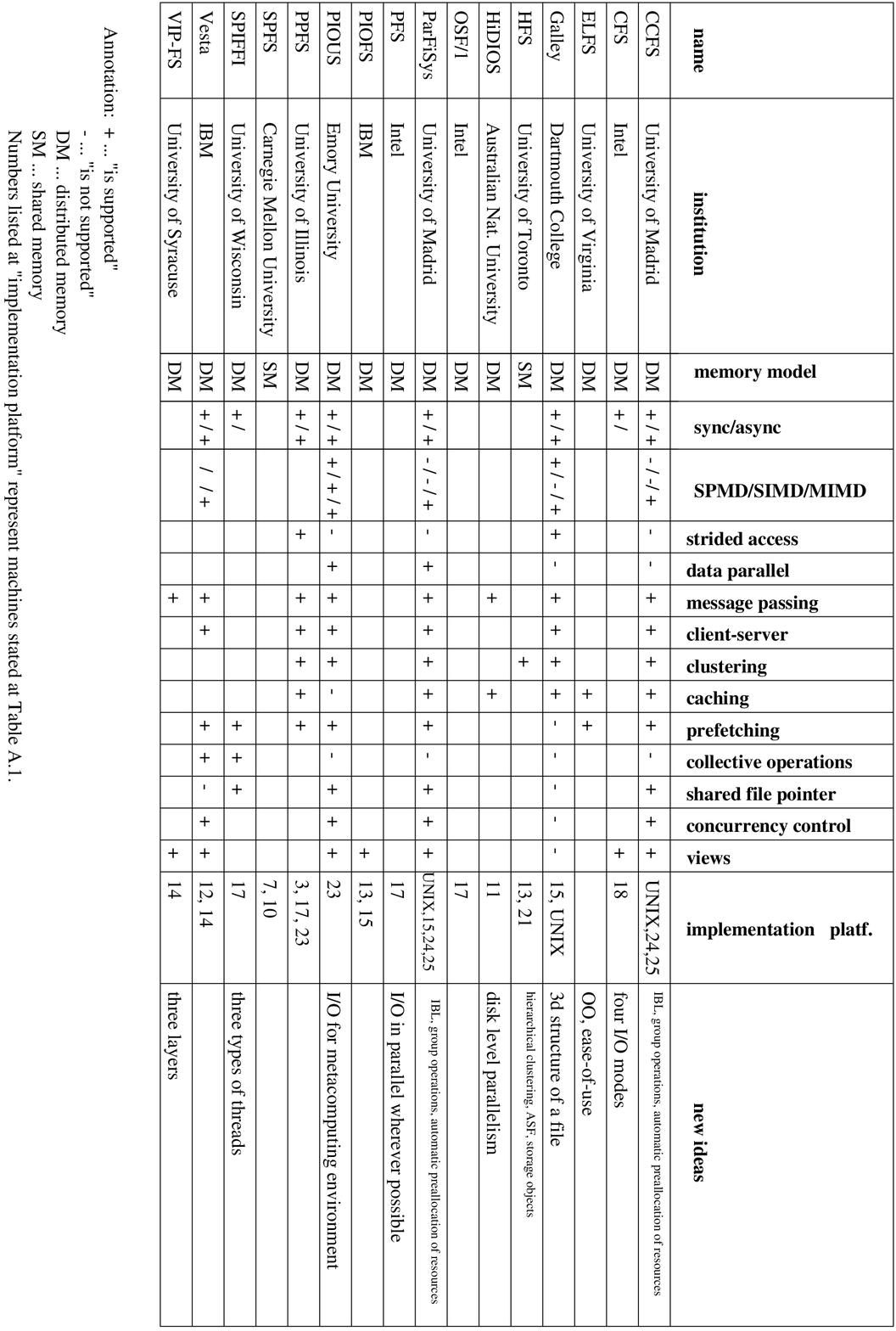, width=14cm,angle=180}
\caption{Parallel I/O products: Parallel File Systems}
\label{fs}
\end{center}
\end{figure}

\begin{figure}
\begin{center}
\epsfig{figure=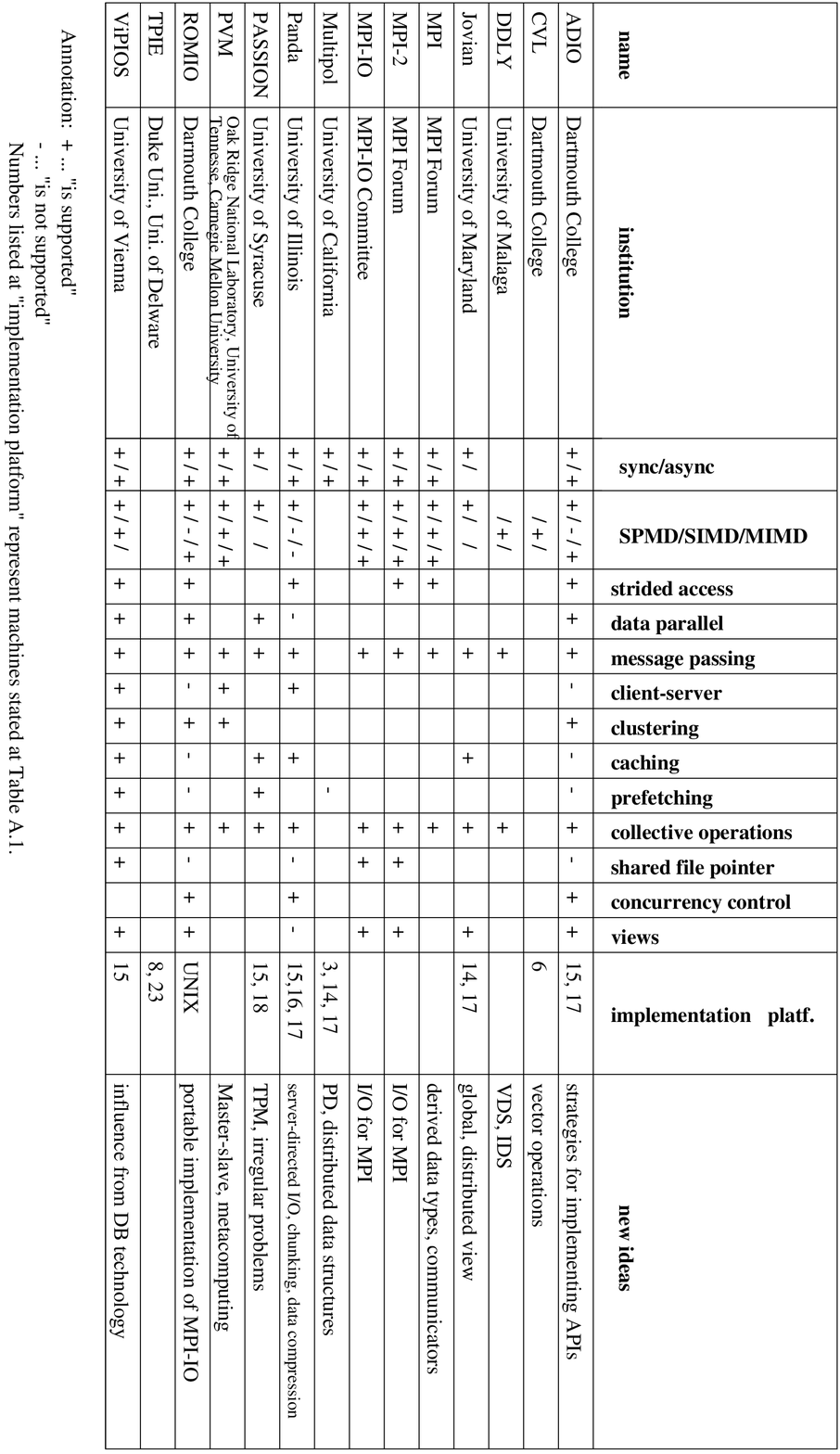, width=12.5cm,angle=180}
\caption{Parallel I/O products: I/O Libraries}
\label{IO}
\end{center}
\end{figure}

The aim of this chapter is to compare different approaches and solutions of the various research teams listed in the dictionary part. (A detailed list of reserach teams is illustrated in Table A.2. Annotation: (et. al.) means that the specific product was produced by more than one institution.) The I/O products can be splitted into three different groups:
\begin{itemize}
\item file systems (see Figure \ref{fs})
\item I/O libraries (see Figure \ref{IO})
\item others, i.e. products that are neither file systems nor I/O libraries (see Figures A.3 and A.4)
\end{itemize}
The platforms which are used by the various approaches are listed in Table A.1.

\begin{figure}
\begin{center}
\epsfig{figure=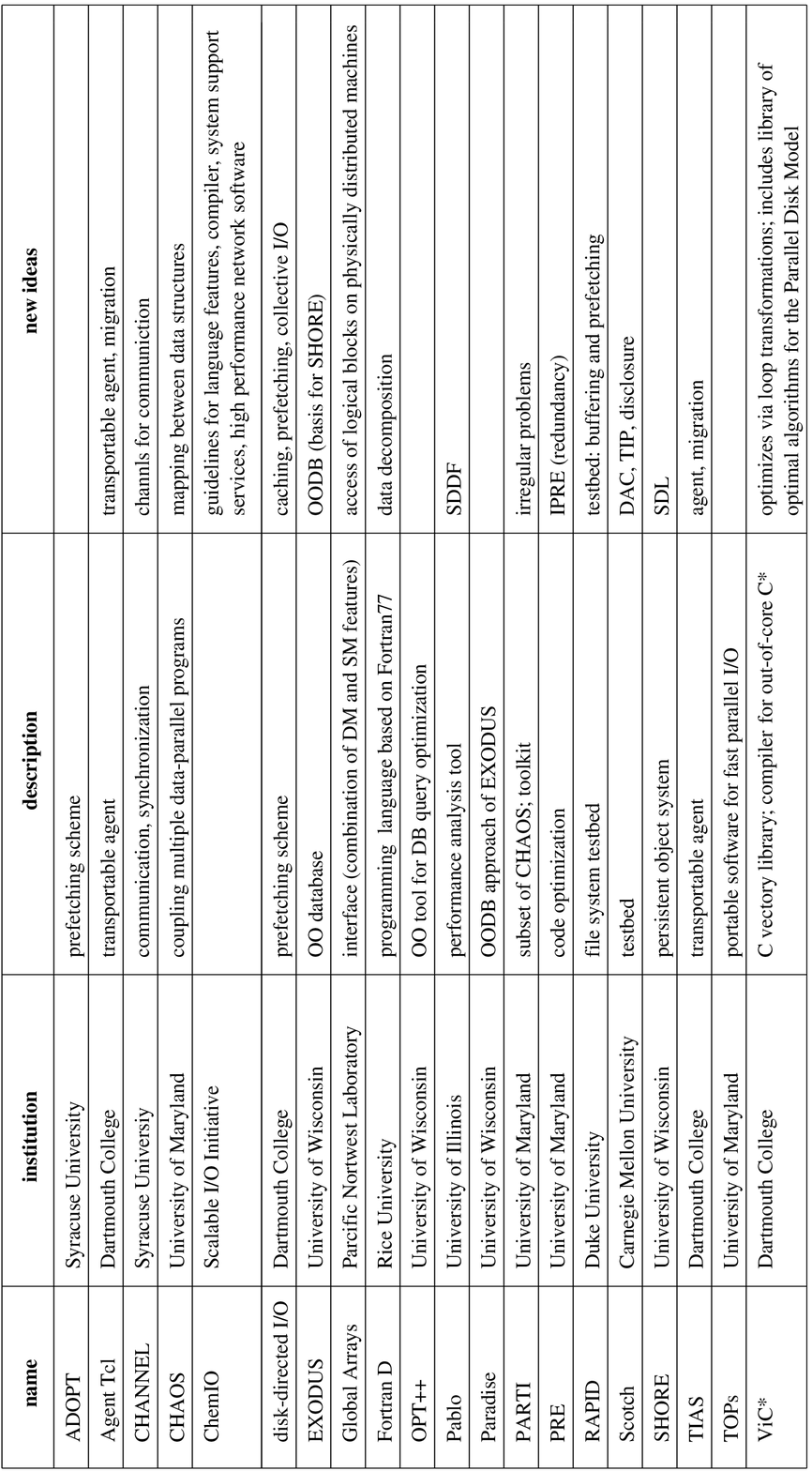, width=12cm}
\caption{Parallel I/O products: others (1)}
\label{others1}
\end{center}
\end{figure}

\begin{figure}
\begin{center}
\epsfig{figure=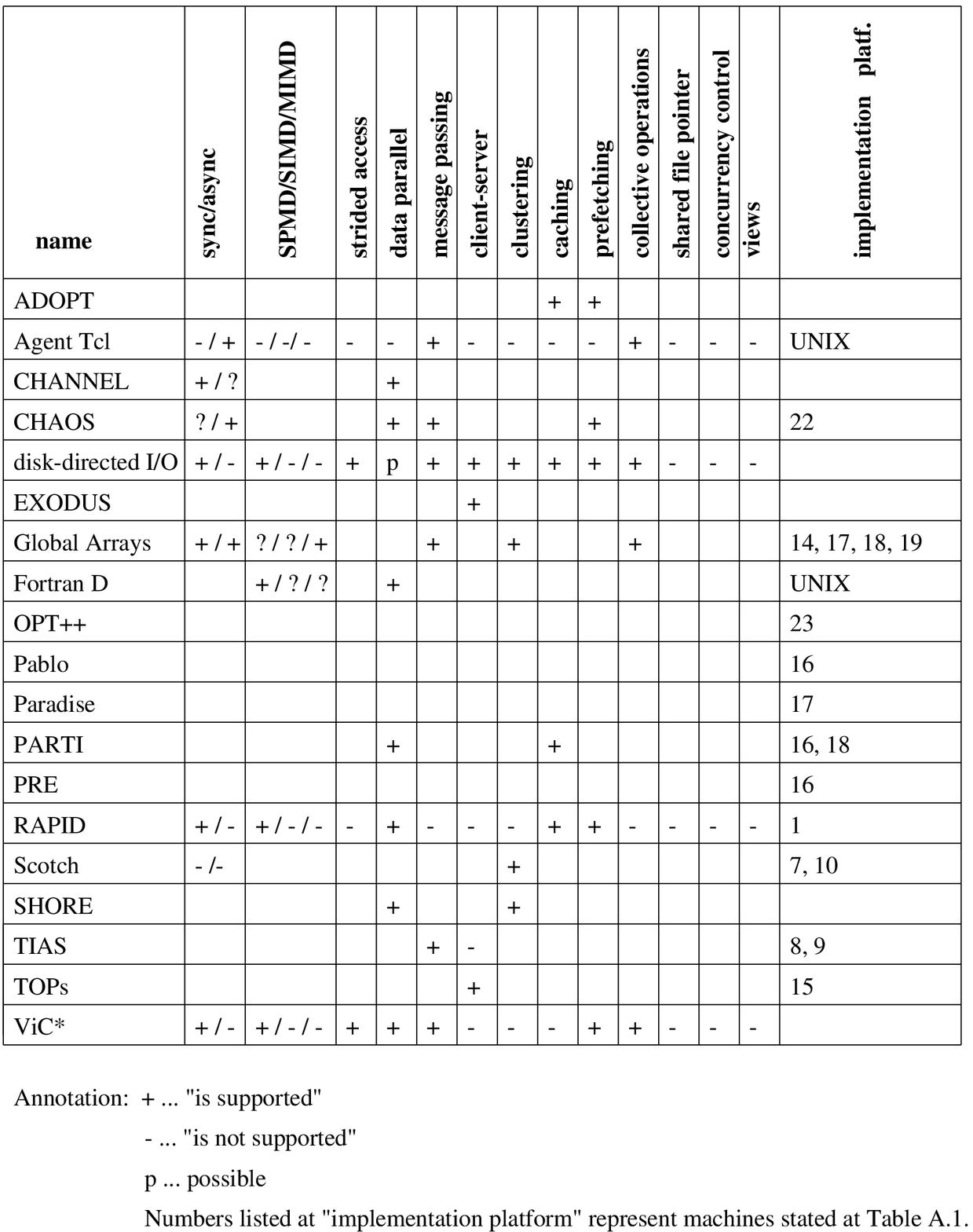, width=14cm}
\caption{Parallel I/O products: others (2)}
\label{others}
\end{center}
\end{figure}

\begin{table}
\begin{center}
\begin{tabular}{|r|l|r|}
\hline
\textbf{No.} &   \textbf{name (usage)} &  \textbf{amount of usage}\\
\hline
1 & Butterfly Plus (RAPID) & 1\\
2 & CM-2 & 0\\
3 &	CM-5 (Multipol, PPFS) & 2\\
4 & Cray C90 & 0\\
5 &	Cray Y-MP &	0\\
6 &	DEC 12000/Sx 2000 (CVL)	& 1\\
7 &	DEC 3000/500  (SPFS) & 	1 \\
8 &	DEC Alpha  (TIAS) &	1\\
9 &	DEC MIPS (TIAS)	& 1\\
10 &	DEC-5000 (STARFISH, SPFS) &	2\\
11 & 	Fujitsu AP1000 (HiDIOS)	& 1\\
12 &	IBM R6000/350 (HFS, Vesta) &	2\\
13 &	IBM RS/6000 (PIOFS) &	1\\
14 &	IBM SP1 (GA, Jovian, Multipol, ROMIO, Vesta, VIP-FS, & \\
   &	ViPIOS) &	7 \\
15 &	IBM SP2 (ADIO, DRA, Galley, Panda, PASSION, ParFiSys, & \\
   &    PIOFS, ROMIO, SIOF, TOPs, ViPIOS) & 11\\
16 &	Intel iPSC/860 hypercube (CHARISMA, Pablo, Panda,  & \\
   &    PARTI, PRE) & 5\\
17 &	Intel Paragon (ADIO, DRA, GA, Multipol, OSF/1, Panda, & \\
   &    Paradise, PFS, PPFS, ROMIO, SPIFFI) & 11\\
18 &	Intel Touchstone Delta (CFS, GA, PARTI, PASSION) & 4\\
19 &	Kendal Square KSR-2 (GA) &	1\\
20 &	MasPar MP-2 &	0\\
21 &	SGI (HFS) &	1\\
22 &	SMP Digital Alpha (CHAOS) &	1\\
23 &	SPARC (OPT++, PIOUS, TPIE) &	3\\
24 &	T800 (ParFiSys) &	1\\
25 &	T9000 (ParFiSys) &	1\\
\hline
\end{tabular}
\caption{The usage of hardware platforms.}
\label{plaforms}
\end{center}
\end{table}

\begin{table}
\begin{center}
\begin{tabular}{|l|l|}
\hline
\textbf{Institution} & \textbf{Product}\\
\hline
Argonne National Laboratories	&	ADIO, ROMIO\\
Australian National University	&	HiDIOS\\
Carnegie Mellon University	&	PVM (et. al.), Scotch\\
Dartmouth College		&	Agent Tcl, CHARISMA, CVL\\
& 					disk-directed I/O,\\
& 					Galley, RAPID, TIAS, ViC*\\
Duke University			&	TPIE\\
Emory University	&		PIOUS, PVM (et. al.)\\
IBM	&				PIOFS, Vesta\\
Intel &					OFS/1, CFS, PFS\\
Message Passing Interface Forum	&	MPI, MPI-2\\
MPI-IO Committee		&	MPI-IO\\
Oak Ridge National Laboratory	&	PVM (et. al.)\\
Parcific Northwest Lab.		&	Global Arrays\\
Rice University			&	Fortran D\\
Scalable I/O Initiative		&		ChemIO\\
University of California	&		Mulipol, RAID,\\
& 						raidPerf, raidSim,\\
&						SIOF\\
University of Delware		&		TPIE (et. al.)\\
University of Illinois		&		Pablo, Panda, PPFS\\
University of Madrid	&			CCFS, ParFiSys\\
University of Malaga 			& DDLY\\
University of Maryland	&		CHAOS,\\
			&			Jovian, PARTI,\\
			&			PRE, TOPs\\
University of Syracuse		&	ADOPT, CHANNEL,\\
				&	PASSION,\\
				&	Two-Phase Method,\\
				&	VIP-FS\\
University of Tennessee	&		PVM (et. al.)\\
University of Toronto		&	HFS, Hector\\
University of Vienna	&		Vienna Fortran, ViPIOS\\
University of Virginia	&		ELFS\\
University of Wisconsin	&		EXODUS, OPT++, Paradise,\\
		&			SHORE, SPIFFI\\
\hline
\end{tabular}
\caption{Research teams and their products.}
\end{center}
\end{table}

\chapter{Parallel Computer Architectures}

Since there appear so many different machines in this dictionary, this part
of the appendix shall give an overview of parallel architectures in
general. Furthermore, some of the machines mentioned in the dictionary part
will be explained explicitly and in more detail.\\

In general, there are two main architectures for parallel machines, SIMD
and MIMD architectures. SIMD machines are supposed to be the cheapest ones,
and the architecture is not as complex as in MIMD machines. In particular,
all the processing elements have to execute the same instructions whereas
in MIMD machines many programs can be executed at the same time. Hence,
they are said to be the "real" parallel machines~\cite{ParArchi}. A major
difference between the two types is the interconnecting network. In a SIMD
architecture this network is a static one while MIMD machines have
different ones depending on the organization of the address space. This
also results in two different communication mechanisms for MIMD machines:
message passing systems (also called distributed memory machines) and
virtual shared memory systems (NUMA: nonuniform memory access). Massively
parallel machines apply UMA architectures that are based on special
crossbar interconnecting networks.\\

\subsection*{SIMD Machines}
These machines are supposed to be "out of date"~\cite{ParArchi}, but they are
still in use.

\begin{itemize}
\item \textbf{Connection Machines CM-200}\\
This machine was built by the Thinking Machines Corporation. CM-200 is the
most modern edition of version 2 (CM-2). The machine consists of 4096 to
65535 microprocessors with a one-bit word length. Moreover, the
bit-architecture of each processing element enables to define different
instructions. In comparison, CM-5 is a MIMD machine.
\item \textbf{MasPar-2}\\
This machine was built by MasPar, an affiliate company from DEC. MasPar-2
consists of up to 16K 32-bit micro processors. Although float comma
operations are micro coded, the performance for float comma operations of
CM-200 is better. Furthermore, the front-end computer is a DECstation 5000.\\
\end{itemize}

\subsection*{Distributed Memory MIMD Machines}
One processor can only directly access its own memory while the memories of
other processors have to be accessed via message passing.\\

MIMD machines have some different topologies like hypercube or grid. The
interconnectivity depends on the amount of links and whether they can be
used concurrently. Systems with a flat topology need four links in order to
establish a 2-dimensional grid (Intel Paragon) whereas systems with 3-d
grids need six links. Hypercubes need most links, and each processing node
has a links to neighboring nodes.

\begin{itemize}
\item \textbf{Hypercube Systems}
  \begin{itemize}
  \item \textbf{Intel iPSC/860}: 60 MFLOPS (60 MHz)
  \item \textbf{nCUBE-2S}: 15 MIPS, 4 MFLOPS
  \item \textbf{nCUBE-3}: 50 MIPS, 50 MFLOPS (floating point), 100 MFLOPS ("multiply-add")
  \end{itemize}
\item \textbf{2-dimensional Topologies}
  \begin{itemize}
  \item \textbf{INMOS Transputer T805}: 5 MIPS, 1,5 MFLOPS
  \item \textbf{Intel Paragon XP/S}: (max. 300 GFLOPS) 50 MHz i860XP: 40 MIPS,     75 MFLOPS
  \end{itemize}
\item \textbf{3-dimensional Topologies}
  \begin{itemize}
  \item \textbf{Parsytec GC} (based on an INMOS T9000 processor): 25 MFLOPS
  \end{itemize}
\item \textbf{Multilevel Interconnecting Network}
  \begin{itemize}
  \item \textbf{Thinking Machines CM-5}\\
  The principles of CM-2/CM-200 and the MIMD principle are combined in the
  CM-5, and the processors are normal SPARC micro processors.  4*32 MFLOPS
  per node
  \item \textbf{IBM SP2}\\
  The SP2 consists of RS/6000 processors. 125 MFLOPS (thin nodes), 266 MFLOPS
  (wide nodes)
  \end{itemize}
\end{itemize}

\subsection*{Shared Memory MIMD Machines}
In contrast to a Cray Y-MP where a uniform memory access is used, in shared
memory machines the amount of processors is much bigger and, hence,
non-uniform memory access is applied.

\begin{itemize}
\item \textbf{Ring Topologies}
  \begin{itemize}
  \item \textbf{Kendal Square Research KSR-2 AllCache}: 80 MIPS, 80 MFLOPS
  Convex Exemplar SPP1000: 200 MFLOPS
  \end{itemize}
\item \textbf{3-dimensional Topologies}
  \begin{itemize}
  \item \textbf{CRAY T3D}\\
  This is a massive parallel machine with up to 2048 processors. 150 MFLOPS
  \end{itemize}
\item \textbf{Multilevel Interconnecting Network}
  \begin{itemize}
  \item \textbf{MANNA}\\
  MANNA is a massively parallel machine for numeric and non-numeric applications.
  50 MHz, 50 MIPS, 100 MFLOPS (32 bit), 50 MFLOPS (64 bit)
  \item \textbf{Meiko CS-2}\\
  Fujitsu vector multiprocessor: 100 MFLOPS, SPARC RISC processor: 40 MFLOPS
  4 Crossbar switch
  \item \textbf{Fujitsu VPP500}\\
  This machine is supposed to be one of the most powerful massively parallel
  systems. 1,6 GFLOPS (vectors), 300 MIPS and 200 MFLOPS (scalars)
  \end{itemize}
\end{itemize}
